\RequirePackage[left]{lineno}

\documentclass{aastex62}

\usepackage{graphicx}
\usepackage{amsmath}
\usepackage{aas_macros}
\usepackage{xcolor}
\usepackage{accents}
\usepackage{xspace}

\graphicspath{{./figures/}}

\newcommand*\ruleline[2][0.97]{\par\noindent\raisebox{1ex}{\makebox[ #1 \linewidth]{\hrulefill\quad\raisebox{-.6ex}{#2}\quad\hrulefill}}}

\newcommand{\fhescenai}{FHES~J1325.3$-$3946}

\newcommand{\fhescenairext}{1.46}
\newcommand{\fhescenairexterr}{0.06}
\newcommand{\fhescenairexterrsys}{0.27}
\newcommand{\fhescenaiindex}{2.22}
\newcommand{\fhescenaiindexerr}{0.14}
\newcommand{\fhescenaiindexerrsys}{0.08}
\newcommand{\fhescenaii}{FHES~J1332.6$-$4130}

\newcommand{\fhescenaiirext}{0.62}
\newcommand{\fhescenaiirexterr}{0.04}
\newcommand{\fhescenaiirexterrsys}{0.10}
\newcommand{\fhescenaiiindex}{2.08}
\newcommand{\fhescenaiiindexerr}{0.12}
\newcommand{\fhescenaiiindexerrsys}{0.04}
\newcommand{\fhescrab}{FHES~J0534.5$+$2201}

\newcommand{\fhescrabtsext}{42.7}
\newcommand{\fhescrabtsextsys}{11.7}
\newcommand{\fhescrabrext}{0.030}
\newcommand{\fhescrabrexterr}{0.003}
\newcommand{\fhescrabrexterrsys}{0.007}

\newcommand{\fhesdmhaloiextdaic}{11.9}

\newcommand{\fhesdmhaloiiextdaic}{29.4}
\newcommand{\fhesfornax}{FHES~J0322.2$-$3710}

\newcommand{\fheshardi}{FHES~J1723.5$-$0501}

\newcommand{\fheshardirext}{0.73}
\newcommand{\fheshardirexterr}{0.10}
\newcommand{\fheshardirexterrsys}{0.01}
\newcommand{\fheshardiindex}{1.97}
\newcommand{\fheshardiindexerr}{0.08}
\newcommand{\fheshardiindexerrsys}{0.06}

\newcommand{\fheshardiii}{FHES~J1741.6$-$3917}

\newcommand{\fheshardiiirext}{1.35}
\newcommand{\fheshardiiirexterr}{0.03}
\newcommand{\fheshardiiirexterrsys}{0.29}
\newcommand{\fheshardiiiindex}{1.80}
\newcommand{\fheshardiiiindexerr}{0.04}
\newcommand{\fheshardiiiindexerrsys}{0.06}
\newcommand{\fheshardiv}{FHES~J2129.9$+$5833}

\newcommand{\fheshardivrext}{1.09}
\newcommand{\fheshardivrexterr}{0.13}
\newcommand{\fheshardivrexterrsys}{0.03}
\newcommand{\fheshardivindex}{2.30}
\newcommand{\fheshardivindexerr}{0.12}
\newcommand{\fheshardivindexerrsys}{0.04}
\newcommand{\fheshardv}{FHES~J2304.0$+$5406}

\newcommand{\fheshardvrext}{1.58}
\newcommand{\fheshardvrexterr}{0.35}
\newcommand{\fheshardvrexterrsys}{0.17}
\newcommand{\fheshardvindex}{1.95}
\newcommand{\fheshardvindexerr}{0.08}
\newcommand{\fheshardvindexerrsys}{0.15}
\newcommand{\fheshlatexti}{3FGL~J2142.2$-$2546}

\newcommand{\fheshlatextiassoc}{PMN~J2142$-$2551}
\newcommand{\fheshlatextits}{81.4}

\newcommand{\fheshlathaloi}{3FGL~J0850.0$+$4855}

\newcommand{\fheshlathaloiassoc}{GB6~J0850$+$4855}
\newcommand{\fheshlathaloits}{1771.4}

\newcommand{\fheshlatunassoci}{3FGL~J0434.3$-$1411c}

\newcommand{\fhesmknfourtwoonetsext}{2.2}

\newcommand{\fhesmthirtyone}{FHES~J0043.2$+$4109}

\newcommand{\fhesmthirtyonetsext}{15.5}

\newcommand{\fhesroph}{FHES~J1626.9$-$2431}

\newcommand{\fhessnrcta}{FHES~J0006.7$+$7314}

\newcommand{\fhessnrctarext}{0.98}
\newcommand{\fhessnrctarexterr}{0.05}
\newcommand{\fhessnrctarexterrsys}{0.04}

\newcommand{\fhessnri}{FHES~J0426.4$+$5529}

\newcommand{\fhessnrii}{FHES~J1208.7$-$5229}

\newcommand{\fhessnriii}{FHES~J1642.1$-$5428}

\newcommand{\fhessnriiirext}{0.57}
\newcommand{\fhessnriiirexterr}{0.02}
\newcommand{\fhessnriiirexterrsys}{0.05}
\newcommand{\fhessnriiiindex}{1.78}
\newcommand{\fhessnriiiindexerr}{0.12}
\newcommand{\fhessnriiiindexerrsys}{0.08}
\newcommand{\fhessofti}{FHES~J0000.2$+$6826}

\newcommand{\fhessoftirext}{0.98}
\newcommand{\fhessoftirexterr}{0.04}
\newcommand{\fhessoftirexterrsys}{0.01}
\newcommand{\fhessoftiindex}{2.72}
\newcommand{\fhessoftiindexerr}{0.11}
\newcommand{\fhessoftiindexerrsys}{0.07}
\newcommand{\fhessoftii}{FHES~J0242.5$+$5229}

\newcommand{\fhessoftiii}{FHES~J0430.5$+$3525}

\newcommand{\fhessoftiiirext}{1.11}
\newcommand{\fhessoftiiirexterr}{0.10}
\newcommand{\fhessoftiiirexterrsys}{0.09}
\newcommand{\fhessoftiiiindex}{2.59}
\newcommand{\fhessoftiiiindexerr}{0.11}
\newcommand{\fhessoftiiiindexerrsys}{0.05}
\newcommand{\fhessoftiv}{FHES~J0940.6$-$6128}

\newcommand{\fhessoftv}{FHES~J1501.0$-$6310}

\newcommand{\fhessoftvrext}{1.29}
\newcommand{\fhessoftvrexterr}{0.13}
\newcommand{\fhessoftvrexterrsys}{0.25}
\newcommand{\fhessoftvindex}{2.44}
\newcommand{\fhessoftvindexerr}{0.09}
\newcommand{\fhessoftvindexerrsys}{0.07}
\newcommand{\fhessoftvi}{FHES~J2208.4$+$6443}

\newcommand{\fhessoftvirext}{0.93}
\newcommand{\fhessoftvirexterr}{0.11}
\newcommand{\fhessoftvirexterrsys}{0.11}
\newcommand{\fhessoftviindex}{2.78}
\newcommand{\fhessoftviindexerr}{0.14}
\newcommand{\fhessoftviindexerrsys}{0.15}
\newcommand{\NUMEXTSOURCES}{24}
\newcommand{\NUMPTSOURCES}{2520}
\newcommand{\NUMUNASSOCPTSOURCES}{70}
\newcommand{\NUMEXTSOURCESSIGNIFICANT}{23}
\newcommand{\NUMCONFUSEDSOURCES}{8}
\newcommand{\NUMHIGHLATUNASSOC}{328}
\newcommand{\NUMHIGHLATASSOC}{1360}
\newcommand{\NUMHIGHLAT}{1688}
\newcommand{\NUMKNOWNEXTSOURCES}{5}
\newcommand{\NUMNEWEXTSOURCES}{19}
\newcommand{\NUMNEWASSOCEXTSOURCES}{5}

\newcommand{\NUMUNASSOCEXTSOURCES}{6}

\newcommand{\NUMTHREEFGLONLYSOURCES}{1112}

\newcommand{\NUMTHREEFHLONLYSOURCES}{218}
\newcommand{\NUMTHREEFXLSOURCES}{1120}

\newcommand{\NUMUNASSOCFXLSOURCES}{70}
\newcommand{\hlatextits}{10.6}
\newcommand{\hlatextisigma}{2.9}
\newcommand{\hlatextisigmaglobal}{-1.5}
\newcommand{\hlathaloits}{16.3}
\newcommand{\hlathaloisigma}{3.6}
\newcommand{\hlathaloisigmaglobal}{0.7}
\newcommand{\hlatunassocts}{9.2}
\newcommand{\hlatunassocsigma}{2.7}
\newcommand{\hlatunassocsigmaglobal}{-0.4}
\newcommand{\tevits}{9.4}
\newcommand{\tevisigma}{2.6}
\newcommand{\tevisigmaglobal}{0.9}

\newcommand{\NUMROIS}{2689}
\newcommand{\NUMTHREEFGLROI}{2469}
\newcommand{\NUMTHREEFHLROI}{220}

\newcommand{\NUMCATLATEXT}{55} 

\newcommand{\FLUXUNASSOCMIN}{$4.3\times10^{-11}\,\mathrm{cm}^{-2}\,\mathrm{s}^{-1}$}
\newcommand{\FLUXUNASSOCMAX}{$1.1\times10^{-9}\,\mathrm{cm}^{-2}\,\mathrm{s}^{-1}$}
\newcommand{\FLUXUNASSOCMEDIAN}{$2.5\times10^{-10}\,\mathrm{cm}^{-2}\,\mathrm{s}^{-1}$}
\newcommand{\FLUXRATIOUNASSOC}{$\sim2$}
\newcommand{\FLUXALLMEDIAN}{$4.7\times10^{-10}\,\mathrm{cm}^{-2}\,\mathrm{s}^{-1}$}

\newcommand{\gr}{$\gamma$-ray}

\newcommand{\Grs}{$\gamma$-rays}
\newcommand{\hess}{H.E.S.S.}
\newcommand{\ee}{$e^+e^-$}
\newcommand{\lext}{\mathcal{L}_{\mathrm{ext}}}

\newcommand{\lhalo}{\mathcal{L}_{\mathrm{halo}}}
\newcommand{\lsrc}{\mathcal{L}_{\mathrm{src}}}
\newcommand{\fhalo}{F_{\mathrm{halo}}}

\newcommand{\deltalnl}{\Delta\ln\mathcal{L}}
\newcommand{\liken}{\mathcal{L}_{\mathrm{n}}}
\newcommand{\likenplusone}{\mathcal{L}_{\mathrm{n+1}}}
\newcommand{\likenplusext}{\mathcal{L}_{\mathrm{n} + \mathrm{ext}}}
\newcommand{\likenplushalo}{\mathcal{L}_{\mathrm{n} + \mathrm{halo}}}
\newcommand{\likenplusm}{\mathcal{L}_{\mathrm{n} + \mathrm{m}}}
\newcommand{\likenplusmplusone}{\mathcal{L}_{\mathrm{n} + \mathrm{m} + 1}}

\newcommand{\daicm}{\Delta_{\rm m}}

\newcommand{\tsmplusone}{\mathrm{TS}_{\rm m + 1}}
\newcommand{\nunplusone}{\nu_{\rm n + 1}}
\newcommand{\nunplusm}{\nu_{\rm n + \rm m}}

\newcommand{\Deltaext}{\Delta_{\mathrm{ext}}}
\newcommand{\Deltahalo}{\Delta_{\mathrm{halo}}}
\newcommand{\tsext}{\mathrm{TS}_{\mathrm{ext}}}
\newcommand{\rext}{\mathrm{R}_{\mathrm{ext}}}
\newcommand{\rinner}{\mathrm{R}_{\mathrm{inner}}}
\newcommand{\rcasc}{\mathrm{R}_{\mathrm{casc}}}
\newcommand{\tshalo}{\mathrm{TS}_{\mathrm{halo}}}
\newcommand{\rhalo}{\mathrm{R}_{\mathrm{halo}}}
\newcommand{\gammahalo}{\Gamma_{\mathrm{halo}}}

\newcommand{\plocal}{p_{\mathrm{local}}}

\renewcommand{\widehat}{\hat}
\newlength{\dhatheight}
\newcommand{\doublehat}[1]{%
    \settoheight{\dhatheight}{\ensuremath{\widehat{#1}}}%
    \addtolength{\dhatheight}{-0.35ex}%
    \widehat{\vphantom{\rule{1pt}{\dhatheight}}%
    \smash{\widehat{#1}}}}

\newcommand{\irf}[1]{\texttt{#1}}

\newcommand{\Sectionref}[1]{Section~\ref{#1}}
\newcommand{\Appendixref}[1]{Appendix~\ref{#1}}
\newcommand{\Tableref}[1]{Table~\ref{tab:#1}}
\newcommand{\Tablerefs}[2]{Tables~\ref{tab:#1} and \ref{tab:#2}}
\newcommand{\Figureref}[1]{Figure~\ref{fig:#1}}
\newcommand{\Equationref}[1]{Equation~\ref{#1}}

\newcommand{\Fermi}{{\textit{Fermi}}}
\newcommand{\fermi}{\Fermi}

\usepackage{color}

\ifdefined\bwfigures
\else
\fi

\ifdefined \fdg \else
\newcommand\fdg{\mbox{$~.\!\!^\circ$}\xspace} 
\fi

\newcommand\dg{\mbox{$~\!\!^\circ$}\xspace}

\begin{document}

\title{Search for Spatial Extension in High-Latitude Sources Detected by the \textit{Fermi} Large Area Telescope}

\date{\today}

\author{M.~Ackermann}
\affiliation{Deutsches Elektronen Synchrotron DESY, D-15738 Zeuthen, Germany}
\author{M.~Ajello}
\affiliation{Department of Physics and Astronomy, Clemson University, Kinard Lab of Physics, Clemson, SC 29634-0978, USA}
\author{L.~Baldini}
\affiliation{Universit\`a di Pisa and Istituto Nazionale di Fisica Nucleare, Sezione di Pisa I-56127 Pisa, Italy}
\author{J.~Ballet}
\affiliation{Laboratoire AIM, CEA-IRFU/CNRS/Universit\'e Paris Diderot, Service d'Astrophysique, CEA Saclay, F-91191 Gif sur Yvette, France}
\author{G.~Barbiellini}
\affiliation{Istituto Nazionale di Fisica Nucleare, Sezione di Trieste, I-34127 Trieste, Italy}
\affiliation{Dipartimento di Fisica, Universit\`a di Trieste, I-34127 Trieste, Italy}
\author{D.~Bastieri}
\affiliation{Istituto Nazionale di Fisica Nucleare, Sezione di Padova, I-35131 Padova, Italy}
\affiliation{Dipartimento di Fisica e Astronomia ``G. Galilei'', Universit\`a di Padova, I-35131 Padova, Italy}
\author{R.~Bellazzini}
\affiliation{Istituto Nazionale di Fisica Nucleare, Sezione di Pisa, I-56127 Pisa, Italy}
\author{E.~Bissaldi}
\affiliation{Dipartimento di Fisica ``M. Merlin" dell'Universit\`a e del Politecnico di Bari, I-70126 Bari, Italy}
\affiliation{Istituto Nazionale di Fisica Nucleare, Sezione di Bari, I-70126 Bari, Italy}
\author{R.~D.~Blandford}
\affiliation{W. W. Hansen Experimental Physics Laboratory, Kavli Institute for Particle Astrophysics and Cosmology, Department of Physics and SLAC National Accelerator Laboratory, Stanford University, Stanford, CA 94305, USA}
\author{E.~D.~Bloom}
\affiliation{W. W. Hansen Experimental Physics Laboratory, Kavli Institute for Particle Astrophysics and Cosmology, Department of Physics and SLAC National Accelerator Laboratory, Stanford University, Stanford, CA 94305, USA}
% ...
\collaboration{The \emph{Fermi}-LAT Collaboration}
\author{J.~Biteau}
\affiliation{Institut de Physique Nucl\'eaire, Universit\'e Paris-Sud, Univ. Paris/Saclay, 15 rue Georges Clemenceau, 91406 Orsay, Cedex, France}

\begin{abstract}
  We present a search for spatial extension in high-latitude
  ($|b|>5\dg$) sources in recent \textit{Fermi} point source catalogs.  
 The result is the  
 \Fermi\ High-Latitude Extended Sources Catalog, which provides source extensions (or upper limits thereof) and
  likelihood profiles for a suite of tested source morphologies.
   We find
  \NUMEXTSOURCES~extended sources,  \NUMNEWEXTSOURCES~of which  were not
  previously characterized as extended.  
These include sources that are potentially associated   
  with supernova remnants and star 
forming regions.
  We also found extended \gr\ emission in the vicinity of the Cen~A radio lobes and---at GeV energies for the first time---spatially coincident with the radio emission of the SNR CTA 1, as well as from the Crab Nebula. 
  We also searched for halos around active galactic nuclei, which are predicted
from electromagnetic cascades that are induced by the  $e^+e^-$~pairs that are deflected in intergalactic magnetic fields. These are produced when   
 $\gamma$-rays interact with background radiation fields. 
We do not find evidence for extension in individual sources or in stacked source samples.
This enables us to place limits on the flux of the extended source components, which are then
used to constrain the intergalactic magnetic field 
  a coherence length $\lambda \gtrsim 10\,$kpc, even when conservative assumptions 
  on the source duty cycle are made. 
  This improves previous limits by several orders of magnitude.
\end{abstract}
\pacs{95.35.+d,95.30.Cq,98.35.Gi}

\section{INTRODUCTION}\label{sec:intro}

Extended \gr\ sources provide a unique probe into a plethora of physics topics, ranging from the acceleration of 
relativistic particles and emission of (very) high energy $\gamma$-rays to searches for new physics. 
Known astrophysical sources from which spatial extension has been observed at \gr\ energies  include 
supernova remnants \citep[SNRs,][]{2006A&A...449..223A, Acero:2015prw}, 
pulsar wind nebulae \citep[PWNe,][]{2013ApJ...774..110G, 2017arXiv170208280H}, 
and molecular clouds \citep{strong1982,2008A&A...481..401A}.
An additional extended $\gamma$-ray source class might be star-forming regions (SFRs), 
one of which has been identified so far at $\gamma$-ray energies, namely the Cygnus Cocoon~\citep{2011Sci...334.1103A}.
Furthermore, spatial extension at \gr\ energies has been detected from nearby galaxies such as the 
Magellanic Clouds~\citep{TheFermi-LAT:2015lxa, Fermi-LAT:2010fcp} 
and M31~\citep{Ackermann:2017nya}, as well as from the lobes of active galactic nuclei (AGNs), such as Cen A~\citep{2010Sci...328..725A}. 

Extended \gr\ emission from
otherwise point-like AGN could be due to electromagnetic 
cascades~\citep{protheroe1993}.
The $\gamma$-rays  interact with photons 
of the extragalactic background light \citep[EBL,][]{hauser2001,kashlinsky2005} 
to form \ee~pairs \citep{nikishov1962,gould1967,gould1967a,dwek2013}. 
The \ee~pairs can, in turn, inverse-Compton (IC) scatter
photons of the cosmic microwave background (CMB),
thereby initiating the cascade. 
The pairs are deflected in the intergalactic magnetic field (IGMF); depending on its strength and coherence length, an extended \gr\  halo may form around AGNs, often referred to as pair halo
beam-broadened cascades~\citep{aharonian1994}.
The cascade emission can also lead to an excess in the GeV regime of \gr\  spectra, and the non-observation of this feature has been used to derive 
lower limits on the IGMF strength---or conversely on the filling factor of the IGMF \citep{neronov2010,tavecchio2011,dolag2011,taylor2011,2012ApJ...747L..14V}. 
These limits depend on the activity time of AGNs~\citep{dermer2011,finke2015} and on their intrinsic spectra~\citep{arlen2014}. 

Apart from the intrinsic extension of astrophysical objects, 
extended emission from unidentified \gr\ emitters that lack a counterpart at other wavelengths 
can be used to probe the nature of dark matter (DM). 
The observed universe includes a significant component of matter that does not interact like any known field in the Standard Model of particle physics. 
Though solid observational evidence exists for the gravitational influence of DM from the earliest moments of the universe's history to the present day, no direct measurements 
have been made~\citep{Zwicky:1933gu,Rubin:1980zd,Olive:2003iq}.
For instance, extended emission should be produced in the case of the annihilation or decay 
of weakly interacting massive particles gravitationally bound in virialized sub-structures of the halo of the Milky Way \citep[e.g.][]
{Pieri:2007ir, Kuhlen:2008aw, Zechlin:2012by, Mirabal:2012em}.

The above searches for source extension profit from the all-sky survey of 
the Large Area Telescope (LAT) on board the \Fermi\ satellite, 
which detects $\gamma$-rays with energies from 20\,MeV to over 300\,GeV~\citep{Atwood:2009ez}. 
It has discovered a wealth of \gr\ sources, culminating in the two most recent \gr\ source catalogs:  the \textit{Fermi} Third Source Catalog~\citep[3FGL,][]{3fgl} and
Third Hard-Source Catalog \citep[3FHL,][]{2017arXiv170200664T}.
Together, these two catalogs contain more than 3000 sources.
With the release of the latest event selection and reconstruction software, and associated analysis tools \citep[\irf{Pass 8,}][]{2013arXiv1303.3514A}, 
the reconstruction of the photon arrival directions has improved significantly; a reduction of the 68\,\%
containment radius of the point spread function (PSF), particularly at high energies ($>$10 GeV), as has been demonstrated by the 3FHL. 
In combination with an eight-year data set, this provides 
an improved sensitivity to search for spatial extensions. 

This work follows several previous searches for spatially extended
sources at GeV energies.  \citet{2012ApJ...756....5L} reported the
first systematic search for spatially extended sources in LAT data; they
identified 21 extended sources, based on an analysis of two years of
\irf{Pass 7} data.  The most recent search for extended sources, the
Fermi Galactic Extended Source Catalog (FGES), looked for new
sources within $7\dg$ of the Galactic Plane using six years of \irf{Pass
  8} data above 10~GeV \citep{2017arXiv170200476T}.  This search
reported 46 extended sources, eight of which were new extended sources with
clear associations.  Counting all FGES associated sources, as
well as sources found in other dedicated analyses, the LAT has detected
\NUMCATLATEXT~extended sources.

We report here on the \Fermi\ High-Latitude Extended Sources Catalog (FHES),
a comprehensive search for spatially extended \gr\ sources above
5\dg\ Galactic latitude using 7.5 years of \irf{Pass 8} data above
1~GeV.  The FHES encompasses a region of the sky complementary to the
FGES, which only considered low Galactic latitudes.  The FHES has a lower energy threshold than 
FGES because we remove the region of the Galactic Ridge 
where the emission coming from the interstellar medium is very large at 1 GeV. 
Due to its lower energy threshold, the FHES is also sensitive to source populations with
softer spectra.

In Section~\ref{sec:data}, we discuss the \textit{Fermi}-LAT instrument and the data set, sources, and background models used for this analysis, 
as well as the methodology developed to build the extended source catalog. 
The catalog and  a study of individual objects are described in Section~\ref{sec:analysis}.
In Section~\ref{sec:stack}, we turn to  sources located at $|b| > 20\dg$ that show weak evidence for extension, and present a source stacking analysis 
of AGN samples in the search of pair-halo emission.
Due to the absence of a clear pair-halo signal, we derive limits on the IGMF in Section~\ref{sec:results}. 
Finally, we conclude in Section~\ref{sec:conclusion}.

\section{Fermi-LAT DATA AND ANALYSIS}\label{sec:data}

The {\it Fermi}-LAT is a pair-conversion telescope. 
Incoming \Grs\ pass through the anti-coincidence detector and convert in the tracker to \ee~pairs. 
The charged particle direction is reconstructed using the information in the tracker, and the energy is estimated from depositions in the calorimeter.  
Detailed descriptions of the LAT and its performance can be found in dedicated papers~\citep{Atwood:2009ez,2013arXiv1303.3514A}.

\subsection{Data Selection}

We analyze 90 months of LAT data (2008 August 4 to 2016 February 4) and select the \irf{P8R2 SOURCE}-class of events, 
which is the recommended class for most analyses and provides good sensitivity for analysis of point sources and extended 
sources.\footnote{\url{https://fermi.gsfc.nasa.gov/ssc/data/analysis/documentation/Cicerone/Cicerone_Data/LAT_DP.html}} 
The \irf{Pass 8} data benefit from an improved PSF, effective area, and energy reach. 
More accurate Monte Carlo simulations of the detector and the environment in low-Earth orbit \citep{2013arXiv1303.3514A} 
have reduced the systematic uncertainty in the LAT instrument response functions (IRFs). 
We have selected events in the energy range from 1~GeV to 1~TeV, which is determined by the angular resolution at lower 
energies and declining acceptance with increasing energy. 
Each source is analyzed with a binned maximum-likelihood analysis, using eight logarithmic bins per decade in energy and a 
region of interest (ROI) of $6\dg \times 6\dg$ with an angular pixelization of $0.025\dg$.  We summarize our data selection 
in Table \ref{tab:data}.

Within the different event classes, \irf{Pass 8} offers \textit{event
  types}, subdivisions based on event-by-event uncertainties in the
directional and energy measurements, which can increase the
sensitivity of likelihood-based analyses.  In this work, we use the
set of four PSF event-type selections that subdivide the events in our
data sample according to the quality of their directional
reconstruction.  Specifically, the data sample is split by event type
into two data selections that are analyzed in a joint likelihood:
\irf{evtype}$=$32 (PSF3, which corresponds to the best quality of
angular reconstruction) and \irf{evtype}$=$28 (PSF0+PSF1+PSF2).  We
choose to combine the three worst PSF event types for computational
efficiency.  In Monte Carlo studies, we found that PSF3 events provide
most of the power for distinguishing between point-like and extended
hypotheses.  The data reduction and exposure calculations are
performed using the LAT \textit{ScienceTools} version
11-05-03,\footnote{\url{http://fermi.gsfc.nasa.gov/ssc/data/analysis/software}}
\textit{fermipy} \citep{2017arXiv170709551W} version
00-15-01,\footnote{\url{http://fermipy.readthedocs.io}} and the
\irf{P8R2\_SOURCE\_V6} IRFs.  We enable the correction for energy
dispersion for all model components except the Galactic diffuse and
isotropic components.

We perform an independent analysis on \NUMTHREEFGLROI~and~
\NUMTHREEFHLROI~ROIs centered on the positions of the sources with $|b| > 5\dg$ listed in the 3FGL and 3FHL, respectively. 
Among the 3FHL sources considered, we exclude
sources that have an association with a 3FGL source or an angular
separation from a 3FGL source that is less than twice its 95\%
positional uncertainty.
The analysis procedure is outlined in Section~\ref{Sec:Optimization}.
The cut on Galactic latitude is chosen
to avoid regions where systematic errors in the diffuse emission model
could bias the measurement of the angular extension or produce spurious
detections.  
We additionally exclude the following 3FGL and 3FHL
sources:
\begin{enumerate}
\item SMC (3FGL~J0059.0$-$7242e)
\item LMC (3FGL~J0526.6$-$6825e) and four sources in the vicinity of
  the LMC (3FGL~J0524.5-6937, 3FGL~J0525.2-6614, 3FGL~J0456.2-6924,
  and 3FHL~J0537.9-6909)
\item Cygnus~Loop (3FGL~J2051.0$+$3040e)
\item Cen~A~Lobes (3FGL~J1324.0$-$4330e)
\end{enumerate}
Those sources have angular sizes that are comparable to, or significantly
larger than, our chosen ROI size of $6\dg \times 6\dg$.  In the case of
the LMC, Cygnus~Loop, and Cen~A~Lobes these sources also have complex
morphologies that are not approximated well by the disk and Gaussian
models that we use in the present work when testing for angular
extension.  Note that, while we exclude the LMC and SMC from our
analysis, we model the emission from these regions using the spatial
templates from the 3FHL.  Our sample includes four sources that were
modeled as point-like objects in the 3FGL but have subsequently been
measured to have angular extension: Fornax~A
\citep{Ackermann:2016arn}, SNR~G295.5+09.7 \citep{Acero:2015prw},
SNR~G150.3+04.5 \citep{Ackermann:2015uya}, and
M31~\citep{Ackermann:2017nya}. 
These sources were handled consistently with all
other potentially extended sources in the fitting procedure.

For the Crab and CTA~1 pulsars (3FGL~J0534.5$+$2201 and
3FGL~J0007.1$+$7303), we use pulsar phase information to constrain
pulsar emission in these regions.  For CTA~1, we use an eight-year
ephemeris derived from \irf{Pass 8} LAT data above 100~MeV
\citep{Kerr:2015tva}.  For the Crab~Pulsar, we use an ephemeris
derived from radio observations with the Jodrell Bank telescope
\citep{1993MNRAS.265.1003L}.\footnote{\url{http://www.jb.man.ac.uk/~pulsar/crab.html}}

\begin{deluxetable*}{cc}
\tablewidth{0pt}
\tablecaption{ \label{tab:data} Summary of {\it Fermi}-LAT data selection criteria. }
\tablehead{Selection & Criteria}
\startdata
Observation Period &  2008 August 4 to 2016 February 4 \\
Mission Elapsed Time (s)\tablenotemark{a}& 239557414 to 476239414 \\ 
Energy Range & 1~GeV-1~TeV\\
Fit Regions & $6.0\dg\times6.0\dg$ ($|b| > 5\dg$; \NUMROIS~ROIs)\\ 
Zenith Range & $\theta_z<$100$\dg$\\
Data Quality Cut\tablenotemark{b} & \texttt{DATA\_QUAL==1} \\
 & \texttt{LAT\_CONFIG==1} \\
\enddata
\tablenotetext{a}{$Fermi$ Mission Elapsed Time is defined as seconds since 2001 January 1, 00:00:00 UTC}
\tablenotetext{b}{Standard data quality selection with the $gtmktime$ Science Tool}
\end{deluxetable*}

\subsection{ROI Model and Optimization}\label{Sec:Optimization}

For each ROI, we start from a baseline model that includes sources
from the 3FGL and standard templates for isotropic and Galactic
diffuse
emission.\footnote{Galactic IEM: gll\textunderscore iem\textunderscore v06.fits, Isotropic: iso\textunderscore P8R2\textunderscore SOURCE\textunderscore V6\textunderscore v06.txt. 
Please see: \url{http://fermi.gsfc.nasa.gov/ssc/data/access/lat/BackgroundModels.html}}
We include 3FGL sources in a $10\dg \times10 \dg$ region centered on
the ROI.  We model each 3FGL source using the same spectral
parameterization as used in the 3FGL.  The 3FGL uses one of three
different spectral parameterizations depending on the source
association and evidence for spectral curvature: power law (PL),
log-parabola (LP), and power law with exponential cutoff (PLE).  We
switch to the LP parameterization for all PL sources detected in our
analysis with Test Statistic~\citep{Chernoff}, (TS) $> 100$.  This ensures that we have an accurate 
model for background sources that may show spectral curvature, and it comes 
without loss of generality because the PL is a special case of an LP.
 For extended sources, we use the
spatial models from the 3FHL \citep{TheFermi-LAT:2017vmf} that include
new or improved spatial templates for some high-latitude extended
sources, including the LMC, Fornax~A, and SNR~G150.3+4.5.

An extended source could be characterized as a
cluster of point sources in the 3FGL because the 3FGL
does not include a criterion for distinguishing between point-like and
extended emission.
Therefore, the baseline model excludes 3FGL sources that are unassociated
and have either $\mathrm{TS} < 100$ or analysis flags indicating
confusion with diffuse emission (flags 5, 6, or 8). 
 Removing unassociated sources
ensures that the characterization of new extended sources is not
biased by the 3FGL sources included in the baseline model.  If the
unassociated sources are genuine point sources, they will be added
back into the model in the course of the ROI optimization (see below).

\begin{figure*}
\centering
\includegraphics[width=.9\linewidth]{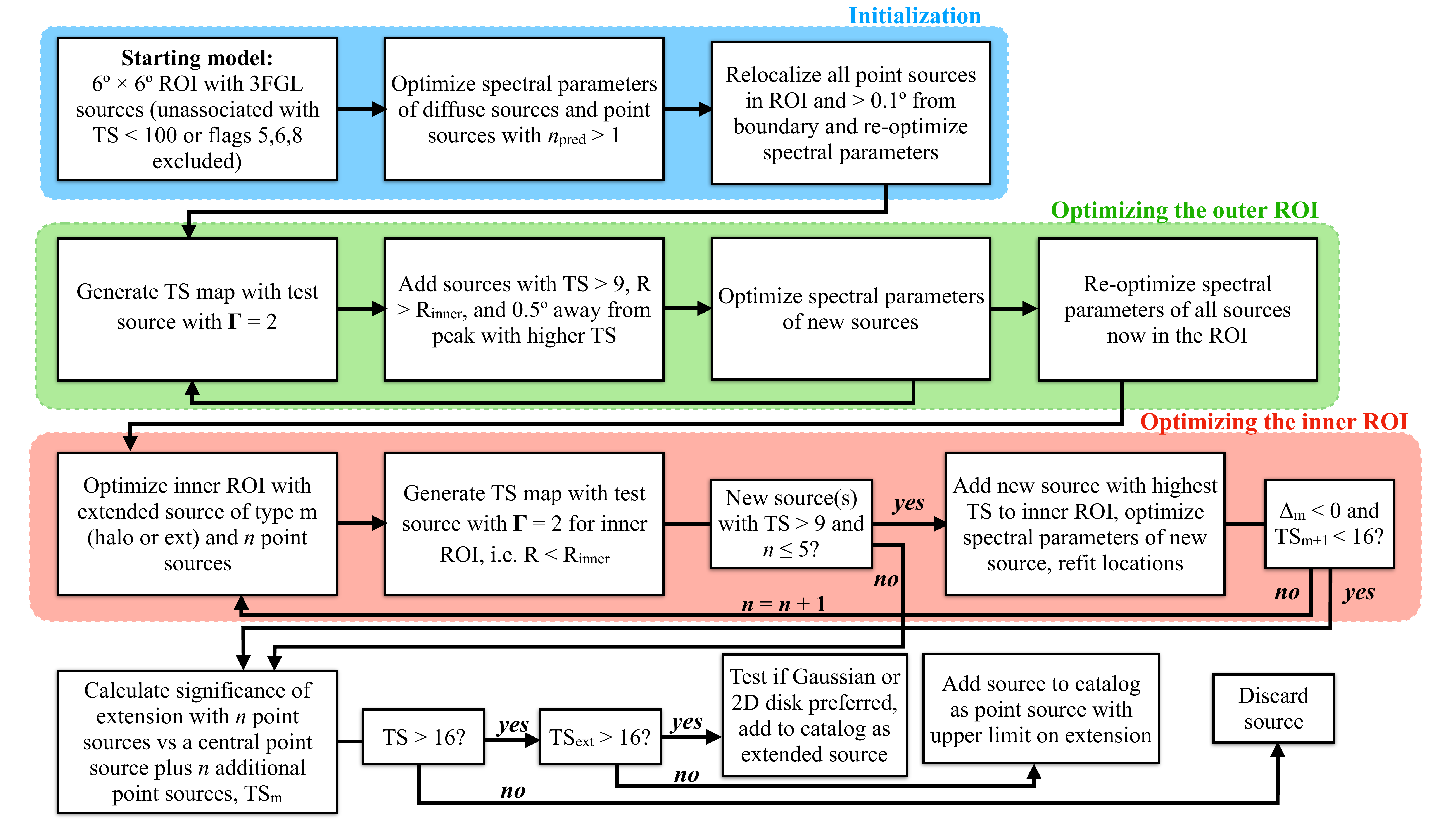}
\caption{\label{fig:flowchart} Flow chart for the analysis procedure. 
See text for further details.}
\end{figure*}

Starting from the baseline model, we proceed to optimize the model by
fitting the spectral and spatial properties of the model components.
We illustrate the analysis procedure in the flow chart in Figure~\ref{fig:flowchart}. 
We first fit the spectral parameters (flux
normalization and spectral shape parameters)
of the Galactic interstellar emission model
 model, and all sources in the model with an amplitude of at
least one expected photon for the initial 3FGL model parameters.  We then individually fit the positions of
all point sources that are inside the ROI and $>0.1\dg$ from the ROI boundary.  When
fitting the position of a source, we fix its spectral shape parameters
but refit its normalization.  After relocalizing point sources, we
re-fit the spectral parameters of all model components.

After optimizing the parameters of the baseline model components, we
further refine the model by identifying and adding new point source
candidates.  The identification of new point sources is
  performed in two successive passes, focusing on the outer
  ($R > \rinner$) and inner ($R \leq \rinner$) ROI, where $R$ is the
  angular distance from the ROI center and $\rinner = 1.0\dg$.
  Sources found to be significantly extended (having a test statistic of extension ,$\tsext, > 16$) are
  reanalyzed with $\rinner = 1.5\dg$ to minimize bias from
  point sources that are confused with the target source.

In the first pass, we use a likelihood-based source-finding algorithm to look for point sources with $R > \rinner$.  We identify candidates
by generating a TS map for a point source that has a PL spectrum with a PL
index $\Gamma=2$.  When generating the TS map, we fix the
  parameters of the background sources and fit only the amplitude of
  the test source. We add a source at every peak in the TS map with
$R > \rinner$ and $\mathrm{TS} > 9$ that is at least $0.5\dg$ from a
peak with higher TS.  New source candidates are modeled with a PL if
the source is detected with TS $< 100$ and an LP otherwise.  Both the
normalization and spectral shape ($\Gamma$ for PL, and index $\alpha$
and curvature $\beta$ for LP) parameters of new source candidates are
fit in this procedure.  We then generate a new TS map after
adding the point sources to the model and repeat the procedure until
no candidates are found satisfying our criteria ($R > \rinner$,
$\mathrm{TS} > 9$).  After completing the search for point sources in
the outer ROI, we re-fit the normalization and spectral shape
parameters of all model components.

In the final pass of the analysis, we look for new point-source
candidates in the inner ROI while simultaneously testing the central
source for extension.  The analysis proceeds iteratively, as follows,
for two independent hypotheses that we denote as extension (extended source) and halo (extended source plus a superimposed point source):
\begin{enumerate}

\item We perform tests for extension (as described in the next
  paragraph) against the null model with $n$ point sources in the
  inner ROI ($n$ includes the source of interest but excludes 3FGL
  point sources included in the baseline model).  
    
\item We derive a model with $n+1$ point sources by searching
  for additional point sources with $\mathrm{TS} > 9$ in the inner ROI
  using the same source-finding algorithm that was applied in the
  outer ROI optimization.  If a peak with $\mathrm{TS} > 9$ is found
  in the TS map, we add a new point source at this location. If more
  than one source candidate is found, we select the one with the
  highest peak TS.  We then individually refit the source
  positions of the central source and any point sources added up to
  this iteration in the inner ROI, starting from the source with the
  highest TS.

\item We repeat steps 1 and 2 until we find that the extension/halo
  hypothesis is preferred over a model with $n+1$ point
    sources (according to the criteria in Equations~\ref{eqn:akaike}
    and \ref{eqn:akaike2}), no point sources with $\mathrm{TS} > 9$
  are found in the source-finding step, or the number of iterations
  exceeds five.

\end{enumerate}

At each iteration $n$, we test for extended emission by comparing the
likelihood of the hypothesis with 
a central point source and 
$n$ additional
point sources ($\liken $) versus
the likelihoods for two alternative hypotheses: replacing the central
source with a symmetric 2D Gaussian ($\likenplusext$), and
superimposing a 2D symmetric Gaussian on the central source
($\likenplushalo$)
(with $n$ additional point sources).
For the extended hypothesis, we replace the
central point source with an extended source that has the same spectral
parameterization.  We then perform a simultaneous fit of the position,
angular size, and spectral parameters (normalization and shape) of the
central source.  In this fit, we free the normalization and spectral shape
parameters of sources within 1.0\dg of the central source and
normalizations of sources within 1.5\dg of the central source.

For the halo hypothesis, we add a new extended source component with
position fixed to that of the central source with a PL spectral
parameterization with index $\gammahalo$ that is independent of the
central source.  The normalization, index, and angular size of the
halo component are left as free parameters.  The normalization of the
central source and all sources within 1.0\dg of the central source are
freed.  We parameterize the angular size of the extended component
with the intrinsic 68\% containment radius, which we denote with
$\rext$ and $\rhalo$ for the case of the extended and halo model,
respectively.

To distinguish an extended source from a cluster of point sources, 
we compare models using the Akaike information criterion
\citep[AIC,][]{Akaike} given by $\mathrm{AIC} = 2k - 2\ln\mathcal{L}$
where $k$ is the number of parameters in the model.  The formulation
of the AIC penalizes models with a larger number of parameters, and
hence minimizes overfitting.  The best model will minimize the AIC. 
Models with fewer parameters are
preferred unless a model with more parameters provides a substantially
better fit.
We define $\Delta_\mathrm{m}$ as the difference between
the AIC of the models with and without extension,

\begin{equation}\label{eqn:akaike}
\Delta_\mathrm{m} = \mathrm{AIC}_{\mathrm{n+m}} - \mathrm{AIC}_{\mathrm{n+1}} = 2\left(\ln\likenplusone - \ln\likenplusm + \nunplusm - \nunplusone\right),
\end{equation}
where $\mathrm{m} =$ ext (halo) for the extension (halo) hypothesis,
$\likenplusone$ is the likelihood for the model with $n+1$ point
sources, $\likenplusm$ is the likelihood for the model with extended
emission, and $\nu_{\rm X}$ is the number of degrees of freedom of the
given model.  If $\Delta_\mathrm{m} > 0$, then a model with an
additional point source is preferred over a model with extension. 

In cases where a bright extended source is superimposed on a fainter
point source, the criterion defined in
\Equationref{eqn:akaike} will tend to prefer the extended source model
with $n$ point sources even when a model with extension and $n+1$
point sources gives a better fit to the data (smaller AIC).
To distinguish this scenario, we define $\tsmplusone$ as twice
  the difference in the log-likelihood of extended source models with
  $n$ and $n+1$ point sources,

\begin{equation}\label{eqn:akaike2}
\tsmplusone = 2\left(\ln\likenplusmplusone - \ln\likenplusm\right),
\end{equation}
where $\likenplusmplusone$ is the likelihood for the extended source
model with $n+1$ point sources.  If no additional point sources are
found in the subsequent iteration, then $\daicm$ and $\tsmplusone$
are undefined.

For the extension and halo hypotheses, we select a best-fit model of
the inner ROI with $n$ point sources, where $n$ is the first
iteration for which $\daicm < 0$ and $\tsmplusone < 16$, or for which no additional point sources are
found (at a level of 3$\sigma$).  Given the best-fit iteration $n$, the evidence for extended
emission is evaluated from the likelihood ratio between models with
and without an extended component,

\begin{equation}\label{eqn:tsext}
\mathrm{TS}_\mathrm{m} = 2\left(\ln\likenplusm - \ln\liken\right).
\end{equation}
Because the hypotheses are nested, we expect the test statistics for the 
extension and halo hypotheses ($\tsext$ and $\tshalo$ respectively) to
be distributed as $\chi^2_{\nu}$ where $\nu$ is the difference in the
number of degrees of freedom ($\nu = 1$ and 3 for the extension and halo
hypotheses, respectively).  We identify a source as extended if
$\tsext > 16$.  Sources that exceed the threshold for extension are
additionally fit with a 2D disk morphology, and the Gaussian or disk
morphology is chosen on the basis of the model with the largest
likelihood.

We find that some extended sources are composites of multiple 3FGL
sources, which results in multiple analysis seeds being associated with the
same source.  Where the same extended source is detected in multiple
analysis seeds (spatial overlap of the 68\% containment circle greater
than 50\%), we merge the analysis seeds into a single seed with
position equal to the average of the seed positions.  We then perform
a new analysis of the source using the merged analysis seed and drop
the original analysis seeds from the catalog.  Six of the
  extended FHES sources were found to be composites of two or more
  3FGL sources.  Merging these seeds resulted in the removal of 15 of
  the original analysis seeds.

If we detect a point source with $\tshalo > 16$, we create a
  new extended source and analyze the ROI with a model that includes
both the point source and an extended component with the same
morphological and spectral parameters as the best-fit halo.  We then
run the analysis pipeline on the extended component, refitting both its
position and extension.  We convert the candidate halo into a separate
  extended FHES source if it is detected with $\mathrm{TS} > 25$.
Nine of the extended FHES sources are found by the search for
  extended halo emission.

\subsection{Diffuse and IRF Systematics}\label{sec:systematics}

The two primary sources of systematic error in our analysis are the
instrument response functions (IRFs) and the Galactic interstellar
emission model (IEM).  We take the total systematic error from the
larger of the errors induced by the IRFs and IEM.  Due to the strong
gradient in IEM intensity with Galactic latitude, IEM uncertainties
are typically subdominant for sources with $|b| > 20\dg$.

Our nominal Galactic IEM is the recommended one for \irf{PASS8} source
analysis, which we denote as IEM-STD.  IEM-STD is based on the IEM
developed with \irf{P7REP} data \citep{2016ApJS..223...26A}.  IEM-STD
has the same spatial distribution as the \irf{P7REP} model, but has
been rescaled with a small, energy-dependent correction to account for
the difference in the influence of energy dispersion in the
\irf{P7REP} and \irf{PASS8} data sets.  To quantify the impact of
diffuse systematics, we repeat our analysis with nine alternative
IEMs: the eight models from \citet{Acero:2015prw} (IEM-A0 to IEM-A7)
and the IEM developed for the study of diffuse emission in the inner
Galaxy \citep[IEM-B][]{TheFermi-LAT:2017vmf}.  Because the models from
\citet{Acero:2015prw} were developed with \irf{P7REP} data, we apply
the same energy dispersion correction that was used for IEM-STD to
obtain models appropriate for \irf{PASS8} analysis.

To evaluate the IEM-induced systematic uncertainty on a fitted
quantity $P$, we follow the method of \citet{Acero:2015prw} by
calculating the dispersion between the nominal value obtained with
IEM-STD and the value obtained with the nine alternative IEMs,

\begin{equation}
\delta P_\mathrm{sys} = \sqrt{\frac{1}{\sum_{i}\sigma_{i}^{-2}}\sum_{i}\sigma_{i}^{-2}\left(P_\mathrm{STD}-P_{i}\right)^2}
\end{equation}
where $P_\mathrm{STD}$ is the measured value obtained with IEM-STD, and
$P_{i}$ and $\sigma_{i}$ are the values and statistical uncertainties
for $P$ obtained with the nine alternative IEMs.

The primary instrumental uncertainty relevant for studies of extension
is the PSF.  To evaluate the systematic uncertainty on the PSF, we consider two
bracketing PSF models based on the recommended systematic error band for
the PSF 68\% containment
radius.\footnote{\url{https://fermi.gsfc.nasa.gov/ssc/data/analysis/LAT_caveats.html}}
We define the following piecewise scaling function for the relative PSF uncertainty versus energy:

\begin{equation}\label{eqn:psfscaling}
f(E) = \left\{
        \begin{array}{ll}
            0.05 & \quad E \leq 10~\mathrm{GeV}\\
            0.05 + 0.1\times\log_{10}\left(E/10~\mathrm{GeV}\right) & \quad E > 10~\mathrm{GeV}
        \end{array}
    \right..
\end{equation}
This function defines a constant 5\% error below 10~GeV that rises to
25\% at 1~TeV. We note that the increase in the systematic
  uncertainty above 10~GeV is driven by the statistical precision of
  the in-flight validation sample, rather than an observed discrepancy
  in the model of the PSF.  We construct bracketing models of the PSF
versus reconstruction angle and energy, $P_{\mathrm{min}}(\theta; E)$
and $P_{\mathrm{max}}(\theta; E)$, by scaling the average PSF,
$P(\theta; E)$, with this function such that
$P_{\mathrm{min}}(\theta; E) = P(\theta \times (1+f(E)); E)(1+f(E))^2$
and
$P_{\mathrm{max}}(\theta; E) = P(\theta \times (1+f(E))^{-1} ;
E)(1+f(E))^{-2}$.  Applying this model to sources detected with
$\tsext > 9$, we find a median systematic error on the 68\%
containment radius of $0.005\dg$.  With the exception of the brightest
LAT sources, the systematic error is much smaller than the statistical
error.

\subsection{Source Associations}\label{sec:associations}

Because our seeds are taken from the 3FGL and 3FHL, we expect the
majority of the FHES sources to have a direct counterpart with a
source from at least one of these two catalogs.  Rather than
performing an independent search for associations, we assign
associations by taking the association of the closest 3FGL or 3FHL
counterpart.  
Positional uncertainties of both 
  FHES
  point sources and 
  extended sources 
  are evaluated by fitting a
  paraboloid to log-likelihood values sampled on a grid centered on
  the best-fit position.  The resulting positional error ellipse is
  parameterized by 68\% uncertainties along the semi-minor and semi-major
  ellipse axes and a position angle.

For the FHES sources that are best-fit by a point source
morphology, we identify the \gr\ counterpart by finding the nearest
3FGL or 3FHL point source with angular separation
$< 1.5\times\sqrt{\theta_{95,\mathrm{FHES}}^2 + \theta_{95,X}^2}$
where $\theta_{95,\mathrm{FHES}}$ and $\theta_{95,\mathrm{X}}$ are the
symmetric 95\% positional uncertainties of the FHES source and 3FGL or 3FHL
source, respectively.  Our association threshold, which is more
inclusive than that used in previous LAT catalogs, is chosen to
achieve a false negative rate $\lesssim 0.1$\%.  The more
inclusive association threshold is motivated by the fact that the data
sets used for the FHES and the 3FGL are largely independent due to the
difference in exposure and the transition from \irf{P7REP} to
\irf{PASS8}.  Where we find both a 3FGL and a 3FHL counterpart, we take
the source association and classification from the 3FGL.

For sources
that have blazar associations, we take the blazar characteristics
(redshift, optical class, synchrotron peak frequency) from the 3LAC
\citep{2015ApJ...810...14A} or 3FHL for sources with a 3FGL or 3FHL
association, respectively.

Associations for the FHES extended sources are performed on a
case-by-case basis by examining positional and morphological
correlations with multiwavelength counterparts.  In several cases, we
find that an extended source may be a composite of 3FGL sources.  We
identify a 3FGL or 3FHL source as a composite counterpart if it is
encompassed within the intrinsic radius of the extended source and has
no point-source counterpart in the best-fit model of the ROI.  The
associations and 3FGL counterparts for FHES extended sources are
discussed further in Section~\ref{sec:IndivSources}.

\subsection{Flux and Extension Likelihood Profiles}\label{sec:like_extraction}

After obtaining the best-fit model for each source, we extract
likelihood profiles that we use for the analysis of stacked samples
(\Sectionref{subsec:stack}) and modeling of pair cascades
(\Sectionref{sec:results}).  The likelihood profiles are evaluated on
a regular grid of parameter values $\mathbf{x}_{i}$ by maximizing the likelihood
with respect to a set of nuisance parameters ($\boldsymbol{\theta}$) at each point in the
coordinate grid. 
The nuisance parameters that maximize the likelihood at each grid point are denoted with $ \hat{\boldsymbol{\theta}}$.
 The tabulated profile likelihood values are included in the
\texttt{LIKELIHOOD} table of the FITS catalog file
(see \Appendixref{sec:fits-format}).  The likelihood profiles
extracted for each source are:
\begin{itemize}
\item $\lext(\rext; \hat{\boldsymbol{\theta}})$: Likelihood versus
  angular extension ($\rext$) of the source of interest
  (\texttt{ext\_dloglike} column in the \texttt{LIKELIHOOD} Table).
  The scan in angular extension is performed on a logarithmic grid
  between $0.00316\dg$ and $1.77\dg$.
\item $\lhalo(\fhalo, \rhalo, \gammahalo; \hat{\boldsymbol{\theta}})$:
  Likelihood for a halo component with a 2D Gaussian morphology and a
  PL spectrum parameterized by flux ($\fhalo$), extension ($\rhalo$),
  and spectral index ($\gammahalo$).  The likelihood is evaluated on a
  logarithmic grid in $\rhalo$ with 15 steps between $0.0316\dg$ and
  $1.77\dg$, a logarithmic grid in $\fhalo$ with 60 steps between
  $10^{-10}~\rm{MeV}~\rm{cm}^{-2}~\rm{s}^{-1}$ and
  $10^{-4}~\rm{MeV}~\rm{cm}^{-2}~\rm{s}^{-1}$, and a grid in
  $\gammahalo$ between 1 and 4 in steps of 0.25.
  (\texttt{halo\_dloglike} column in the \texttt{LIKELIHOOD} Table)
\item ${\lhalo}_{,i}(\fhalo, \rhalo ; \hat{\boldsymbol{\theta}})$:
  Likelihood for a halo component with flux ($\fhalo$) and extension
  ($\rhalo$) in energy bin $i$ (\texttt{halo\_sed\_dloglike} column in
  the \texttt{LIKELIHOOD} Table).  The likelihood is evaluated on a
  logarithmic grid in $\rhalo$ with 15 steps between $0.0316\dg$ and
  $1.77\dg$.  Likelihood evaluation points in $\fhalo$ are chosen
  individually for a given $\rhalo$ and energy bin $i$ to sample points
  around the peak of the likelihood function.
\item ${\lsrc}_{,i}(F; \hat{\boldsymbol{\theta}})$: Likelihood versus
  source flux in energy bin $i$ (\texttt{src\_sed\_dloglike} column in
  the \texttt{LIKELIHOOD} Table). Likelihood evaluation points in $F$
  are chosen individually for a given energy bin to sample points
  around the peak of the likelihood function.
\end{itemize}
For all likelihood profiles, the nuisance parameters include the
normalizations of both diffuse components and all sources in the inner
ROI.  In the case of the likelihood versus extension, we also
simultaneously fit the normalization and spectral shape parameters of
the source of interest.  Following the approach developed for DM
analyses of the SMC and LMC
\citep{2015PhRvD..91j2001B,2016PhRvD..93f2004C}, when evaluating the
likelihood profiles versus flux in a given energy bin $i$
(${\lhalo}_{,i}(\fhalo,\rhalo; \hat{\boldsymbol{\theta}})$ and
${\lsrc}_{,i}(F; \hat{\boldsymbol{\theta}})$), we fit the nuisance
parameters while applying a prior on their values derived from the
broadband (full energy range) fit.  The profile likelihood is given by

\begin{equation}\label{eqn:nuisance}
\mathcal{L}_{i}(\mathbf{x};\hat{\boldsymbol{\theta}}_{i}) = 
\max_{\boldsymbol{\theta}}\mathcal{L}_{i}(\mathbf{x};\boldsymbol{\theta}) \prod_{j}N(\theta_{j}-\tilde{\theta}_{j},5\sigma_{j}),
\end{equation}
where $\mathbf{x}$ represents the parameters of interest, $N$ is the normal
distribution, $\tilde{\theta}_{j}$ and $\sigma_{j}$ are the value and
uncertainty on $\theta_{j}$ obtained from the broadband fit.  This
prior constrains the amplitude of each nuisance parameter to lie
within $5\sigma$ of its value from the broadband fit.

\section{Extension Catalog}\label{sec:analysis}

As described in Section~\ref{sec:data}, this analysis searches for
source extension using 3FGL and 3FHL point sources as targets.  There
are {\NUMCATLATEXT}
 known extended sources in these catalogs, which include the
most current compilation of spatially extended LAT
sources.\footnote{\url{https://fermi.gsfc.nasa.gov/ssc/data/access/lat/3FHL/}}
Most of these sources are Galactic SNRs and PWNe and are well within the 
Galactic plane ($|b|<5\dg$). 
At higher latitudes, extended sources are generally galaxies: for example the Magellanic Clouds, the lobes of Centaurus A, and Fornax A.

From our analysis of {\NUMROIS} seed positions, we identify
{\NUMEXTSOURCES} extended sources and {\NUMPTSOURCES} sources
consistent with a point-like morphology.  The extended source list
includes {\NUMEXTSOURCESSIGNIFICANT} with statistically significant
extension ($\tsext\geq16$), as well as M31, which falls slightly below our
detection threshold ($\tsext=\fhesmthirtyonetsext$).  M31 was
previously detected as extended \citep{Ackermann:2017nya} and the
measured extension from that work is in good agreement with this
analysis.

Using the procedure outlined in Section~\ref{sec:associations} we find
a \gr\ association for all but {\NUMUNASSOCPTSOURCES} of the
{\NUMPTSOURCES} FHES point sources.  From the {\NUMTHREEFHLROI} seeds
that are initialized with a 3FHL source, 
only five sources are not detected in our analysis or do not have a 3FHL association
(note that, if there is a 3FGL counterpart for a 3FHL source, we use the 3FGL
source position).  The unassociated sources have integrated fluxes
between {\FLUXUNASSOCMIN} and {\FLUXUNASSOCMAX}, with a median
{\FLUXUNASSOCMEDIAN} which is a factor of {\FLUXRATIOUNASSOC} lower than
the median of the full catalog ({\FLUXALLMEDIAN}).
Table~\ref{tab:summary} summarizes the number of sources with a 3FGL
or 3FHL association. FHES point sources without a 3FGL or
  3FHL association are excluded from the search for angular extension
  and are not included in the online FHES data products.

A summary of the results of the spatial analyses for the extended sources is shown in Table~\ref{tab:extended_sources}.  
In Table~\ref{tab:extended_sources_syst}, we show the measured
properties of these sources (position, size, flux, and spectral index) along
with their statistical errors and systematic errors obtained from the
nine alternative IEMs and the two bracketing PSF models.
Of the {\NUMEXTSOURCES} extended sources reported in this work,
{\NUMNEWEXTSOURCES} are newly detected.  Nine of the newly detected
sources were found via the halo test (indicated with a dagger in
\Tablerefs{extended_sources}{extended_sources_syst}) and do not have a
direct counterpart in the 3FGL or 3FHL.  The characteristics of the
five previously detected extended sources obtained in this study are
in agreement with those found in previous publications (references are
provided in Section~\ref{subsec:knownsrcs}).  Of the new sources,
{\NUMNEWASSOCEXTSOURCES} have potential associations and the remaining
are classified as unassociated.  We have separated the unassociated
sources into two categories, based on the spectral index of their PL
spectrum: $\Gamma<2.3$ (hard) and $\Gamma>2.3$ (soft) for the 2D
Gaussian extension.  The distinction between hard and soft sources is
made because the soft sources might resemble a mismodeling of the
Galactic diffuse emission, which also has a soft spectrum.

We identify {\NUMCONFUSEDSOURCES} of the {\NUMNEWEXTSOURCES} newly
identified extended sources as ``confused'', indicating sources that
may be spurious, that could be affected by systematic
uncertainties in the IEM, 
or that are seen in the direction of HII regions. The ionized gas is not accounted for in the current IEM, although it can 
significantly contribute to the diffuse $\gamma$-ray emission in the case of massive HII regions \citep{remy2017}.
These sources are grouped into a separate
section at the bottom of
\Tablerefs{extended_sources}{extended_sources_syst}.  We categorize a
source as confused if $\tsext$ falls below our detection threshold
when analyzed with at least one of the alternative IEMs, or if the
fractional systematic uncertainty on the source flux exceeds 50\%.
We also categorize {\fhessoftiii} as confused based on a
  separate analysis with an IEM based on \textit{Planck} dust maps~\citep{2014A&A...571A..11P}.
Finally, we characterize {\fhessofti} as confused after our inspection of the
   velocity-integrated map of H$_{\alpha}$ emission from the Wisconsin H-Alpha Mapper (WHAM) Sky Survey~\citep{0067-0049-149-2-405, whaminprep}. 
   All but one of the confused sources are unassociated, and
four of them ({\fhessofti}, {\fhessoftii}, {\fhessoftiii}, and
{\fhessoftiv}) have soft spectral indices similar to that expected
from Galactic diffuse emission.

The format of the extended source catalog follows the previous
\Fermi-LAT catalogs.  A FITS file with analysis results for all \NUMPTSOURCES\ point sources and \NUMEXTSOURCES\ extended sources
is provided in the online supplementary
material.\footnote{\href{http://www-glast.stanford.edu/pub_data/1261/}{http://www-glast.stanford.edu/pub\_data/1261/} and \url{https://zenodo.org/record/1324474}}
The format of this file is described in \Appendixref{sec:fits-format}.

\begin{figure*}[ht]
\centering
\includegraphics[width=0.99\linewidth]{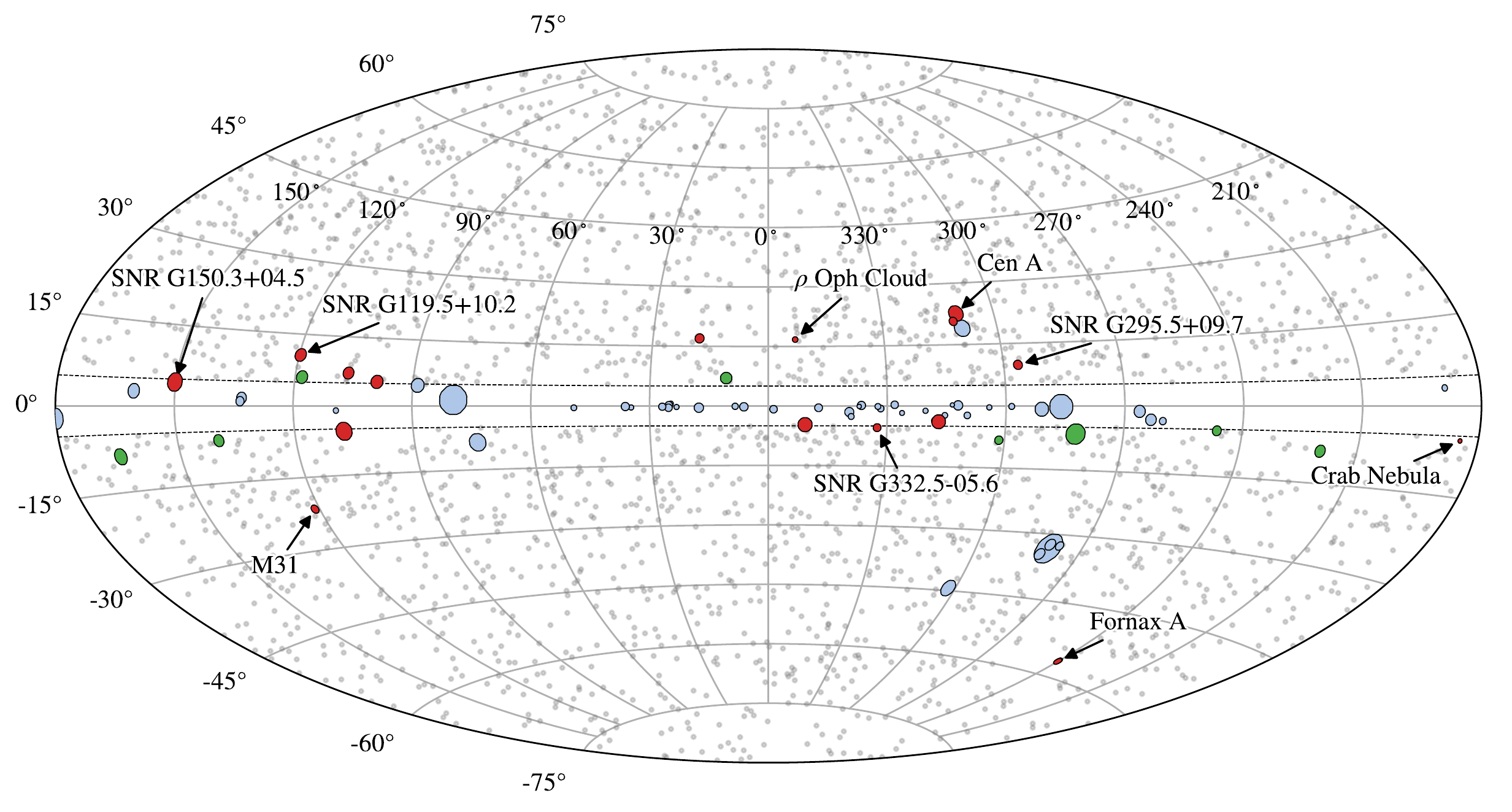}
\caption{Distribution of FHES sources in Galactic coordinates.  Light
  gray markers indicate FHES sources that are fit best by a
  point-source morphology.  Red and green circles with black outlines indicate
  the~\NUMNEWEXTSOURCES~FHES sources that are fit best by an
  extended morphology.  Green circles indicate
  the~\NUMCONFUSEDSOURCES~sources identified as confused
  based on the analysis with alternative IEMs. Two of the confused sources 
  have already been identified as spurious in previous publications~\citep{remy2017}.  
  The size of the marker
  is drawn to the scale of the intrinsic 68\% containment radius of the
  source.  Labeled sources are those with a previously published
  detection of extension or an association with a multiwavelength
  counterpart. Blue circles indicate the position and angular size of
  the 53 known LAT extended sources that fell outside our
  latitude selection or were explicitly excluded from the analysis.
  The dashed lines indicate the boundary of the latitude
  selection. \label{fig:results_allsky} }
\end{figure*}

\begin{figure*}[ht]
\centering
\includegraphics[width=0.9\columnwidth]{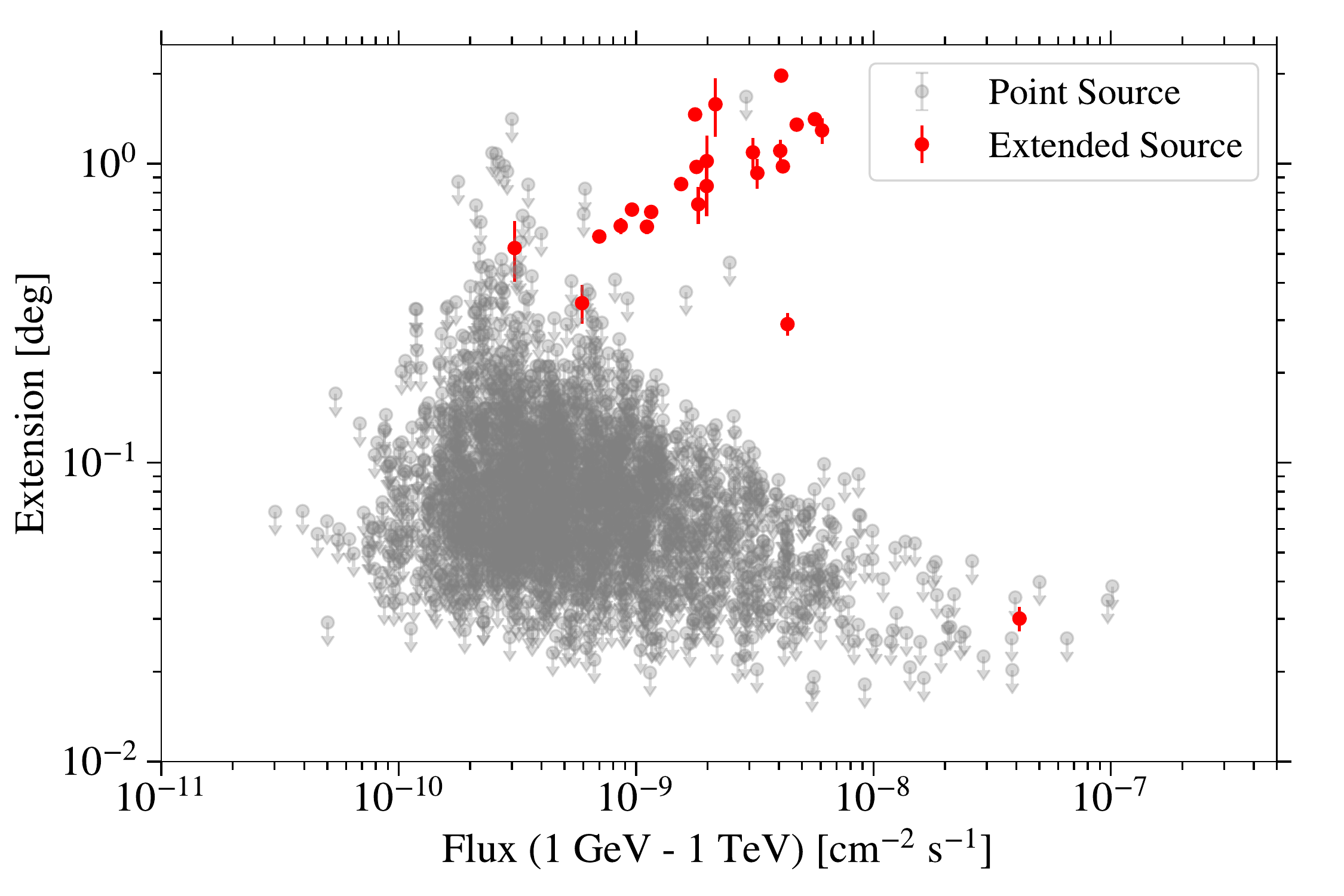}
\caption{Extension versus flux above 1 GeV for sources in the
  FHES catalog.  Gray points show the 95\% CL upper limit
  on the angular extension for point sources (TS$_{\mathrm{ext}}<$ 16).
  Red points show the best-fit value and 1$\sigma$ errors on the
  extension (68\% containment radius) for extended sources (TS$_{\mathrm{ext}}>$ 16). 
  \label{fig:results_extension} }
\end{figure*}

\Figureref{results_allsky} shows the distribution of FHES sources
in Galactic coordinates. We note that all of the new extended sources are found
at low latitudes ($4\dg \lesssim | b | \lesssim 20 \dg$),\footnote{
We find three source at latitudes $|b|$ slightly below our cut value of $5\dg$. 
The reason is that we consider seed positions $|b| > 5\dg$ but our ROIs have 
 sizes $6\dg\times6\dg$. Thus, the FHES source positions can have latitudes as low as $|b| = 2\dg$.
} implying potential
Galactic origin. We found the sources were generally associated with
either SNRs or SFRs, the two exceptions being the $\rho$ Oph Cloud,
which was originally discovered by \textit{COS~B} at \gr\ energies~\citep{1980NYASA.336..211M}
and was previously found as an
extended object in \Grs\ \citep{2012ApJ...756....5L,Abrahams:2016chw},
and the Crab Nebula, which did not pass the extension criteria
threshold in~\citet{2017arXiv170200476T}.   Fig.~\ref{fig:results_extension} shows the detected extension and extension
upper limits for all the sources investigated in this analysis.  
We see that 
the extension upper limit is generally correlated with the flux.
The outlier with the small extension and high flux is the Crab Nebula.

\begin{deluxetable*}{ll}
\tablewidth{0pt}
\tabletypesize{ \small }
\tablecaption{ \label{tab:summary} Summary of analysis seeds and FHES sources. }
\tablehead{
Category & Number
}
\startdata
\multicolumn{2}{c}{\ruleline[0.4]{Analysis Seeds}}\\
3FGL & \NUMTHREEFGLROI \\ 
3FHL & \NUMTHREEFHLROI \\ 
Total & \NUMROIS \\ 
\multicolumn{2}{c}{\ruleline[0.4]{Point Sources}}\\
3FGL Association & \NUMTHREEFGLONLYSOURCES \\ 
3FHL Association & \NUMTHREEFHLONLYSOURCES \\ 
3FGL and 3FHL Association & \NUMTHREEFXLSOURCES \\ 
Unassociated\tablenotemark{a} & \NUMUNASSOCFXLSOURCES \\ 
Total & \NUMPTSOURCES \\ 
\multicolumn{2}{c}{\ruleline[0.4]{Extended Sources}}\\
Known & \NUMKNOWNEXTSOURCES \\ 
Associated & \NUMNEWASSOCEXTSOURCES \\ 
Unassociated & \NUMUNASSOCEXTSOURCES \\ 
Confused & \NUMCONFUSEDSOURCES \\ 
Total & \NUMEXTSOURCES \\ 
\enddata
\tablenotetext{a}{FHES point sources without a 3FGL or 3FHL
    association are excluded from further analysis.}
{\footnotesize \tablecomments{ Number of unassociated extended sources excludes sources classified
as confused. }}
\end{deluxetable*}

\begin{deluxetable*}{lrrllrrrr}
\tablewidth{0pt}
\tabletypesize{ \small }
\tablecaption{ \label{tab:extended_sources} FHES extended sources. }
\tablehead{
Name                      & $l$ [$\dg$] & $b$ [$\dg$] & Association          & Class      & TS         & Model  & $\tsext$   & $\rext$ [$\dg$]     
}
\startdata
FHES~J0006.7$+$7314\tablenotemark{$\dagger$} &     119.67 &      10.65 & SNR~G119.5$+$10.2    & snr        &       38.0 & D      &       37.3 (37.3) & 0.98 $\pm$ 0.05 $\pm$ 0.04 \\ 
FHES~J0043.2$+$4109\tablenotemark{$*$} &     121.27 &     -21.68 & M31                  & gal        &       72.9 & G      &       15.5 (13.2) & 0.52 $\pm$ 0.12 $\pm$ 0.02 \\ 
FHES~J0322.2$-$3710\tablenotemark{$*$} &     240.12 &     -56.78 & Fornax~A             & rdg        &       70.5 & G      &       25.7 (24.6) & 0.342 $\pm$ 0.051 $\pm$ 0.007 \\ 
FHES~J0426.4$+$5529\tablenotemark{$*$} &     150.21 &       4.45 & SNR~G150.3$+$04.5    & snr        &      377.2 & G      &      366.2 (255.6) & 1.41 $\pm$ 0.06 $\pm$ 0.05 \\ 
FHES~J0534.5$+$2201       &     184.55 &      -5.78 & Crab~Nebula          & PWN        &     7879.4 & G      &       42.7 (11.7) & 0.030 $\pm$ 0.003 $\pm$ 0.007 \\ 
FHES~J1208.7$-$5229\tablenotemark{$*$} &     296.36 &       9.84 & SNR~G295.5$+$09.7    & snr        &       84.6 & D      &       76.9 (70.9) & 0.70 $\pm$ 0.03 $\pm$ 0.02 \\ 
FHES~J1325.3$-$3946\tablenotemark{$\dagger$} &     309.99 &      22.63 & Cen~A~Lobes          & rdg        &       38.9 & D      &       35.5 (35.5) & 1.46 $\pm$ 0.06 $\pm$ 0.27 \\ 
FHES~J1332.6$-$4130       &     311.17 &      20.70 & Cen~A~Lobes          & rdg        &       56.6 & D      &       30.0 (30.0) & 0.62 $\pm$ 0.04 $\pm$ 0.10 \\ 
FHES~J1501.0$-$6310\tablenotemark{$\dagger$} &     316.95 &      -3.89 &                      &            &      148.4 & G      &       95.9 (35.7) & 1.29 $\pm$ 0.13 $\pm$ 0.25 \\ 
FHES~J1626.9$-$2431\tablenotemark{$*$} &     353.06 &      16.73 & $\rho$~Oph~Cloud     & mc         &      411.7 & G      &       79.9 (77.6) & 0.29 $\pm$ 0.03 $\pm$ 0.01 \\ 
FHES~J1642.1$-$5428       &     332.48 &      -5.43 & SNR~G332.5$-$05.6    & snr        &       45.2 & D      &       26.4 (21.8) & 0.57 $\pm$ 0.02 $\pm$ 0.05 \\ 
FHES~J1723.5$-$0501       &      17.90 &      16.96 &                      &            &       89.5 & G      &       52.9 (47.4) & 0.73 $\pm$ 0.10 $\pm$ 0.01 \\ 
FHES~J1741.6$-$3917\tablenotemark{$\dagger$} &     350.73 &      -4.72 &                      &            &      189.1 & D      &      188.2 (137.2) & 1.35 $\pm$ 0.03 $\pm$ 0.29 \\ 
FHES~J2129.9$+$5833       &      99.13 &       5.33 &                      &            &       87.7 & G      &       49.4 (42.6) & 1.09 $\pm$ 0.13 $\pm$ 0.03 \\ 
FHES~J2208.4$+$6443       &     106.62 &       7.15 &                      &            &      136.1 & G      &       65.2 (37.0) & 0.93 $\pm$ 0.11 $\pm$ 0.11 \\ 
FHES~J2304.0$+$5406\tablenotemark{$\dagger$} &     107.50 &      -5.52 &                      &            &       46.1 & G      &       43.3 (34.1) & 1.58 $\pm$ 0.35 $\pm$ 0.17 \\ 
\multicolumn{9}{c}{\ruleline{Confused Sources}}\\
FHES~J0000.2$+$6826       &     118.24 &       6.05 & NGC~7822             & sfr        &      194.7 & D      &      149.7 (113.5) & 0.98 $\pm$ 0.04 $\pm$ 0.01 \\ 
FHES~J0242.5$+$5229\tablenotemark{$\dagger$} &     139.54 &      -6.76 &                      &            &       95.0 & G      &       26.9 (26.9) & 0.84 $\pm$ 0.18 $\pm$ 0.32 \\ 
FHES~J0430.5$+$3525\tablenotemark{$\ddagger$} &     165.28 &      -8.86 &                      &            &      153.6 & G      &      100.1 (100.1) & 1.11 $\pm$ 0.10 $\pm$ 0.09 \\ 
FHES~J0631.5$-$0940       &     219.36 &      -8.79 &                      &            &       42.3 & D      &       19.7 (12.7) & 0.86 $\pm$ 0.04 $\pm$ 0.08 \\ 
FHES~J0737.3$-$3205\tablenotemark{$\dagger$} &     246.44 &      -5.30 &                      &            &       63.6 & D      &       61.1 (61.1) & 0.69 $\pm$ 0.03 $\pm$ 0.36 \\ 
FHES~J0940.6$-$6128\tablenotemark{$\dagger$} &     282.10 &      -6.58 &                      &            &       56.8 & D      &       54.2 (8.2) & 1.97 $\pm$ 0.08 $\pm$ 0.56 \\ 
FHES~J1232.9$-$7105\tablenotemark{$\ddagger$} &     301.42 &      -8.28 &                      &            &       58.6 & D      &       25.8 (0.0) & 0.62 $\pm$ 0.03 $\pm$ 0.31 \\ 
FHES~J1743.7$-$1609\tablenotemark{$\dagger$} &      10.72 &       7.01 &                      &            &       33.8 & G      &       30.5 (15.6) & 1.02 $\pm$ 0.22 $\pm$ 0.37 \\ 
\enddata

    \tablenotetext{\dagger}{Detected via halo test (no 3FGL or 3FHL counterpart).}
    \tablenotetext{*}{Detected as extended in previous publication.}
    \tablenotetext{\ddagger}{Identified as spurious in previous publication.}

{\footnotesize \tablecomments{ The TS column gives the test statistic for detection (likelihood
    ratio of models with and without the source).  The $\tsext$ column
    gives the value of $\tsext$ obtained under the primary analysis, and in
    parentheses, the smallest value obtained under the bracketing PSF
    models or alternative IEMs.  The class column gives the class
    designator (snr - Supernova Remnant, rdg - Radio Galaxy, pwn -- Pulsar
    Wind Nebula, mc -- Molecular Cloud, sfr -- Star-Forming Region, gal --
    Galaxy).  The model column indicates the best-fit spatial model for
    each source (G -- Gaussian, D -- Disk).  Here, $\rext$ is the 68\% containment radius of the best-fit spatial model (for the disk model $\rext=0.82\rm R$ where $\rm R$ is the disk radius). The first and second errors
    on $\rext$ are statistical and systematic, respectively. }}

\end{deluxetable*}

\begin{deluxetable*}{lrrrrrrr}
\tablewidth{0pt}
\tabletypesize{ \scriptsize }
\tablecaption{ \label{tab:extended_sources_syst} Measured properties of
    FHES extended sources with their statistical and systematic errors.
     }
\tablehead{
Name                      & $l$ [$\dg$] & $b$ [$\dg$] & $\delta\theta_{\rm stat} [\dg]$ & $\delta\theta_{\rm sys} [\dg]$ & $\rext$ [\dg]        & Index                & Flux (1~GeV -- 1~TeV)\\
                          &            &            &            &            &                      &                      & [$\times 10^{-10}$ cm$^{-2}$ s$^{-1}$]
}
\startdata
FHES~J0006.7$+$7314\tablenotemark{$\dagger$} &     119.67 &      10.65 &       0.13 &       0.18 & 0.98 $\pm$ 0.05 $\pm$ 0.04 & 2.24 $\pm$ 0.16 $\pm$ 0.02 &       18.0 $\pm$        3.5 $\pm$        3.4 \\ 
FHES~J0043.2$+$4109\tablenotemark{$*$} &     121.27 &     -21.68 &       0.12 &       0.03 & 0.52 $\pm$ 0.12 $\pm$ 0.02 & 2.66 $\pm$ 0.21 $\pm$ 0.01 &        7.6 $\pm$        1.0 $\pm$        0.3 \\ 
FHES~J0322.2$-$3710\tablenotemark{$*$} &    240.117 &    -56.784 &      0.078 &      0.003 & 0.342 $\pm$ 0.051 $\pm$ 0.007 & 2.16 $\pm$ 0.13 $\pm$ 0.00 &        5.9 $\pm$        1.0 $\pm$        0.1 \\ 
FHES~J0426.4$+$5529\tablenotemark{$*$} &     150.21 &       4.45 &       0.10 &       0.25 & 1.41 $\pm$ 0.06 $\pm$ 0.05 & 1.81 $\pm$ 0.04 $\pm$ 0.12 &       56.7 $\pm$        4.3 $\pm$       21.1 \\ 
FHES~J0534.5$+$2201       &    184.552 &     -5.781 &      0.002 &      0.000 & 0.030 $\pm$ 0.003 $\pm$ 0.007 & 1.79 $\pm$ 0.04 $\pm$ 0.00 &      412.3 $\pm$        8.7 $\pm$        1.4 \\ 
FHES~J1208.7$-$5229\tablenotemark{$*$} &     296.36 &       9.84 &       0.06 &       0.06 & 0.70 $\pm$ 0.03 $\pm$ 0.02 & 1.81 $\pm$ 0.09 $\pm$ 0.05 &        9.6 $\pm$        1.6 $\pm$        1.3 \\ 
FHES~J1325.3$-$3946\tablenotemark{$\dagger$} &     309.99 &      22.63 &       0.16 &       0.67 & 1.46 $\pm$ 0.06 $\pm$ 0.27 & 2.22 $\pm$ 0.14 $\pm$ 0.08 &       17.7 $\pm$        3.0 $\pm$        6.5 \\ 
FHES~J1332.6$-$4130       &     311.17 &      20.70 &       0.10 &       0.10 & 0.62 $\pm$ 0.04 $\pm$ 0.10 & 2.08 $\pm$ 0.12 $\pm$ 0.04 &        8.6 $\pm$        1.3 $\pm$        2.5 \\ 
FHES~J1501.0$-$6310\tablenotemark{$\dagger$} &     316.95 &      -3.89 &       0.15 &       0.33 & 1.29 $\pm$ 0.13 $\pm$ 0.25 & 2.44 $\pm$ 0.09 $\pm$ 0.07 &       60.7 $\pm$        5.2 $\pm$       10.6 \\ 
FHES~J1626.9$-$2431\tablenotemark{$*$} &     353.06 &      16.73 &       0.03 &       0.05 & 0.29 $\pm$ 0.03 $\pm$ 0.01 & 2.55 $\pm$ 0.07 $\pm$ 0.03 &       43.4 $\pm$        2.6 $\pm$        4.9 \\ 
FHES~J1642.1$-$5428       &     332.48 &      -5.43 &       0.06 &       0.10 & 0.57 $\pm$ 0.02 $\pm$ 0.05 & 1.78 $\pm$ 0.12 $\pm$ 0.08 &        7.0 $\pm$        1.9 $\pm$        2.3 \\ 
FHES~J1723.5$-$0501       &      17.90 &      16.96 &       0.13 &       0.15 & 0.73 $\pm$ 0.10 $\pm$ 0.01 & 1.97 $\pm$ 0.08 $\pm$ 0.06 &       18.3 $\pm$        2.5 $\pm$        2.1 \\ 
FHES~J1741.6$-$3917\tablenotemark{$\dagger$} &     350.73 &      -4.72 &       0.07 &       0.26 & 1.35 $\pm$ 0.03 $\pm$ 0.29 & 1.80 $\pm$ 0.04 $\pm$ 0.06 &       47.5 $\pm$        4.6 $\pm$       17.3 \\ 
FHES~J2129.9$+$5833       &      99.13 &       5.33 &       0.15 &       0.43 & 1.09 $\pm$ 0.13 $\pm$ 0.03 & 2.30 $\pm$ 0.12 $\pm$ 0.04 &       31.1 $\pm$        3.8 $\pm$        2.3 \\ 
FHES~J2208.4$+$6443       &     106.62 &       7.15 &       0.12 &       0.13 & 0.93 $\pm$ 0.11 $\pm$ 0.11 & 2.78 $\pm$ 0.14 $\pm$ 0.15 &       32.4 $\pm$        2.9 $\pm$        9.9 \\ 
FHES~J2304.0$+$5406\tablenotemark{$\dagger$} &     107.50 &      -5.52 &       0.29 &       0.12 & 1.58 $\pm$ 0.35 $\pm$ 0.17 & 1.95 $\pm$ 0.08 $\pm$ 0.15 &       21.6 $\pm$        3.7 $\pm$        7.8 \\ 
\multicolumn{8}{c}{\ruleline{Confused Sources}}\\
FHES~J0000.2$+$6826       &     118.24 &       6.05 &       0.09 &       0.22 & 0.98 $\pm$ 0.04 $\pm$ 0.01 & 2.72 $\pm$ 0.11 $\pm$ 0.07 &       41.5 $\pm$        3.1 $\pm$        3.4 \\ 
FHES~J0242.5$+$5229\tablenotemark{$\dagger$} &     139.54 &      -6.76 &       0.14 &       0.20 & 0.84 $\pm$ 0.18 $\pm$ 0.32 & 2.59 $\pm$ 0.17 $\pm$ 0.29 &       19.8 $\pm$        2.2 $\pm$       20.9 \\ 
FHES~J0430.5$+$3525\tablenotemark{$\ddagger$} &     165.28 &      -8.86 &       0.13 &       0.12 & 1.11 $\pm$ 0.10 $\pm$ 0.09 & 2.59 $\pm$ 0.11 $\pm$ 0.05 &       40.5 $\pm$        3.4 $\pm$        4.3 \\ 
FHES~J0631.5$-$0940       &     219.36 &      -8.79 &       0.11 &       0.57 & 0.86 $\pm$ 0.04 $\pm$ 0.08 & 2.21 $\pm$ 0.12 $\pm$ 0.10 &       15.5 $\pm$        2.6 $\pm$        1.4 \\ 
FHES~J0737.3$-$3205\tablenotemark{$\dagger$} &     246.44 &      -5.30 &       0.07 &       0.03 & 0.69 $\pm$ 0.03 $\pm$ 0.36 & 1.85 $\pm$ 0.08 $\pm$ 0.07 &       11.6 $\pm$        2.1 $\pm$        3.6 \\ 
FHES~J0940.6$-$6128\tablenotemark{$\dagger$} &     282.10 &      -6.58 &       0.15 &       1.07 & 1.97 $\pm$ 0.08 $\pm$ 0.56 & 2.45 $\pm$ 0.11 $\pm$ 0.35 &       40.9 $\pm$        5.5 $\pm$       28.0 \\ 
FHES~J1232.9$-$7105\tablenotemark{$\ddagger$} &     301.42 &      -8.28 &       0.09 &       0.74 & 0.62 $\pm$ 0.03 $\pm$ 0.31 & 2.31 $\pm$ 0.14 $\pm$ 0.57 &       11.1 $\pm$        1.6 $\pm$       10.4 \\ 
FHES~J1743.7$-$1609\tablenotemark{$\dagger$} &      10.72 &       7.01 &       0.24 &       0.32 & 1.02 $\pm$ 0.22 $\pm$ 0.37 & 2.07 $\pm$ 0.11 $\pm$ 0.13 &       19.8 $\pm$        3.8 $\pm$       24.4 \\ 
\enddata

    \tablenotetext{\dagger}{Detected via halo test (no 3FGL or 3FHL counterpart).}
    \tablenotetext{*}{Detected as extended in previous publication.}
    \tablenotetext{\ddagger}{Identified as spurious in previous publication.}

{\footnotesize \tablecomments{ Here, $\delta\theta_{\rm stat}$ and $\delta\theta_{\rm sys}$ are the
    statistical and systematic 68\% positional uncertainties.  The first
    and second errors on $\rext$, Index, and Flux are statistical and
    systematic.  The systematic error is the larger of the IRF and IEM
    systematics.  No systematic errors are given for the Crab Nebula
    position because no measurable change in the best-fit position was
    observed for either the bracketing PSF models or alternative IEMs.
    We define $\rext$ as the 68\% containment radius of the best-fit spatial model
    (for the disk model $\rext=0.82R$ where $R$ is the disk radius).  The
    Index column gives the spectral index for sources parameterized with a
    PL spectrum and the spectral slope at 1~GeV for sources parameterized
    with an LP or PLE spectrum. }}

\end{deluxetable*}

\subsection{Known extended sources}
\label{subsec:knownsrcs}

The {\NUMKNOWNEXTSOURCES} sources in our analysis that are 3FGL point sources but have already observed extensions are: 
\begin{itemize}
\item \fhesfornax: Fornax A  \citep{Ackermann:2016arn},  
\item \fhesmthirtyone: M31 \citep{Ackermann:2017nya}, 
\item \fhesroph: $\rho$ Oph cloud  \citep{2012ApJ...756....5L,Abrahams:2016chw}, 
\item \fhessnri: SNR~G150.3$+$04.5  \citep{Acero:2015prw,Ackermann:2015uya}, 
\item \fhessnrii: SNR~G295.5$+$09.7  \citep{Acero:2015prw}.  
\end{itemize}

\noindent These sources are included in Table~\ref{tab:extended_sources} and their spectral and spatial 
properties are in agreement with the published results.

\subsection{Individual Sources of Interest}\label{sec:IndivSources}

Previously unidentified extended sources are discussed in further
detail in Sections \ref{subsec:indvsrc_cta1}-\ref{subsec:softsources}.
These extended objects often encompass multiple 3FGL sources.  We
performed searches in archival radio, infrared, optical, UV, and X-ray
data to look for potential associations.  These surveys were accessed
using SkyView.\footnote{\url{https://skyview.gsfc.nasa.gov}} Data
include the IR band from the Digital Sky Survey (DSS and DSS2); both
the Low Frequency Instrument (LFI) on the \textit{Planck} satellite at 30 GHz,
44 GHz, and 70 GHz, and the High Frequency Instrument (HFI) at 353 GHz;
the K and K$_\mathrm{a}$ frequencies (23 and 33 GHz respectively) on the
Wilkinson Microwave Anisotropy Probe (\textit{WMAP}); the Sydney University
Molonglo Sky Survey (SUMMS) at a frequency of 843 MHz; and finally, the
Westerbork Northern Sky Survey (WENSS) at a frequency of 325 MHz.
We looked for potential associations to known sources in the
  TeV energy band with TeVCat.\footnote{\url{http://tevcat.uchicago.edu/}}

For sources that we suspect to be associated with cosmic-ray interactions
with the Interstellar Medium (ISM), we perform comparisons with maps of dust optical depth
at 353~GHz ($\tau_{353}$) from \textit{Planck} Public Data Release 1
\citep{2014A&A...571A..11P}. 
Thermal dust emission has been shown to be correlated with components of the ISM, and the \textit{Planck} $\tau_{353}$ map provides much better information than the ISM tracers used for the official \emph{Fermi} IEM \citep{2016ApJS..223...26A}.

In this search for counterparts of the {\NUMNEWEXTSOURCES} sources previously not known to be 
extended, we found {\NUMNEWASSOCEXTSOURCES} sources with potential associations:
two sources in regions of SNRs that were previously undetected by the LAT (Sections~\ref{subsec:indvsrc_cta1} and~\ref{subsec:indvsrc_snrg332}); two 
sources near the Cen A Lobes which extend beyond the current model based on \textit{WMAP} data (Section~\ref{subsec:indvsrc_cenalobes}); 
and one in the direction of the Crab Nebula (Section~\ref{subsec:indvsrc_crab}), which 
is the only source with an extension comparable to the systematic uncertainty on the IRFs. 
Three of the more tentative associations are found in SFRs, and are
discussed in Section~\ref{subsec:indvsrc_sfr}.

Three unassociated extended sources have a spectral and spatial
morphology that is consistent with SNRs or PWNe.  These are further
discussed in Section~\ref{subsec:snrysources}.  The sources
  with soft spectra consistent with the Galactic diffuse emission are
discussed in Section~\ref{subsec:softsources}.  With the
  exception of the possible SFR source \fhessoftiii,
  sources identified as confused are not discussed further.

\subsubsection{CTA 1: SNR~G119.5$+$10.2 (\fhessnrcta)}
\label{subsec:indvsrc_cta1}

The SNR CTA 1 is located about 1400~pc away in the constellation of Cepheus, 
and has an estimated age of 1.3$\times$10$^4$ years~\citep{Slane:2003du}. 
The pulsar, PSR J0007+7303, located within the SNR CTA~1, is the first \gr\
only pulsar discovered with the \Fermi-LAT~\citep{Abdo:2008nz}. The associated PWN has been detected at very-high \gr\ energies 
with VERITAS~\citep{2013ApJ...764...38A}. 
In the first two years of LAT observations, extended emission that could have been related to the PWN was detected at the $\sim$2$\sigma$ level.
 A subsequent LAT analysis of
PSR~J0007+7303 with over seven years of \irf{Pass 8} data found no
evidence for extended \gr\ emission
over the 0.3\dg region encompassing the TeV source VER~J0006$+$729
\citep{2016ApJ...831...19L}.  

We perform this analysis in the off-pulse of the pulsar \gr\
emission using an eight-year \gr\ ephemeris and the phase
interval $\phi \in [0.55, 1.05]$.  We include a point-source component
at the location of PSR~J0007+7303 (TS=153) to model the off-peak
emission from the pulsar.  The best-fit model also includes a new
point source to the west of the PSR location. 
\citet{2016ApJ...831...19L} identified this object as a variable
source and found a probable association with the quasar S5~0016$+$73.

\begin{figure}[!ht]
\centering
\includegraphics[width=0.48\columnwidth]{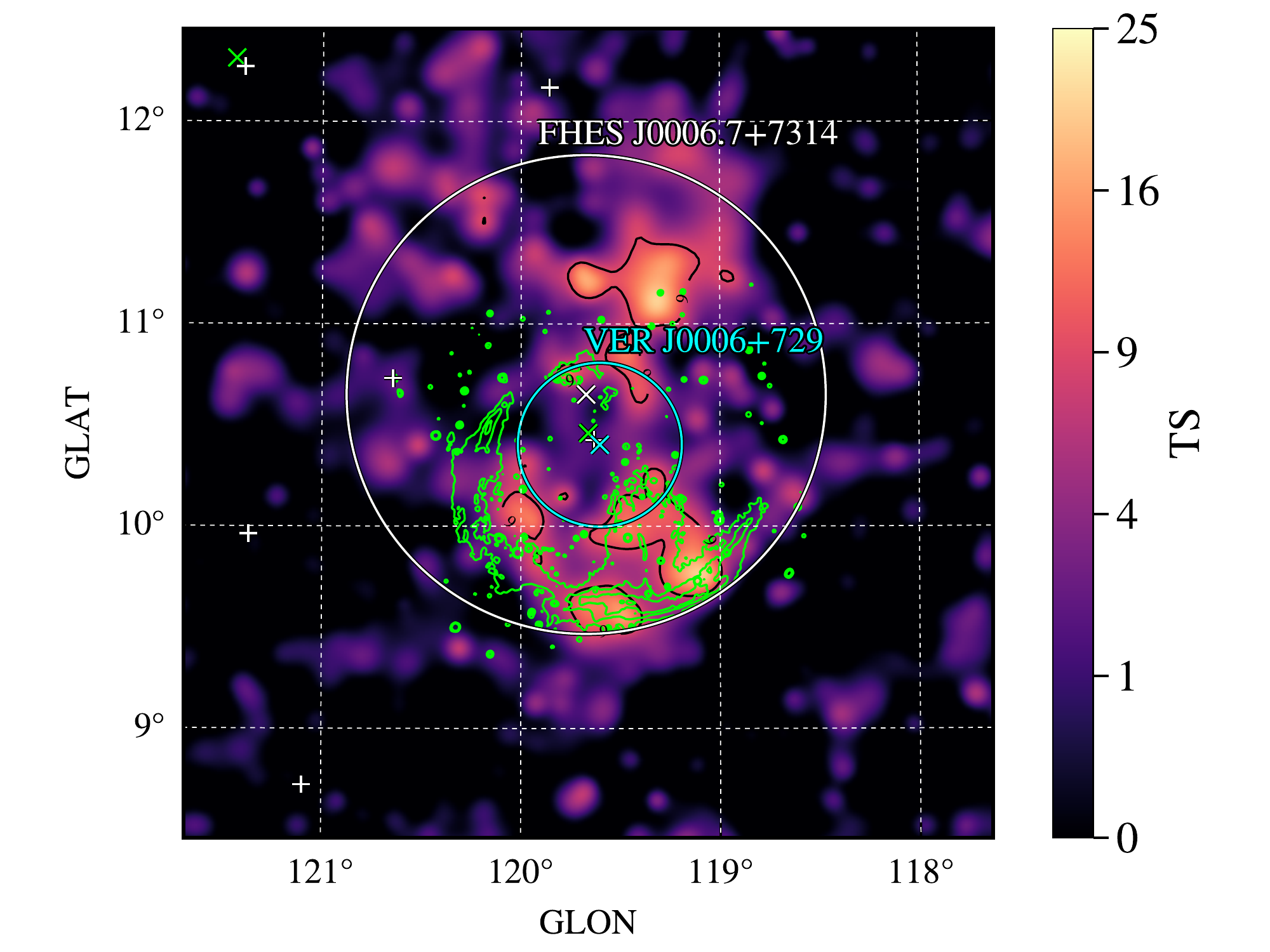}
\includegraphics[width=0.48\columnwidth]{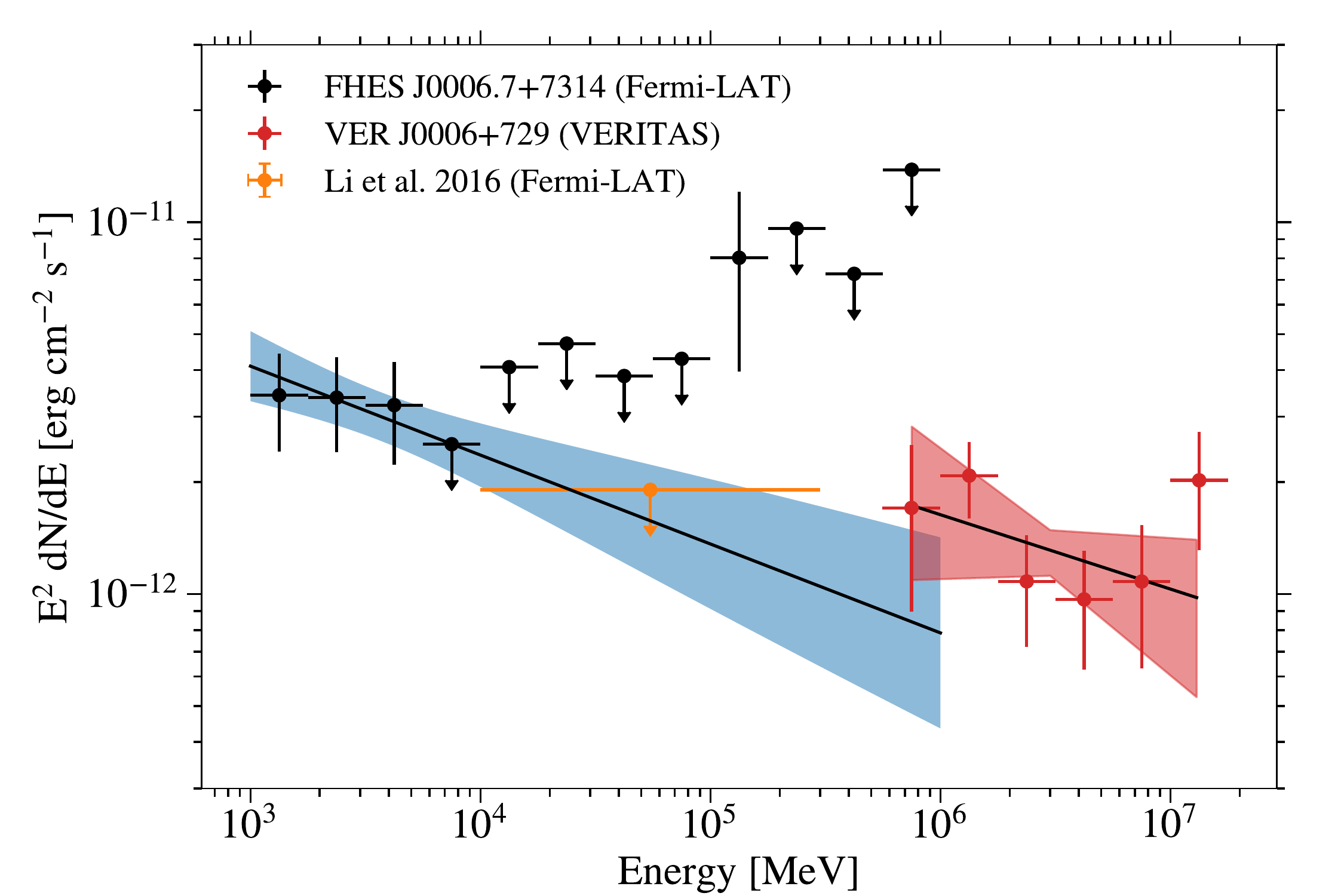}
\caption{\textit{Left:} TS map of {\fhessnrcta} which is associated
  with the CTA~1 SNR (SNR~G119.5$+$10.2).  The white circle with
  central marker $\times$ indicates the best-fit disk extension
  and centroid of the FHES source.  White crosses indicate the positions
  of point-source candidates with TS $> 9$ from the best-fit model for
  the region.  Green crosses indicate the positions of sources in the
  3FGL catalog. Green contours show the map of radio continuum
  emission from the CTA~1 SNR measured at 1420~MHz
  \citep{1997A&A...324.1152P}.  The cyan circle and cross indicate the
  angular extent (68\% containment) and centroid of the TeV source
  VER~J0006$+$729 \citep{2013ApJ...764...38A}.  \textit{Right:}
  spectral energy distributions of {\fhessnrcta} from this analysis
  and the VERITAS spectrum of VER~J0006$+$729.  Upper
    limit points for {\fhessnrcta} are computed at 95\% C.L.  The orange
  marker shows the 99\% upper limit from \citet{2016ApJ...831...19L}
  on the energy flux between 10~GeV and 300~GeV measured within the
  0.3\dg angular extent of
  VER~J0006$+$729.  \label{fig:tsmap0}}
\end{figure}

We find evidence for an extended \gr\ source {\fhessnrcta} that is
correlated with the radio emission at
1420\,MHz~\citep{1997A&A...324.1152P}, 
which is evident from the TS map (Figure~\ref{fig:tsmap0}, left), where the source is shown overlaid
with the radio emission contours from the CTA~1 SNR.  
The map is generated with the central source from 
the ROI removed and a point source added instead at each pixel 
(modeled with a power law with index $\Gamma = 2$). 
The extension is fit best by a disk with
$\rext=\fhessnrctarext\dg\pm\fhessnrctarexterr\dg\pm\fhessnrctarexterrsys\dg$. 
The \gr\ emission is somewhat larger in angular extent than the radio shell
($D\sim1.5\dg$), with a suggestion of elongation beyond the northern edge of
the shell.  The TeV \gr\ emission is located farther north, inside the incomplete radio shell,
and is also shown in the figure.  
There is an obvious difference in angular size between the TeV and GeV \gr\ emission. 
A morphology similar to the GeV emission is seen in \textit{ROSAT} PSPC X-ray
images of the region \citep{1995ApJ...453..284S,1997ApJ...485..221S}.
In the right panel of Figure~\ref{fig:tsmap0}, we compare the \Fermi-LAT spectrum of \fhessnrcta~to the one of the VERITAS source, VER\,J0006+729.
There is evidence for mismatch in the flux normalization observed between the two spectra,
even when taking into account the difference in angular size. This could indicate a 
spectral break at higher energies, or the observation of two separate sources. 
However, the spectral indices agree well with each
other.

\subsubsection{SNR~G332.5$-$05.6 (\fhessnriii)}
\label{subsec:indvsrc_snrg332}

\begin{figure}[ht]
\centering
\includegraphics[width=0.48\columnwidth]{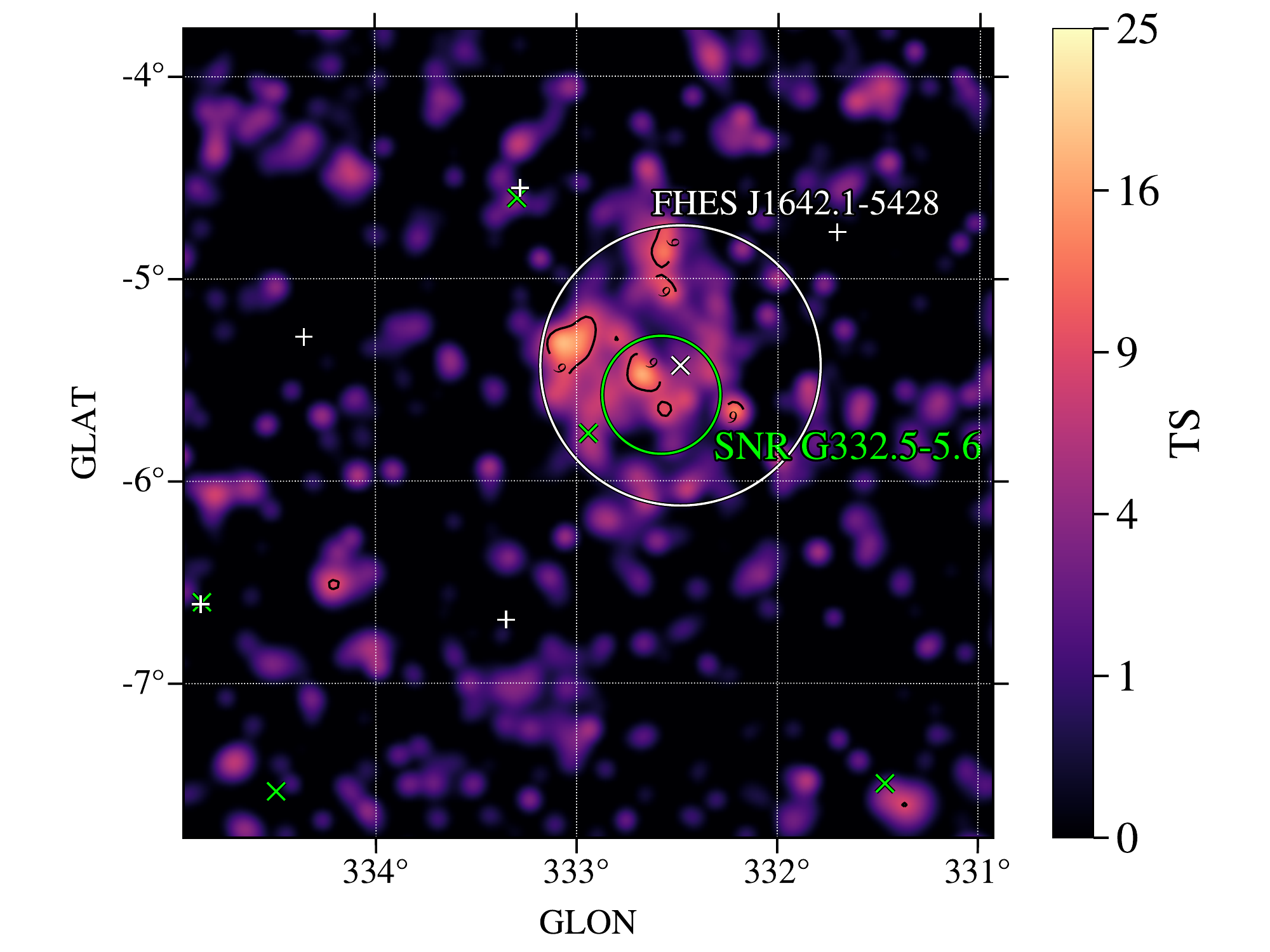}
\caption{ TS map of {\fhessnriii}, which is associated with
  SNR~G332.5$-$05.6.  The white circle with central marker
  $\times$ indicates the best-fit disk extension and centroid of
  the FHES source.  White crosses indicate the positions of point-source
  candidates with TS $> 9$ from the best-fit model for the region.
  The green circle indicates the angular extent of the radio
    SNR from \citet{Reynoso:2006zn}. Green crosses indicate the
  positions of sources in the 3FGL catalog. \label{fig:tsmapsnriii}}
\end{figure}

SNR~G332.5-5.6, located in the constellation Norma, is between 7000-9000 years old and is $\sim$3.4~kpc
away \citep{Reynoso:2006zn}.  It has been detected in radio and in
X-ray wavelengths as an extended object by
$XMM Newton$~\citep{Suarez:2015hga}, and $Suzaku$~\citep{Zhu:2015fwa}, as well
as by $ATCA$ and $ROSAT$~\citep{Reynoso:2006zn}.  It was not detected in the
first LAT SNR Catalog~\citep{Acero:2015prw}; however, in the 3FGL
(3FGL~J1645.9$-$5420) it was classified as having a potential
association with a SNR or PWN. 
X-ray observations show strong X-ray emission from the center of the
remnant, which has similar morphology to that of the central radio
emission.  No radio, X-ray, or a \gr\ pulsars have been found
in the vicinity of SNR~G332.5-5.6.  Fig.~\ref{fig:tsmapsnriii} shows
the TS map of the extended \gr\ emission in the region.  We find the
disk radius of {\fhessnriii} to be
$\fhessnriiirext\dg\pm\fhessnriiirexterr\dg\pm\fhessnriiirexterrsys\dg$,
with a spectral index
$\Gamma=\fhessnriiiindex\pm\fhessnriiiindexerr\pm\fhessnriiiindexerrsys$,
making it one of the hardest sources in the catalog.

\subsubsection{Cen A Lobes ({\fhescenai} and {\fhescenaii})}
\label{subsec:indvsrc_cenalobes}

Cen A is one of the brightest radio sources in the sky. It was 
first identified as a \gr\ source by $COS~B$~\citep{1981ApJ...243L..69S}, and later by $OSSE$~\citep{1995ApJ...449..105K} and $EGRET$~\citep{1995ApJS..101..259T}.
It was also one of the first \gr\ sources to be identified
with a galaxy (NGC 5128) outside of our Milky Way~\citep{Israel1998}. 
Extending from the bright central source is a pair of radio lobes with a total angular extent of $\sim$10$^{\circ}$, which makes Cen A
the largest non-thermal extragalactic radio source visible from the Earth. 
At a distance of 3.7 Mpc, it is also the closest radio-loud galaxy. The radio lobes are approximately 600\,kpc across. 
Extended \gr\ emission, coming from the lobes as well as the radio core, has been detected at \gr\ energies with the LAT~\citep{2010Sci...328..725A}.
Very high energy \gr\ emission has been observed with \hess\, which is only consistent with the core and inner jets~\citep{2009ApJ...695L..40A}. 
The LAT \gr\ emission from the lobes is consistent with the morphology found with \textit{WMAP} as well as  
the 30 GHz \textit{Planck} data~\citep{Sun:2016ibh}. 

\begin{figure}[ht]
\centering
\includegraphics[width=0.49\columnwidth]{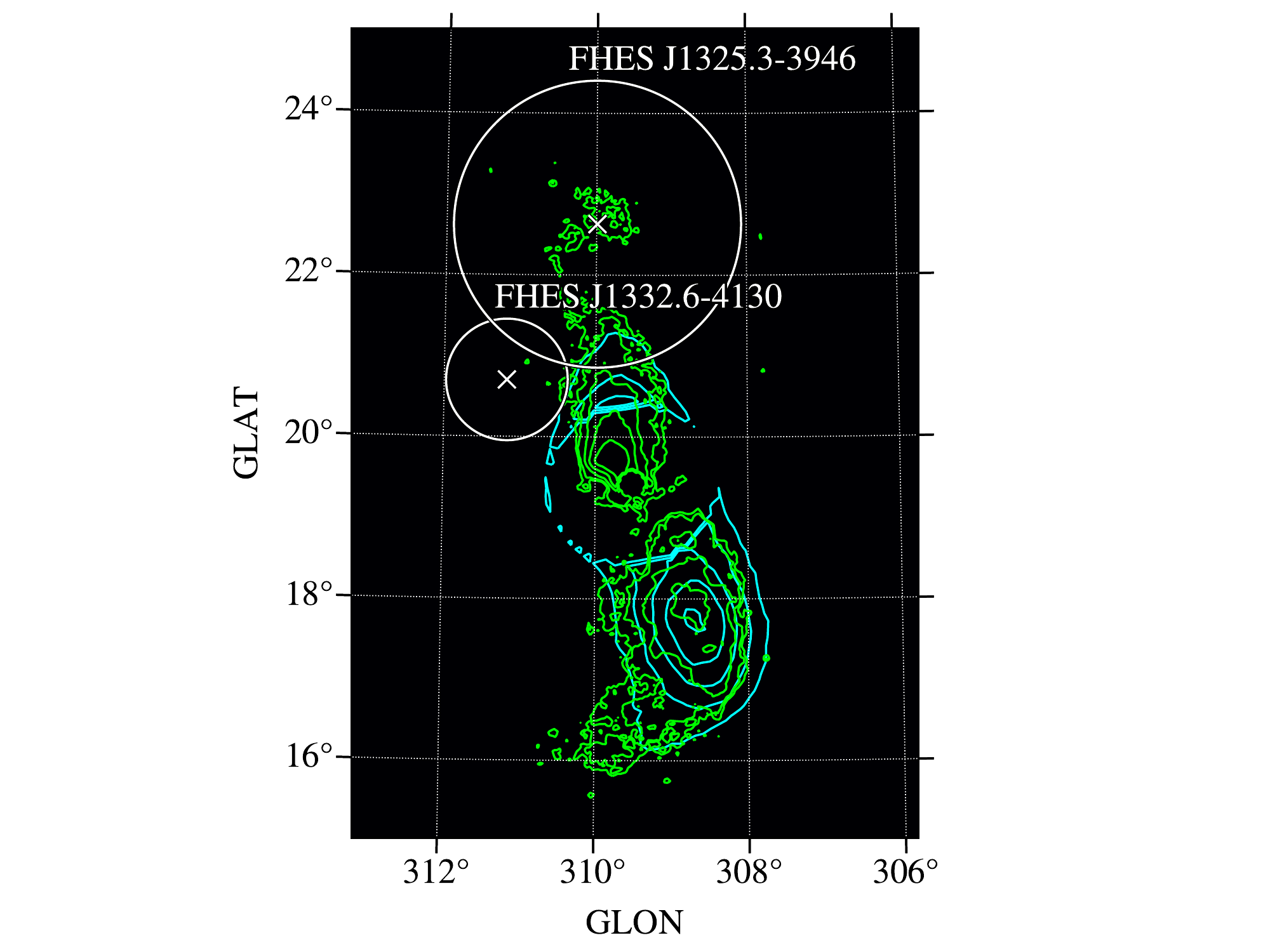}
\caption{ Map of the Cen~A region, showing contours for the LAT
  \gr\ Cen A Lobes template (cyan) and Parkes radio continuum
  map at 5~GHz (green).  The white circles with central marker $\times$
  indicate the best-fit disk radius and centroid of the two FHES
  sources associated with Cen A: {\fhescenai} and
  {\fhescenaii}.\label{fig:cena}}
\end{figure}

In addition to the $\gamma$-ray emission, which follows the 3FHL
template of the lobes based on \textit{WMAP}, there appear to be additional
extended $\gamma$-ray components beyond the edge of the northern Cen~A
Lobe.  \Figureref{cena} shows a map of the Cen~A region with the
position and extension of the two FHES sources overlaid.  We note that
the analysis of these two sources was performed independently and the
background models do not include the neighboring FHES extended source.  
However, the optimization procedure partially compensates for excess emission
outside the search region via the inclusion of point-source
components.  Given that the best-fit disk models of these two sources
partially overlap, it is likely that these two sources belong to a
single diffuse emission component associated with Cen~A.

\begin{figure}[ht]
\centering
\includegraphics[width=0.49\columnwidth]{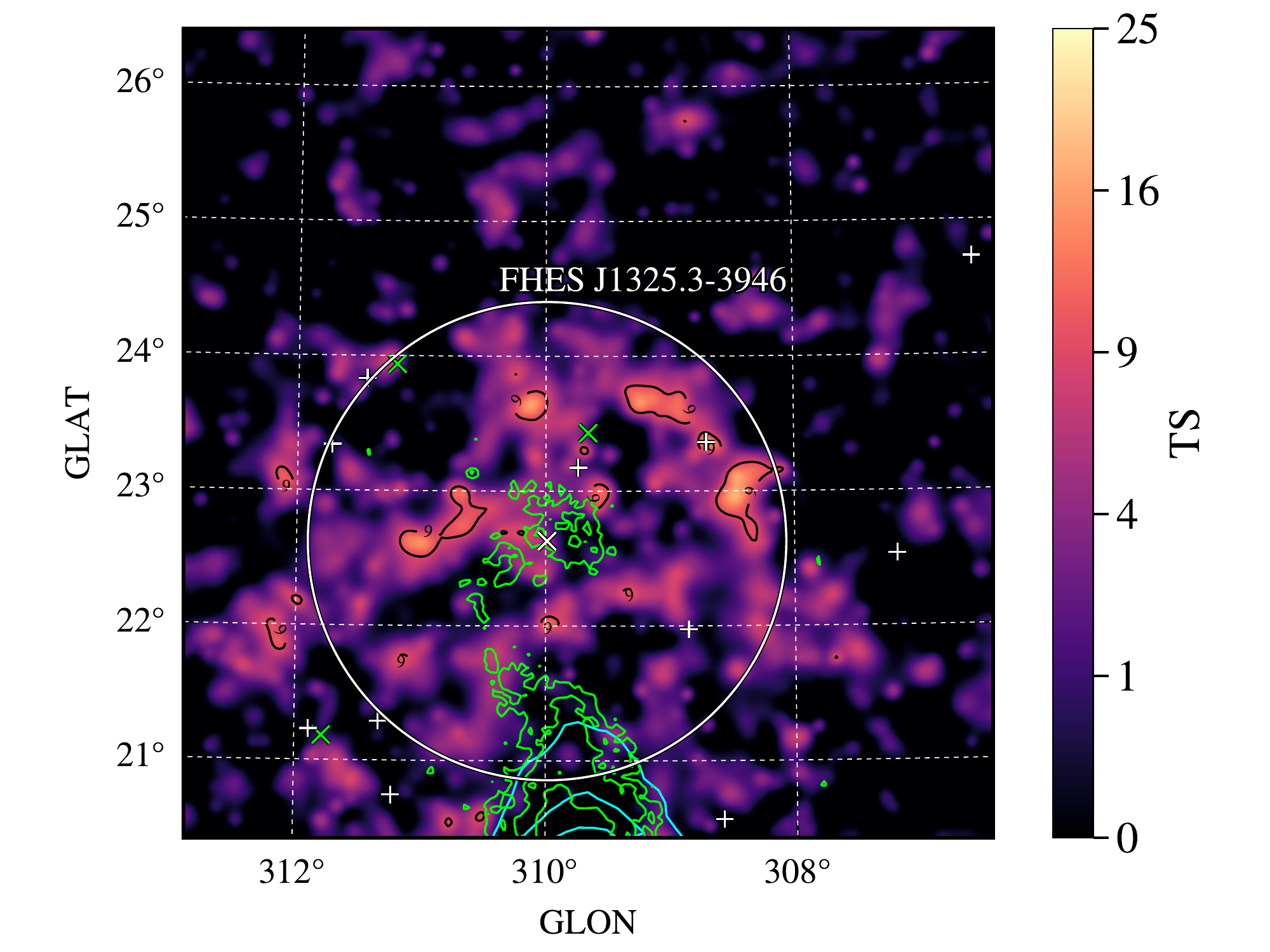}
\includegraphics[width=0.49\columnwidth]{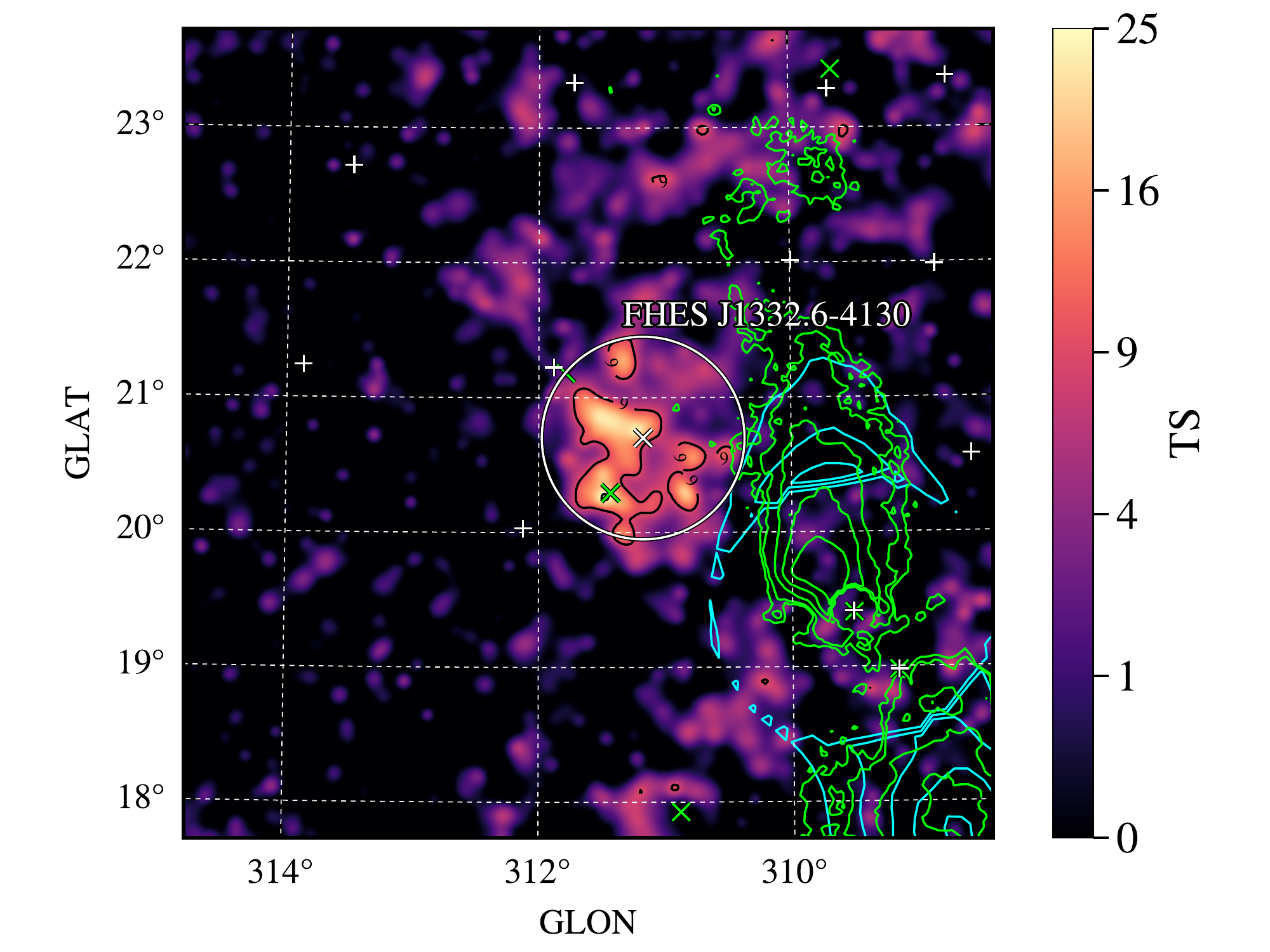}
\caption{ TS maps of {\fhescenai} and {\fhescenaii}, which are
  associated with the Cen A lobes.  The white circle with central
  marker $\times$ indicates the best-fit disk extension and centroid of the
  FHES source.  White crosses indicate the positions of point-source
  candidates with TS $> 9$ from the best-fit model for the region.
  Green crosses indicate the positions of sources in the 3FGL catalog.
  Overlaid are contours of the LAT \gr\ Cen A Lobes template
  (cyan) and Parkes radio continuum map at 5~GHz
  (green).\label{fig:cena_tsmap}}
\end{figure}

\Figureref{cena_tsmap} shows the individual TS maps for the
two sources with the two distinct regions around the north lobe: one
directly north (\fhescenai) and one west (\fhescenaii).  We find the
extension of the northern (western) source to be
$\fhescenairext\dg\pm\fhescenairexterr\dg\pm\fhescenairexterrsys\dg$
($\fhescenaiirext\dg\pm\fhescenaiirexterr\dg\pm\fhescenaiirexterrsys\dg$)
and the spectral index to be
$\Gamma=\fhescenaiindex\pm\fhescenaiindexerr\pm\fhescenaiindexerrsys$
($\Gamma=\fhescenaiiindex\pm\fhescenaiiindexerr\pm\fhescenaiiindexerrsys$).
These sources, the western one in particular,  are harder than both the north and south lobes, which have spectral indices of $\Gamma=2.52^{+0.16}_{-0.19}$ and
$\Gamma=2.60^{+0.14}_{-0.15}$, respectively.  The origin of this
emission beyond the edge of the radio contours is unclear so far.

\subsubsection{Crab Nebula (\fhescrab)}
\label{subsec:indvsrc_crab}

The Crab Nebula is a PWN associated with the young pulsar
PSR~J0534$+$2200, which is the compact remnant of a supernova explosion that
occurred in the year 1054\,AD, at a distance of
$\sim2\,$kpc~\citep[see, e.g.][for a review]{2008ARA&A..46..127H}.  In
the 3FGL, the \gr\ emission from the Crab Nebula was decomposed
into three components: an Inverse Compton component (IC;
3FGL~J0534.5$+$2201i), a synchrotron component (3FGL~J0534.5$+$2201s),
and the Crab pulsar (3FGL~J0534.5$+$2201).  The point-like emission of
the Crab Pulsar dominates the nebula at energies below 10~GeV, while
the IC component dominates above 10~GeV.  Due to the strong degeneracy
between the IC and pulsar components, it is not possible to obtain a
stable fit to both components simultaneously.  To constrain the
contribution of the Crab pulsar, we perform an independent phased
analysis of the region using a joint fit to on-
($\phi \in [0.0, 0.68]$) and off-pulse ($\phi \in [0.68, 1.0]$)
selections in which we set the amplitude of the pulsar to zero in the
off-pulse interval.  With this analysis, we obtain a best-fit PLE
parameterization for the on-pulse pulsar emission with
$N_{0} = 6.06\times10^{-9}~\rm{cm}^{-2}~\rm{s}^{-1}\rm{MeV}^{-1}$ at 0.635~GeV,
$\Gamma = 2.24$, and $E_{c} = 15.4~\rm{GeV}$.  When fitting the extension of the Crab
Nebula, we fix the spectral model of the pulsar to the one obtained
from the phased analysis and remove the synchrotron component from the
model.

\begin{figure*}[ht]
\includegraphics[width=0.48\columnwidth]{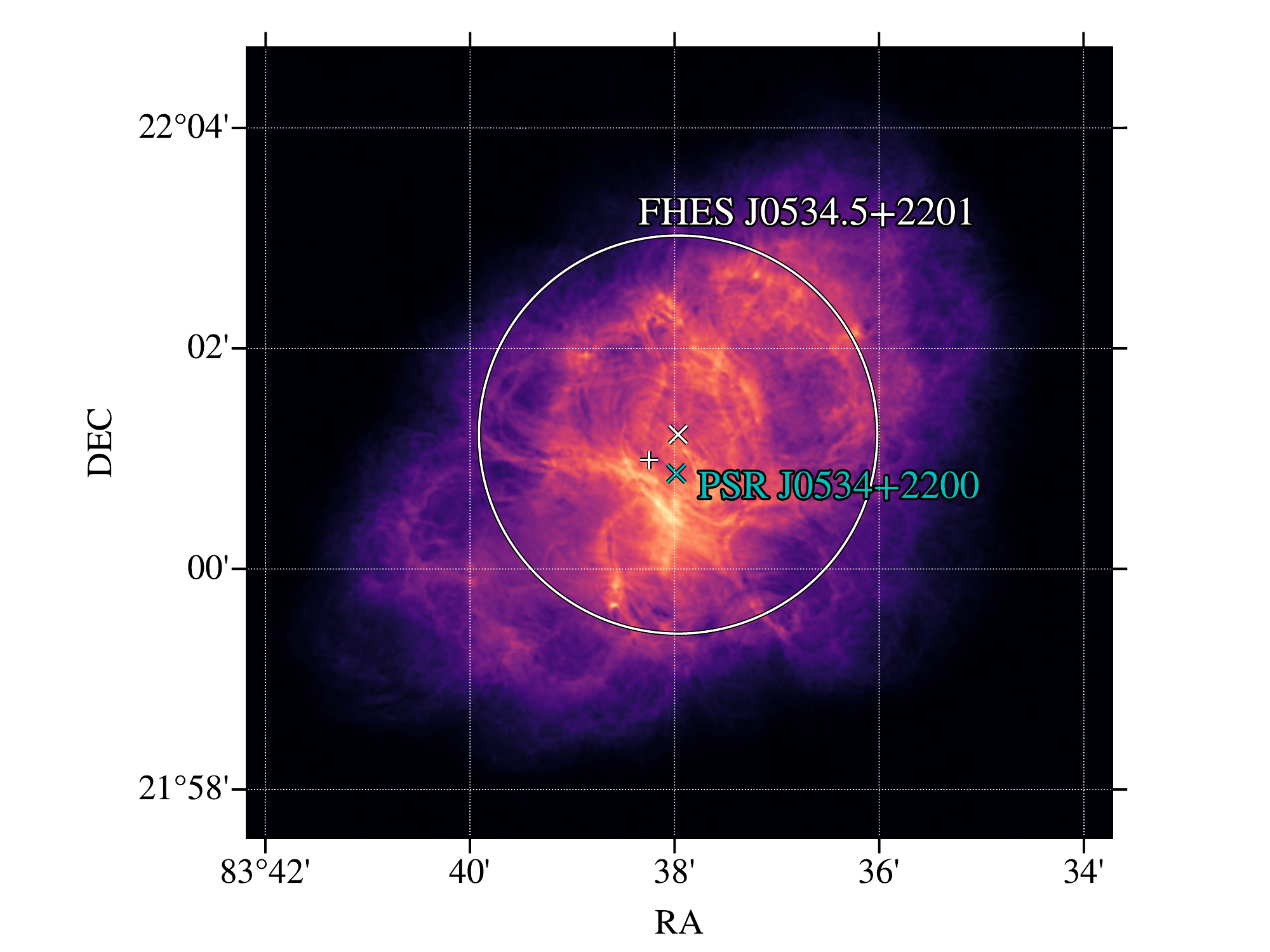}
\includegraphics[width=0.48\columnwidth]{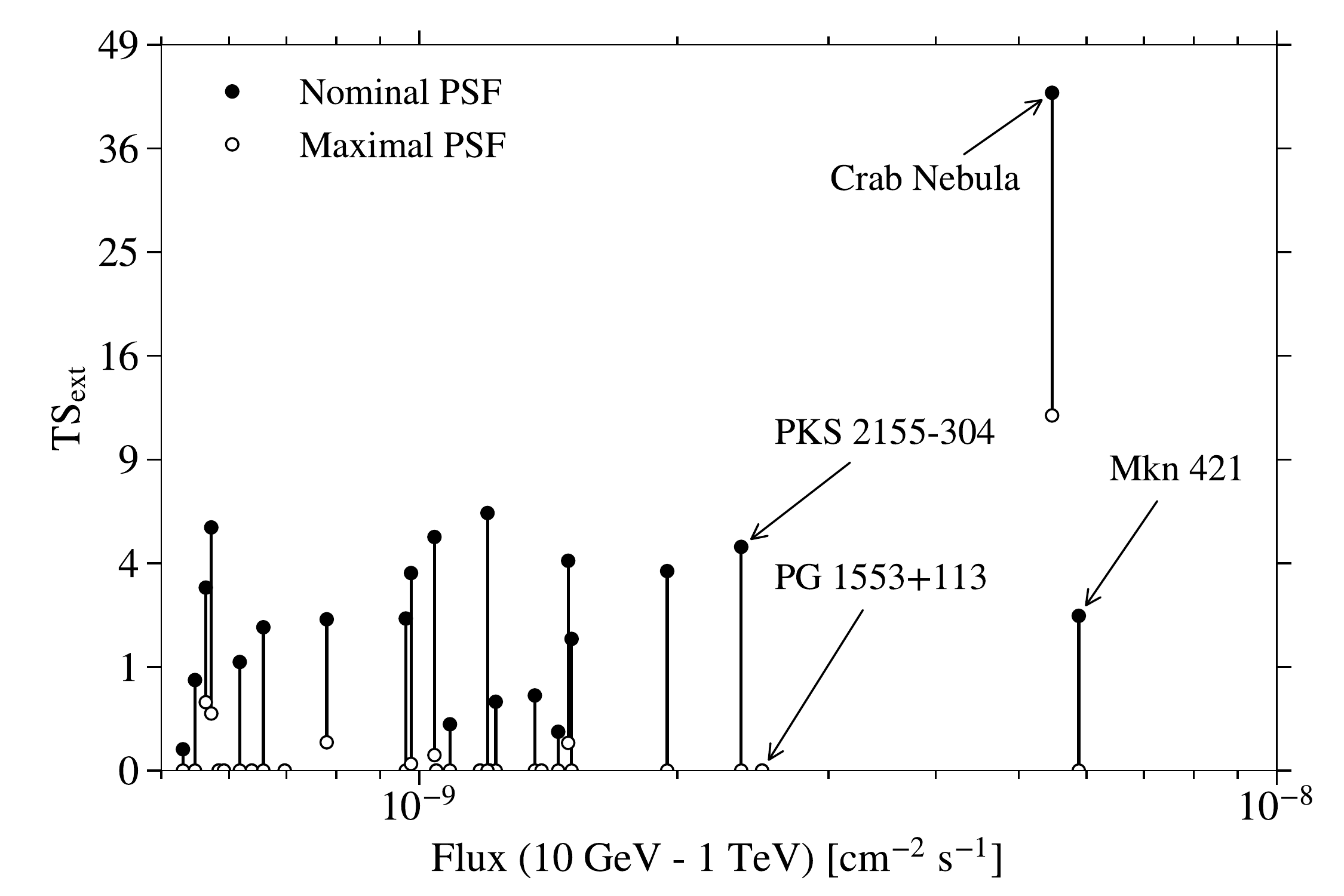}
\caption{\textit{Left:} VLA radio image of the Crab Nebula at 3~GHz
  \citep{2017arXiv170402968D}, overlaid with the position and 68\%
  containment radius of {\fhescrab} (white $\times$ marker and circle).  The
  cyan marker indicates the location of PSR~J0534$+$2200 as determined
  from optical/radio measurements. \textit{Right:} $\tsext$ versus
  photon flux above 10~GeV.  Filled and open circles show the value of
  $\tsext$ obtained with the nominal and bracketing models of the
  PSF.\label{fig:mapcrab}}
\end{figure*}

Our analysis detects an extension of
$\fhescrabrext\dg \pm \fhescrabrexterr\dg \pm \fhescrabrexterrsys\dg$ in {\fhescrab} which is associated
with the IC component of the Crab Nebula
(3FGL~J0534.5$+$2201i).  
The left panel of Fig.~\ref{fig:mapcrab} shows a VLA radio image
of the Crab Nebula overlaid with the 68\% containment radius of
{\fhescrab}.
The nebula spectrum is fit with an LP that has a spectral index $\alpha = 1.79\pm0.04$ and curvature $\beta = (1.67 \pm 0.70)\times10^{-2}$.

The extension of {\fhescrab} is comparable to the LAT angular resolution (68\% containment radius)
for the best-reconstructed events at high energy ($\sim0.03\dg$ for
PSF3 events with $E>30$~GeV), and is therefore particularly
sensitive to systematic uncertainties of the LAT PSF model.
Bracketing models for the PSF systematic uncertainty discussed in
Section~\ref{sec:systematics} were developed by comparing the nominal
PSF model derived from Monte Carlo simulations of the detector against
the angular distribution of high-latitude blazars.

Using a model that increases the size of the PSF according to Eq.~\eqref{eqn:psfscaling}, we find that $\tsext$ drops from {\fhescrabtsext}
to {\fhescrabtsextsys}.  In the right panel of Fig.~\ref{fig:mapcrab}, we show the
value of $\tsext$ obtained for the sources with photon flux above
10~GeV larger than $5\times10^{-10}$ cm$^{-2}$ s$^{-1}$.  If the extension of
{\fhescrab} arises from systematic errors in the PSF, we would expect to
see a trend toward increasing $\tsext$ in higher flux objects; however, this was not observed.  
The BL~Lac object Mkn~421, which has comparable flux to the
Crab Nebula above 10~GeV, has $\tsext$ of {\fhesmknfourtwoonetsext} and 0.0 for the nominal
and bracketing models of the PSF, respectively.  Given the absence of
significant extension in high-latitude sources of comparable flux, we conclude that the measured extension is 
probably intrinsic to the Crab Nebula rather than the result of an instrumental artifact.

Furthermore, the measured extension of {\fhescrab} agrees well with predictions
from simple synchrotron-self-Compton models when the spatial extension
of the photon densities is modeled with two-dimensional Gaussian
distributions that emit synchrotron radiation in an homogeneous
magnetic field~\cite[e.g.][]{1998ApJ...503..744H,2010A&A...523A...2M}.
In addition, the result is consistent with recent results from the \hess\ Collaboration who measured an extension of $0.022\dg\pm0.001\dg\pm0.003\dg$ of the IC component of the nebula above energies of 700\,GeV~\citep{Holler:2017dtw}.\footnote{We note that the \hess\ results are quoted in terms of the width of a 2D Gaussian $\sigma = 0.0145\dg$, whereas  our results are given in terms of the 68\,\% confidence radius. The two quantities are related through $r_{68} = \sqrt{-2\sigma^2\ln(1-0.68)}$.}

\subsubsection{FHES sources in SFR regions: \fhessoftiii, \fhessofti, \fheshardiv}
\label{subsec:indvsrc_sfr}

SFRs are found in giant molecular clouds. 
These clouds collapse and produce stars of all spectral types, some of which are massive O- and B-type stars. 
Because of their relatively short life spans, higher densities of the latter are found in and near their parent SFRs. 
Those stars produce strong radiation fields, stellar winds, and supernova explosions that create large bubbles in the clouds. 
The density of SNRs in those regions is larger than the Galactic average. SFRs are thus expected to be sites of efficient cosmic-ray acceleration through different processes~\citep{2014A&ARv..22...77B}. 
Models include diffusive acceleration by the shockwaves of SNRs~\citep{2015ICRC...34....8C} 
and by the termination shock of massive stellar winds~\citep{2005AJ....130.2185L}, 
as well as stochastic acceleration by the magnetic turbulence induced by all those shockwaves \citep{2001AstL...27..625B, 2016A&A...591A..71M}. 
The Cygnus Cocoon is the only SFR firmly associated with an extended \gr\ source seen by the LAT~\citep{2011Sci...334.1103A}. 
It may be associated with the ARGO~J2031+4157 source at TeV energies~\citep{2014ApJ...790..152B}. 
Other SFRs have potential associations with GeV point sources, such as the G25.0+0.0 region~\citep{Katsuta:2017ctk}, 
NGC 3603~\citep{2017A&A...600A.107Y}, 
and Westerlund~2~\citep{2017arXiv171002803Y}, 
but it is difficult to estimate the contribution from unresolved sources unrelated to cosmic-ray production in 
such complex regions, as was demonstrated for 30 Doradus in the LMC~\citep{2010A&A...512A...7A,2015Sci...347..406H}. 
Other \gr\ sources detected beyond TeV energies are also tentatively associated with SFRs, such as Westerlund~1~\citep{2013MNRAS.434.2289O} 
and HESS J1848-018~\citep{2008AIPC.1085..372C,2013AdSpR..51..258D}. 
Our analysis finds three extended sources spatially consistent with the directions of SFR regions. 
They are described in more detail below.

SFRs present unique challenges for modeling the ISM and associated
diffuse \gr\ emission.  The intense radiation fields near OB
associations give rise to sharp gradients in both dust properties and
temperature.  Both our standard and alternative IEMs use dust
corrections derived from the Schlegel-Finkbeiner-Davis (SFD) map of
\citet{1998ApJ...500..525S}.  Generally, we have found a correlation
between the sources listed in this section and the SFD maps, which
trace the interstellar reddening related to the color excess, E(B-V).
The SFD map uses a relatively coarse correction for dust temperature,
with an angular resolution of 0.7\dg.  In the vicinity of SFRs, where
dust temperature can vary on much smaller angular scales, 
IEM models including SFD information have localized biases that can induce spurious sources \citep[see, e.g., Fig.~11 in][]{1fgl} or suppress real sources.

Because all of the alternative IEMs considered in
\Sectionref{sec:systematics} use the same SFD-based corrections, we
are not able to evaluate the systematic uncertainties associated with
these corrections.  Comparison with IEMs derived from \textit{Planck} dust maps
would test this hypothesis directly.  We have analyzed the
  three extended FHES sources associated with SFRs via IEM with
  \textit{Planck}-derived dust corrections that was fit to 8~years of \irf{Pass 8} LAT data.  
As demonstrated in \cite{remy2017}, improved treatment of the IEM rules out 
  some of these (\fhessoftiii~for example) as extended sources. 
Additionally, we used results from the WHAM Sky Survey to 
  see if any of these sources were spatially coincident with ionized gas missing from the IEM. 
  We found that one source (\fhessofti), which partially overlaps with NGC~7822, a SFR at a distance of
1~kpc with a diameter of $\sim$0.4\dg \citep{Quireza:2006sn},
  is also spatially coincident with a large region of H$_{\alpha}$ emission.

\begin{figure}[ht]
\centering
\includegraphics[width=0.48\columnwidth]{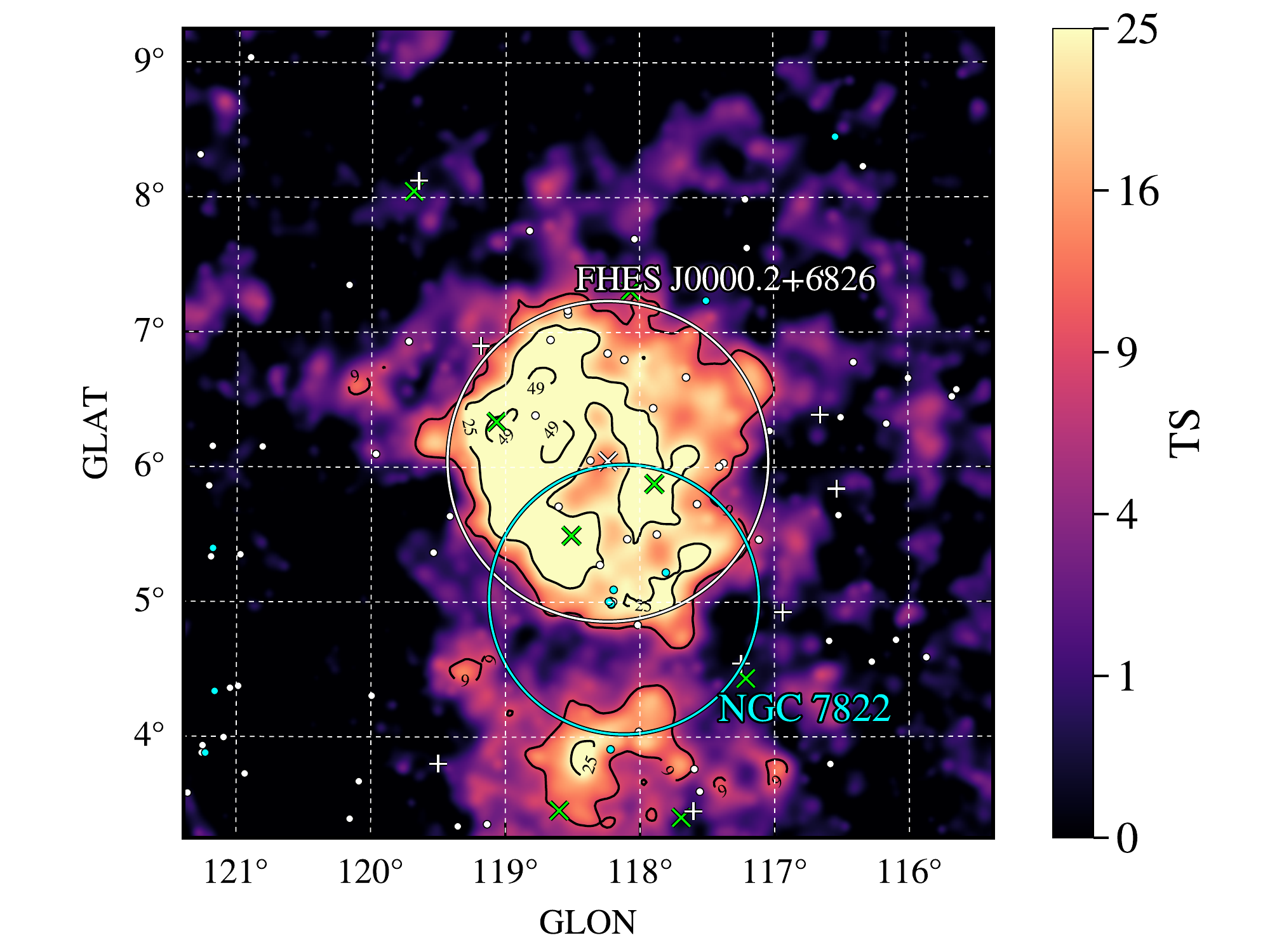}
\includegraphics[width=0.48\columnwidth]{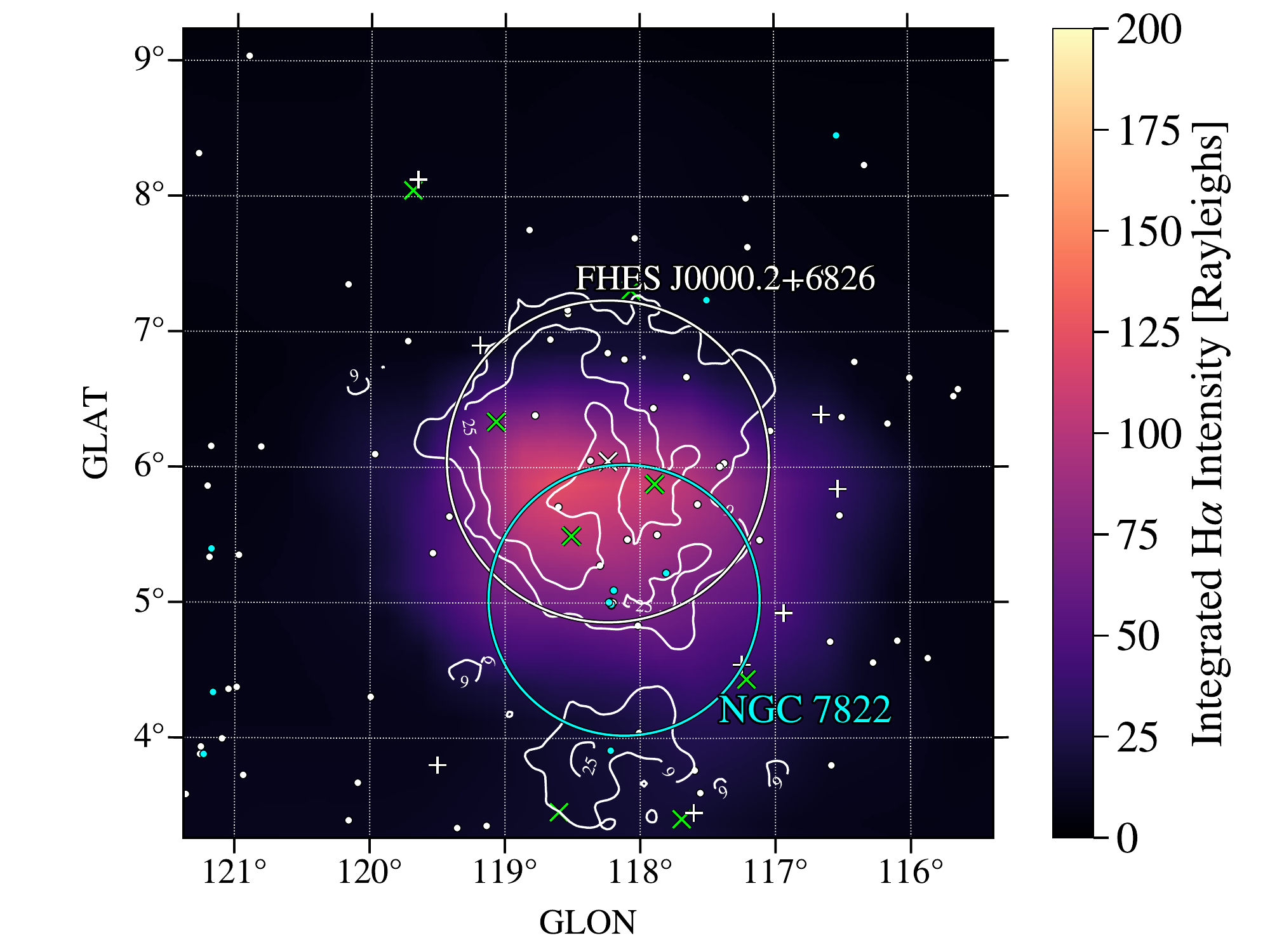}\\
\includegraphics[width=0.48\columnwidth]{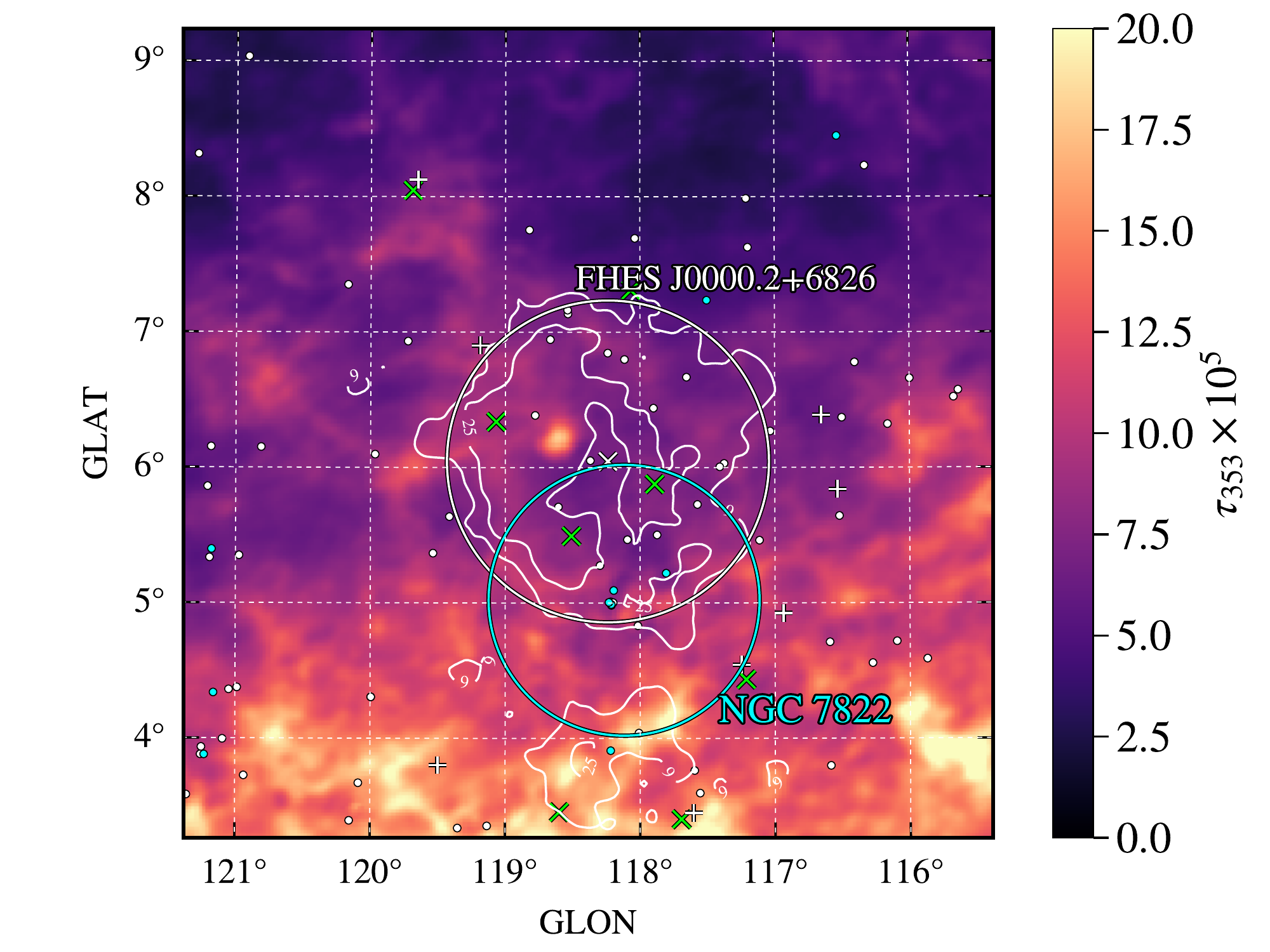}
\includegraphics[width=0.48\columnwidth]{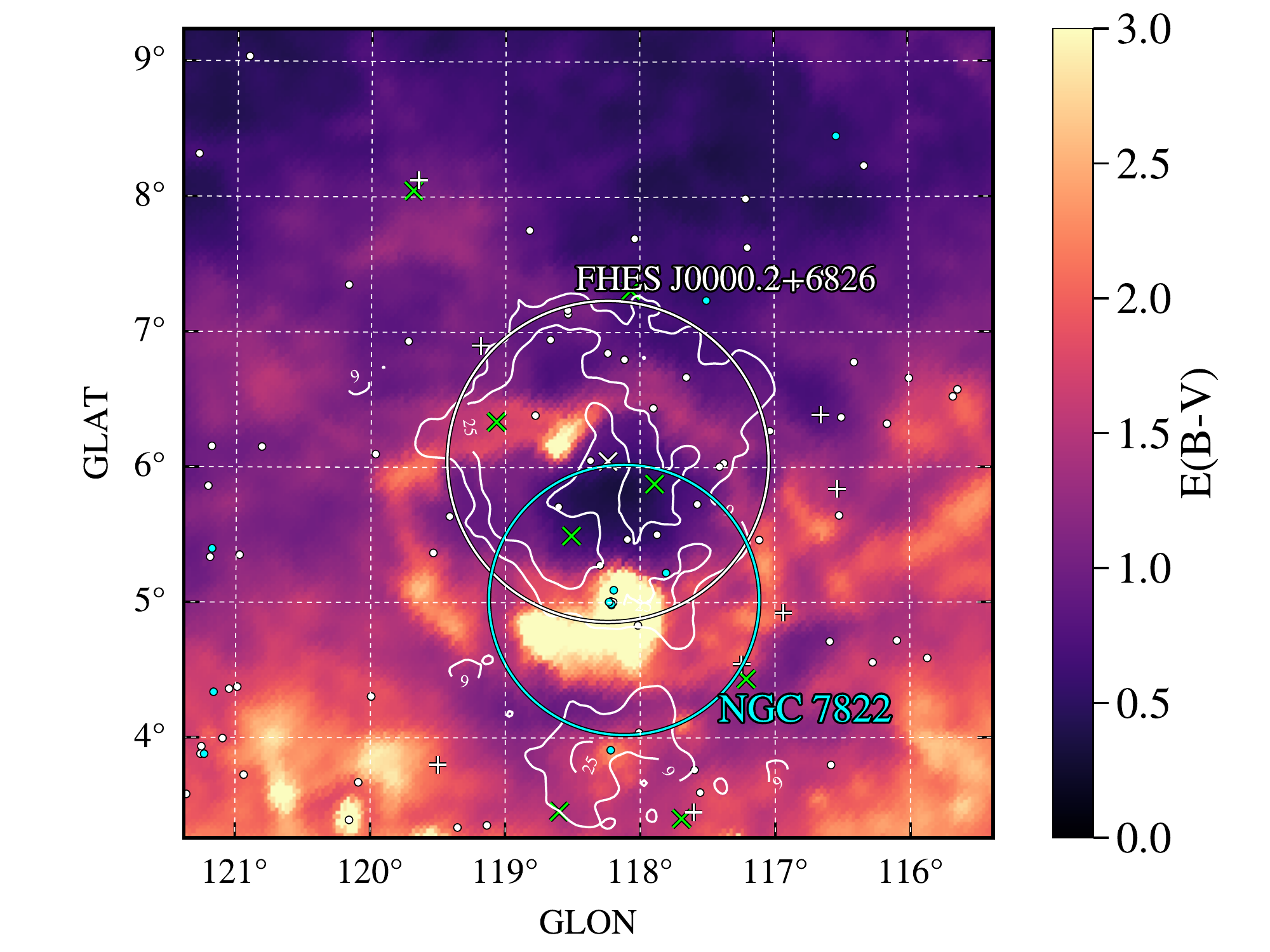}
\caption{ TS map of {\fhessofti} (top).   The top right plot shows the velocity-integrated 
map of H$_{\alpha}$ from the WHAM Sky Survey,
with the LAT TS isocontours overlaid.
Bottom panels show maps of the \textit{Planck} dust optical depth at 353 GHz (left) 
and SFD dust reddening (right), with LAT TS isocontours overlaid.
The white circle with central marker $\times$ indicates the best-fit disk
extension and centroid of the FHES source.  White crosses indicate the 
positions of point-source candidates with TS $> 9$ from the best-fit
model for the region. The LAT TS isocontours are also shown in white.  
Green crosses indicate the positions of sources
in the 3FGL catalog. Filled white and cyan markers indicate the positions
of B and O stars from the SIMBAD database.  The cyan circle indicates
the location of the HII region NGC~7822.\label{fig:tsmap_fhessofti}}
\end{figure}

{\fhessofti} is a soft-spectrum source
($\Gamma = \fhessoftiindex \pm \fhessoftiindexerr \pm
\fhessoftiindexerrsys$) that is modeled best by a disk with
$\rext = \fhessoftirext\dg \pm \fhessoftirexterr\dg \pm
\fhessoftirexterrsys\dg$.  The best-fit model encompasses four 3FGL
sources.\footnote{3FGL sources in the region of \fhessofti:
  3FGL~J2356.9$+$6812, 3FGL~J0004.2$+$6757, 3FGL~J0008.5$+$6853, and
  3FGL~J2355.4$+$6939}. All four sources are unassociated and were
measured in the 3FGL with indices between 2.4 and 2.7.  The spectral
indices of the 3FGL sources are consistent, within one standard deviation, with the index measured for
{\fhessofti}.
 \Figureref{tsmap_fhessofti}
shows a comparison of the LAT TS map of the region to the H$_{\alpha}$ emission, SFD, and
\textit{Planck} dust maps. Although there is no correlation with the cold dust (\textit{Planck}), 
a large deficit in the SFD map is observed in the
southern part of {\fhessofti}.  This feature is not observed in the
\textit{Planck} map and is likely attributable to dust temperature variations
within NGC~7822. 
The \gr\ map is correlated best with  H$_{\alpha}$ emission, coming from regions of ionized gas, which is not accounted for in the IEM.
For comparison, we have also indicated the location of the dozens of O-
and B-type stars in the region in \Figureref{tsmap_fhessofti}.  There
appears to be an over-density of O- and B-type stars inside
{\fhessofti}, particularly toward the southern edge of the source.
As we can not rule out the possibility that the \gr\ emission is due 
to the ionized gas not accounted for in the IEM,
we mark this source as confused.

{\fhessoftiii} is located near NGC~1579, an SFR at a distance of 700~pc~\citep{2013A&A...558A..53K}. 
It is a soft-spectrum source
($\Gamma = \fhessoftiiiindex \pm
\fhessoftiiiindexerr\pm\fhessoftiiiindexerrsys$ ) that is 
modeled best by a disk with
$\rext = \fhessoftiiirext\dg \pm \fhessoftiiirexterr \pm
\fhessoftiiirexterrsys\dg$.  The
best-fit model encompasses three 3FGL sources that do not have
point-source counterparts.\footnote{The 3FGL sources in the region of
  \fhessoftiii: 3FGL~J0431.7$+$3503, 3FGL~J0426.3+3510, and
  3FGL~J0429.8+3611c} {\fhessoftiii} is a composite of these three
sources, which are unassociated and also have spectral indices measured
in the 3FGL between 2.4 and 2.7.  \cite{remy2017} found that this excess is 
due to dark neutral gas, and when combining HI, CO, and DNM gas components, 
the excess toward NGC~1579 disappears (see Fig. 7 of \cite{remy2017}). 

\begin{figure*}[ht]
\centering
\includegraphics[width=0.48\columnwidth]{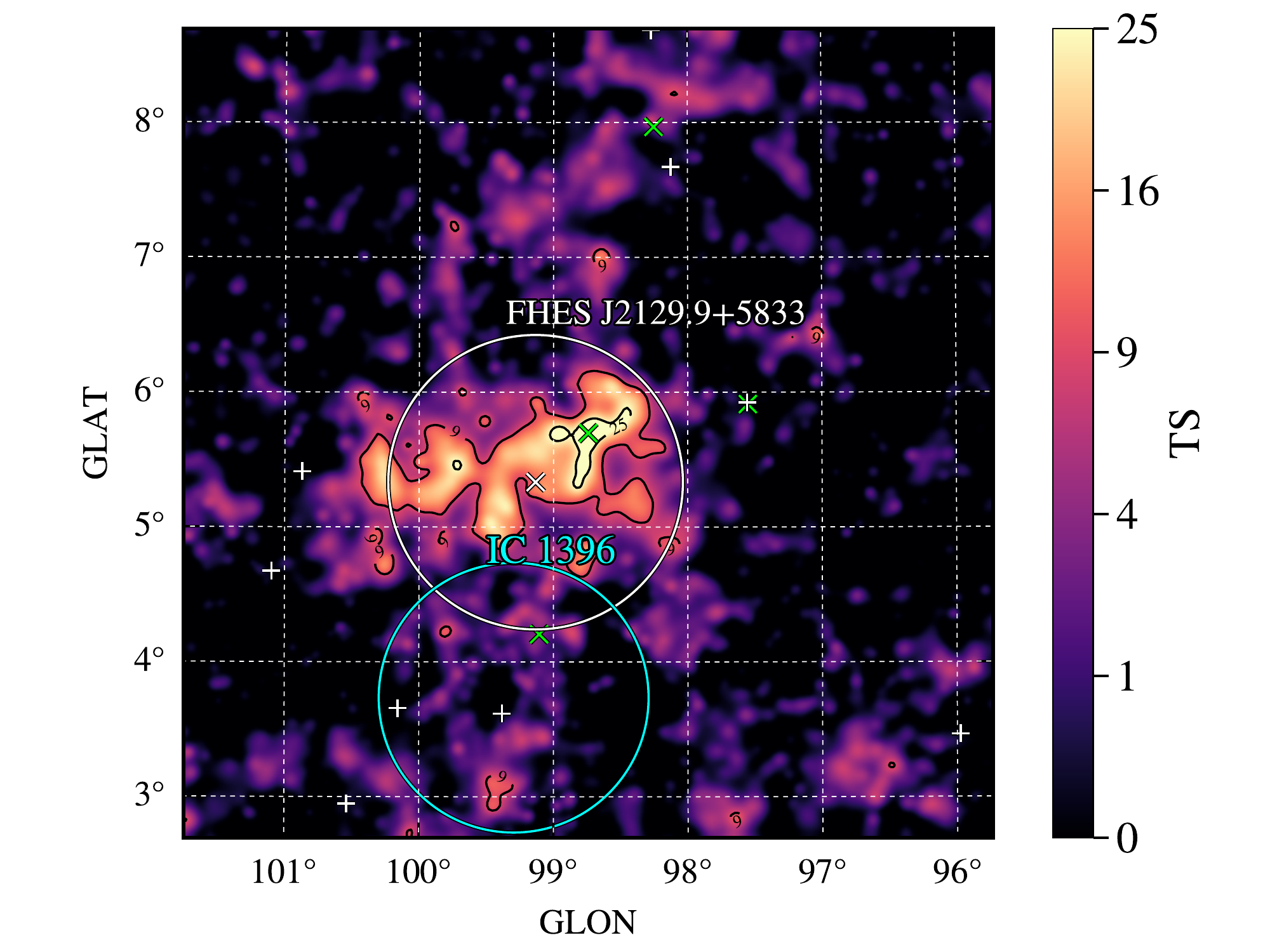}
\includegraphics[width=0.48\columnwidth]{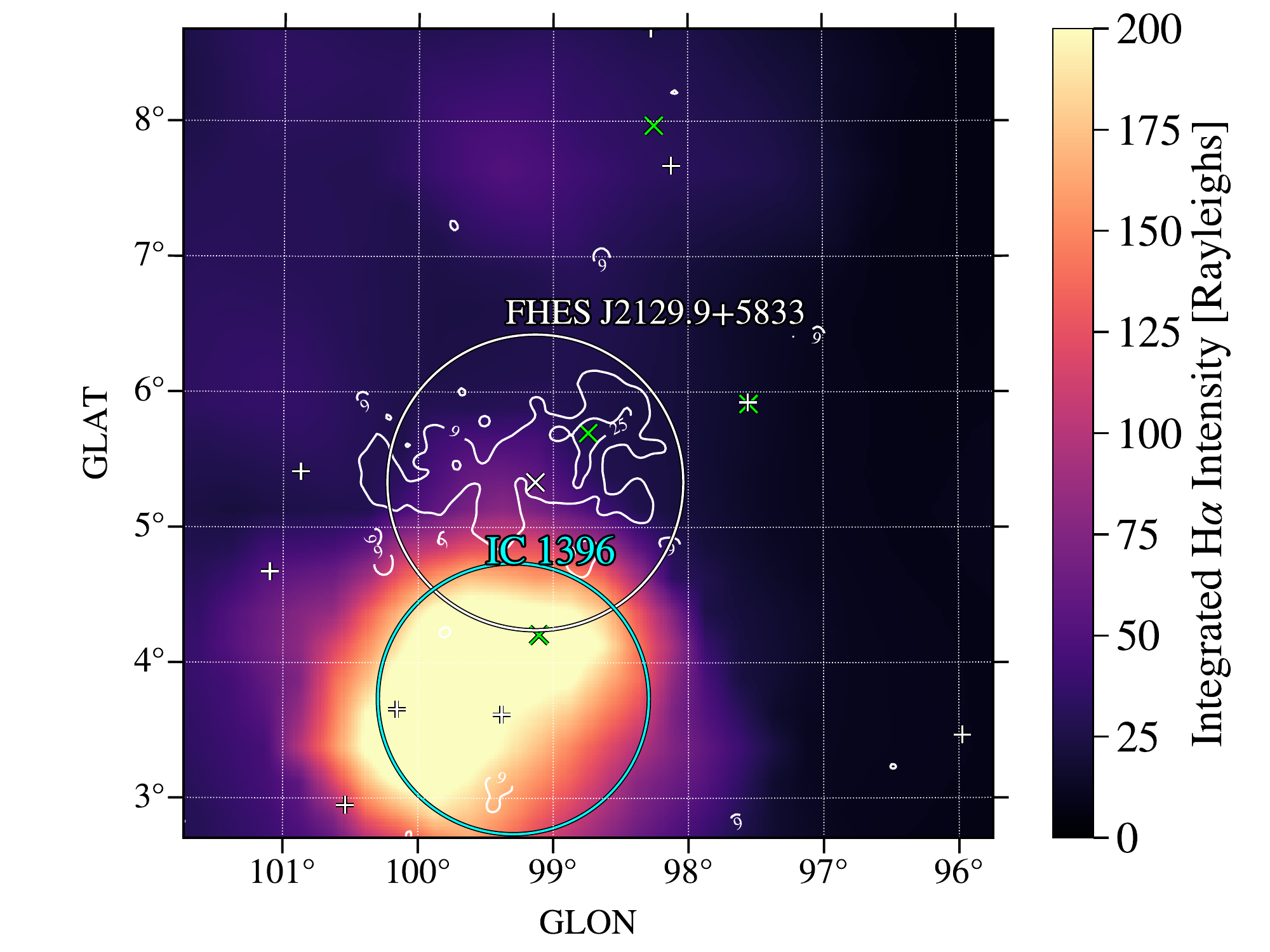}\\
\includegraphics[width=0.48\columnwidth]{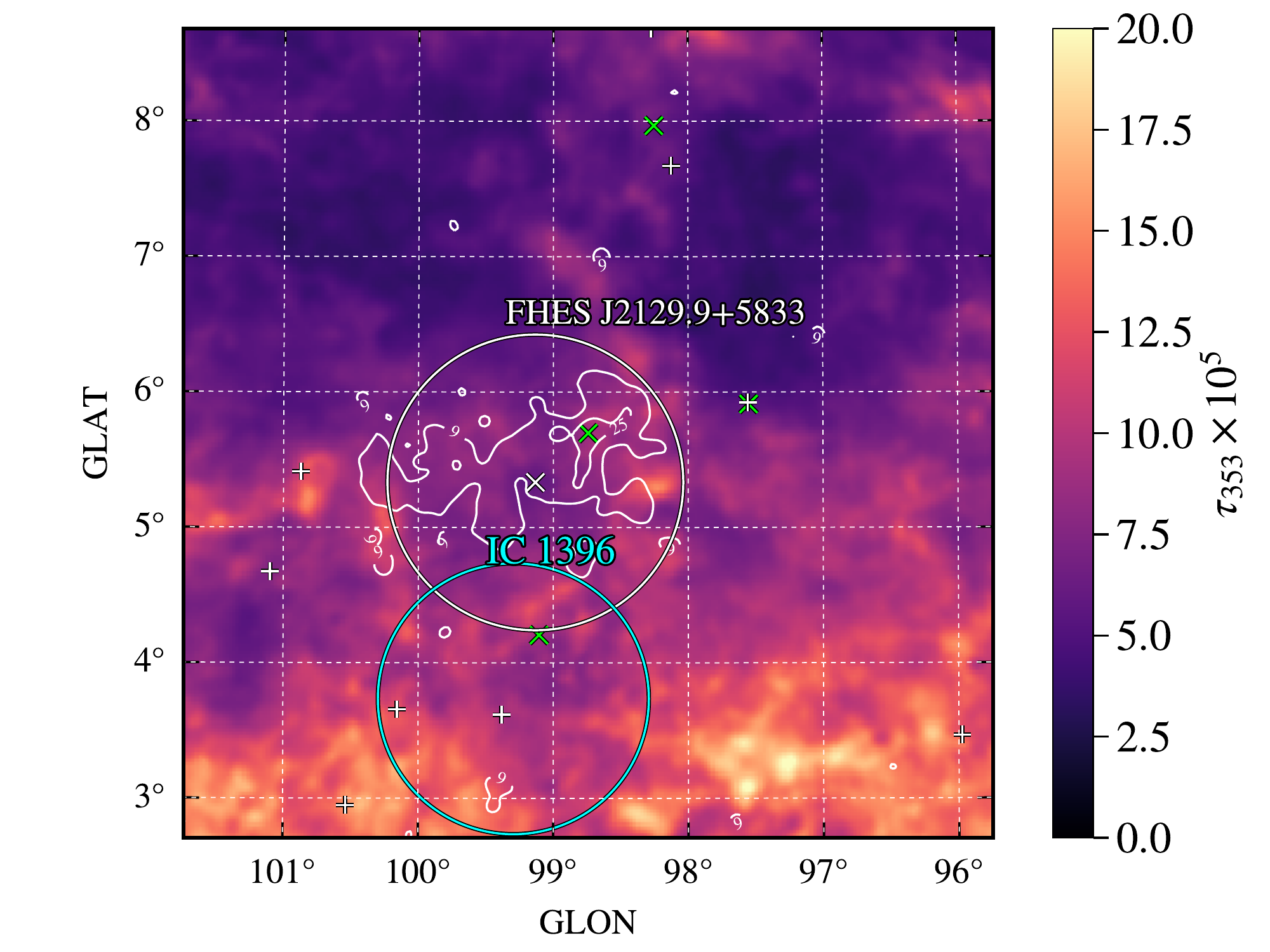}
\includegraphics[width=0.48\columnwidth]{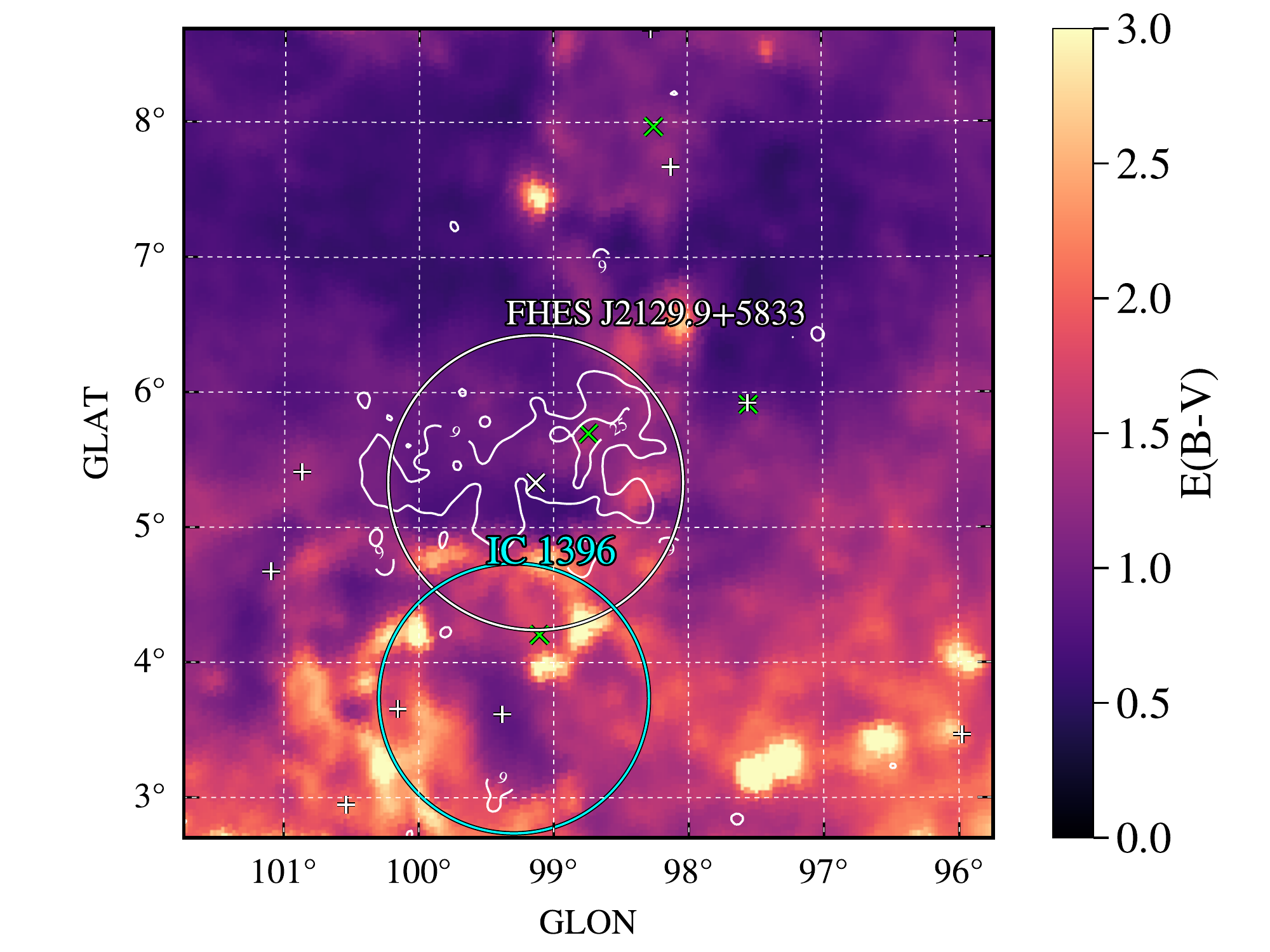}
\caption{ 
TS map of {\fheshardiv} (top).  The top right plot shows the velocity-integrated 
map of H$_{\alpha}$ from the WHAM Sky Survey,
with the LAT TS isocontours overlaid.
Bottom panels show maps of \textit{Planck} dust optical depth at 353 GHz (left) 
and SFD dust reddening (right), with LAT TS isocontours overlaid.
The white circle with central marker $\times$ indicates the best-fit disk
extension and centroid of the FHES source.  White crosses indicate the
positions of point-source candidates with TS $> 9$ from the best-fit
model for the region.  
 The LAT TS isocontours are also shown in white.  
Green crosses indicate the positions of sources
in the 3FGL catalog. 
The cyan circle indicates
the location of the HII region IC~1396.\label{fig:tsmap_fheshardiv}}
\end{figure*}

{\fheshardiv} is located near IC~1396, which is a large and
comparatively faint-emission nebula and SFR over 30 pc across, located
about 735~pc away.  It has an intermediate spectral hardness
($\Gamma = \fheshardivindex \pm
\fheshardivindexerr\pm\fheshardivindexerrsys$) and is best-modeled spatially
by a disk with
$\rext = \fheshardivrext\dg \pm \fheshardivrexterr\dg
\pm\fheshardivrexterrsys\dg$.
The $\gamma$-ray emission appears to be located primarily in a
comparatively low-density region of dust, gas, and stars as seen in
\Figureref{tsmap_fheshardiv}.  
No obvious features are visible in either the SFD dust reddening or \textit{Planck} dust optical depth maps of the region.
There is a possibility that this is a newly found source belonging to a more common class of 
extended \gr\ emitters, such as SNRs or PWNe, and not necessarily emission from the SFR itself.

\subsubsection{FHES sources potentially associated with SNR/PWN:  \fheshardi, \fheshardiii, \fheshardv} \label{subsec:snrysources}

There are over 30 SNRs and PWNe with known $\gamma$-ray emission
generally found at lower latitudes, near the Galactic
plane~\citep{Acero:2015prw}.  Extragalactic SNRs were also detected in
the Magellanic Clouds.  In addition to the previously detected SNRs CTA~1
and SNR~G332.5$-$05.6 discussed in Sections~\ref{subsec:indvsrc_cta1}
and~\ref{subsec:indvsrc_snrg332}, we find two additional sources, one
close to the Galactic plane at $b = -4.8\dg$ (\fheshardiii) and the
other (\fheshardi)  at a higher latitude, $b=17.9\dg$, which is coincident
with an unclassified radio shell.  Furthermore, we identify one more
source as a potential SNR candidate that, however, lacks a
multiwavelength counterpart: {\fheshardv} at $b = -5.5\dg$.

\begin{figure*}[ht]
\includegraphics[width=0.48\columnwidth]{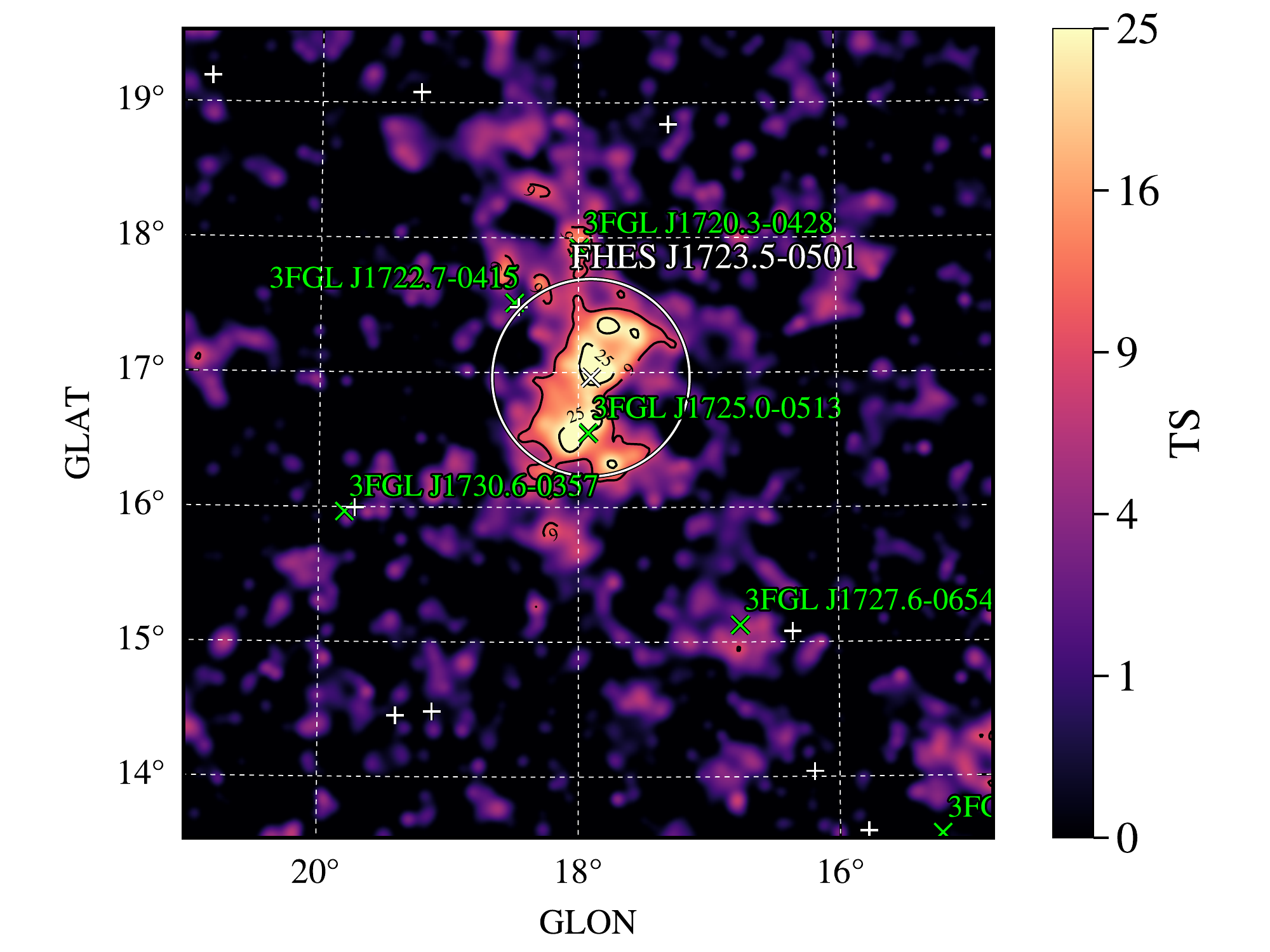}
\includegraphics[width=0.48\columnwidth]{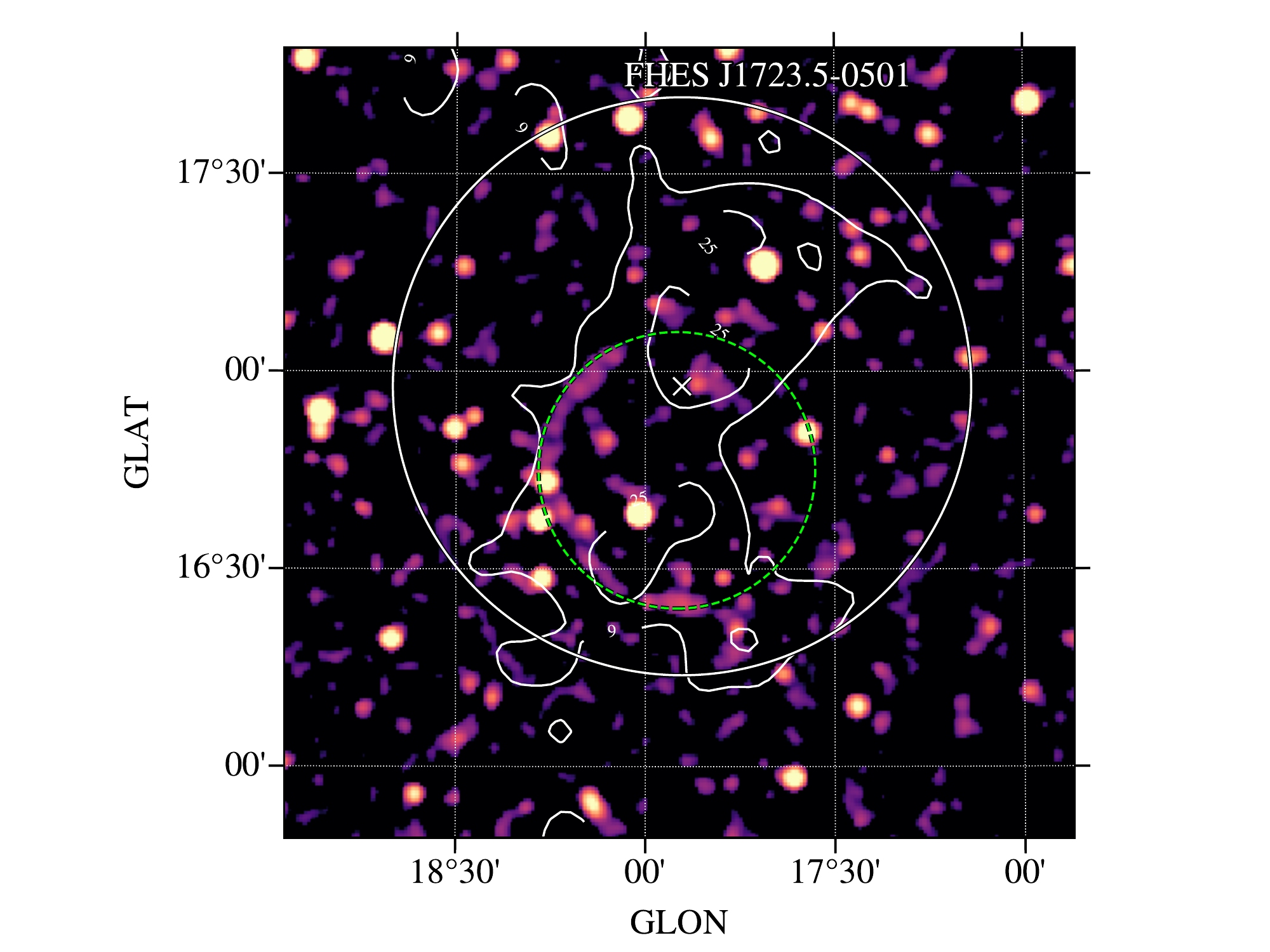}

\caption{\textit{Left:}  TS map of {\fheshardi}.
  The white circle with central marker $\times$ indicates the extension (68\%
  containment radius) and centroid of the FHES source.  White crosses
  indicate the positions of point-source candidates with TS $> 9$ from the
  best-fit model for the region.  Green crosses indicate the positions
  of sources in the 3FGL catalog. \textit{Right:} Map of continuum
  emission at 1.4~GHz from NVSS \citep{1998AJ....115.1693C}, smoothed
  with a Gaussian kernel of radius $0.012\dg$.  White contours show
  the  TS map of {\fheshardi}.  The dashed green circle traces
  the circular feature observed in the radio
  map. \label{tsmap_fheshardi}}
\end{figure*}

{\fheshardi} is the highest-latitude unassociated candidate, and its TS
map is shown in Figure~\ref{tsmap_fheshardi} (left). 
 It encompasses a
shell-like structure in the NVSS (1.4 GHz) image
(Figure~\ref{tsmap_fheshardi}, right) and has an angular extent of
$\rext=\fheshardirext\dg\pm\fheshardirexterr\dg\pm\fheshardirexterrsys\dg$ and a hard spectral index
($\Gamma=\fheshardiindex\pm\fheshardiindexerr\pm\fheshardiindexerrsys$).  The size of the radio shell
($D\sim0.7\dg$), seen best along the southwestern edge of the
\gr\ emission, is comparable to the size of the FHES
source. 
There are no previously known SNRs at this location.
 {\fheshardi} encompasses the unassociated source
3FGL~J1725.0$-$0513, which does not have a point-source counterpart in
our model of the region.  
Given its high latitude, we suggest that this source could be 
associated with a type Ia SNR because these are not necessarily 
located close to the regions of star formation. SN 1006 
represents an example of a remnant of a type Ia supernova
explosion detected in \Grs\ at high Galactic latitude~\citep{2017ApJ...851..100C}. 

\begin{figure}[ht]
\centering
\includegraphics[width=0.48\columnwidth]{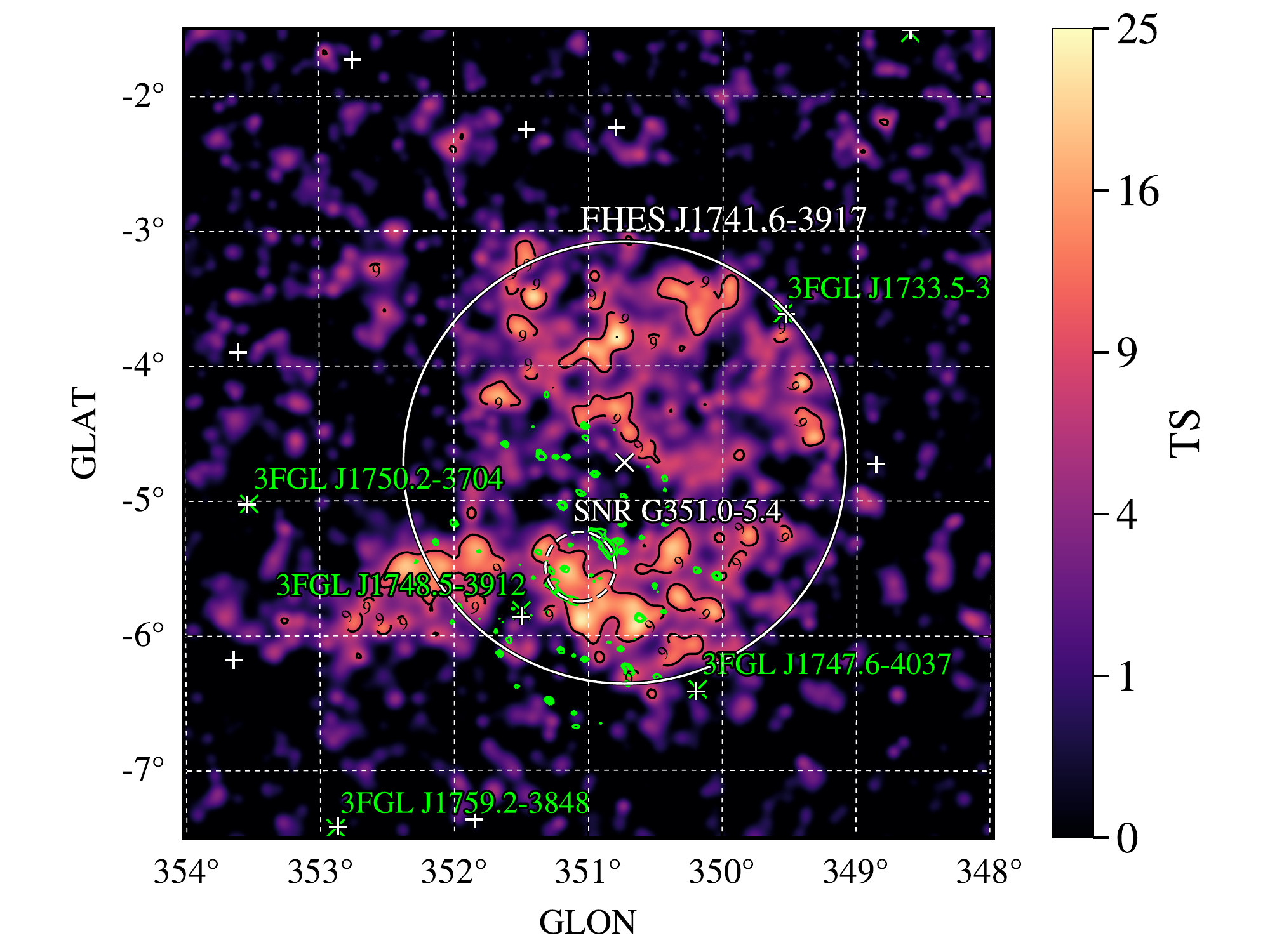}
\caption{ TS map ($\Gamma=2$) of {\fheshardiii}.  The white
  circle with central marker $\times$ indicates the best-fit disk extension and
  centroid of the FHES source.  White crosses indicate the positions of
  point-source candidates with TS $> 9$ from the best-fit model for
  the region.  Green crosses indicate the positions of sources in the
  3FGL catalog.  Green contours show the GMRT radio map of
  SNR~G351.0$-$5.4 at 325~MHz from
  \citet{deGasperin:2014caa}.\label{fig:tsmap_fheshardiii}}
\end{figure}

{\fheshardiii} has a large angular extent
($\rext=\fheshardiiirext\dg\pm\fheshardiiirexterr\dg\pm\fheshardiiirexterrsys\dg$) and encompasses the known,
radio-detected SNR~G351.0$-$5.4~\citep{deGasperin:2014caa}. 
However, the $\gamma$-ray emission appears to be much larger than the radio SNR. 
The TS map is shown in \Figureref{tsmap_fheshardiii}.  It has a hard
spectral index ($\Gamma=\fheshardiiiindex\pm\fheshardiiiindexerr\pm\fheshardiiiindexerrsys$), which suggests that it
may be associated with a young, shell-type SNR similar to, e.g., Tycho's SNR or Cas~A \citep[][]{2010ApJ...710L..92A,2017ApJ...836...23A}.  
{\fheshardiii} is
near to, or encompasses, three point sources that have
direct 3FGL counterparts: 3FGL~J1748.5$-$3912, 3FGL~J1733.5$-$3941,
and 3FGL~J1747.6$-$4037.  Sources 3FGL~J1748.5$-$3912 and 3FGL~J1733.5$-$3941
are both unassociated.  Source 3FGL J1747.6$-$4037 is located on the
southern edge of {\fheshardiii} and is associated with the millisecond
pulsar PSR~J1747$-$4036.  We note that the characteristics of
{\fheshardiii} match well with the new \gr\ source
G350.6$-$4.7 reported by \citet{Araya:2016cbn}, based on an analysis of
eight years of LAT data. Source G350.6$-$4.7 is found at the same location
($l=350.6\dg$, $b=-4.7\dg$), with similar angular extent and spectrum
($\Gamma= 1.68\pm0.04\pm0.14$, $R=1.7\dg\pm0.2\dg$).

In addition to the previous sources, we also found one new
unassociated hard-spectrum source.  The hardness of the spectrum for
 {\fheshardv}
($\Gamma = \fheshardvindex\pm\fheshardvindexerr\pm\fheshardvindexerrsys$) may imply an association with an SNR or PWN. However,
there is no clear overlap with known objects in the TeV, X-ray, or
radio wavelengths in the considered multiwavelength surveys and catalogs.
For this extended object, there are 3FGL and 3FHL sources
within the 68\% containment radius; however, both 3FGL/3FHL sources
have point-source counterparts in our model.  They are hence
presumably unrelated to the FHES sources.\footnote{ The sources are
  3FHL~2308.8+5424 with an angular separation of $0.77\dg$ and spectral index $\Gamma = 2.06\pm0.53$, and 
  3FGL~2309.0+5428 with a separation of $0.82\dg$ and $\Gamma = 1.70\pm0.25$. 
   The 3FHL source is associated with 1RXS\,J2300852.2+542559, an AGN 
  of unknown class.  }

\begin{figure}[ht]
\centering
\includegraphics[width=0.48\columnwidth]{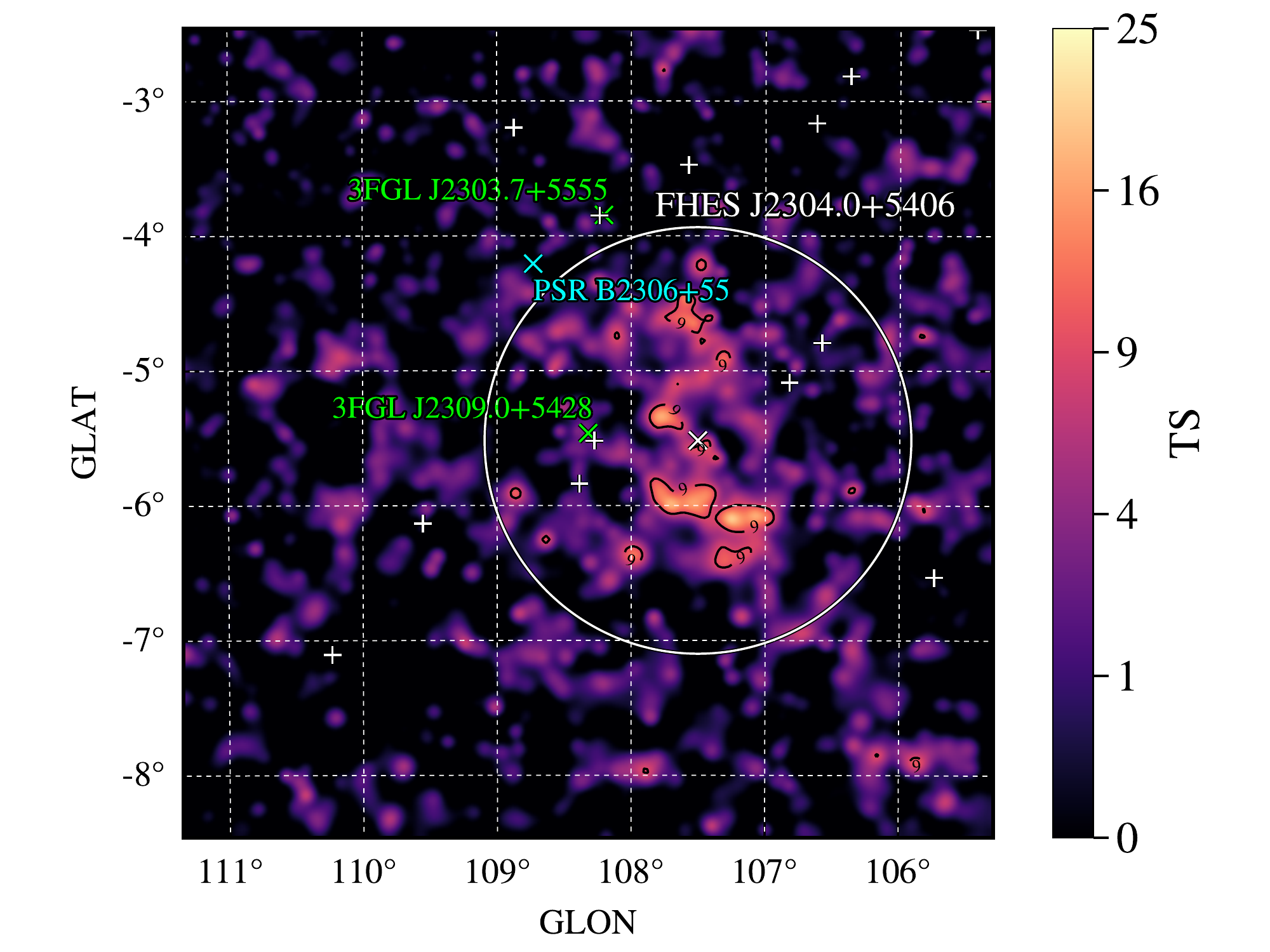}
\caption{ 
   TS map of the unassociated, hard-spectrum source {\fheshardv}. 
  The map is generated with
  a point-source morphology and a PL spectrum ($\Gamma=2$).  The white
  circle with central marker $\times$ indicates the extension (68\% containment
  radius) and centroid of the FHES source.  White crosses indicate the 
  positions of point-source candidates with TS $> 9$ from the best-fit
  model for the region.  Green crosses indicate the positions of
  sources in the 3FGL catalog. \label{tsmap_hard}}
\end{figure}

\fheshardv~has a large angular extent ($\rext=\fheshardvrext\pm\fheshardvrexterr\pm\fheshardvrexterrsys\dg$), as seen in the TS map shown in Fig.~\ref{tsmap_hard} (right).   
There is a nearby pulsar PSR~B2306$+$55 ($\sim$2 kpc away) at the northwest edge of the source.  
However, it is quite old ($\sim$10 Myr), so any associated SNR would be too old to drive particle acceleration.
Additionally, the pulsar has a relatively low spin-down power (7.3$\times$10$^{31}$ erg s$^{-1}$), which would be too low to power a $\gamma$-ray bright PWN~\citep{2013ApJ...773...77A}.

\subsubsection{Unassociated Soft-spectrum Sources: \fhessoftv, \fhessoftvi} \label{subsec:softsources}
\label{sec:indvsrclast}

The remaining  
two soft-spectrum candidates have spectral indices which are similar to that expected for 
Galactic diffuse emission ($\Gamma\sim$2.7). 

\begin{figure*}[ht]
\centering
\includegraphics[width=0.48\columnwidth]{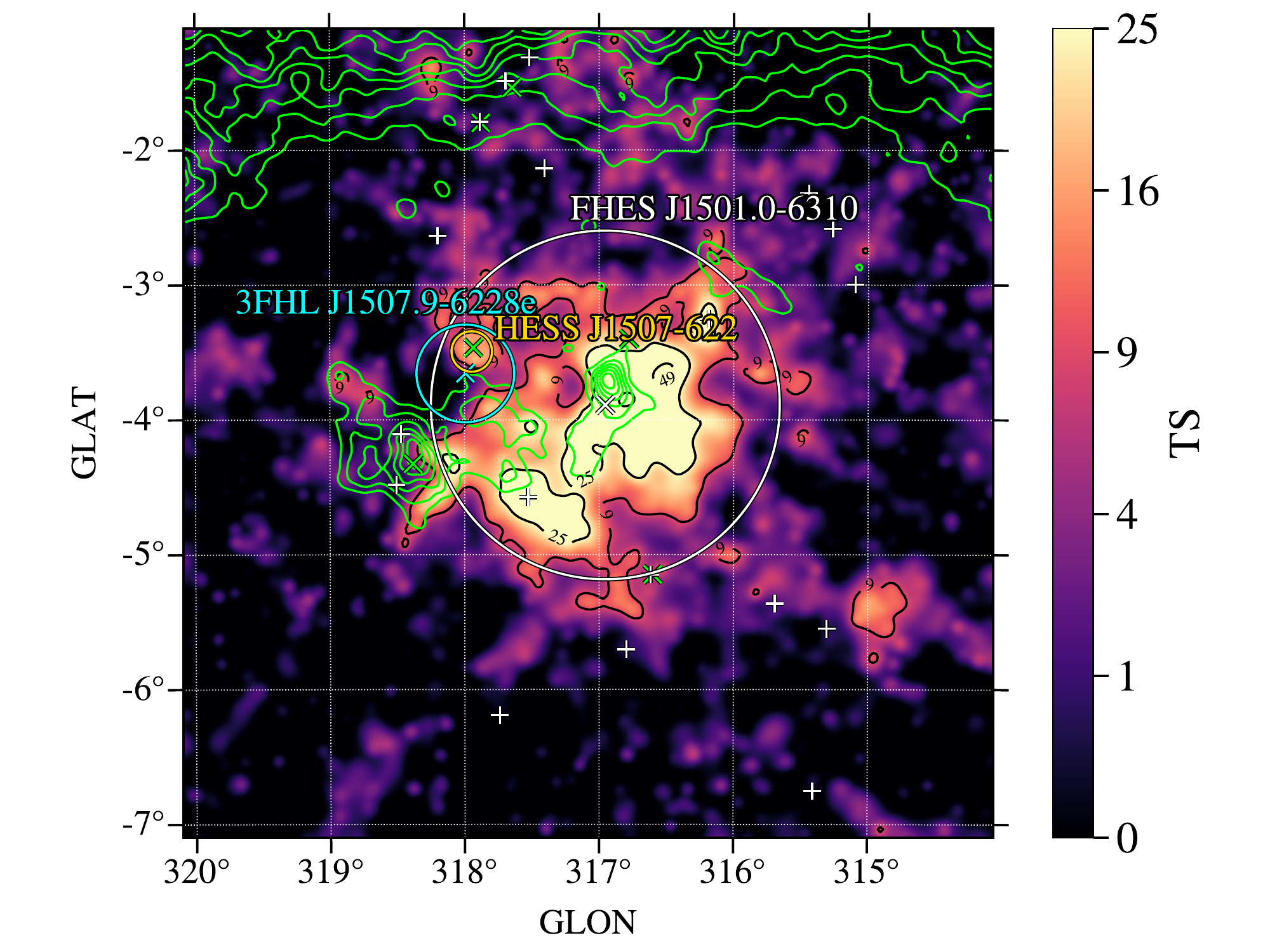}
\includegraphics[width=0.48\columnwidth]{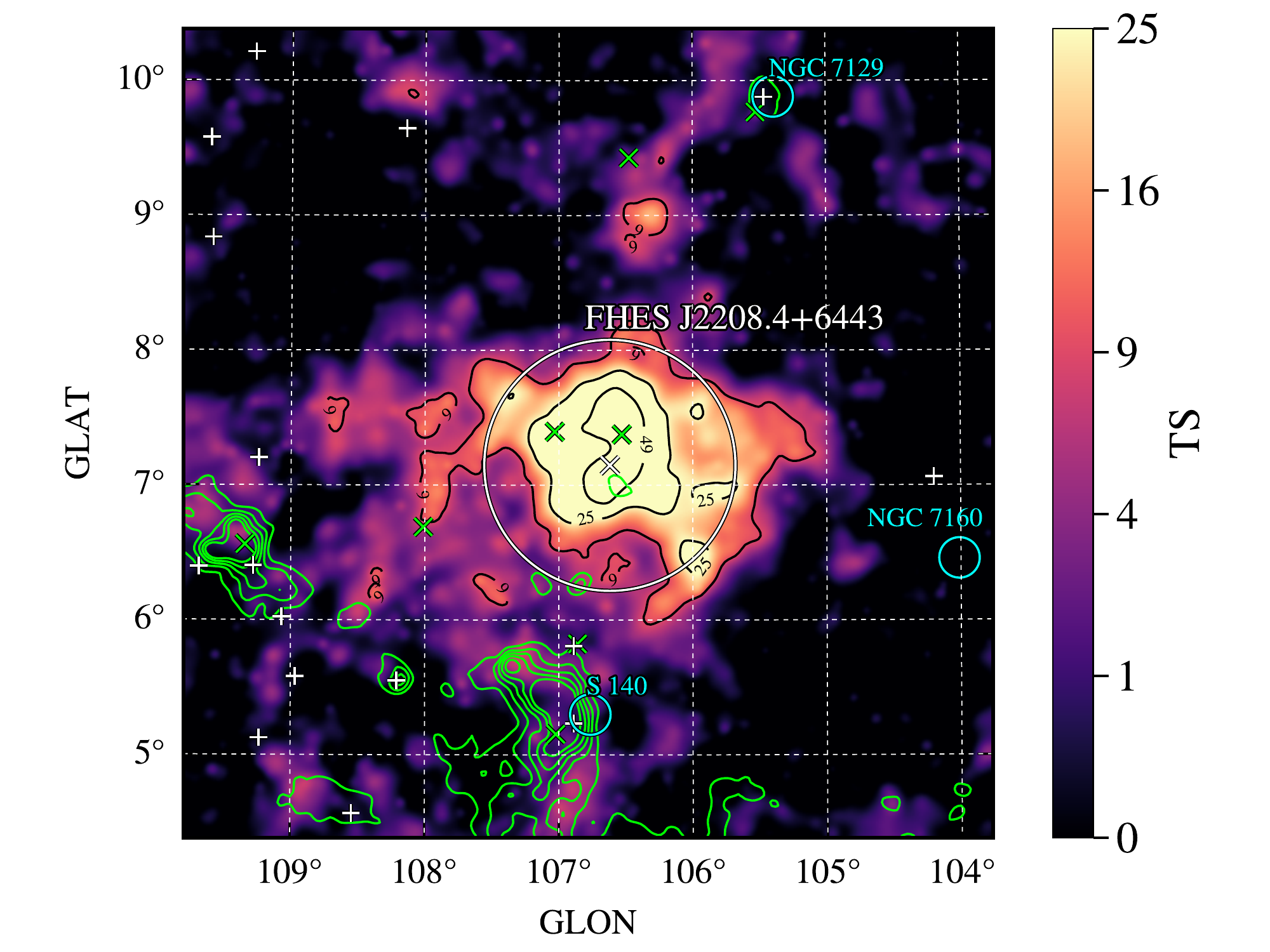}
\caption{ TS maps of unassociated, soft-spectrum sources {\fhessoftv}
  (left) and {\fhessoftvi} (right).  Green contours show the \textit{Planck}
  dust optical depth at 353~GHz.  The white circle with central marker $\times$
  indicates the extension (68\% containment radius) and centroid of
  the FHES source.  White crosses indicate the positions of point-source
  candidates with TS $> 9$ from the best-fit model for the region.
  Green crosses indicate the positions of sources in the 3FGL
  catalog. The 3FHL and \hess\ source extensions are shown as cyan and yellow contours (left). Cyan circles indicate the locations of HII
  regions (right).  
  \label{tsmap_soft_1}}
\end{figure*}

{\fhessoftv} is fit best with an extension of size
$\rext=\fhessoftvrext\dg\pm\fhessoftvrexterr\dg\pm\fhessoftvrexterrsys\dg$
and a spectral index of
$\Gamma=\fhessoftvindex\pm\fhessoftvindexerr\pm\fhessoftvindexerrsys$.
The TS map is shown in Figure~\ref{tsmap_soft_1} (left).  Three
3FGL/3FHL sources have an angular separation $d < \rext$, namely
3FGL~J1457.6-6249 ($d = 0.53\dg$), 3FGL~J1503.7-6426 ($d = 0.94\dg)$,
and 3FHL~J1507.9-6228e ($d = 1.40\dg$); they have spectra
$\Gamma = 2.45\pm0.12$, $\Gamma = 2.33\pm0.07$, and
$\Gamma = 1.86\pm0.15$, respectively.  The source 3FGL~J1503.7-6426 is
classified as a blazar of unknown type, while the other sources do not
have a multiwavelength counterpart.  In our model, 3FHL~J1507.9-6228e is an
  extended source that replaces 3FGL~J1506.6-6219; it is
  represented spatially as a disk of radius $0.36\dg$.  This source
  may be associated with the unidentified \hess\ source HESS~J1507-622
  \citep{2011A&A...525A..45H}, which is located at the same position but
  has a smaller spatial extent ($R=0.15\pm0.02\dg$).
The 3FHL/3FGL sources have harder spectra than the FHES source, yet
the measured spectral index of the latter fits well with the spectral
index of the \hess\ source
($\Gamma = 2.24 \pm 0.16_\mathrm{stat} \pm 0.20_\mathrm{sys}$) and the
one found in a dedicated \emph{Fermi} analysis of TeV detected PWNe
that gave $\Gamma = 2.33\pm0.48$ for energies above 10\,GeV
\citep{2013ApJ...773...77A}.  The 3FHL and \hess\ source extensions
are shown as cyan and yellow contours, respectively
(Figure~\ref{tsmap_soft_1}; left).  We also show the \textit{Planck}
dust optical depth contours (green contours).  The FHES source
encompasses the regions with high dust optical depth that are in the
direction of the Circinus molecular cloud complex.

{\fhessoftvi} comprises the two unassociated 3FGL sources
(3FGL~J2206.5$+$6451 with $d = 0.25\dg$ and  $\Gamma = 2.84 \pm 0.25$ as well as 
3FGL~J2210.2$+$6509 with  $d = 0.48\dg$ and  $\Gamma = 2.48 \pm 0.16$). It has an angular extent
$\rext=\fhessoftvirext\dg\pm\fhessoftvirexterr\dg\pm\fhessoftvirexterrsys\dg$
and a spectral index of
$\Gamma=\fhessoftviindex\pm\fhessoftviindexerr\pm\fhessoftviindexerrsys$,
making it the softest source in our analysis.  Both 3FGL sources are unassociated. 
The FHES source is located within
the Cepheus Bubble, which is a large region ($D\sim 10^{\circ}$)
containing several SFRs~\citep{2000A&A...354..645A,
  2008hsf1.book..136K}.  Although not located within an SFR,
{\fhessoftvi} is in the vicinity of several, the nearest being S140
($\sim 2\dg$ south at the peak of the dust map), NGC 7129
($\sim 2\dg$ north), and NGC 7160 ($\sim 2\dg$ east).  We
note that IC~1396, which is tentatively associated with \fheshardiv,
is an SFR that also lies in the Cepheus Bubble.

\section{Search for extended emission from Extragalactic Sources}
\label{sec:stack}

All of the unassociated extended sources in our analysis are detected
at Galactic latitudes $|b| < 20\dg$, indicating a Galactic origin. 
We now turn to sources at higher latitudes.
These are most probably of extragalactic origin. The most common extragalactic sources observed at \gr\ energies are blazars
(radio-loud AGNs with their jets orientated closely to the line of sight).
As discussed in \Sectionref{sec:intro} and further below, extended emission of blazars
could be caused by secondary \Grs\ from electromagnetic cascades.
Interestingly, some authors found evidence for extended emission around AGNs in analyses of 
\textit{Fermi}-LAT data \citep{Kotelnikov:2014vda,Chen:2014rsa}, while others did not detect any 
significant extension with combined \textit{Fermi}-LAT and \hess\ observations~\citep{hess2014ph}, 
VERITAS observations~\citep{2017ApJ...835..288A}, or
\textit{Fermi}-LAT data only \citep{2011A&A...526A..90N,Ackermann:2013yma}.

Alternatively, extended \gr\ emission from unassociated sources could be due to a DM annihilation signal from sub-halos of the Milky Way. Searches performed on the unassociated 3FGL sources have yielded upper limits~\citep{Buckley:2010vg, Ackermann:2012nb} or are inconclusive~\citep{Mirabal:2012em}.
Recently, however, two possible DM sub-halos were identified~\citep{2016JCAP...05..049B,2017PhRvD..95j2001X}, on which we comment below. 

\subsection{High-latitude Extended Candidates}
\label{sec:hlat_candidates}

To search for sub-threshold extended extragalactic candidates, we
examine a sample of {\NUMHIGHLAT} high-latitude point sources 
listed in the 3FGL and 3FHL with
  $|b| > 20\dg$ containing {\NUMHIGHLATASSOC} AGN and
  {\NUMHIGHLATUNASSOC} unassociated sources.  In composing this sample,
we exclude high-latitude sources detected as extended ({\fhescenai},
{\fhescenaii}, {\fhesmthirtyone}, and {\fhesfornax}) and sources with
pulsar associations.  At these high Galactic latitudes, the intensity of the
Galactic diffuse emission is much lower and 
the effect of systematic uncertainties
from the IEM should be less severe.  Tables
\ref{tab:hlat_ext_sources} and \ref{tab:hlat_halo_sources} present the five
most significant sub-threshold candidates when ordering sources by
$\tsext$ and $\tshalo$.

In the absence of systematic effects, we expect the null distributions
of $\tsext$ and $\tshalo$ to follow a mixture of $\chi^2$
distributions.
However, modeling uncertainties can cause deviations from a purely
statistical distribution.  These uncertainties could include
systematic uncertainties in the IEM and IRFs or the
contribution of unresolved sources.  Rather than model these
systematic effects directly, we derive an empirical model for the null
distribution by fitting a function to the measured distributions of
$\tsext$ and $\tshalo$ in our sample.  We model the tail of the
distributions with an exponential function:

\begin{equation}\label{eqn:tsfunction}
f(\mathrm{TS_{ext/halo}};p_{0}) = \frac{p_{0}}{2}e^{-(\mathrm{TS_{ext/halo}}-9)/2},
\end{equation}
where $p_{0}$ is the $p$-value for $\mathrm{TS}_{ext/halo} > 9$.  
We restrict our fit to
the range of the TS distribution between 4 and 9 where the upper bound
is chosen to avoid biasing the fit with the distribution of genuinely
extended sources that may be detected with $\mathrm{TS_{ext/halo}} > 9$.

\begin{figure*}[ht]
\includegraphics[width=0.48\columnwidth]{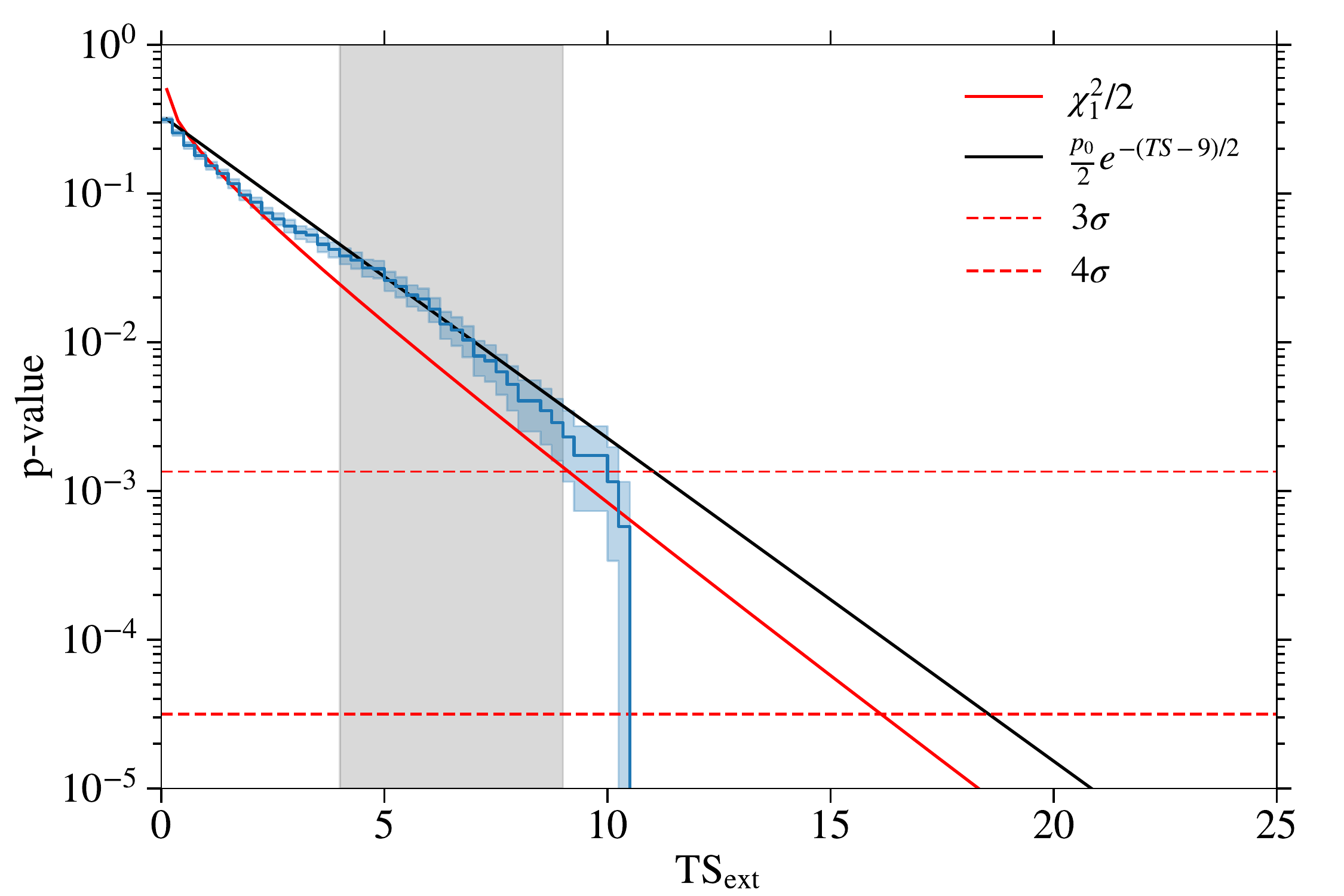}
\includegraphics[width=0.48\columnwidth]{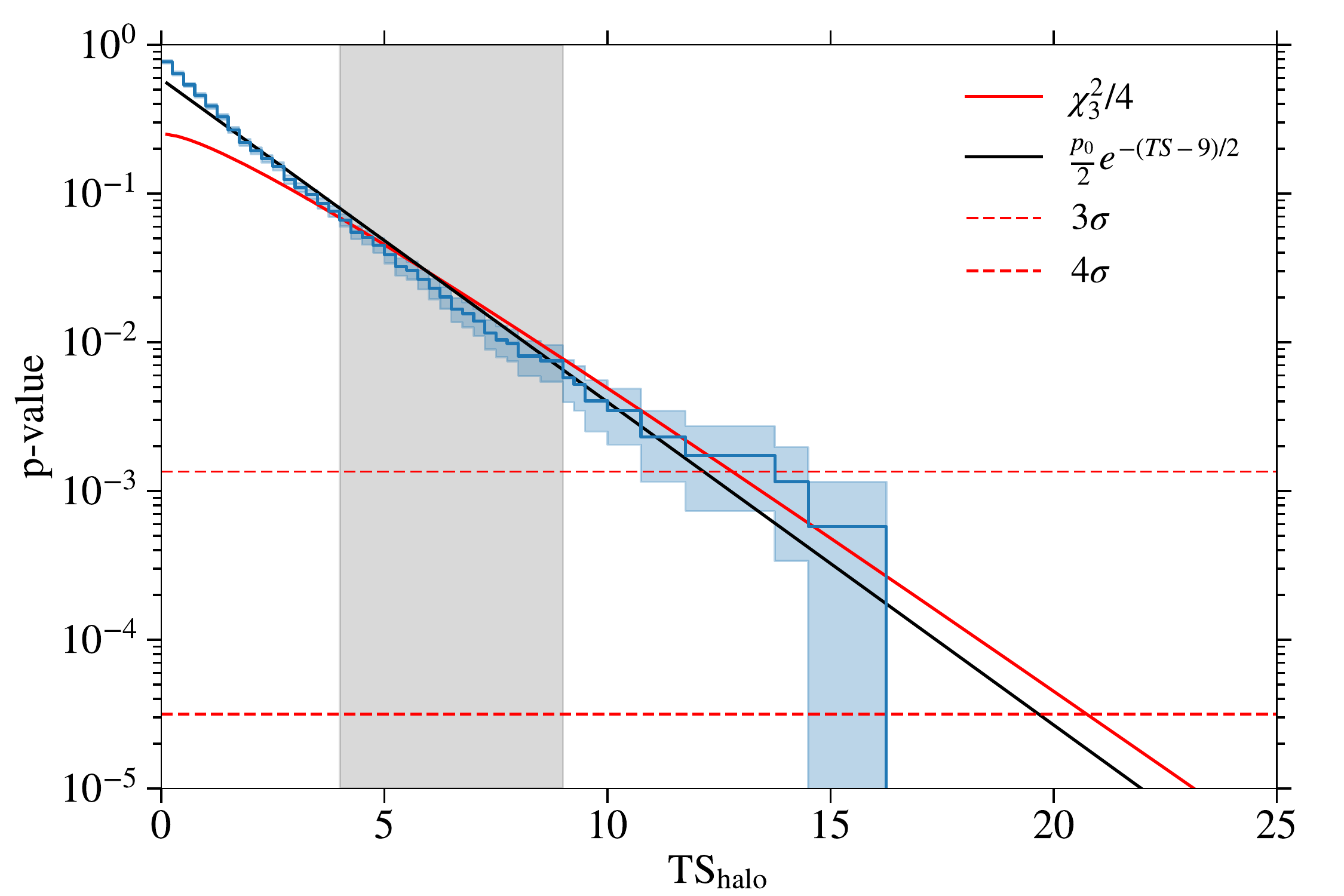}
\caption{Cumulative distributions of $\tsext$ (left) and $\tshalo$
  (right) derived from {\NUMHIGHLAT} FHES sources in the high-latitude
  sample ($|b| > 20\dg$).  The solid black line shows the best fit to
  the distribution derived with parameterization in
  Equation~\ref{eqn:tsfunction}.  The shaded region indicates the
  range of the distribution that was used to fit the parameterization.
  The red curve is the theoretical distribution for a likelihood ratio
  with the number of degrees of freedom of the given
  test. \label{fig:tshist}}
\end{figure*}

In Fig.~\ref{fig:tshist}, we show the cumulative distribution of
$\tsext$ and $\tshalo$ against the best-fit parameterization derived
with Equation~\ref{eqn:tsfunction}.  We find best-fit values of
$p_{0} = 3.51\times10^{-3}$ and $p_{0} = 6.14\times10^{-3}$ for $\tsext$
and $\tshalo$, respectively.  In the case of $\tsext$, we expect the
distribution to follow $\chi^2_{1}/2$ (one bounded degree of freedom),
which has a tail probability $1.35\times10^{-3}$ for
$\mathrm{TS_{ext/halo}} > 9$.  The larger than expected tail probability implies
that IRF or modeling uncertainties are skewing the distribution toward
higher $\tsext$ values.  In the case of $\tshalo$, we expect the
distribution to follow $\chi^2_{3}/4$ (three degrees of freedom with
two bounded parameters).  Here, we observe a better match with the
theoretical expectation.
Using these parameterizations, we derive the values of
$\plocal$ shown in Tables \ref{tab:hlat_ext_sources} and
\ref{tab:hlat_halo_sources}.  

We note that the parameterization of Equation~\ref{eqn:tsfunction}
ignores the potential influence of source properties (e.g. latitude,
flux, or spectral hardness) on the distribution of $\tshalo$ and
$\tsext$.  Distributions of $\tshalo$ and $\tsext$ for different
subpopulations did not show a strong relationship with source
properties, and accounting for these differences in the
parameterization would have had a small effect on the implied local
significance.  When examining all FHES sources, the largest effect was
seen when comparing low- and high-latitude sources ($|b| < 20\dg$ and
$|b| \geq 20\dg$) where the distribution of low-latitude sources was
found to be more skewed toward large TS values.  This effect could be
attributed to genuine sub-threshold sources of Galactic origin or the
influence of residuals in the IEM.  A similar effect was observed in
the $\tsext$ distributions of fainter ($\rm TS < 100$) versus brighter
($\rm TS > 100$) sources.  This behavior is consistent with the
greater susceptibility of faint sources to source confusion due to
their higher spatial density.

The two most significant candidates are {\fheshlathaloi}
with $\tshalo = \hlathaloits~(\hlathaloisigma\sigma)$, and
\fheshlatexti~
with $\tsext = \hlatextits~(\hlatextisigma\sigma)$.  The former 
is associated with Low-synchrotron-frequency peaked (LSP) BL~Lac object {\fheshlathaloiassoc} at
unknown redshift, and the latter is associated with
{\fheshlatextiassoc}, an active galaxy of uncertain type (bcu class)
with unknown redshift.  Both objects are consistent with an
intermediate extension ($0.14\dg$ and $0.5\dg$) that could suggest
confusion with a nearby sub-threshold point source.  Although both
sources have similar spectral indices ($\Gamma \sim 2.4$),
{\fheshlatexti} (the extension candidate) 
is detected with a much lower significance
 than {\fheshlathaloi} (the halo candidate)
($\mathrm{TS}=\fheshlatextits$ versus $\mathrm{TS}=\fheshlathaloits$).

We evaluate the global significance of the two highest TS candidates
by treating every object in the high-latitude sample as an independent
trial, such that the probability of observing an object with
$\plocal>p$ is $1-(1-p)^{N}$, where $N$ is the trials factor
corresponding to the number of objects in our sample.  Using a trial
factor of {\NUMHIGHLAT} for the number of sources in the high-latitude
sample, we derive global significances for {\fheshlatexti} and
{\fheshlathaloi} of $\hlatextisigmaglobal\sigma$ and
$\hlathaloisigmaglobal\sigma$, respectively.  
The most significant
unassociated source is {\fheshlatunassoci}, wchich has
$\tsext=\hlatunassocts~(\hlatunassocsigma\sigma)$ and a global
significance of $\hlatunassocsigmaglobal\sigma$.
We conclude that both
are consistent with being drawn from our parameterizations for the
null distributions of $\tsext$ and $\tshalo$.

\subsection{High-latitude Unassociated Sources}
\label{sec:hlat_unass}

There are {\NUMHIGHLATUNASSOC} unassociated objects in the high-latitude sample. 
With no obvious counterparts at other
wavelengths, the \gr\ emission of these sources could be due to
annihilation of DM particles in DM subhalos of the Milky Way.  Due to
the proximity of such subhalos to Earth, the emission could likely be
extended.  We find no evidence for an individual unassociated source
with statistically significant extension.  

Several other recent works have identified possible DM subhalo
candidates among the sample of unassociated 3FGL sources.
\citet{2016JCAP...05..049B} identify 3FGL~J2212.5$+$0703 as an
unassociated source that shows evidence for spatial extension of
$0.25\dg$ with a statistical significance of $5.1\sigma$, although they
also find that a model with a second nearby point source provides an
equally good fit to the data.  \citet{2017PhRvD..95j2001X} identify
3FGL~J1924.8$-$1034 as another potential DM subhalo candidate and
report a significance for spatial extension of $5.4\sigma$ for a
best-fit extension radius of $0.15\dg$.  Our analysis finds no
evidence for significant extension in either of these sources
($\tsext \sim 0$).  In both cases, a model with two close point sources
is strongly preferred over one with angular extension
($\Deltaext=\fhesdmhaloiextdaic$ and $\Deltaext=\fhesdmhaloiiextdaic$ for 3FGL~J2212.5$+$0703 and
3FGL~J1924.8$-$1034, respectively).

\begin{deluxetable*}{lrrlrrrrrrr}
\tablewidth{0pt}
\tabletypesize{ \scriptsize }
\tablecaption{ \label{tab:hlat_ext_sources} Analysis results for the five most significant high-latitude extended source candidates ($|b| > 20^\circ$). }
\tablehead{
Name                      & $l$ [deg]  & $b$ [deg]  & Association               & Class      & $z$        &$\tsext$   & $p_\mathrm{local}$ & $\rext$ [deg]        & $\Deltaext$
}
\startdata
3FGL~J2142.2$-$2546       &      23.41 &     -47.98 & PMN~J2142$-$2551          & bcu        & \nodata    &      10.6 & $3.0\times10^{-3}$ (2.7$\sigma$) & 0.14 $\pm$ 0.03      & \nodata   \\ 
3FGL~J0002.2$-$4152       &     334.25 &     -71.99 & 1RXS~J000135.5$-$415519   & bcu        & \nodata    &      10.4 & $3.3\times10^{-3}$ (2.7$\sigma$) & 0.12 $\pm$ 0.03      & \nodata   \\ 
3FGL~J2103.9$-$6233       &     332.69 &     -38.95 & PMN~J2103$-$6232          & bcu        & \nodata    &      10.1 & $3.7\times10^{-3}$ (2.7$\sigma$) & 0.07 $\pm$ 0.01      & \nodata   \\ 
3FGL~J0107.0$-$1208       &     137.64 &     -74.63 & PMN~J0107$-$1211          & bcu        & \nodata    &       9.5 & $4.8\times10^{-3}$ (2.6$\sigma$) & 0.64 $\pm$ 0.12      & -1.9      \\ 
3FGL~J0434.3$-$1411c      &     210.82 &     -36.55 & CO~cloud                  &            & \nodata    &       9.2 & $5.4\times10^{-3}$ (2.5$\sigma$) & 0.77 $\pm$ 0.17      & -0.1      \\ 
\enddata

{\footnotesize \tablecomments{ Here, $\Deltaext$ is the difference in the Akaike Information Criterion
between the best-fit extended source model and a model with one
additional point source (see Equation \ref{eqn:akaike}).   }}

\end{deluxetable*}

\begin{deluxetable*}{lrrlrrrrrrr}
\tablewidth{0pt}
\tabletypesize{ \scriptsize }
\tablecaption{ \label{tab:hlat_halo_sources} Analysis results for the five most significant high-latitude halo candidates ($|b| > 20^\circ$). }
\tablehead{
Name                      & $l$ [deg]  & $b$ [deg]  & Association               & Class      & $z$        &$\tshalo$  & $p_\mathrm{local}$ & $\rhalo$ [deg]       & $\gammahalo$         & $\Deltahalo$
}
\startdata
3FGL~J0850.0$+$4855       &     170.47 &      39.27 & GB6~J0850$+$4855          & bll        & \nodata    &16.3       & $1.2 \times 10^{-4}$ (3.7$\sigma$) & 0.52 $\pm$ 0.13      & 2.15 $\pm$ 0.22      & -7.8      \\ 
3FHL~J0901.5$+$6712       &     147.19 &      37.28 & 1RXS~J090140.8$+$671158   &            & \nodata    &14.7       & $2.7 \times 10^{-4}$ (3.5$\sigma$) & 0.78 $\pm$ 0.25      & 4.00 $\pm$ 0.28      & \nodata   \\ 
3FGL~J0107.0$-$1208       &     137.64 &     -74.63 & PMN~J0107$-$1211          & bcu        & \nodata    &13.8       & $4.4 \times 10^{-4}$ (3.3$\sigma$) & 0.75 $\pm$ 0.15      & 2.10 $\pm$ 0.21      & -2.5      \\ 
3FGL~J0626.6$-$4259       &     251.05 &     -22.44 & 1RXS~J062635.9$-$425810   & bcu        & \nodata    &10.8       & $1.9 \times 10^{-3}$ (2.9$\sigma$) & 0.66 $\pm$ 0.21      & 2.13 $\pm$ 0.22      & -1.1      \\ 
3FGL~J2318.6$+$1912       &      94.49 &     -38.40 & TXS~2315$+$189            & bcu        & \nodata    &10.8       & $2.0 \times 10^{-3}$ (2.9$\sigma$) & 0.57 $\pm$ 0.18      & 2.62 $\pm$ 0.32      & -5.6      \\ 
\enddata

{\footnotesize \tablecomments{ Here, $\Deltahalo$ is the difference in the Akaike Information Criterion
between the best-fit halo model and a model with one additional point
source (see Equation \ref{eqn:akaike}). }}

\end{deluxetable*}

\subsection{Individual TeV-selected AGNs}

AGNs with strong TeV emission are among the best candidates for
secondary cascade emission because the amplitude of the cascade
component is expected to be proportional to the fraction of the
primary emission that is absorbed by the EBL.  We consider the sample
of 38~TeV-selected AGNs compiled in \citet{2015ApJ...812...60B}, which
have all been detected above $\sim100\,$GeV.  The cascade component could
appear
as an extended component superimposed on the point-like emission
of the AGN.

Table \ref{tab:tev_sources} shows the analysis results for all objects
in the TeV-selected AGN sample.  No source shows evidence for
extension with $\tsext > 9$, and upper limits on the angular extension
lie between $0.02\dg$ and $0.09\dg$.  PKS~1510$-$08 shows
a hint of a halo component with
$\tshalo = \tevits~(\tevisigma \sigma)$ where the significance
is quoted prior to trial penalization.  However, the halo
model is only marginally preferred over a model with an additional
point source ($\Deltahalo \sim 0$).  Using the model for the null
distribution of $\tshalo$ derived in \Sectionref{sec:hlat_candidates}
and a trials factor of 38, we find a global significance for halo
emission associated with PKS~1510$-$08 of $\tevisigmaglobal\sigma$.

\begin{deluxetable*}{lrrlrrrrrrr}
\tablewidth{0pt}
\tabletypesize{ \footnotesize }
\tablecaption{ \label{tab:tev_sources} Analysis results for the sample of 38 TeV-selected AGN.  }
\tablehead{
Name                      & $l$        & $b$        & Association               & $z$      &Class      & Energy Flux          &$\tsext$   & $\rext$              & $\tshalo$  & $\Deltahalo$\\ 
                          & [deg]      & [deg]      &                           &          &           & [eV cm$^{-2}$ s$^{-1}$] &           & [deg]                &            &           
}
\startdata
3FGL~J0013.9$-$1853       &      74.53 &     -78.09 & RBS~0030                  &    0.094 &bll        &        1.1 $\pm$        0.4 &       0.0 & $< 0.05$             &        0.1 & \nodata   \\
3FGL~J0152.6$+$0148       &     152.38 &     -57.54 & PMN~J0152$+$0146          &    0.080 &bll        &        4.0 $\pm$        0.8 &       0.0 & $< 0.04$             &        0.8 & \nodata   \\
3FGL~J0222.6$+$4301       &     140.15 &     -16.77 & 3C~66A                    &    0.444 &BLL        &       42.1 $\pm$        2.5 &       0.1 & $< 0.02$             &        0.4 & \nodata   \\
3FGL~J0232.8$+$2016       &     152.94 &     -36.59 & 1ES~0229$+$200            &    0.139 &bll        &        2.0 $\pm$        0.6 &       0.0 & $< 0.05$             &        0.9 & \nodata   \\
3FGL~J0303.4$-$2407       &     214.62 &     -60.18 & PKS~0301$-$243            &    0.260 &BLL        &       18.4 $\pm$        1.9 &       0.0 & $< 0.02$             &        0.5 & \nodata   \\
3FGL~J0316.6$+$4119       &     150.19 &     -13.71 & IC~310                    &    0.019 &rdg        &        1.6 $\pm$        0.6 &       0.0 & $< 0.05$             &        1.9 & \nodata   \\
3FGL~J0319.8$+$1847       &     165.10 &     -31.70 & RBS~0413                  &    0.190 &bll        &        4.1 $\pm$        0.9 &       7.7 & $< 0.09$             &        6.7 & \nodata   \\
3FGL~J0349.2$-$1158       &     201.93 &     -45.71 & 1ES~0347$-$121            &    0.185 &bll        &        2.7 $\pm$        0.8 &       1.3 & $< 0.08$             &        1.2 & \nodata   \\
3FGL~J0416.8$+$0104       &     191.81 &     -33.16 & 1ES~0414$+$009            &    0.287 &bll        &        3.7 $\pm$        0.9 &       0.0 & $< 0.05$             &        0.0 & \nodata   \\
3FGL~J0449.4$-$4350       &     248.80 &     -39.92 & PKS~0447$-$439            &    0.205 &bll        &       37.2 $\pm$        2.6 &       0.0 & $< 0.03$             &        0.2 & \nodata   \\
3FGL~J0521.7$+$2113       &     183.60 &      -8.71 & TXS~0518$+$211            &    0.108 &bll        &       39.6 $\pm$        2.7 &       0.4 & $< 0.03$             &        3.4 & \nodata   \\
3FGL~J0648.8$+$1516       &     198.98 &       6.33 & RX~J0648.7$+$1516         &    0.179 &bll        &       11.8 $\pm$        1.6 &       0.0 & $< 0.04$             &        0.3 & \nodata   \\
3FGL~J0710.3$+$5908       &     157.41 &      25.42 & 1H~0658$+$595             &    0.125 &bll        &        4.8 $\pm$        0.9 &       0.0 & $< 0.02$             &        0.0 & \nodata   \\
3FGL~J0721.9$+$7120       &     143.98 &      28.02 & S5~0716$+$71              &    0.127 &BLL        &       34.0 $\pm$        1.8 &       0.0 & $< 0.03$             &        4.1 & \nodata   \\
3FGL~J0809.8$+$5218       &     166.25 &      32.91 & 1ES~0806$+$524            &    0.138 &bll        &       15.0 $\pm$        1.5 &       0.0 & $< 0.03$             &        0.3 & \nodata   \\
3FGL~J1010.2$-$3120       &     266.91 &      20.05 & 1RXS~J101015.9$-$311909   &    0.143 &bll        &        5.8 $\pm$        1.1 &       0.0 & $< 0.02$             &        0.0 & \nodata   \\
3FGL~J1015.0$+$4925       &     165.53 &      52.71 & 1H~1013$+$498             &    0.212 &bll        &       34.5 $\pm$        2.4 &       0.0 & $< 0.02$             &        0.5 & \nodata   \\
3FGL~J1103.5$-$2329       &     273.18 &      33.08 & 1ES~1101$-$232            &    0.186 &bll        &        3.8 $\pm$        0.9 &       0.1 & $< 0.07$             &        2.8 & \nodata   \\
3FGL~J1104.4$+$3812       &     179.83 &      65.04 & Mkn~421                   &    0.031 &BLL        &      222.3 $\pm$        6.5 &       2.2 & $< 0.02$             &        1.6 & \nodata   \\
3FGL~J1136.6$+$7009       &     131.90 &      45.65 & Mkn~180                   &    0.045 &bll        &        6.1 $\pm$        0.9 &       0.0 & $< 0.02$             &        0.5 & \nodata   \\
3FGL~J1217.8$+$3007       &     188.85 &      82.06 & 1ES~1215$+$303            &    0.130 &bll        &       20.5 $\pm$        1.9 &       0.0 & $< 0.03$             &        2.0 & \nodata   \\
3FGL~J1221.3$+$3010       &     186.40 &      82.74 & PG~1218$+$304             &    0.182 &bll        &       22.9 $\pm$        2.2 &       3.1 & $< 0.04$             &        3.1 & \nodata   \\
3FGL~J1221.4$+$2814       &     201.69 &      83.29 & W~Comae                   &    0.103 &bll        &        5.1 $\pm$        0.9 &       5.9 & $< 0.07$             &        4.1 & \nodata   \\
3FGL~J1224.9$+$2122       &     255.06 &      81.66 & 4C~$+$21.35               &    0.435 &FSRQ       &       13.5 $\pm$        1.3 &       1.1 & $< 0.03$             &        1.6 & \nodata   \\
3FGL~J1256.1$-$0547       &     305.10 &      57.06 & 3C~279                    &    0.536 &FSRQ       &       11.4 $\pm$        1.1 &       5.5 & $< 0.05$             &        4.3 & \nodata   \\
3FGL~J1314.7$-$4237       &     307.56 &      20.05 & MS~13121$-$4221           & \nodata  &bcu        &        2.8 $\pm$        0.8 &       0.0 & $< 0.03$             &        0.8 & \nodata   \\
3FGL~J1427.0$+$2347       &      29.49 &      68.20 & PKS~1424$+$240            & \nodata  &BLL        &       45.4 $\pm$        2.8 &       4.1 & $< 0.03$             &        3.2 & \nodata   \\
3FGL~J1428.5$+$4240       &      77.50 &      64.90 & H~1426$+$428              &    0.129 &bll        &        6.2 $\pm$        1.1 &       2.4 & $< 0.04$             &        1.0 & \nodata   \\
3FGL~J1512.8$-$0906       &     351.28 &      40.14 & PKS~1510$-$08             &    0.360 &FSRQ       &       20.6 $\pm$        1.5 &       2.2 & $< 0.04$             &        9.4 & -1.8      \\
3FGL~J1555.7$+$1111       &      21.91 &      43.96 & PG~1553$+$113             & \nodata  &BLL        &       84.8 $\pm$        3.9 &       0.0 & $< 0.02$             &        5.3 & \nodata   \\
3FGL~J1653.9$+$3945       &      63.60 &      38.85 & Mkn~501                   &    0.034 &BLL        &       78.5 $\pm$        3.7 &       3.7 & $< 0.03$             &        3.5 & \nodata   \\
3FGL~J2000.0$+$6509       &      98.01 &      17.67 & 1ES~1959$+$650            &    0.047 &bll        &       36.8 $\pm$        2.3 &       3.6 & $< 0.03$             &        3.4 & \nodata   \\
3FGL~J2009.3$-$4849       &     350.39 &     -32.60 & PKS~2005$-$489            &    0.071 &BLL        &       13.8 $\pm$        1.6 &       0.1 & $< 0.03$             &        1.7 & \nodata   \\
3FGL~J2158.8$-$3013       &      17.73 &     -52.24 & PKS~2155$-$304            &    0.116 &bll        &       78.1 $\pm$        3.9 &       4.7 & $< 0.03$             &        2.1 & \nodata   \\
3FGL~J2202.7$+$4217       &      92.60 &     -10.45 & BL~Lacertae               &    0.069 &BLL        &       15.3 $\pm$        1.3 &       1.9 & $< 0.04$             &        6.7 & \nodata   \\
3FGL~J2250.1$+$3825       &      98.25 &     -18.56 & B3~2247$+$381             &    0.119 &bll        &        5.6 $\pm$        1.0 &       0.6 & $< 0.04$             &        3.3 & \nodata   \\
3FGL~J2347.0$+$5142       &     112.89 &      -9.91 & 1ES~2344$+$514            &    0.044 &bll        &       13.9 $\pm$        1.4 &       0.0 & $< 0.02$             &        0.9 & \nodata   \\
3FGL~J2359.3$-$3038       &      12.82 &     -78.03 & H~2356$-$309              &    0.165 &bll        &        2.6 $\pm$        0.7 &       0.0 & $< 0.03$             &        0.5 & \nodata   \\
\enddata

{\footnotesize \tablecomments{ The class column gives the class designator following the
convention of the 3FGL: bll = BL Lac type of blazar, rdg = radio
galaxy, FSRQ = flat spectrum radio quasar, bcu = active galaxy of
uncertain type. As in the 3FGL, designations shown in capital letters are firm identifications; lower-case letters indicate associations.
Energy flux is integrated from the spectral model
between 10~GeV and 1~TeV. }}

\end{deluxetable*}

\subsection{Stacking analysis}\label{subsec:stack}

Interest in stacking {\it Fermi}-LAT data to search for IGMF induced pair halos  
was partly triggered by the initial study of~\citet{2010ApJ...722L..39A}, who found a hint of extension in the stacked images of 170~AGNs observed over 11~months. 
In their analysis, AGNs detected above 10~GeV at large Galactic latitudes, $|b|>10\dg$, were compared to an early version of the PSF based on ground-based beam tests as well as Monte-Carlo simulations. 
A comparison to the profile of the Crab Nebula by~\citet{2011A&A...526A..90N} nonetheless suggested an instrumental effect. 
This was further investigated by the {\it Fermi}-LAT collaboration~\citep{Ackermann:2013yma} with an updated PSF based on on-orbit data, and using bright pulsars (Vela and Geminga) as control point-like sources. 
The AGN and pulsar extensions relative to the PSF proved to be consistent with zero in the 3-30~GeV energy range, where both samples contain ample statistics. 
More recently,~\citet{Chen:2014rsa} selected, a priori,  a subset of 24 nearby high-synchrotron-peaked BL Lacs (HSP). It was searched for potential IGMF-induced extension  and compared with reference samples of 26 flat-spectrum radio quasars (FSRQs), as well as the Geminga and Crab pulsars. 
This yielded a $2.3\sigma$ indication for extension in the HSP sample around 1~GeV.

We search for extended emission by stacking 3FHL and 3FGL samples of AGN  from which significant cascade emission could be expected. The considered samples are:
\begin{enumerate}
\item\textit{HSPs.} We select HSPs with a synchrotron 
peak $\nu_{\rm sync} > 10^{15}\,$Hz. Such sources are promising emitters of very-high-energy $\gamma$ rays necessary to 
induce the cascade. This selection leaves us with 299 sources.
\item\textit{Non-variable HSPs.} In this sub-sample of the HSP sample, we further demand that the variability index in the 
3FGL to be smaller than $100$, which corresponds to a significance of less than 4.2$\,\sigma$ that the source flux is time-variable. 
This reduces the sample to 258 sources. 
Cascade photons can arrive with a significant 
time delay~\citep{plaga1995}, and thus we exclude sources whose average flux might 
be dominated by strong flaring activity from which the cascade photons might not yet have reached Earth.
\item\textit{Close-by HSPs with $z< 0.2$.} We also limit the first sample to close-by AGNs, as the cascade emission 
leads to a broader angular emission profile from closer sources~\citep{2010ApJ...722L..39A}. 
This additional cut reduces the sample size to 72 objects.
\item\textit{Extreme HSPs.} As a further subset of the HSP sample, we only consider extreme HSP (XHSPs) with 
$\nu_{\rm sync} > 10^{17}\,$Hz that additionally show a large ratio of X-ray to radio flux, $F_{\rm X} / F_{\rm R} > 10^4$. 
The radio and X-ray fluxes are extracted from the 3LAC catalog. 
This criterion was identified in~\citet{Bonnoli:2015yia} as a promising tracer of sources that have a hard spectral index 
and thus are likely emitters of $\gamma$ rays beyond multiple TeV.
In total, there are 24 sources in this sample.
\end{enumerate}
We also consider the same sample as in \citet{Chen:2014rsa} (24 sources) and all TeV detected AGN listed in \citet{2015ApJ...812...60B} (38 sources).

Additionally, we define two control samples for which we do not expect to find any evidence for extension:
\begin{enumerate}
\item\textit{Low synchrotron peak blazars (LSPs).} 
This control sample contains a subset of low synchrotron peak blazars (FSRQs, BL Lacs, and blazars of unknown type with 
$\nu_{\rm sync} < 10^{14}\,$Hz) with $\nu_{\rm sync} < 10^{13}\,$Hz.
 We infer from predictions of the blazar sequence~\citep{Ghisellini:2017ico} that such sources are unlikely to emit a significant amount of \Grs\ at the highest energies.
The sample consists of 246 sources.
\item\textit{Pulsars.} As a second control sample, we consider a population of pulsars.
We exclude the pulsars CTA1 and the Crab, for which we have identified the PWN as extended or possibly extended,  as well as sources in the 3FGL with latitudes $|b|<$5$^{\circ}$.
This leaves us with 89 pulsars.
\end{enumerate}

For each sample, we sum the likelihoods of individual sources, assuming common parameters for all sources. 
In the case of the extension hypothesis, the common parameter is $\rext$ and we use 2D Gaussian spatial profiles (no disks) for all sources. 
For the halo, the common parameters are $\rhalo$, $\gammahalo$ and the ratio $f_{\rm halo}$ between the point source and halo energy flux integrated between $1\,$GeV and $1\,$TeV. 
We find the best-fit parameters for $\rext$ ($\rhalo$, $\gammahalo$, and $f_{\rm halo}$) for the extension (halo) hypothesis from the summed likelihoods.
For the extension hypothesis, we repeat the procedure also with the likelihoods obtained from the bracketing IRFs. 
In the case of a non-detection of extension or halo, we report upper limits on $\rext$ and $f_\mathrm{halo}$, respectively. For the halo case, we do so by fixing $\gammahalo = 2$, which is the spectral shape generally expected for the cascade \citep[e.g.][]{protheroe1993}, and fixing $\rhalo$ to values of $0.1\dg,0.316\dg$, and $1.0\dg$. 
Thus, we are left with one free parameter each for both extension and halo.  
We calculate one-sided 95\,\% confidence limits on these quantities by stepping over them and profiling over the parameters of the other sources in the ROI until the summed likelihood changes by $2\Delta\ln\mathcal{L} = 2.71$.

We present the stacked $\rm{TS}$ values for a halo and extended emission for each sample in Table~\ref{tab:stack_samples}, along with the combined best-fit values 
of $\rext$ and the limit values for $f_{\rm Halo}$, as well as the number of sources in each sample, $\mathrm{N}_\mathrm{obj}$.

\begin{deluxetable*}{lrrrrrrr}
\tablewidth{0pt}
\tabletypesize{ \small }
\tablecaption{  \label{tab:stack_samples} Analysis results for stacked object samples testing hypotheses of extension and halo emission. }
\tablehead{
  &&&&&\multicolumn{3}{c}{$f_\mathrm{halo}$ ($\gammahalo=2$)} \\ \cline{6-8}
Name                      & N$_\mathrm{obj}$ & $\tsext$   & $\rext$ [deg]        &$\tshalo$  & $\rhalo=0.1\dg$      & $\rhalo=0.316\dg$    & $\rhalo=1.0\dg$     
}
\startdata
HSPs                           &        300 &       30.5 (1.1) & 0.015 $\pm$ 0.001 $\pm$ 0.013 &       0.0 & $< 0.032$            & $< 0.005$            & $< 0.005$            \\ 
Non-Variable HSPs              &        258 &       24.3 (0.3) & 0.017 $\pm$ 0.002 $\pm$ 0.013 &       2.0 & $< 0.040$            & $< 0.006$            & $< 0.008$            \\ 
HSPs ($z < 0.2$)               &         72 &       15.6 (0.4) & 0.016 $\pm$ 0.002 $\pm$ 0.013 &       2.2 & $< 0.017$            & $< 0.006$            & $< 0.004$            \\ 
XHSPs                          &         24 &       13.1 (0.1) & 0.018 $\pm$ 0.003 $\pm$ 0.014 &       4.4 & $< 0.063$            & $< 0.015$            & $< 0.009$            \\ 
Sample of \citet{Chen:2014rsa} &         24 &        0.6 (0.1) & $< 0.030$            &       0.1 & $< 0.043$            & $< 0.008$            & $< 0.013$            \\ 
TeV-Selected AGN               &         38 &       18.4 (0.7) & 0.015 $\pm$ 0.002 $\pm$ 0.013 &       0.0 & $< 0.040$            & $< 0.013$            & $< 0.010$            \\ 
\hline
LSPs                           &        247 &        1.5 (0.2) & $< 0.040$            &       1.8 & $< 0.004$            & $< 0.008$            & $< 0.008$            \\ 
PSRs                           &         88 &       26.3 (0.2) & 0.030 $\pm$ 0.003 $\pm$ 0.027 &       3.8 & $< 0.004$            & $< 0.004$            & $< 0.006$            \\ 
\enddata

{\footnotesize \tablecomments{  The last two samples are control samples for which we do not
expect to find any intrinsic extension. In the second column,
$\mathrm{N}_\mathrm{obj}$ denotes the number of sources in the sample.
The value of $\tsext$ in parentheses is the minimum obtained with the
two bracketing models of the PSF (see Section~\ref{sec:systematics}).
Here, $\rext$ and $\rhalo$ are the best-fit intrinsic 68\% containment radii
obtained when fitting the sample with a Gaussian morphology and a
Gaussian halo component, respectively.  The $\rext$ column includes
the statistical and systematic (IRF) errors. We provide 90\%
C.L. limits on the Halo flux ratio $f_{\rm Halo}$ because the halo is
not detected in any sample. For easier comparison between the samples,
the limits are provided for a fixed spectral index $\gammahalo = 2$
and different values of the halo extension.   }}

\end{deluxetable*}

We find the highest $\tsext$ values for the samples 
encompassing all HSPs and non-variable HSPs with best-fit values 
of $\rext = 0.015\dg\pm0.001\dg\pm0.013\dg$ and $\rext = 0.017\dg\pm0.002\dg\pm0.013\dg$, respectively. 
The second uncertainty represents half the difference between the best-fit values when the different bracketing PSFs are used to estimate the systematic uncertainty (cf. Section~\ref{sec:systematics}). 
The extension is found to be consistent with uncertainties 
in the PSF.\footnote{One should note, however, that the bracketing PSFs 
were derived by considering samples of blazars that were assumed to be pointlike;
see \url{https://fermi.gsfc.nasa.gov/ssc/data/analysis/LAT_caveats.html}}
The interpretation is further supported by the fact that samples containing 
bright sources show larger $\tsext$ values. 
Indeed, the pulsar sample yields $\tsext = 26.4$, indicating again that 
the high $\tsext$ value of the HSP sample is connected to the systematic uncertainty in 
the PSF modeling. 
In terms of a pair halo, the sources should not only be bright but 
also have a hard spectrum that extends well into the TeV range. 
However, the XHSP and TeV-selected AGN sample show lower $\tsext$ values 
than the ``pure'' HSP samples.

Similarly, we do not find any evidence for halo emission in any of the stacked 
samples. 
In contrast to the extension model, none of the control samples show evidence for a halo component.
We did not compute systematic uncertainties in the halo case because---in contrast to the extension case---the likelihood depends on $\rhalo$ and $\gammahalo$,  which would make it extremely computationally expensive. 
We expect these systematic effects to be subdominant
in the halo-hypothesis case, compared to the statistical uncertainties. The reasons are the small flux of the halo 
component and the fact that most sources are located at high Galactic latitudes where uncertainties 
on the diffuse emission are less pronounced.

We cannot confirm the evidence for halo emission reported by \citet{Chen:2014rsa}. The stacked analysis for the low-redshift TeV blazars used in their sample 
results in the lowest values for $\tsext$ and $\tshalo$ of all samples considered.

\section{Limits on the intergalactic magnetic field}\label{sec:results}

With no clear evidence for an extension 
of individual blazars or stacked samples of BL Lac objects, 
we use the FHES to derive constraints 
on the coherence length, $\lambda$, and field strength, $B$, of the IGMF.
We use both spectral and spatial information from the catalog, as well 
as spectra from imaging air Cherenkov telescopes (IACTs),
to derive these constraints. 
A significant source detection at very high \gr\ energies with IACTs is essential for this study in order to probe the intrinsic spectrum in the regime where it is strongly affected by EBL absorption.

\subsection{Source Selection}
\label{sec:tevsrc}
We again use the list of VHE-emitting sources compiled in~\cite{2015ApJ...812...60B} to select sources 
detected both with the \emph{Fermi} LAT and IACTs.
We set aside five objects with uncertain redshifts: S5~0716+714, 3C~66A, PKS~0447-439, PG~1553+113, and PKS~1424+240. 
We further limit the sample to IACT spectra with a well-measured EBL cutoff, i.e. significant spectral points up to an optical depth $\tau > 2$, assuming the 
EBL model of~\citet{Dominguez:2010bv}. 
In this way, we guarantee that we have sufficient statistics in the very high energy 
part of the spectra, which is most important to model the contribution from the 
cascade.
Moreover, we exclude sources that show evidence for variability beyond the 4.2$\,\sigma$ level, corresponding to a variability index larger than $100$ in the 3FGL catalog. 
In this way, we also exclude sources whose flux level is dominated by flaring events and whose quiescent state is much lower than the average flux level. 
In the case that the same emission mechanism is responsible for \Grs\ at energies probed with the \fermi\ LAT and IACTs,
this cut implies that the IACT spectra are also a good representation for the average flux level. 
We further discard H\,1426+428 because the \textit{Fermi}-LAT measurement does not match that recorded with HEGRA during their 
2000 observation, which implies that the source was in a different emission state in the past.
We note that the HEGRA Collaboration reported two spectra for H~1426+428, one corresponding to observations in 1999-2000 and one to observations 
in 2002 \citep{2003A&A...403..523A}. These two spectra show a flux mismatch by a factor of 2.5, similar to that observed in X-rays, indicative of source variability. 
H~1426+428 has been detected again with VERITAS~\citep{2017ApJ...835..288A}, but  the spectrum is not yet published.

This selection leaves us with 9 BL~Lac objects, for which we have 15 IACT spectra for the IGMF analysis, as shown in Table~\ref{tab:VHEdata}. All objects listed in this table are high frequency peak blazars with redshifts ranging from 0.105 to 0.287. Most of these sources
have already been used in the past to set constraints on the IGMF~\citep[e.g.][]{neronov2010,dermer2011,tavecchio2011,2011A&A...529A.144T,arlen2014,hess2014ph,finke2015}.

\begin{deluxetable*}{lcccccccr}
\tablewidth{0pt}
\tabletypesize{ \small }
\tablecaption{\label{tab:VHEdata}Spectra from ground-based instruments used in the IGMF study ordered by increasing redshift.}
\tablehead{
Source &  $z$ & R.A. [$^{\rm o}$] & Dec. [$^{\rm o}$] & 3FGL name & 3FGL var. index & Experiment  &  Obs. Period  &  Ref. \\
}
\startdata
	1ES~1312-423  & 0.105 &	198.76	&	-42.61	&	J1314.7-4237 	&	45.0 & 	H.E.S.S. & 	2004-2010 &   (1)\\
	RGB~J0710+591  & 0.125 &	107.63	&	59.14	&	J0710.3+5908	&	55.5 & 	VERITAS & 	2008-2009 &   (2)\\
	1ES~0229+200  & 0.14 &	38.20	&	20.29	&	J0232.8+2016	&	49.2 & 	H.E.S.S. & 	2005-2006 &  (3) \\
	& & &	&   &   & 	VERITAS & 	2009-2012 &  (4) \\
	1RXS J101015.9-311909& 0.143& 152.57 &	-31.32& J1010.2-3120 &	86.3&	H.E.S.S.&	2006-2010	& (5) \\
	& & &	&   &   & 	VERITAS & 	2009-2012 &   (6) \\
	H~2356-309  & 0.165 &	359.78	&	-30.63	&	J2359.3-3038 	&	41.0 & 	H.E.S.S. & 	2004 &  (7) \\
	& & &	&   &   & 		H.E.S.S. & 	2005 &  (8) \\
	& & &	&   &   & 		H.E.S.S. & 	2006 &  (9)\\
	1ES 1218+304 &0.182& 185.34 &	30.18& J1221.3+3010 & 92.5 &	VERITAS&	2007	& (10) \\
	 & & & & & &	VERITAS&	2008-2009	& (11)\\
	1ES~1101-232  & 0.186 &	165.91	&	-23.49	&	J1103.5-2329	&	36.5 & 	H.E.S.S. & 	2004-2005 &   (12) \\
	1ES~0347-121  & 0.185 &	57.35	&	-11.99	&	J0349.2-1158	&	44.3  & 	H.E.S.S. & 	2006  &   (13) \\
	1ES~0414+009  & 0.287&	64.22	&	1.09	&	J0416.8+0104	&	55.8  & 	H.E.S.S. & 	2005-2009 &   (14) \\
	& & &	&   &   & 	VERITAS & 	2008-2011 &  (15) \\ 
\enddata
{\footnotesize \tablecomments{ From left to right: source name, redshift, right ascension and declination (J2000), name of the source, and variability index from the 3FGL catalog, experiment, observation period, and reference for the VHE spectra.}
\tablereferences{
(1)~\citet{2013MNRAS.434.1889H},
(2)~\citet{2010ApJ...715L..49A},
(3)~\citet{2007AA...475L...9A},
(4)~\citet{2014ApJ...782...13A},
(5)~\citet{2012AA...542A..94H},
(6)~\citet{2014ApJ...782...13A},
(7)~\citet{2010AA...516A..56H},
(8)~\citet{2010AA...516A..56H},
(9)~\citet{2010AA...516A..56H},
(10)~\citet{2009ApJ...695.1370A},
(11)~\citet{2010ApJ...709L.163A},
(12)~\citet{2007AA...470..475A},
(13)~\citet{2007AA...473L..25A},
(14)~\citet{2012AA...538A.103H},
(15)~\citet{2012ApJ...755..118A}.
}}
\end{deluxetable*}

\subsection{Modeling of the Cascade Emission}\label{sec:simulations}

In order to model the expected cascade emission from these sources, we generate a library of cascade templates for different IGMF configurations
using the ELMAG Monte Carlo code \citep{2012CoPhC.183.1036K}.
This open-source code computes the 
observed photon flux (primary and cascade photons) 
by sampling an input intrinsic \gr\ spectrum of a source assumed to be viewed on-axis, i.e. $\theta_{\mathrm{obs}} = 0^\circ$,
using a weighted sampling procedure~\citep[see][for details]{2012CoPhC.183.1036K}. 
Interactions with the CMB and EBL are taken into account, 
and we choose to trace all secondary particles with an 
energy $\epsilon \geq \epsilon_\mathrm{thr}, $ where $\epsilon_\mathrm{thr} = 100\,$MeV.
Energy losses due to inverse-Compton scattering and synchrotron radiation
are integrated out for $\epsilon < \epsilon_\mathrm{thr}$. 
The energy $\epsilon$, observation angle $\vartheta$, and time delay $\Delta t$ for
the final \Grs\ reaching the observer are recorded in a multidimensional 
histogram. 
 
ELMAG adopts a simplified description of the IGMF,
namely that its field strength is constant in cells that have a   
size equal to the coherence length. 
The $e^+e^-$~pairs are deflected in a coherent manner in each cell. 
ELMAG uses the small-angle approximation~\citep[see][for details]{2012CoPhC.183.1036K}, i.e.
the total deflection angle $\beta$ can be
 accumulated following a random-walk approximation, such
 that 
 the emission angle $\alpha$ is related to the observation angle $\vartheta$ through $\alpha = \beta - \vartheta$~\citep[see Fig.~1 in][]{2009ApJ...703.1078D}. 
If the total squared deflection angle exceeds $\pi^2 / 4$,
the deflection angle $\beta$ of the cascade photons is randomized. 
This occurs when~\citep{2009PhRvD..80l3012N,meyer2012}

\begin{equation}
B / \mathrm{G}\gtrsim 
\begin{cases}
2\times10^{-15} (1+z_r)^4(\epsilon / \mathrm{GeV})
& \lambda\gg1\,\mathrm{kpc},\\
8\times10^{-15} (1+z_r)^4(\epsilon / \mathrm{GeV})^{3/4}(\lambda/\mathrm{kpc})^{-1/2} & \lambda\lesssim\mathrm{\,kpc},
\end{cases}\label{eq:small-angle}
\end{equation}
where $B$ is the IGMF strength  at $z = 0$, and $z_r$ is the redshift where the pair production takes place, producing secondary $\gamma$ rays of energy $\epsilon$. We note that this formula includes 
the $(1+z)^3$ dependence of the IGMF, which is neglected in the ELMAG implementation used here.
Importantly, if $\beta - \vartheta > \theta_\mathrm{jet}$, where $\theta_\mathrm{jet}$ is the jet opening angle, the small-angle approximation breaks down and the photon is dismissed. 
Taking $\epsilon = 1\,$GeV and $z_r \approx z$, more and more photons are randomized---and consequently, are likely to be dismissed for deflection angles larger than $\theta_{\rm jet}$ for magnetic fields larger than $\sim 10^{-15}\,\mathrm{G}$.
Because $\theta_{\mathrm{obs}} = 0^\circ$ is assumed, the simulated cascades are symmetric in surface brightness and do not show the elongated features seen in 3D Monte Carlo simulations~\citep{2010ApJ...719L.130N}.
As shown by \cite{arlen2014}, an increasing viewing angle should increase the cascade contribution if the observed point-source spectrum is held constant. 
The rejection of high-angle photons is thus expected to yield conservative results.

We simulate the full cascade flux over a grid of 
redshifts and in bins of injected \gr\ energy $\Delta E$ between 100\,MeV and 32\,TeV (using eight bins per decade) using the EBL model of~\citet{Dominguez:2010bv}.
In each injected energy bin, we 
assume a power-law intrinsic spectrum with index $\Gamma_\mathrm{int} = 2$. 
We use an $(8\times8)$ logarithmic grid over the magnetic field and coherence length with $B/\mathrm{G} \in [10^{-20};
10^{-12}]$ and $\lambda/\mathrm{Mpc} \in [10^{-4};10^4]$.
We thereby probe IGMF values for which hints have been claimed~\citep{Chen:2014rsa,Chen:2014qva} and that are in a relevant range 
for astrophysical or primordial generation of the IGMF~\citep{Durrer:2013pga}.
We also study different jet opening angles
$\theta_\mathrm{jet} = 1^\circ, 3^\circ, 6^\circ, 10^\circ$. 
The corresponding bulk Lorentz factors for a conical jet,
 $\Gamma_\mathrm{L} = \theta_\mathrm{jet}^{-1} \sim 60, 20, 10, 6$, are broadly consistent with typical values inferred from broadband emission modeling of AGN.
We assume that the sources have been active for a particular 
time $t_{\rm max}$ and all cascade photons arriving with a time delay $\Delta t > t_{\rm max}$ are discarded. We test $t_{\rm max} = 10,\ 10^4,\ 10^7$\,years, where the first case 
corresponds to the conservative case in which blazars have only been active during a timescale comparable to the 
observation time with the \fermi\ LAT. 
AGN activity times are nonetheless estimated to 
lie between $10^6$ and $10^8$\,years~\citep{parma2002}, which is reflected by the maximum $t_{\rm max}$ value tested, whereas $t_{\rm max} = 10^4$\,years is our choice for an intermediate case.

In this way, we end up with a multidimensional cube for the cascade flux $dN/d\epsilon d\Omega$ (in units per energy and per solid angle) in bins of observed energy $\epsilon$, of 
observation angle $\vartheta$ that corresponds to the solid angle $\Omega$, and of injected energy $E$ for a source at redshift $z$ with parameters 
$\mathcal{S} = (\theta_\mathrm{jet},t_\mathrm{max})$ and IGMF parameters $\mathcal{B} = (B,\lambda)$. 
We simulate $N_{\mathrm{inj}, j}$ photons and calculate the yield $y_j$ of cascade flux per injected particle for narrow bins of injected energy $\Delta E_j$,

\begin{equation}
y_j(\epsilon,\vartheta;z,\mathcal{S},\mathcal{B}) = 
\frac{1}{N_{\mathrm{inj},j}}
\frac{dN}{d\epsilon d\Omega}(\epsilon,\vartheta; E_{j},z,\mathcal{S},\mathcal{B}),
\end{equation}
where $E_{j}$ denotes the central energy of $\Delta E_j$.\footnote{
The number of injected particles is given by the sum over the initial Monte Carlo weights calculated by ELMAG in the $j$th energy bin. We inject 600 particles for each energy bin.
For $2 \leq \tau < 4$, we increase the number of particles by a factor of $w_{\mathrm{init},j}  = \lfloor\tau\rfloor$. For $\tau \geq 4$, we increase the number by a factor of 4. The initial Monte Carlo weights are adjusted accordingly by $w_{\mathrm{init},j}^{-1}$.}
We obtain the cascade energy flux per solid angle in an observed energy bin $\Delta\epsilon_i$ for an arbitrary injected \gr\ spectrum $\phi(E,\mathbf{p})$ with parameters $\mathbf{p}$ 
 by reweighting the yields $y_j$ with a weight $w_j$, and summing the cascade flux over all injected energy bins $j$,

\begin{equation}
F_{i}(\vartheta; z,\mathcal{B},\mathcal{S},\mathbf{p}) = \Delta\epsilon_i \sum\limits_{j} w_j(\mathbf{p}) y_j(\epsilon, \vartheta; z,\mathcal{B},\mathcal{S}),
\end{equation}

\noindent where $\epsilon$ lies within the $i$th observed energy bin and the weights are given by 

\begin{equation}
w_j(\mathbf{p}) = \int\limits_{\Delta E_j} dE\, \phi(E,\mathbf{p}).
\end{equation}

\noindent The final cascade flux $F_{\mathrm{casc},i}$ in the observed energy bin $\Delta\epsilon_i$ 
(the same energy binning is used as in Section~\ref{sec:data}) is then obtained by integrating over the 
entire solid angle filled by the cascade, $\Omega_\mathrm{casc}$:

\begin{equation}
F_{\mathrm{casc},i}(z,\mathcal{B},\mathcal{S},\mathbf{p}) = \int\limits_{\Omega_\mathrm{casc}}F_{i}(\vartheta; z,\mathcal{B},\mathcal{S},\mathbf{p})\,d\Omega.\label{eq:casc-flux}
\end{equation}

\noindent The dependence of $F_{\mathrm{casc},i}$ on $\mathbf{p}$ is introduced through the weights $w_j$.
Additionally, the 68\,\% containment radius ${\rcasc}_{,i}(z,\mathcal{B},\mathcal{S},\mathbf{p})$ of the cascade is given through the relation 

\begin{equation}
0.68 = \frac{2\pi}{F_{\mathrm{casc},i}}\int_{0}^{R_{\mathrm{casc},i}}F_{i}(\vartheta; z,\mathcal{B},\mathcal{S},\mathbf{p})\sin\vartheta d\vartheta
\end{equation}

\noindent Note that ${\rcasc}_{,i}$ is, in general, also a function of observed energy, as well as the source and IGMF parameters.

\subsection{IGMF Constraints}
\label{sec:igmf-constraints}

With the simulated cascades in hand, we are in a position
to compare the theoretical cascade spectra and their spatial extension versus the results of the extended catalog for the case of 
a source with a halo.
For the analysis, we make the following assumptions in addition to those made when calculating the cascade flux $F$ (discussed in the previous section):
\begin{enumerate}
\item The source flux does not vary over the observation time and the IACT spectra are good representatives of the average flux level of the sources.  
\item The intrinsic spectrum for each source over the whole \fermi-LAT and IACT energy range can be parameterized with a single LP function with 
exponential cutoff. The observed spectrum is then obtained by multiplying the intrinsic spectrum by the EBL absorption, which is parameterized 
through $\exp(-\tau(E,z))$, where $\tau(E,z)$ is the optical depth, which we assume to follow the model of \citet{Dominguez:2010bv}.
The optical depth is a function of primary \gr\ energy and source redshift and is given by the same EBL model that we use for the ELMAG simulation. 
The observed spectrum is then given by the function
\begin{equation}
\phi_\mathrm{obs}(E,\mathbf{p},z) = N (E/E_0)^{-(\alpha + \beta \ln(E/E_0))}\exp\left[-\left(E/E_\mathrm{cut}+\tau(E,z)\right)\right],
\label{eq:spec}
\end{equation}
which has four free fit parameters,  $\mathbf{p} = (N_0,\alpha,\beta,E_{\rm cut})$. 
We only assume concave spectra, i.e. $\beta \geq 0$ and set $E_0 = 1\,$TeV throughout. 
Enforcing $\beta \geq 0$ should lead to conservative results for the cascade contribution, as it will 
decrease the intrinsic source flux at high energies.

\item Accounting for the cascade contribution does not change the best-fit spectrum of the central point source in the entire \textit{Fermi}-LAT 
energy band by more than $5\,\sigma$ (see Section~\ref{sec:like_extraction}).
\item In each energy bin $i$, the spatial morphology of the cascade can be approximated using a 2D Gaussian halo component with a 68\% 
containment radius equal to that of the cascade, i.e. ${\rhalo}_{,i} = {\rcasc}_{,i}$. 
\item The cascade is not suppressed by the dissipation of energy 
of the $e^+e^-$ beam into plasma instabilities
 \citep[the efficiency of these instabilities is a matter of ongoing debate; see, e.g.][]{broderick2012,sironi2014,menzler2015,chang2016}.
\end{enumerate} 

Given these assumptions, we can use the computed source likelihoods of the extended source catalog to constrain the IGMF parameters. 
The extraction of the likelihoods is described in \Sectionref{sec:like_extraction}.
We use the SED likelihoods for the halo, 
${\lhalo}_{,i}(F_\mathrm{halo}, \rhalo)\equiv{\lhalo}_{,i}(F_\mathrm{halo}, \rhalo;\hat{\boldsymbol{\theta}} | \mathbf{D})$ 
 (given in the \texttt{halo\_sed\_dloglike} column 
in the catalog fits file, see Appendix~\ref{sec:fits-format})
which are provided as a function of the halo flux $F_\mathrm{halo}$, the 68\,\% containment radius $\rhalo$, for each observed energy bin $\Delta\epsilon_i$.
Here, $\mathbf{D}$ denotes the data from the considered source, with the parameters $\boldsymbol{\theta}$ of the other sources in the ROI having already been profiled over.  

The likelihood for a cascade with flux $F_{\mathrm{casc},i}$ (calculated through Eq.~\eqref{eq:casc-flux}) 
and containment radius ${\rcasc}_{,i}$ is then simply given by

\begin{equation}
\mathcal{L}_\mathrm{casc}(z,\mathcal{B},\mathcal{S},\mathbf{p}) = 
\prod\limits_{i}{\lhalo}_{,i}(F_{\mathrm{casc},i}(z,\mathcal{B},\mathcal{S},\mathbf{p}), {\rcasc}_{,i}(z,\mathcal{B},\mathcal{S},\mathbf{p})).
\end{equation}

The catalog also provides the likelihood for the central point source in each energy bin, ${\lsrc}_{,i}(\phi(E_i,\mathbf{p},z))\equiv{\lsrc}_{,i}(\phi(E_i,\mathbf{p},z);\hat{\boldsymbol{\theta}}|\mathbf{D})$ (in the \texttt{src\_sed\_dloglike} column in the catalog fits file; see Appendix~\ref{sec:fits-format}).
For the likelihood of the IACT spectrum, $\mathcal{L}_\mathrm{IACT}(\phi(E_i,\mathbf{p},z))\equiv\mathcal{L}_\mathrm{IACT}(\phi(E_i,\mathbf{p},z)|\mathbf{D}_\mathrm{IACT})$, 
we assume a normal distribution centered on the reported flux and a width equal to the flux uncertainty of the measured IACT spectrum ($\mathbf{D}_\mathrm{IACT}$). 
We neglect any contribution of the reprocessed cascade flux to 
the IACT spectrum, which is generally well-justified, given the 
source spectra and IGMF parameters under consideration. 
In the case that multiple IACT spectra are available for the same source (cf.~Table~\ref{tab:VHEdata}), 
we test whether the observed spectra are compatible with one another by fitting 
them with a simple power law. For sources where this is the case within 2\,$\sigma$ statistical uncertainties, we use all the IACT data points simultaneously 
(1ES\,0229+200, 1ES\,1218+304, H\,2356-309, 1ES\,0414+009).

In order to find the best-fit intrinsic parameters $\mathbf{p}$ for a given IGMF and source, 
we maximize the product of the cascade and point source likelihoods, 

\begin{equation}
\mathcal{L}(z,\mathcal{B},\mathcal{S},\mathbf{p}) = 
\mathcal{L}_\mathrm{casc}(z,\mathcal{B},\mathcal{S},\mathbf{p}) \times
\left(\prod\limits_i{\lsrc}_{,i}(\phi_\mathrm{obs}(E_i,\mathbf{p},z)) \right) 
\times\left(\prod\limits_j\mathcal{L}_{\mathrm{IACT},j}(\phi(E_j,\mathbf{p},z))\right).
\label{eq:likelihood}
\end{equation}

\begin{figure*}
\centering
\includegraphics[width = .49\linewidth]{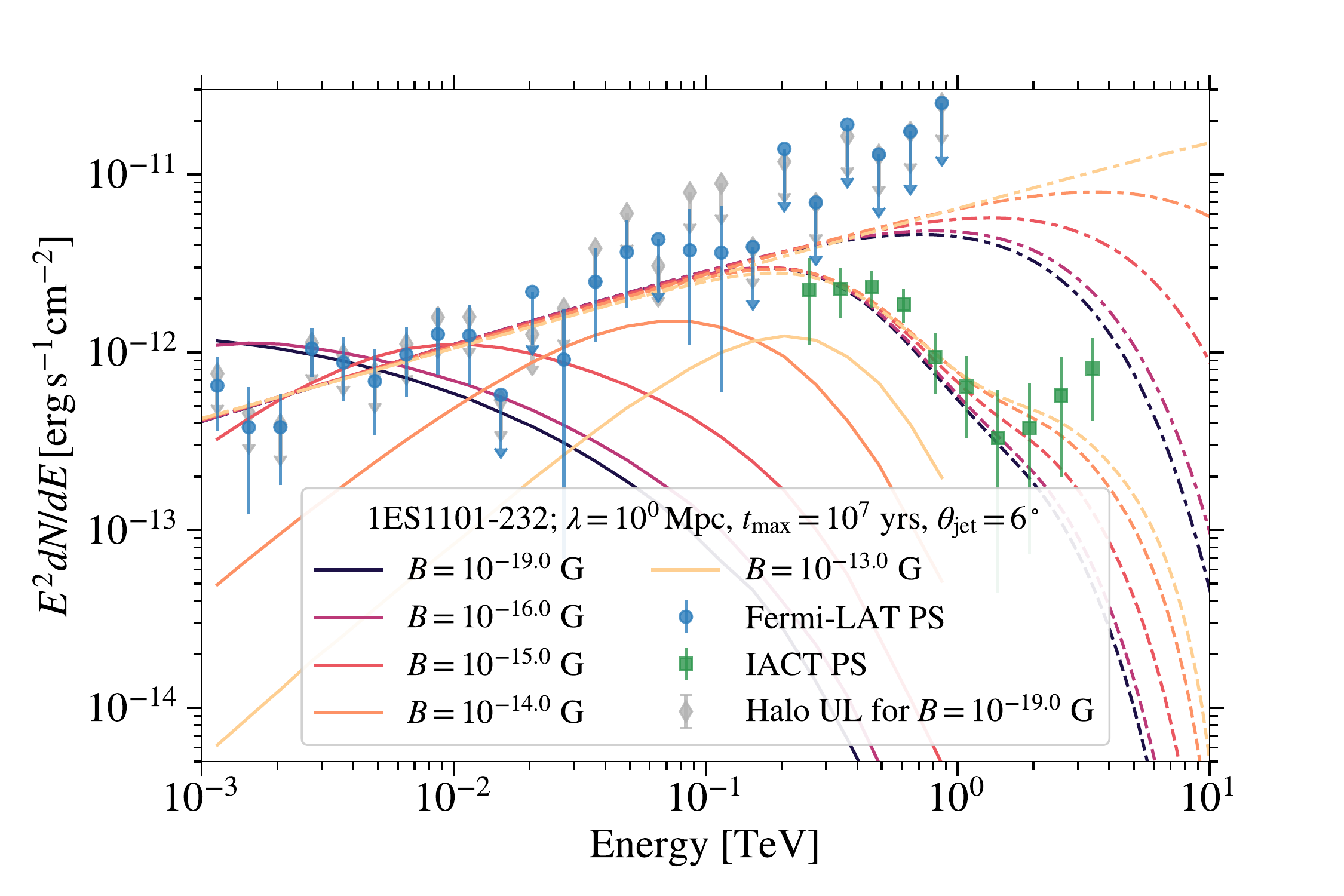}
\includegraphics[width = .49\linewidth]{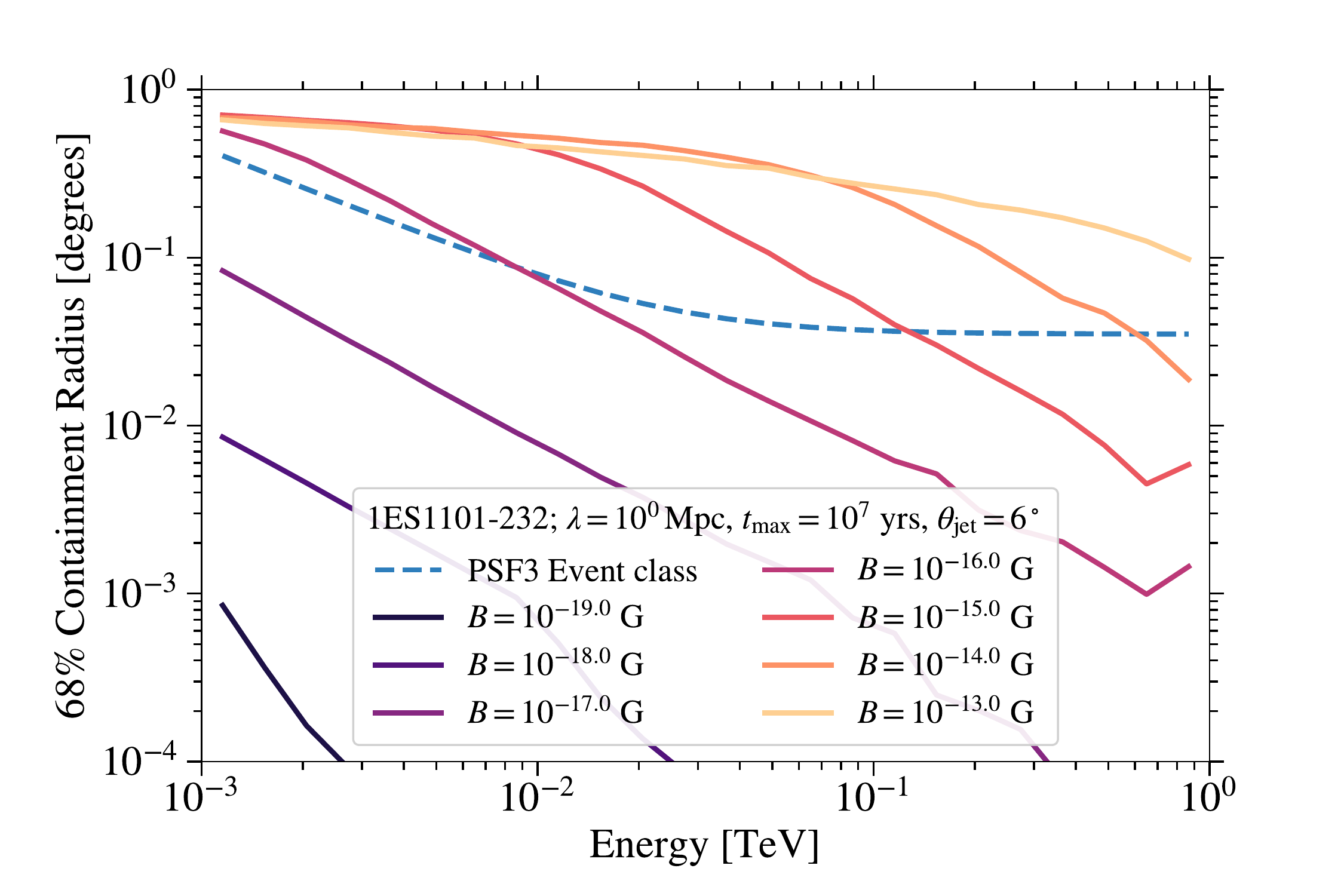}
\caption{\textit{\textbf{Left}}: Fit of the intrinsic spectrum and cascade component to the IACT and \fermi-LAT data of 1ES\,1101-232 ($z = 0.186$) for different IGMF strengths. A source activity time of $t_\mathrm{max} = 10^7$\,years and a jet opening angle of $\theta_{\rm jet} = 6^\circ$ are assumed, along with a coherence length of 1\,Mpc.
The best-fit intrinsic spectra multiplied with EBL absorption are shown as dashed lines with colors matching those of the cascade component (solid lines). 
The intrinsic spectra are shown as dash-dotted lines. Upper limits on the halo energy flux for widths equal to that of the cascade for $B = 10^{-19}$\,G are shown as gray diamonds.
\textit{\textbf{Right}}: Containment radii for the cascade ($\rcasc$) for different $B$-field strengths and the PSF (PSF3 event class) as a function of energy for the same source and parameters as the right panel. 
We show the containment radii for two additional $B$-field strengths ($10^{-18}$\,G and $10^{-17}$\,G) compared to the left panel to better illustrate the IGMF dependence on this quantity. The spectra for these values of $B$ would be very similar to the ones shown for  $10^{-19}$\,G or $10^{-16}$\,G. 
\label{fig:sed-igmf}
}
\end{figure*}

As an example, we show the best-fit spectrum and cascade contribution 
for various magnetic-field strengths in the left panel of Figure~\ref{fig:sed-igmf} for 1ES\,1101-232. 
In this example, we have assumed that the source has been active for $t_{\rm max} = 10^7\,$years. 
This maximum value of $t_\mathrm{max}$ yields the largest cascade contribution, and the differences 
in the fit for the different IGMF values are most pronounced.
As the magnetic field decreases, the contribution from the cascade becomes larger at lower energies. 
To compensate for this, the fit of the intrinsic spectrum (dotted lines) prefers lower values of the cutoff energy, $E_{\rm cut}$.
For high $B$-field values, the fit is insensitive to the cutoff at the highest energies. 
In the right panel of Figure~\ref{fig:sed-igmf}, we show the containment radii $\rcasc$ and the 68\,\% containment radius for the \fermi-LAT PSF for the event class PSF3.
Only for the largest tested IGMF strengths does the halo size increase beyond the PSF. 
For $B \lesssim 10^{-16}\,$G, the halo appears point-like over the entire \fermi-LAT energy range. 
For this reason, the constraints are driven primarily by spectral features of the cascade.
We show the same figure for the other considered blazars in Appendix~\ref{sec:igmf-spec} for the 
minimum and maximum considered activity times, along with the best-fit parameters of the sources yielding constraints on the IGMF.

For each tested IGMF realization and selected source (fixing $z$ and $\mathcal{S}$), we 
maximize the likelihood of Eq.~\eqref{eq:likelihood} by profiling over
the intrinsic spectral parameters $\mathbf{p}$. We then calculate the 
likelihood ratio test statistic:

\begin{equation}
\mathrm{TS}(B,\lambda) = -2 \ln\left(\frac{\mathcal{L}(B,\lambda,{\widehat{\mathbf{p}}}(B,\lambda))}{\mathcal{L}(\doublehat{B},\doublehat{\lambda})}\right).
\label{eq:teststat}
\end{equation}
In the numerator, ${\widehat{\mathbf{p}}}$ denotes the best-fit nuisance parameters for fixed values of
$(B,\lambda)$, and the denominator
gives the unconditionally maximized likelihood with maximum likelihood estimators $\doublehat{B}$ and $\doublehat{\lambda}$. 

\begin{figure*}
\centering
\includegraphics[width = 0.49\linewidth]{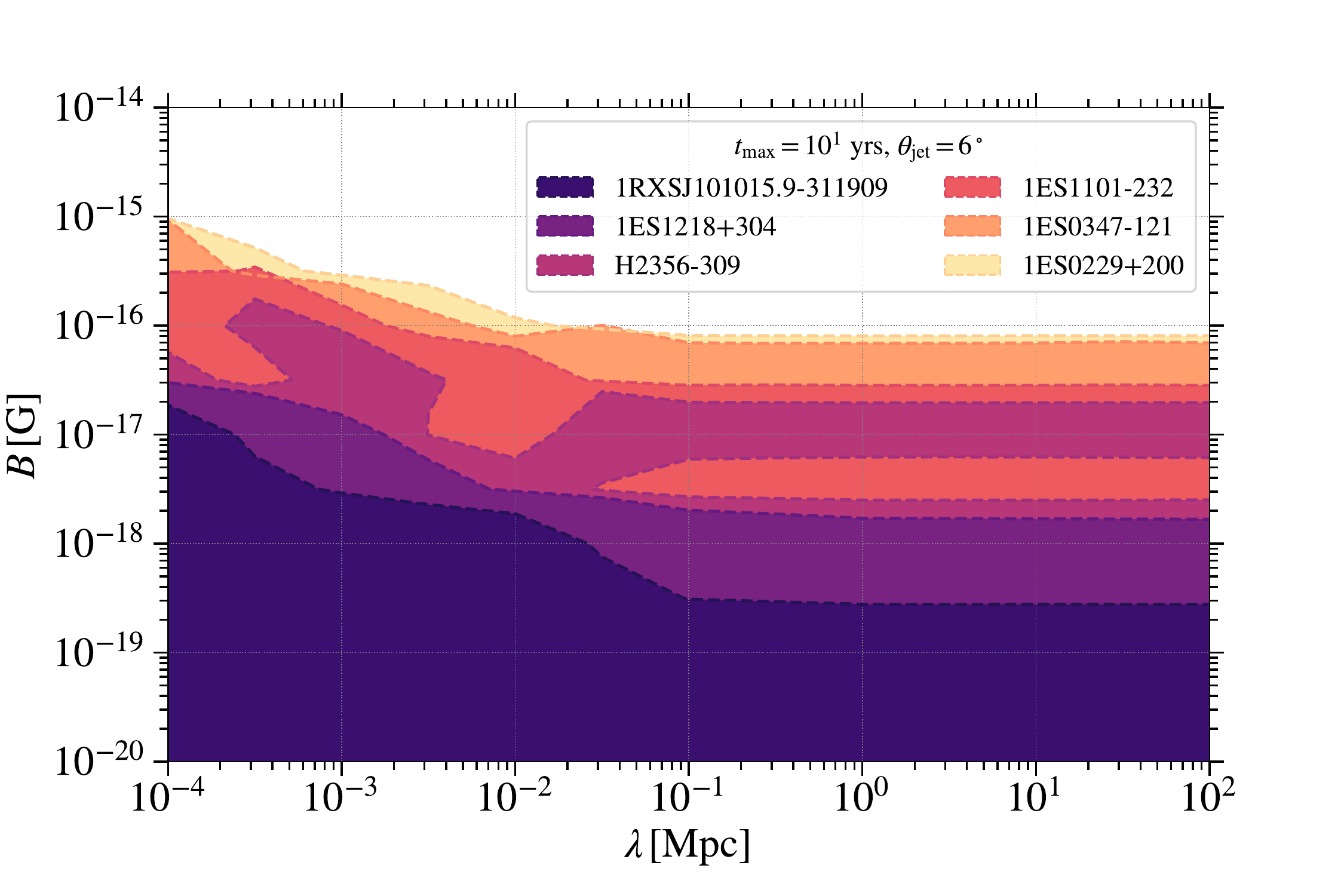}
\includegraphics[width = 0.49\linewidth]{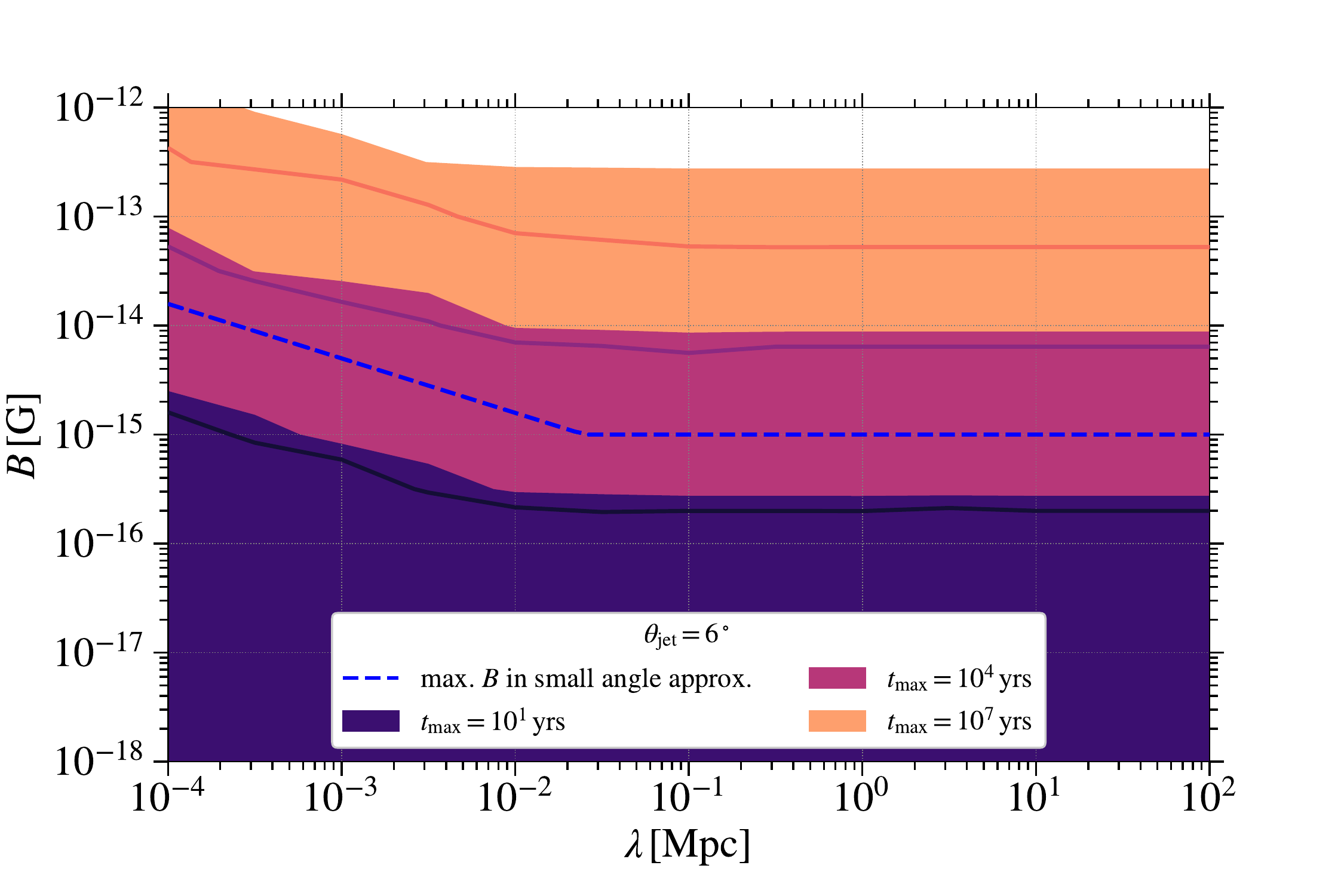}
\caption{\label{fig:like} 95\,\% lower limits on the field strength of the  IGMF for $\theta_{\rm jet} = 6^\circ$. \textit{Left:} Exclusions for $t_{\rm max} = 10\,$years for individual sources. \textit{Right:} Combined exclusion limits for different blazar activity times. 
The solid lines indicate the combined limits 
if the sources 1ES\,0229+200 and 1ES\,1218+304 are excluded from the sample. Above the blue dashed line, the small angle approximation adopted by 
ELMAG breaks down for an increasing number of cascade photons (cf. Eq.~\eqref{eq:small-angle} where an energy of 1\,GeV has been assumed for the cascade photons).}
\end{figure*}

For all tested sources, we find that the best-fit parameters $\doublehat{B}$ and $\doublehat{\lambda}$ coincide with IGMF parameters that lead 
to a strong deflection of the $e^+e^-$ pairs and a consequent suppression of the cascade flux. We therefore derive $95\,\%$ confidence lower limits on the IGMF by excluding parameters for which $\mathrm{TS}(B,\lambda) \geq 5.99$, corresponding to a $\chi^2$ distribution with two degrees of freedom ($B$-field strength and coherence length).  
The limits for the individual sources are shown in the left panel of Fig.~\ref{fig:like} for $\theta_{\rm jet} = 6^\circ$ and a conservative choice of $t_{\rm max} = 10$\,years.

Clearly, a number of spectra yield strong constraints and the lower limit of the IGMF lies between $10^{-17}$\,G and $10^{-16}$\,G.
These constraints are driven by the $\mathcal{L}_\mathrm{casc}$ term in Eq.~\eqref{eq:likelihood} as it gives the largest contribution to the $\mathrm{TS}(B,\lambda)$ values.
The strongest constraints come from the observations of 1ES\,0229+200, as well as the \hess\ observations of 1ES\,0347-121, H\,2356-309, and 1ES\,1101-232.
The non-monotonic behavior of the limits of H\,2356-309 can be explained with the fit stability, in particular the best-fit value for $E_\mathrm{cut}$. 
 Less than 5\,\% of the tested parameter space is excluded for 
 the combined VERITAS and \hess\ spectrum of 1ES\,0414+009, 
as well as for 1ES\,1312-423 and RGB\,0710+591, 
  and we do not show those results here.

We derive combined limits on the IGMF by stacking the individual IGMF likelihoods of the individual sources. We consider only the six sources that yield strong constraints by themselves.
The results for different choices of $t_{\rm max}$ are shown in the right panel of Fig.~\ref{fig:like}.\footnote{The limits on the IGMF are available in plain ASCII files at 
\url{http://wwwglast.
stanford.edu/pub_data/1261/} and \url{https://zenodo.org/record/1324474}}
Even for the most conservative case of $t_{\rm max} = 10$\,years,
we are able to exclude magnetic fields below $\sim3\times10^{-16}$\,G
for $\lambda > 10^{-2}$\,Mpc. 
If we additionally exclude the sources 1ES\,1218+304 and 1ES\,0229-200, for which evidence for variability has been found~\citep{2014ApJ...782...13A}, the limits weaken only marginally for short activity times---but by almost a factor of 5 for $t_\mathrm{max} = 10^7$\,years (solid lines in the right panel of Fig.~\ref{fig:like}).
For such long activity times, the limits improve by three orders of magnitude compared to $t_\mathrm{max}=10$\,years, limiting the $B$ field to be above $3\times10^{-13}\,$G. 
For such high $B$ fields, however, 
one can see from Eq.~\eqref{eq:small-angle} that the small angle approximation adopted by ELMAG breaks down, 
as indicated by the blue dashed line for cascade photons of $\epsilon = 1\,$GeV.
Due to the fact that ELMAG randomizes the deflection angles 
for large deflections and discards the photons when $\beta > \theta_\mathrm{jet}$, the results for long activity times
also depend on the assumed opening angle. 
Assuming $\theta_{\rm jet} = 1^\circ$ instead of $\theta_{\rm jet} = 6^\circ$ decreases the limits by a factor of $1/2$, as more photons are discarded. 
For $\theta_{\rm jet} = 10^\circ$ the results are comparable to $\theta_{\rm jet} = 6^\circ$.
This effect is not observed for $t_\mathrm{max}=10$\,years where the limits are independent of $\theta_{\rm jet}$.

We do not test the impact of different EBL models, as we expect the difference in the limits to be negligible. 
This has been shown in a sensitivity study by~\citet{Meyer:2016xvq} for the Cherenkov Telescope Array (CTA) 
that also utilized the ELMAG code and compared results for the EBL model of \citet{Dominguez:2010bv} 
and \citet{Finke:2009xi}. The slightly larger photon density of the \citeauthor{Finke:2009xi} model  
gives rise to more electron-positron pairs, estimated to be on the order of $5\,\%$, when comparing the two EBL models above~\citep{Meyer:2016xvq}. 
The resulting difference of the limits should consequently be small, compared to the effect of the uncertain blazar activity time scales. 

A larger impact on the limits is given by the systematic uncertainty of the energy scale of IACTs.
This is commonly assumed to be of the order of $\pm15\,\%$, but a cross-calibration between the 
\emph{Fermi}~LAT and IACTs, using the spectrum of the Crab Nebula, found the uncertainty to be on the 
order of $5\,\%$~\citep{2010A&A...523A...2M}.
Nevertheless, recalculating the limits for $\theta_\mathrm{jet} = 6\dg$ and $t_\mathrm{max} = 10\,$years 
with a rescaling of the IACT energy scale by $+15\,\%$ and $-15\,\%$ results in 
$B \gtrsim 4\times10^{-16}$\,G and $B \gtrsim 10^{-16}$\,G, respectively, for $\lambda > 10^{-2}$\,Mpc.

\subsection{Discussion of IGMF Constraints}
\label{sec:igmf-discussion}
Even for the extremely conservative choice of $t_\mathrm{max} = 10\,$years, our results limit the IGMF to be larger than $3\times10^{-16}\,$G for $\lambda \gtrsim 10^{-2}$\,Mpc. Theregore, our results improve the limits derived by~\cite{finke2015} by more than three orders of magnitude, even though we have used a similar source sample and assumptions  (\cite{finke2015} tested $t_\mathrm{max} = 3\,$years and a maximum primary \gr\ energy equal to the highest-energy data point of the IACT spectrum). 
One major difference is that \cite{finke2015}
use a semi-analytic calculation of the cascade~\citep{dermer2011,meyer2012} that only considers the first generation of the cascade. Repeating our analysis using the semi-analytic model in~\citet{meyer2012}, our limits weaken by a factor of five.
The remaining differences can be explained by the very different analysis techniques used. 
In the present analysis, we simultaneously fit the intrinsic spectrum 
and the cascade contribution to the data, profiting from the results of the FHES derived in bins of energy. In contrast to that, \citet{finke2015} exclude IGMF models that lead to an integrated cascade flux larger than the measured flux between 0.1 and 300\,GeV. Therefore, the present analysis uses more information (spectral and spatial) and is consequently more sensitive to the cascade emission. 

In contrast to~\cite{arlen2014}, we are able to rule out $B = 0$ with high significance. 
As noted in~\cite{finke2015}, \cite{arlen2014} use EBL models with low photon densities, 
partly incompatible with lower limits on the EBL from galaxy number counts. 
More importantly, they allow for a spectral break at lower \gr\ energies and very hard 
spectral indices below a few tens of GeV. 
As a result, the point-source flux at these energies is strongly suppressed and the entire GeV flux is dominated by the cascade. 
Such extreme assumptions are in tension with the assumption adopted here that the spectral parameters of the point 
sources are allowed to vary from the broadband energy fit by a maximum of $5\,\sigma$ in each energy bin. 
However, we do allow for curvature in the spectra by using a log-parabola in addition to the exponential cutoff. 
As can be seen from Figures \ref{fig:sed-igmf}, \ref{fig:all-igmf-spec}, and \ref{fig:all-igmf-spec2}, 
large values of $\beta$ are not preferred. 

Under the assumption that the considered blazars have been active for more than 10\,years, 
our limits agree with the values found in a recent study by the VERITAS collaboration~\citep{2017ApJ...835..288A}. 
That study places a lower limit on  the IGMF strength, which lies between $\sim 5\times10^{-15}$ and 
$\sim 7\times10^{-14}$\,G (for coherence lengths larger than the inverse-Compton cooling length) 
due to the absence of a broadening of the angular distribution of \Grs\  from the source 1ES\,1218+304.
The limits also agree with \hess\ measurements from PKS\,2155-305 that ruled out IGMF strengths of 
$(0.3$-$3)\times 10^{-15}\,$G for $\lambda = 1\,$Mpc~\citep{hess2014ph}. 
Both of these studies assumed blazar activity times long enough for the pair halo to be observable with IACTs.

For an activity time $t > 10^4$\,years, our analysis also excludes $B$ field values suggested in~\citet{Chen:2014qva}, where hints for a helical IGMF were found 
from correlations of arrival directions of diffuse \Grs. 
It should be noted, however, that the cascade flux and spatial extension depend on the helicity of the IGMF, which is not included in the ELMAG 1D simulation~\citep{2016PhRvD..94h3005A}.
We cannot confirm hints for pair halos as found in~\citet{Chen:2014rsa} with our stacking analysis nor with our dedicated IGMF analysis, which rules out the values suggested therein.
Likely reasons for this discrepancy are the use of the updated \irf{Pass 8} instrumental response and the usage of the dedicated PSF event classes in the present analysis (cf. Sec.~\ref{sec:data}). Furthermore, we run dedicated source-finding algorithms, providing a complete modeling of each ROI, while the analysis in~\citet{Chen:2014rsa} relied on the two-year LAT point source catalog. 

The obtained limits are on the same order of magnitude as the projected exclusion limits for the future CTA presented in~\citet{Meyer:2016xvq}, which, however, only took the 
spectral features of the cascade into account and only used simulated observations from four blazars.

\section{Conclusions}\label{sec:conclusion}
We have presented the first \fermi\ catalog of high-latitude ($|b| > 5\dg$) extended sources (FHES)
for energies between 1~GeV and 1~TeV.
Using the improved \irf{Pass 8} event reconstruction and data analysis, we are able to identify 
{\NUMEXTSOURCES} extended sources, {\NUMNEWEXTSOURCES} of which are identified as such for the first time.

We are able to associate {\NUMNEWASSOCEXTSOURCES} of the
{\NUMNEWEXTSOURCES} new sources with counterparts from multiwavelength
catalogs.  We identify two SNRs (SNR~G119.5$+$10.2 and
SNR~G332.5$-$05.6) and emission beyond the 
\textit{WMAP} template in the radio
lobes of Cen~A.  We also find evidence for extension of
$\rext = \fhescrabrext\dg \pm \fhescrabrexterr\dg \pm
\fhescrabrexterrsys\dg$ from the Crab Nebula.  
Even though the detection is
not significant when systematic uncertainties of the PSF are taken into account, it should be noted that
 the measured extension
agrees well with predictions from synchrotron self-Compton emission
scenarios and is not observed in blazars with a similar flux above
10~GeV (Mkn~421, PG~1553$+$113, and PKS~2155$-$304). 
It is also in accordance with the extension recently reported by the \hess\ collaboration~\citep{Holler:2017dtw}.
Furthermore, we have found evidence for extended \gr\ emission toward three SFRs (NGC~7822, NGC~1579, and IC~1396). 
However, NGC~7822 and NGC~1579 have been identified as spurious via limitations in the IEM. 
IC~1396 remains as a tentative association.

Three of the five unassociated newly discovered extended sources have hard spectra with $\Gamma \lesssim 2$, suggesting an association with an SNR or PWN.  
However, our search for radio, X-ray, or TeV counterparts in archival data was inconclusive. 
Among these objects, we identify {\fheshardi} and {\fheshardiii} as the two most promising SNR candidates.
Follow-up observations at other wavelengths are encouraged in order to identify the origin of the \gr\ emission.

None of the newly discovered extended sources are located at a Galactic
latitude $|b| > 20\dg$ and the only extragalactic sources
reported here have been previously identified as extended (including
M\,31, Fornax\,A, and the Cen\,A lobes).  After correcting for trials,
we do not find evidence of extended emission in high-latitude
sources whether considered individually or as a population.  This
is also true for the sample of 38 IACT-detected blazars in the sample
of \citet{2015ApJ...812...60B}.  Among the sources in this sample,
PKS~1510$-$08 shows the strongest evidence for halo emission
($\tshalo = \tevits$) which corresponds to a local significance of
$\tevisigma\,\sigma$. However, in this case, the model including a
halo is only marginally preferred over the model with an
additional point source in the ROI.  The rather large TS values found
for extension in stacked source samples of high-synchrotron peaked BL
Lac objects are consistent with systematic uncertainties in the PSF.
None of the unassociated sources above  $|b| > 20\dg$ show evidence for 
extension, and we cannot confirm the DM sub-halo interpretation 
of two 3FGL sources~\citep{2016JCAP...05..049B,2017PhRvD..95j2001X}.
 
Using the results of the extended source catalog, we are able to
derive strong limits on the IGMF, limiting
$B \gtrsim 3\times10^{-16}$\,G for $\lambda \gtrsim 10\,$kpc, for a
conservative assumption of the activity time of the considered blazars
of 10\,years.  
The modeling of the extension performed here makes the results more 
conservative, as we did not assume that the cascade emission is point-like.
Compared to previous studies of the IGMF,
our analysis uses both spatial and spectral information in
the \textit{Fermi}-LAT energy range and simultaneously fits the
intrinsic source spectrum and cascade contribution.
Even though the constraints are driven mostly by the spectral features 
caused by the cascade, 
the detection of pair halos remains a ``smoking gun'' signature of the IGMF that can only
be addressed with a full modeling of the spectrum and the spatial source morphology.
Using longer activity times of $10^4$ ($10^7$) years
improves the limits to $B \gtrsim 9\times10^{-15}$\,G
($B \gtrsim 3\times10^{-13}$\,G).  For such large fields, however, the
actual jet opening and viewing angle of the blazar become important to
accurately model the halo.  The influence of these effects in the
limit of large field strength ($B \gtrsim 10^{-15}$\,G) is not considered
in the simplified 1D Monte-Carlo calculation used by ELMAG.

In the future, dedicated 3D Monte Carlo codes should be used in the future to 
search for the cascade emission at higher values of the IGMF, in order to 
accurately model the source extension and take into account 
the viewing angle of the blazar~\citep{2010ApJ...719L.130N,2016PhRvD..94h3005A,2017MNRAS.466.3472F}. 
Such an analysis should also re-examine our assumption that the point-source spectrum does not change by 
more than 5\,$\sigma$ when the halo component is derived. 
Further extensions could include more realistic 
models of the intergalactic field, including a full treatment of its turbulence spectrum~\citep{caprini2015} 
and its helicity~\citep{chen2015helical}.

\begin{acknowledgments}
The authors would like to thank Ievgen Vovk for helpful discussions
and Gloria Dubner for providing the VLA high resolution map of the Crab Nebula. 
The \textit{Fermi}-LAT Collaboration acknowledges generous ongoing support
from a number of agencies and institutes that have supported both the
development and the operation of the LAT as well as scientific data analysis.
These include the National Aeronautics and Space Administration and the
Department of Energy in the United States, the Commissariat \`a l'Energie Atomique
and the Centre National de la Recherche Scientifique / Institut National de Physique
Nucl\'eaire et de Physique des Particules in France, the Agenzia Spaziale Italiana
and the Istituto Nazionale di Fisica Nucleare in Italy, the Ministry of Education,
Culture, Sports, Science and Technology (MEXT), High Energy Accelerator Research
Organization (KEK) and Japan Aerospace Exploration Agency (JAXA) in Japan, and
the K.~A.~Wallenberg Foundation, the Swedish Research Council and the
Swedish National Space Board in Sweden.
 
Additional support for science analysis during the operations phase is gratefully acknowledged from the Istituto Nazionale 
di Astrofisica in Italy and the Centre National d'\'Etudes Spatiales in France.
This work performed in part under DOE Contract DE-AC02-76SF00515.

This research has made use of the SIMBAD database,
operated at CDS, Strasbourg, France and the NASA/IPAC Extragalactic Database (NED) which is 
operated by the Jet Propulsion Laboratory, California Institute of Technology, under contract with the 
National Aeronautics and Space Administration.

The Wisconsin H$\alpha$ Mapper and its H$\alpha$ Sky Survey have been funded primarily by the National Science Foundation. The facility was designed and built with the help of the University of Wisconsin Graduate School, Physical Sciences Lab, and Space Astronomy Lab. NOAO staff at Kitt Peak and Cerro Tololo provided on-site support for its remote operation.

\end{acknowledgments}

\software{
ELMAG~\citep{2012CoPhC.183.1036K}, fermipy~\citep[version 00-15-01,][]{2017arXiv170709551W}
}

\begin{appendix}
\section{Blazar Spectra for IGMF Constraints}
\label{sec:igmf-spec}
\begin{figure*}
\centering
\includegraphics[width = .49\linewidth]{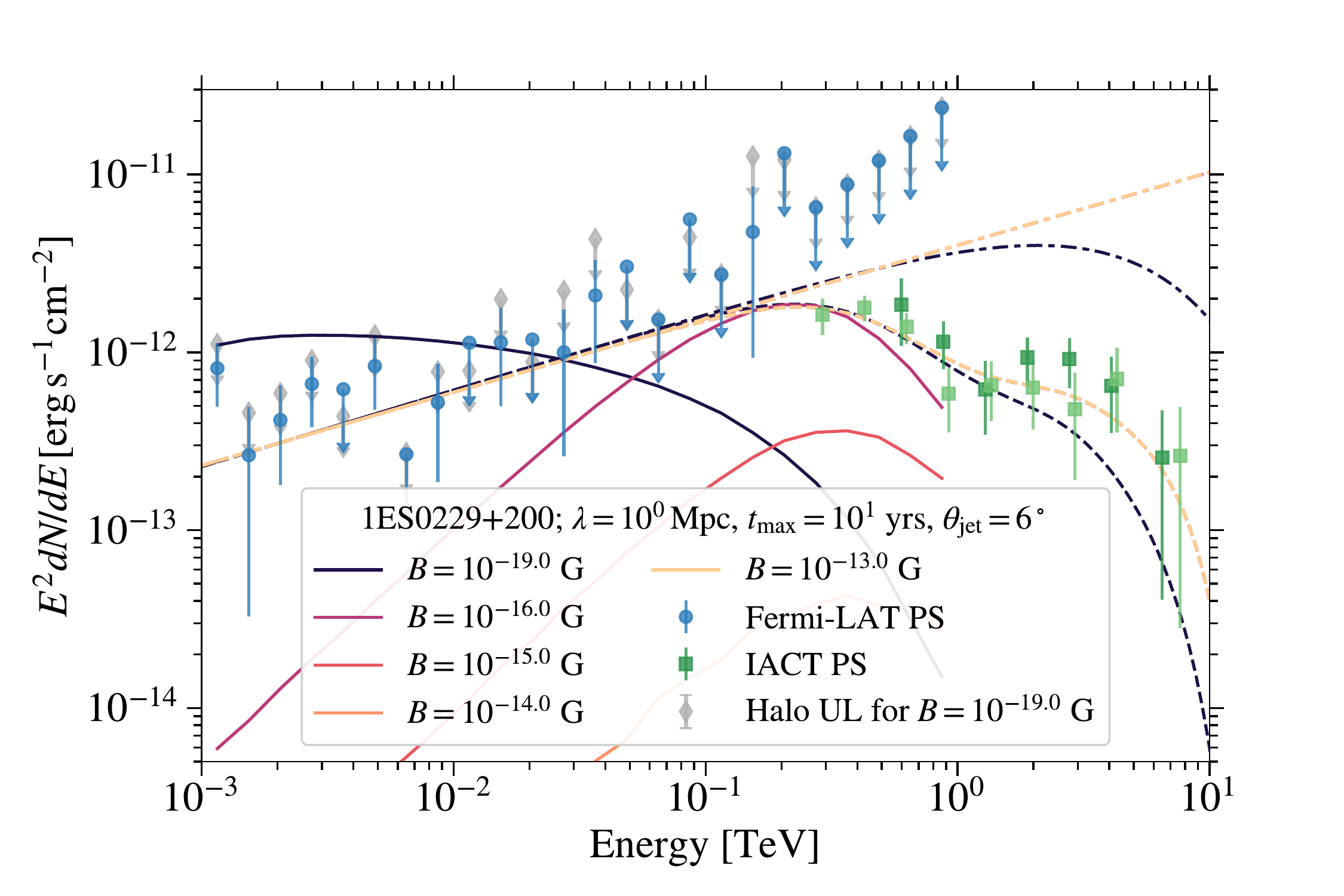}
\includegraphics[width = .49\linewidth]{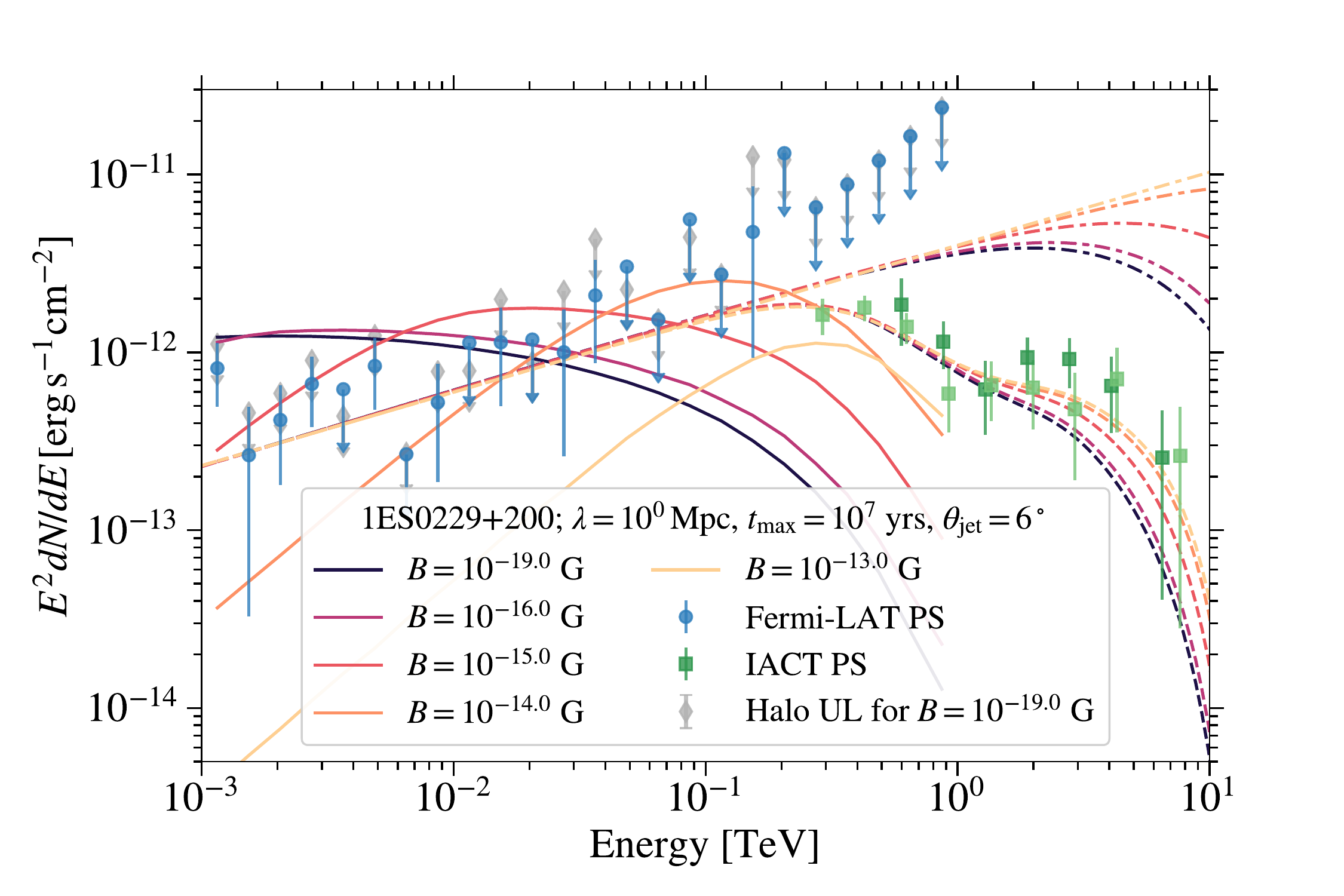}
\includegraphics[width = .49\linewidth]{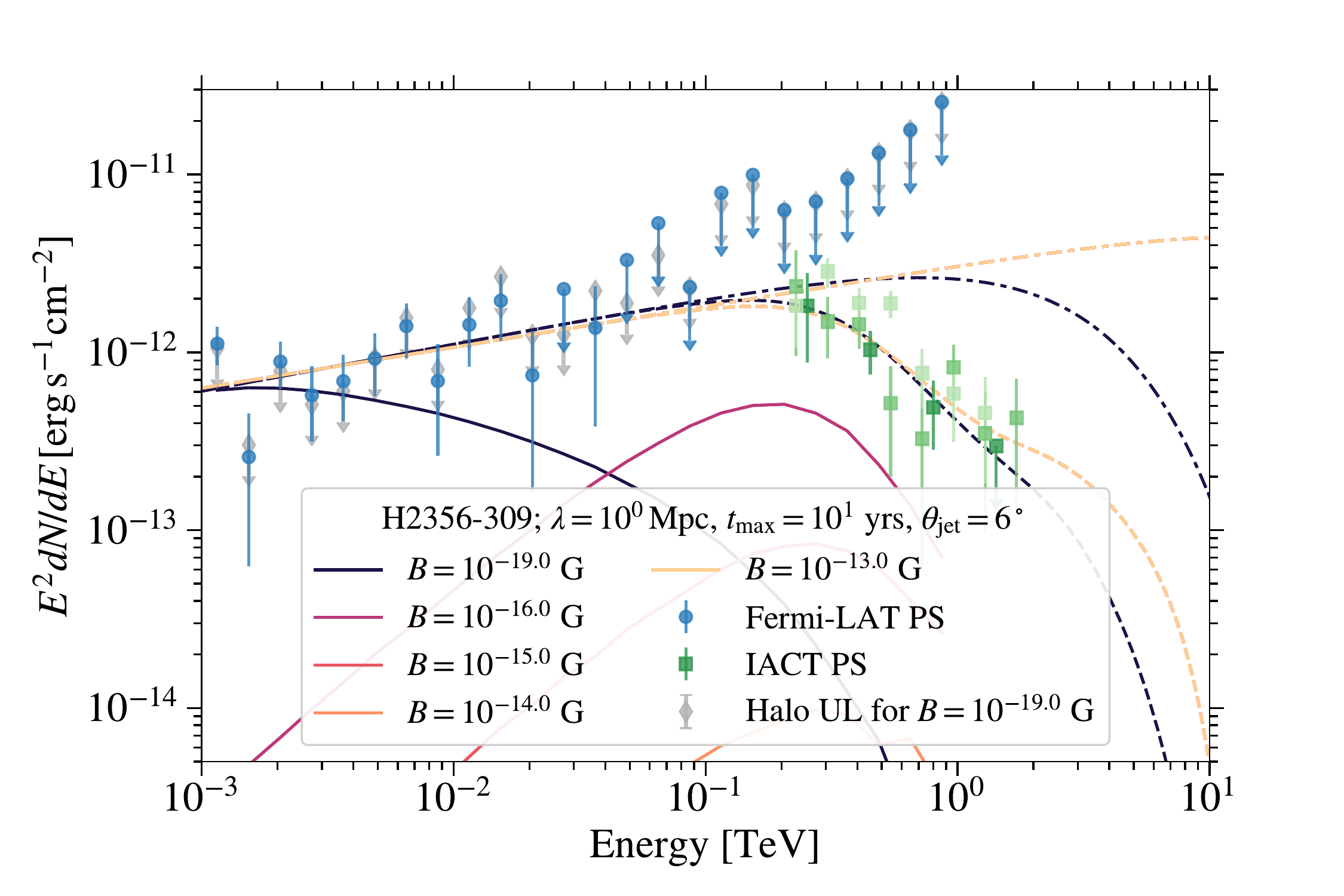}
\includegraphics[width = .49\linewidth]{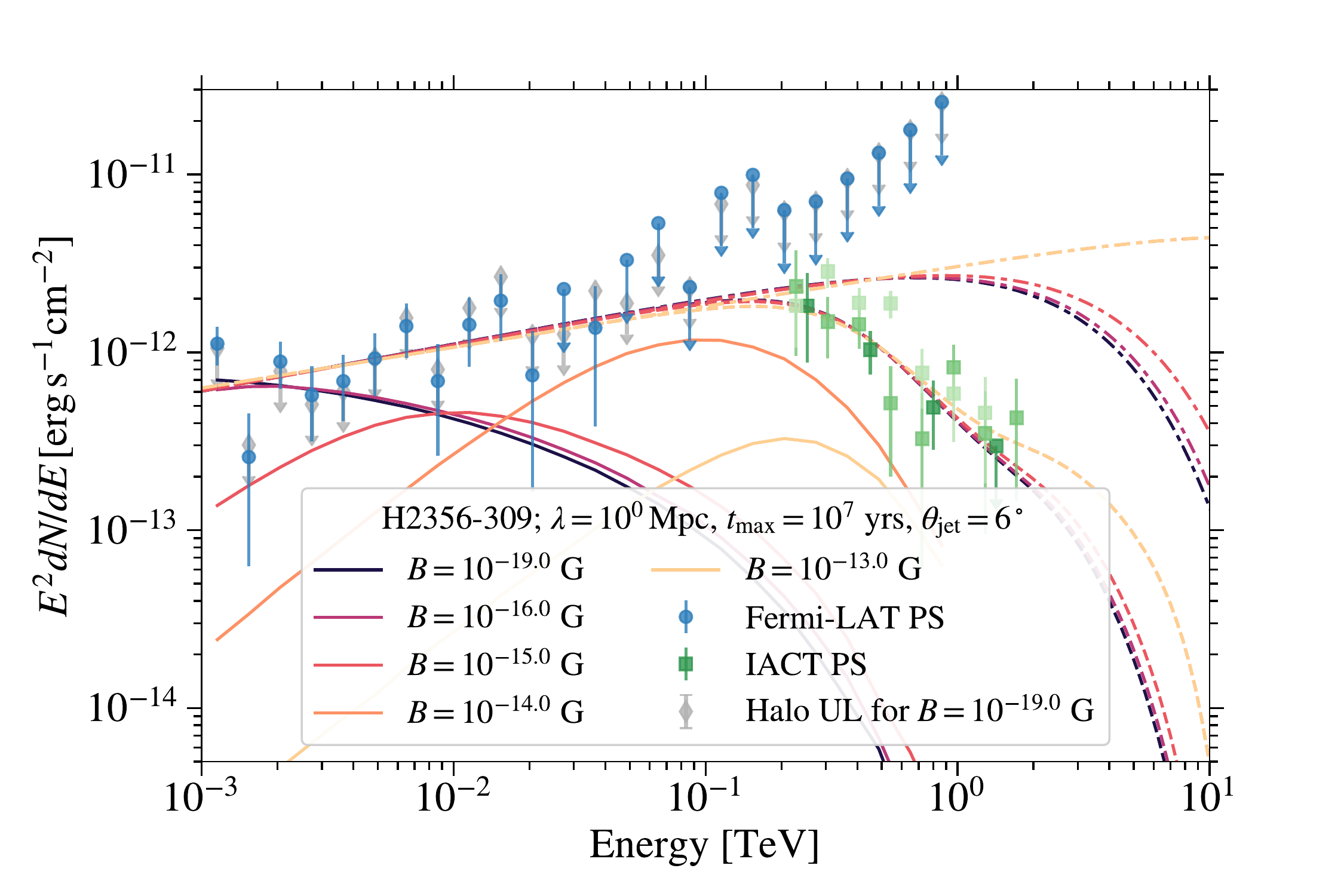}
\includegraphics[width = .49\linewidth]{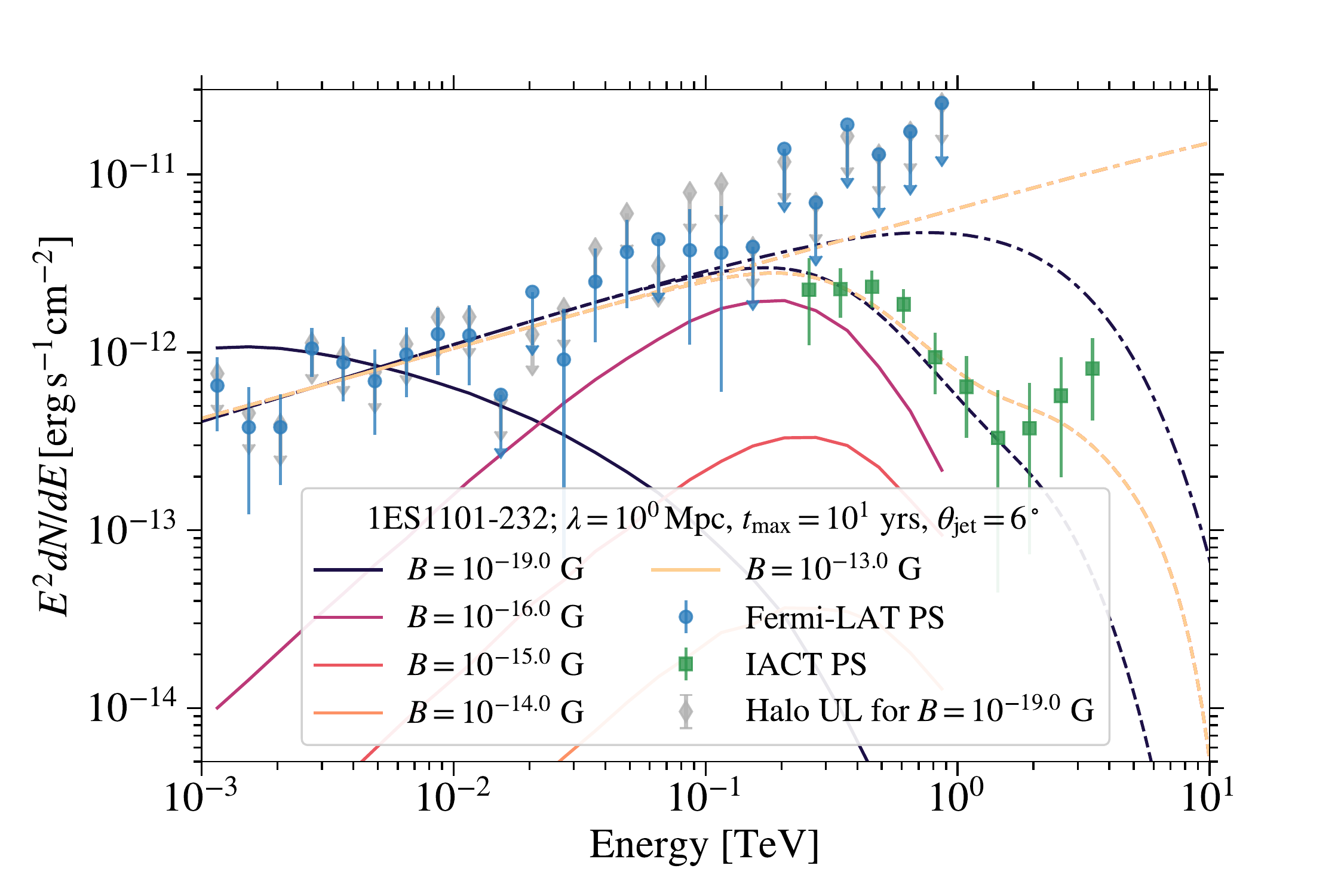}
\includegraphics[width = .49\linewidth]{./sed_1ES1101-232_HESS_2004-2005_idL8_tmax1e+07_thj6_elmag.pdf}
\caption{Same as Figure~\ref{fig:sed-igmf} (left) for the TeV detected
blazars 1ES\,0229+200, H\,2356-309, and 1ES\,1101-232 that yield constraints on the IGMF. 
The left column shows the constraints for  $t_\mathrm{max} = 10$\,years, while the right column shows results for  $t_\mathrm{max} = 10^7$\,years.
\label{fig:all-igmf-spec}}
\end{figure*}

\begin{figure*}
\centering
\includegraphics[width = .49\linewidth]{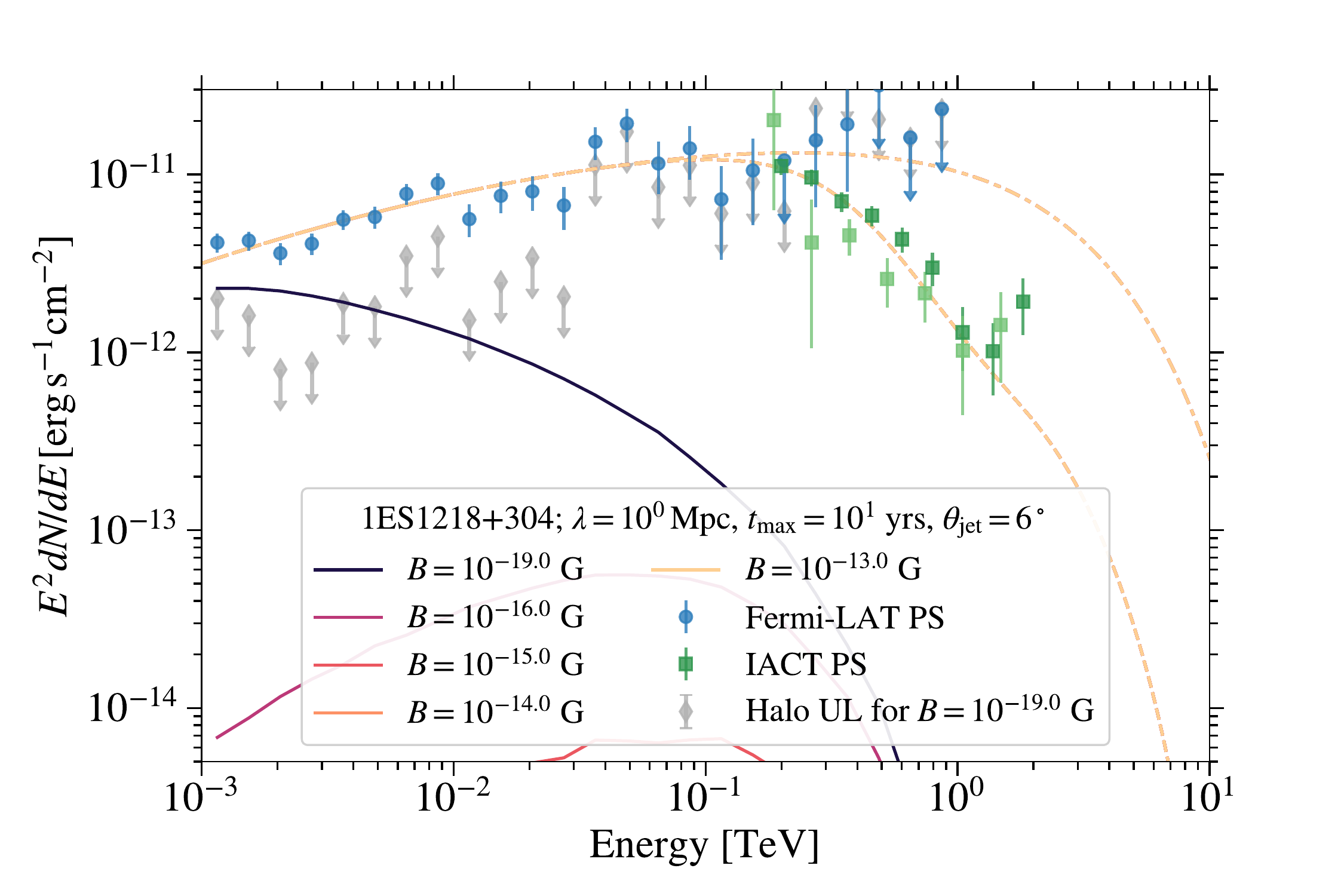}
\includegraphics[width = .49\linewidth]{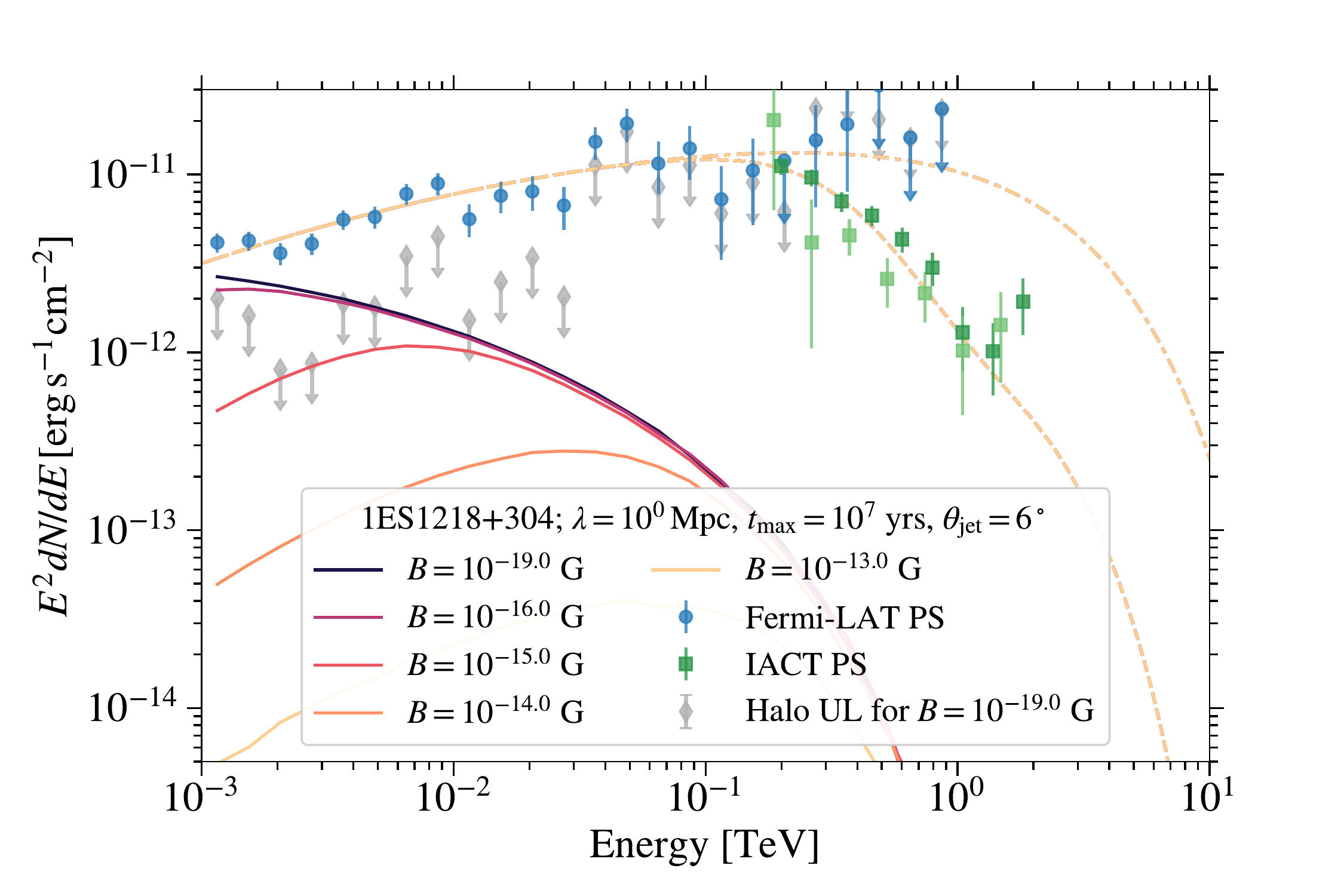}
\includegraphics[width = .49\linewidth]{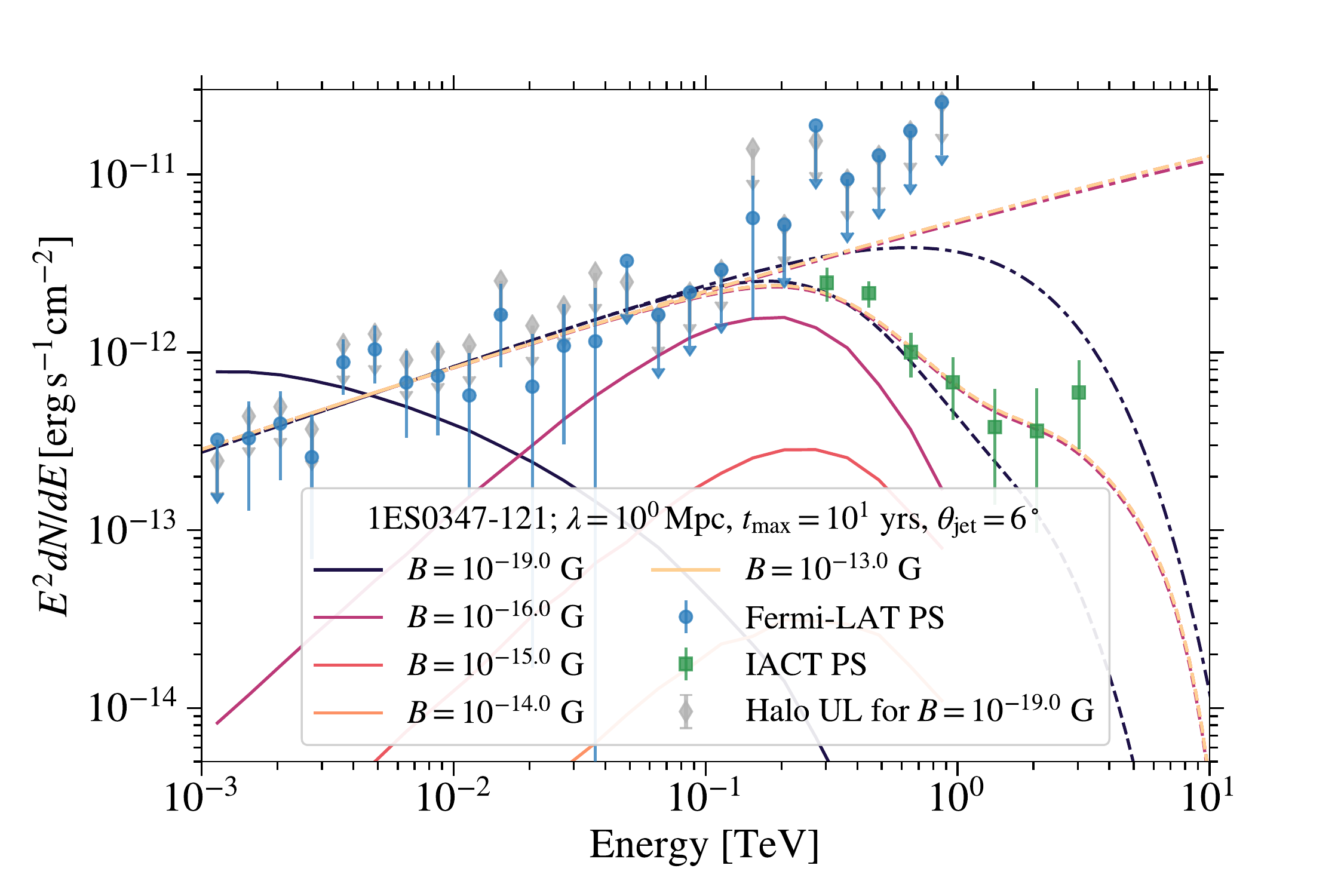}
\includegraphics[width = .49\linewidth]{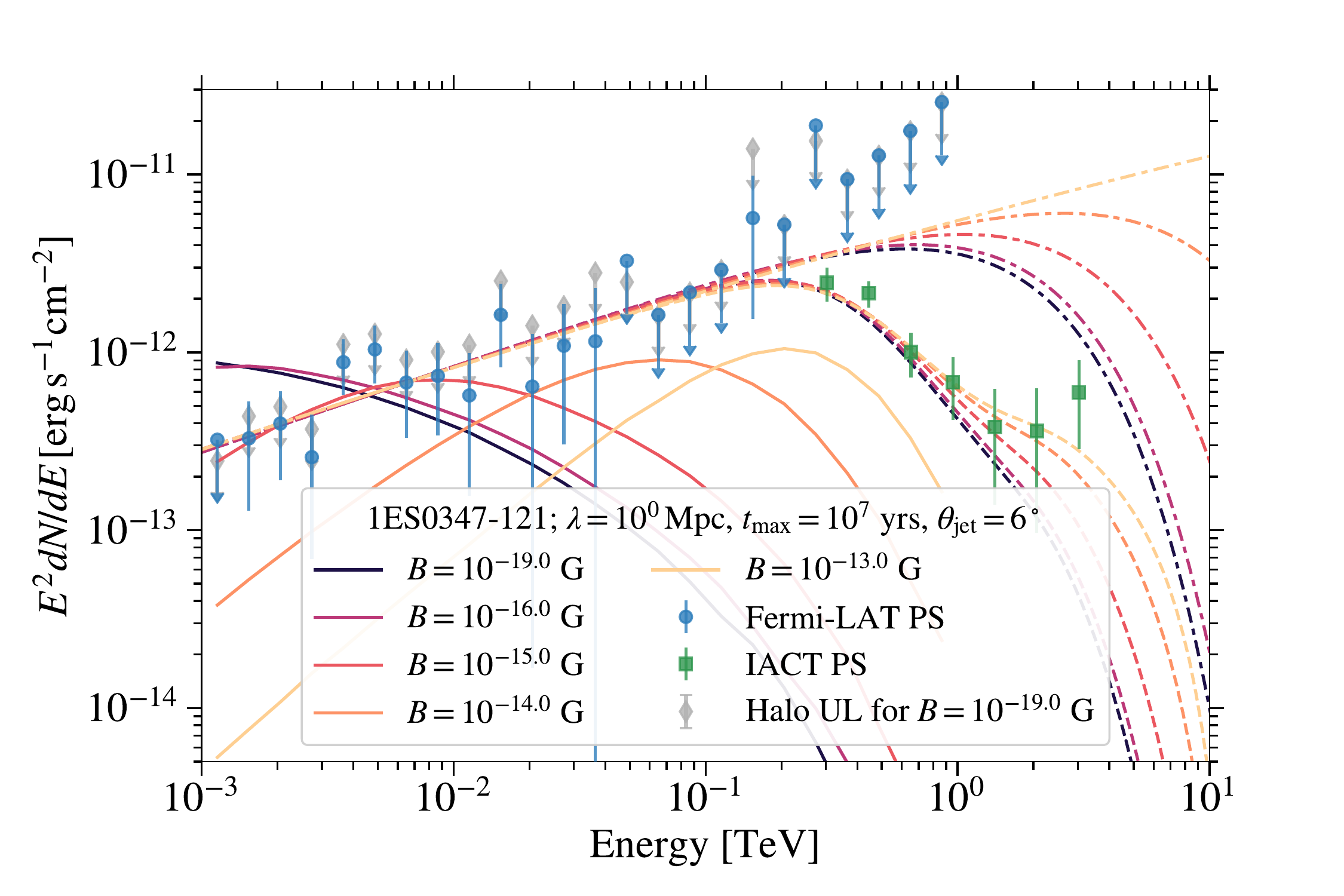}
\caption{Same as Figure~\ref{fig:all-igmf-spec} for the TeV-detected
blazars 1ES\,1218+304 and 1ES0347-121 that yield constraints on the IGMF. 
\label{fig:all-igmf-spec2}}
\end{figure*}

In Figures~\ref{fig:all-igmf-spec} and~\ref{fig:all-igmf-spec2}, we show all blazar spectra obtained with \emph{Fermi} LAT and IACTs
that yield constraints on the IGMF as described in Section~\ref{sec:igmf-constraints}. 
We show the two extreme cases for the assumed activity time, $t_\mathrm{max} = 10$\,years and $10^7$\,years. 
For larger magnetic fields, it is obvious that the cascade flux is increased for the larger values of $t_\mathrm{max}$.
For 1ES\,0229+200, 1ES\,1218+304, and H\,2356-309, the IACT spectra are fitted simultaneously ,
as they are not significantly different from each other. 

\begin{figure*}
\centering
\includegraphics[width = .49\linewidth]{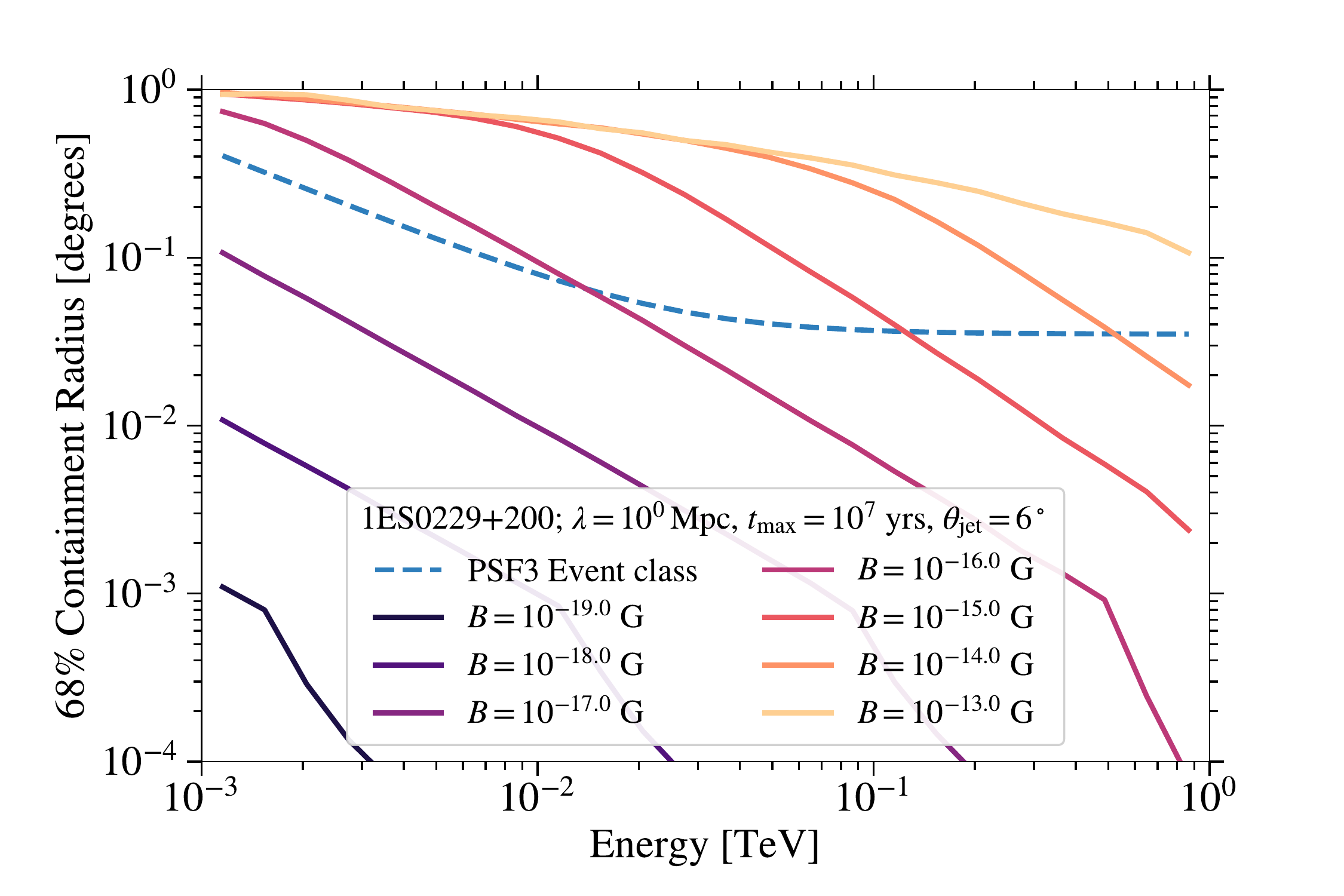}
\includegraphics[width = .49\linewidth]{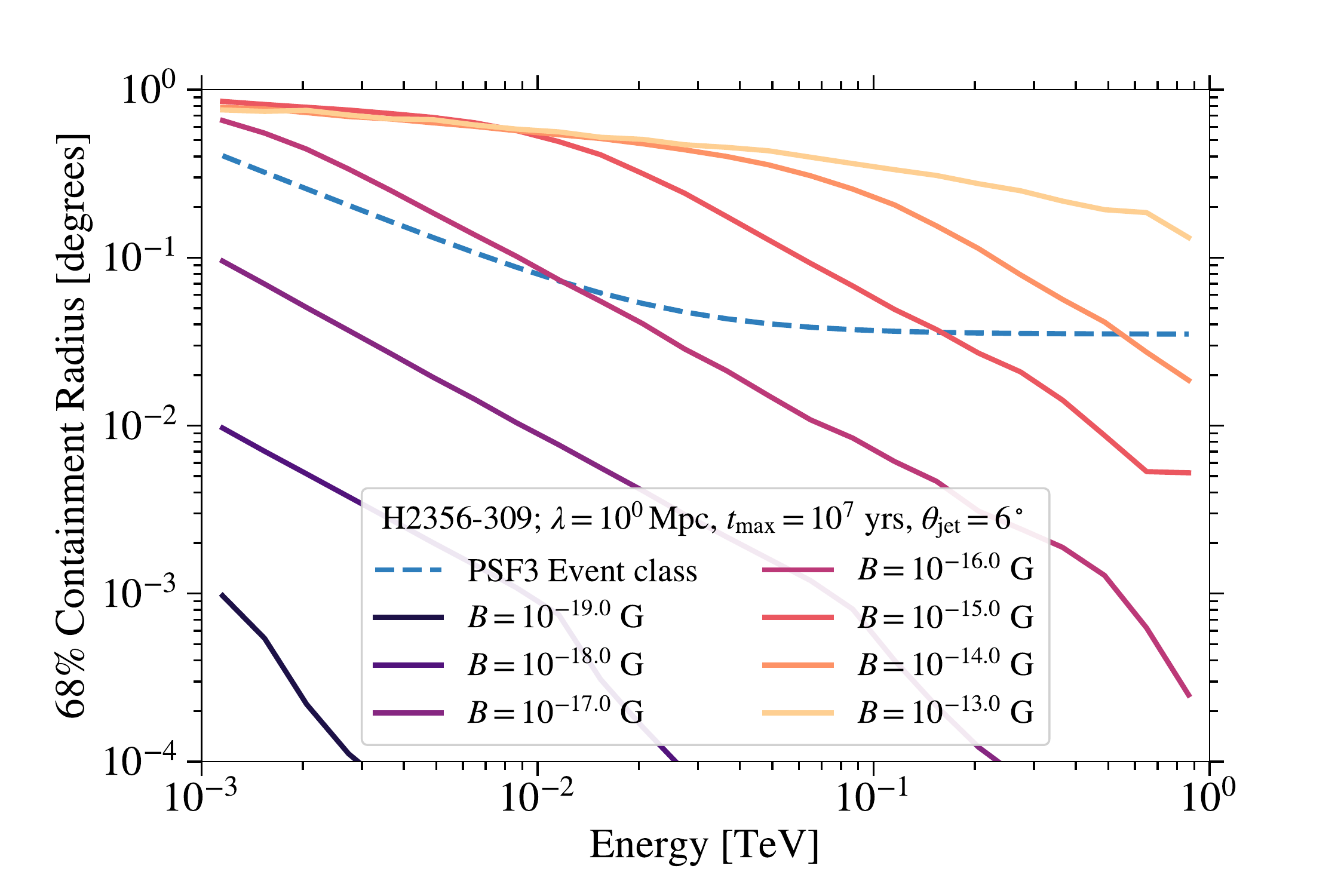}
\includegraphics[width = .49\linewidth]{./th68_1ES1101-232_HESS_2004-2005_idL8_tmax1e+07_thj6_elmag.pdf}
\includegraphics[width = .49\linewidth]{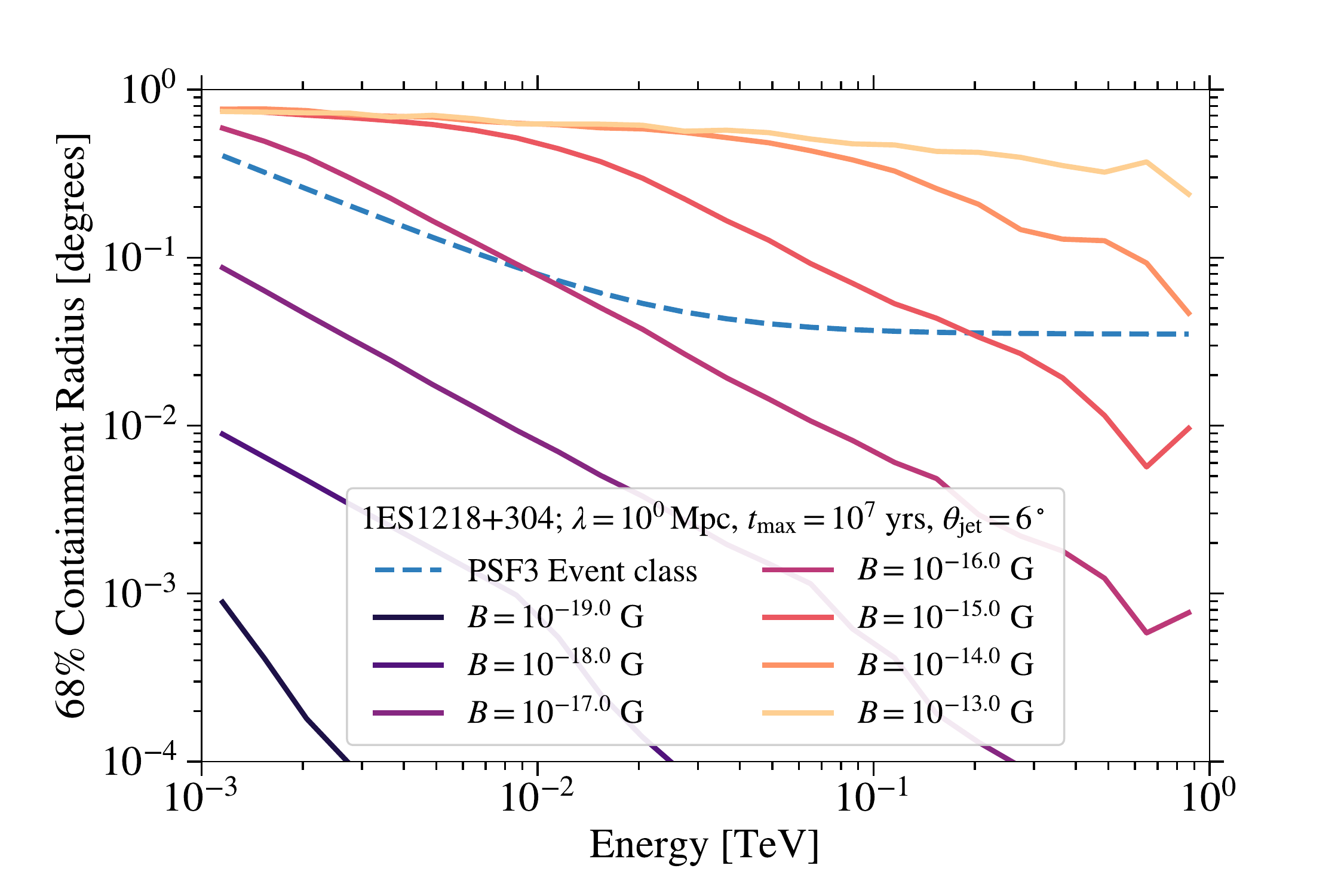}
\includegraphics[width = .49\linewidth]{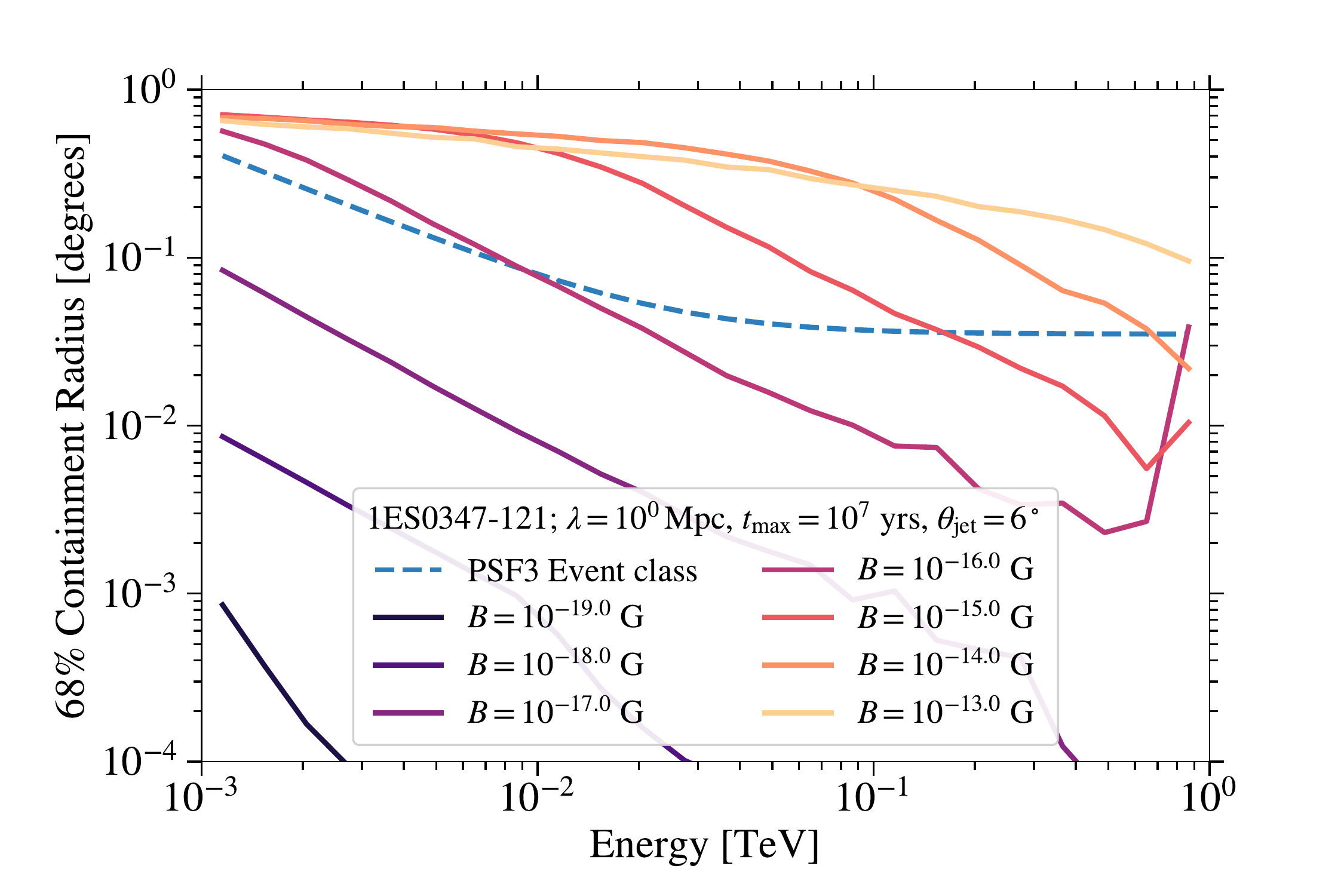}
\caption{Same as Figure~\ref{fig:sed-igmf} (right) for the TeV-detected
blazars that yield constraints on the IGMF, assuming a maximum activity time of $t_\mathrm{max} = 10^7$\,years. 
\label{fig:all-igmf-spec3}}
\end{figure*}

In Figure~\ref{fig:all-igmf-spec3} we show the $r_{68}$ extension of the halo as a function of energy for the case of  $t_\mathrm{max} = 10^7$\,years. For the smaller values of $t_\mathrm{max}$ considered here, too many cascade photons are lost to lead to an extension beyond the \emph{Fermi}-LAT PSF (shown by the dashed line).

Finally, in Table~\ref{tab:bfpars}, we show the best-fit parameters of all six spectra that 
lead to constraints on the IGMF for the $B$-field values plotted in Figures~\ref{fig:all-igmf-spec} and \ref{fig:all-igmf-spec2} for $\lambda = 1\,$Mpc, $t_\mathrm{max} = 10\,$years, and $\theta_\mathrm{jet} = 6\dg$. The table shows that, for the high $B$-field values, the best-fit parameters 
do not change but are very different for the $B = 10^{-20}$\,G, the lowest $B$~field considered. 

\begin{deluxetable*}{rcccc}
\tablewidth{0pt}
\tabletypesize{\scriptsize}
\tablecaption{\label{tab:bfpars}Best-fit parameters for the intrinsic spectra of the 
sources used to derive IGMF constraints.
}
\tablehead{
$\log_{10}(B / \mathrm{G})$ & $N_0$ & $\alpha$ & $\beta$ & $E_\mathrm{cut}$ \\
 & $[10^{-13} \mathrm{TeV}^{-1} \mathrm{s}^{-1}\mathrm{cm}^{-2}]$&  & $[10^{-3}]$ & [TeV] \\
}
\startdata
    \multicolumn{5}{c}{\ruleline[0.5]{1ES0229+200}}\\
$-19.0$ & $1.422$ & $-1.569$ & $0.000$ & $4.834$ \\
$-16.0$ & $1.440$ & $-1.587$ & $0.000$ & \nodata \\
$-15.0$ & $1.439$ & $-1.586$ & $0.000$ & \nodata \\
\multicolumn{5}{c}{\ruleline[0.5]{1ES0347-121}}\\
$-19.0$ & $1.714$ & $-1.515$ & $0.000$ & $1.319$ \\
$-16.0$ & $1.773$ & $-1.524$ & $7.483$ & \nodata \\
$-15.0$ & $1.773$ & $-1.523$ & $6.867$ & \nodata \\
\multicolumn{5}{c}{\ruleline[0.5]{1ES0414+009}}\\
$-19.0$ & $5.537$ & $-1.769$ & $0.000$ & $0.265$ \\
\multicolumn{5}{c}{\ruleline[0.5]{1RXSJ101015.9-311909}}\\
$-19.0$ & $3.801$ & $-1.500$ & $43.932$ & $7.680$ \\
$-16.0$ & $3.801$ & $-1.500$ & $43.932$ & $7.680$ \\
\multicolumn{5}{c}{\ruleline[0.5]{1ES1101-232}}\\
$-19.0$ & $2.551$ & $-1.565$ & $0.000$ & $1.725$ \\
$-16.0$ & $2.652$ & $-1.605$ & $0.000$ & \nodata \\
\multicolumn{5}{c}{\ruleline[0.5]{1ES1218+304}}\\
$-19.0$ & $19.740$ & $-1.528$ & $35.404$ & $2.589$ \\
$-16.0$ & $19.740$ & $-1.528$ & $35.404$ & $2.589$ \\
\multicolumn{5}{c}{\ruleline[0.5]{1ES1312-423}}\\
$-19.0$ & $1.768$ & $-1.500$ & $34.197$ & $2.537$ \\
$-16.0$ & $1.770$ & $-1.500$ & $34.983$ & $3.044$ \\
\multicolumn{5}{c}{\ruleline[0.5]{H2356-309}}\\
$-19.0$ & $3.778$ & $-1.735$ & $0.000$ & $2.627$ \\
$-16.0$ & $3.925$ & $-1.769$ & $0.000$ & \nodata \\
\enddata
{\footnotesize \tablecomments{
The same $B$-field strengths as in Fig.~\ref{fig:sed-igmf} are used here. 
If they are not displayed in the table, the best-fit parameters are the same as for a lower $B$-field strength.
Parameters are shown for $t_\mathrm{max} = 10\,$years, 
$\lambda = 1\,$Mpc, and $\theta_\mathrm{jet} = 6\dg$.
If no value for $E_\mathrm{cut}$ is given, this
fit parameter is unconstrained, i.e. the spectrum does not show an exponential cutoff.}}
\end{deluxetable*}

\section{Catalog File Format}
\label{sec:fits-format}

Complete analysis results for all FHES sources are provided
in the gll\_hes\_v20.fits catalog file available at \href{https://www-glast.stanford.edu/pub_data/1261/}{https://www-glast.stanford.edu/pub\_data/1261/} and \url{https://zenodo.org/record/1324474}.  The catalog includes all
analysis seeds with a detection TS $>25$.  The catalog file contains
five FITS tables:

\begin{itemize}
\item A \texttt{CATALOG} table with best-fit parameters and likelihood
  ratios (TS, $\tsext$, and $\tshalo$) for each source (see
  \Tableref{catalog_table}). 
\item A \texttt{SED} table with the likelihood profiles for the SEDs for each source  (see
  \Tableref{sed_table}).
\item Two tables  \texttt{LIKELIHOOD\_FHES\_SOURCES} and \texttt{LIKELIHOOD\_IGMF\_SOURCES} with likelihood profiles versus flux
  and angular extension for the Gaussian, halo, and disk model for the sources with detected extension and the blazars used to derive limits on the IGMF, respectively
  (see \Tableref{likelihood_table}). The likelihood profiles for other sources can be obtained from the authors upon request. 
\item A \texttt{SCAN\_PARS} table defining the grid values used in the  \texttt{LIKELIHOOD\_FHES\_SOURCES} and \texttt{LIKELIHOOD\_IGMF\_SOURCES} tables (see \Tableref{scan_pars_table}).
\item An \texttt{EBOUNDS} table defining the energy bins as well as reference flux values used in the \texttt{LIKELIHOOD\_FHES\_SOURCES}, \texttt{LIKELIHOOD\_IGMF\_SOURCES}, and \texttt{SED} tables (see \Tableref{ebounds_table}).
\end{itemize}

\begin{deluxetable*}{lll}
\tablewidth{0pt}
\tabletypesize{ \small }
\tablecaption{ \label{tab:catalog_table} FHES source table format (\texttt{CATALOG} HDU) }
\tablehead{
Column & Unit & Description
}
\startdata
\texttt{name}\tablenotemark{a} & \nodata & Source designation \\
\texttt{assoc\_3fgl} & \nodata & 3FGL Source association \\
\texttt{assoc\_3fhl} & \nodata & 3FHL Source association \\
\texttt{assoc\_fl8y} & \nodata & Name of associated source in preliminary LAT eight-year point source list (FL8Y) \\
\texttt{sep\_fl8y} & deg & Angular separation of between FHES source and associated source in FL8Y \\
\texttt{assoc} & \nodata & Source association from 3FGL (column \texttt{ASSOC1}) or 3FHL (column \texttt{ASSOC}) \\
\texttt{class} & \nodata & Source class from 3FGL (column \texttt{CLASS1}) or 3FHL (column \texttt{CLASS}) \\
\texttt{redshift} & \nodata & Redshift from 3FHL or 3LAC \\
\texttt{nupeak} & Hz & Synchrotron peak frequency from 3FHL or 3LAC \\
\texttt{RAJ2000}\tablenotemark{b} & deg & Right ascension of 3FGL or 3FHL source \\
\texttt{DEJ2000}\tablenotemark{b} & deg & Declination of 3FGL or 3FHL source \\
\texttt{GLON}\tablenotemark{b} & deg & Galactic longitude of 3FGL or 3FHL source \\
\texttt{GLAT}\tablenotemark{b} & deg & Galactic longitude of 3FGL or 3FHL source \\
\texttt{pos\_r68}\tablenotemark{b} & deg & Symmetric statistical error (68\%) on position \\
\texttt{pos\_r68\_sys}\tablenotemark{b,d} & deg & Symmetric systematic error (68\%) on position \\
\texttt{pos\_r95}\tablenotemark{b} & deg & Symmetric statistical error (95\%) on position \\
\texttt{index} & \nodata & Spectral slope at 1 GeV \\
\texttt{index\_err} & \nodata & Statistical error on \texttt{index} \\
\texttt{index\_sys\_err}\tablenotemark{d} & \nodata & Systematic error ($1\sigma$) on \texttt{index} \\
\texttt{SpectrumType} & \nodata & Spectral type (PowerLaw, LogParabola, PLSuperExpCutoff) \\
\texttt{ts} & \nodata & Detection test statistic (TS) \\
\texttt{ts\_ext} & \nodata & Extension test statistic $\tsext$ \\
\texttt{ts\_ext\_sys}\tablenotemark{d} & \nodata & Minimum extension test statistic ($\tsext$) under bracketing models for systematics \\
\texttt{ts\_halo} & \nodata & Halo test statistic $\tshalo$ \\
\texttt{flux} & cm$^{-2}$~s$^{-1}$ & Integrated photon flux (1~GeV -- 1~TeV) \\
\texttt{flux\_err} & cm$^{-2}$~s$^{-1}$ & Statistical error ($1\sigma$) on \texttt{flux} \\
\texttt{flux\_sys\_err}\tablenotemark{d} & cm$^{-2}$~s$^{-1}$ & Systematic error ($1\sigma$) on \texttt{flux} \\
\texttt{eflux} & MeV~cm$^{-2}$~s$^{-1}$ & Integrated energy flux (1~GeV -- 1~TeV) \\
\texttt{eflux\_err} & MeV~cm$^{-2}$~s$^{-1}$ & Statistical error ($1\sigma$) on \texttt{eflux} \\
\texttt{ext\_model} & \nodata & Best-fit spatial morphology (RadialGaussian, RadialDisk, PointSource) \\
\texttt{ext\_r68}\tablenotemark{c} & deg & Best-fit extension ($\rext$) \\
\texttt{ext\_r68\_err} & deg & Statistical error ($1\sigma$) on \texttt{ext\_r68} \\
\texttt{ext\_r68\_sys\_err}\tablenotemark{d} & deg & Systematic error ($1\sigma$) on \texttt{ext\_r68} \\
\texttt{ext\_r68\_ul95} & deg & 95\% CL upper limit on \texttt{ext\_r68} \\
\enddata

\tablenotetext{a}{The source designation is the FHES name for extended sources and the name of the associated 3FGL or 3FHL source for point sources.}
\tablenotetext{b}{If the source was found to be extended, the measured position and position uncertainties from the FHES are given. If the source was found to be point-like, position and position uncertainties are taken from the 3FGL or 3FHL.}
\tablenotetext{c}{For sources best-fit with the disk model, $\rext = 0.82R$, where $R$ is the disk radius.}
\tablenotetext{d}{Systematic uncertainties are only given for sources with detected extension listed in Tables \ref{tab:extended_sources} and \ref{tab:extended_sources_syst}.}

{}

\end{deluxetable*}

\begin{deluxetable*}{llll}
\tablewidth{0pt}
\tabletypesize{ \footnotesize }
\tablecaption{ \label{tab:sed_table} FHES source SED table format (\texttt{SED} HDU) }
\tablehead{
Column & Format & Unit & Description
}
\startdata
\texttt{name}\tablenotemark{a} & \nodata & \nodata & Source designation \\
\texttt{norm} & 24D & \nodata & Normalization in each bin, in units of the reference model \\
\texttt{norm\_err} & 24D & \nodata & Symmetric error on the normalization in each bin, in units of the reference model \\
\texttt{norm\_errp} & 24D & \nodata & Upper $1\sigma$  error on the normalization in each bin, in units of the reference model \\
\texttt{norm\_errn} & 24D & \nodata & Lower $1\sigma$ error on the normalization in each bin, in units of the reference model \\
\texttt{norm\_ul} & 24D & \nodata & Upper limit on the normalization in each bin, in units of the reference model \\
\texttt{norm\_scan}\tablenotemark{b} & 24 $\times$ 9D & \nodata & Array of 24$\times$9 normalization values for the profile likelihood scan \\
\texttt{dloglike\_scan} & 24 $\times$ 9D & \nodata & Array of 24$\times$9 delta-loglikelihood values for the profile likelihood scan in 24 energy bins and 9 scan points \\
\enddata

\tablenotetext{a}{The source designation is the FHES name for extended sources and the name of the associated 3FGL or 3FHL source for point sources.}
\tablenotetext{b}{A row-wise multiplication with any of \texttt{ref} columns in the \texttt{EBOUNDS} HDU can be used to convert this matrix to the respective unit.}

{}

\end{deluxetable*}

\begin{deluxetable*}{llll}
\tablewidth{0pt}
\tabletypesize{ \small }
\tablecaption{ \label{tab:likelihood_table} FHES likelihood table format (\texttt{LIKELIHOOD} HDUs) }
\tablehead{
Column & Format & Unit & Description
}
\startdata
\texttt{name}\tablenotemark{a} & \nodata & \nodata & Source designation \\
\texttt{ext\_gauss\_dloglike} & 31E  & \nodata & $\deltalnl$ values for extension likelihood scan of 2D Gaussian ($\rext$) \\
\texttt{ext\_disk\_dloglike} & 31E  & \nodata & $\deltalnl$ values for extension likelihood scan of 2D Disk ($\rext$) \\
\texttt{halo\_dloglike} & 15 $\times$ 13 $\times$ 61E & \nodata & $\deltalnl$ values for halo likelihood grid scan ($\rhalo$, $\gammahalo$, $\fhalo$) \\
\texttt{halo\_sed\_dloglike} & 15 $\times$ 24 $\times$ 61E & \nodata & $\deltalnl$ values for halo SED likelihood grid scan ($\rhalo$, $E_\mathrm{ref}$, $\fhalo$) \\
\texttt{halo\_src\_sed\_dloglike} & 24 $\times$ 9E & \nodata & $\deltalnl$ values for \texttt{halo\_src\_sed\_eflux} \\
\texttt{halo\_src\_sed\_eflux} & 24 $\times$ 9E & MeV~cm$^{-2}$~s$^{-1}$ & Energy flux evaluation points for source SED likelihood scan \\
\enddata

\tablenotetext{a}{The source designation is the FHES name for extended sources and the name of the associated 3FGL or 3FHL source for point sources.}

{}

\end{deluxetable*}

\begin{deluxetable*}{llll}
\tablewidth{0pt}
\tabletypesize{ \small }
\tablecaption{ \label{tab:scan_pars_table} FHES parameter table format (\texttt{SCAN\_PARS} HDU) }
\tablehead{
Column & Format & Unit & Description
}
\startdata
\texttt{ext\_r68} & 30D & deg & Evaluation points for likelihood scan versus source extension ($\rext$) \\
\texttt{halo\_r68} & 15D & deg & $\rhalo$ values for halo likelihood scan (\texttt{halo\_dloglike}) \\
\texttt{halo\_index} & 13D & \nodata & $\gammahalo$ values for halo likelihood scan (\texttt{halo\_dloglike}) \\
\texttt{halo\_eflux} & 61D & MeV~cm$^{-2}$~s$^{-1}$ & $\rhalo$ values for halo likelihood scan (\texttt{halo\_dloglike}) \\
\enddata

{}

\end{deluxetable*}

\begin{deluxetable*}{lll}
\tablewidth{0pt}
\tabletypesize{ \small }
\tablecaption{ \label{tab:ebounds_table} FHES SED parameter table format (\texttt{EBOUNDS} HDU) }
\tablehead{
Column & Unit & Description
}
\startdata
\texttt{e\_min} & MeV & Lower edge of energy bin \\
\texttt{e\_max} & MeV & Upper edge of energy bin \\
\texttt{e\_ref} & MeV & Central energy of bin \\
\texttt{ref\_flux} & cm$^{-2}$~s$^{-1}$ & Flux of the reference model in each bin \\
\texttt{ref\_eflux} & MeV~cm$^{-2}$~s$^{-1}$ & Energy flux of the reference model in each bin \\
\texttt{ref\_dnde} & MeV$^{-1}$~cm$^{-2}$~s$^{-1}$ & Differential flux of the reference model evaluated at the bin center \\
\enddata

{}

\end{deluxetable*}

\end{appendix}

\newpage

\bibliography{extension_main}

\newcommand{\noop}[1]{}
\begin{thebibliography}{}
\expandafter\ifx\csname natexlab\endcsname\relax\def\natexlab#1{#1}\fi

\bibitem[{{Abdalla} {et~al.}(2017){Abdalla}, {Abramowski}, {Aharonian}, {Ait
  Benkhali}, {Akhperjanian}, {Andersson}, {Ang{\"u}ner}, {Arrieta}, \&
  et~al.}]{2017arXiv170208280H}
{Abdalla}, H., {Abramowski}, A., {Aharonian}, F., {et~al.} 2017, ArXiv
  e-prints, arXiv:1702.08280

\bibitem[{{Abdo} {et~al.}(2008){Abdo}, {Ackermann}, {Atwood}, {Baldini},
  {Ballet}, {Barbiellini}, {Baring}, {Bastieri}, {Baughman}, {Bechtol},
  {Bellazzini}, {Berenji}, {Blandford}, {Bloom}, {Bogaert}, {Bonamente},
  {Borgland}, {Bregeon}, {Brez}, {Brigida}, {Bruel}, {Burnett}, {Caliandro},
  {Cameron}, {Caraveo}, {Carlson}, {Casandjian}, {Cecchi}, {Charles},
  {Chekhtman}, {Cheung}, {Chiang}, {Ciprini}, {Claus}, {Cohen-Tanugi},
  {Cominsky}, {Conrad}, {Cutini}, {Davis}, {Dermer}, {de Angelis}, {de Palma},
  {Digel}, {Dormody}, {do Couto e Silva}, {Drell}, {Dubois}, {Dumora},
  {Edmonds}, {Farnier}, {Focke}, {Fukazawa}, {Funk}, {Fusco}, {Gargano},
  {Gasparrini}, {Gehrels}, {Germani}, {Giebels}, {Giglietto}, {Giordano},
  {Glanzman}, {Godfrey}, {Grenier}, {Grondin}, {Grove}, {Guillemot}, {Guiriec},
  {Harding}, {Hartman}, {Hays}, {Hughes}, {J{\'o}hannesson}, {Johnson},
  {Johnson}, {Johnson}, {Johnson}, {Kamae}, {Kanai}, {Kanbach}, {Katagiri},
  {Kawai}, {Kerr}, {Kishishita}, {Kiziltan}, {Kn{\"o}dlseder}, {Kocian},
  {Komin}, {Kuehn}, {Kuss}, {Latronico}, {Lemoine-Goumard}, {Longo}, {Lonjou},
  {Loparco}, {Lott}, {Lovellette}, {Lubrano}, {Makeev}, {Marelli}, {Mazziotta},
  {McEnery}, {McGlynn}, {Meurer}, {Michelson}, {Mineo}, {Mitthumsiri},
  {Mizuno}, {Moiseev}, {Monte}, {Monzani}, {Morselli}, {Moskalenko}, {Murgia},
  {Nakamori}, {Nolan}, {Nuss}, {Ohno}, {Ohsugi}, {Okumura}, {Omodei},
  {Orlando}, {Ormes}, {Ozaki}, {Paneque}, {Panetta}, {Parent}, {Pelassa},
  {Pepe}, {Pesce-Rollins}, {Piano}, {Pieri}, {Piron}, {Porter}, {Rain{\`o}},
  {Rando}, {Ray}, {Razzano}, {Reimer}, {Reimer}, {Reposeur}, {Ritz},
  {Rochester}, {Rodriguez}, {Romani}, {Roth}, {Ryde}, {Sadrozinski}, {Sanchez},
  {Sander}, {Parkinson}, {Schalk}, {Sellerholm}, {Sgr{\`o}}, {Siskind},
  {Smith}, {Smith}, {Spandre}, {Spinelli}, {Starck}, {Strickman}, {Suson},
  {Tajima}, {Takahashi}, {Takahashi}, {Tanaka}, {Thayer}, {Thayer}, {Thompson},
  {Thorsett}, {Tibaldo}, {Torres}, {Tosti}, {Tramacere}, {Usher}, {Van Etten},
  {Vilchez}, {Vitale}, {Wang}, {Watters}, {Winer}, {Wood}, {Yasuda}, {Ylinen},
  \& {Ziegler}}]{Abdo:2008nz}
{Abdo}, A.~A., {Ackermann}, M., {Atwood}, W.~B., {et~al.} 2008, Science, 322,
  1218

\bibitem[{{Abdo} {et~al.}(2010{\natexlab{a}}){Abdo}, {Ackermann}, {Ajello},
  {Baldini}, {Ballet}, {Barbiellini}, {Bastieri}, {Bechtol}, {Bellazzini},
  {Berenji}, {Blandford}, {Bloom}, {Bonamente}, {Borgland}, {Bouvier},
  {Brandt}, {Bregeon}, {Brez}, {Brigida}, {Bruel}, {Buehler}, {Buson},
  {Caliandro}, {Cameron}, {Caraveo}, {Carrigan}, {Casandjian}, {Cecchi}, {{\c
  C}elik}, {Charles}, {Chekhtman}, {Cheung}, {Chiang}, {Ciprini}, {Claus},
  {Cohen-Tanugi}, {Conrad}, {Dermer}, {de Palma}, {Digel}, {Silva}, {Drell},
  {Dubois}, {Dumora}, {Favuzzi}, {Fegan}, {Fukazawa}, {Funk}, {Fusco},
  {Gargano}, {Gasparrini}, {Gehrels}, {Germani}, {Giglietto}, {Giordano},
  {Giroletti}, {Glanzman}, {Godfrey}, {Grenier}, {Grondin}, {Grove}, {Guiriec},
  {Hadasch}, {Harding}, {Hayashida}, {Hays}, {Horan}, {Hughes}, {Jean},
  {J{\'o}hannesson}, {Johnson}, {Johnson}, {Kamae}, {Katagiri}, {Kataoka},
  {Kerr}, {Kn{\"o}dlseder}, {Kuss}, {Lande}, {Latronico}, {Lee},
  {Lemoine-Goumard}, {Llena Garde}, {Longo}, {Loparco}, {Lovellette},
  {Lubrano}, {Makeev}, {Martin}, {Mazziotta}, {McEnery}, {Michelson},
  {Mitthumsiri}, {Mizuno}, {Monte}, {Monzani}, {Morselli}, {Moskalenko},
  {Murgia}, {Nakamori}, {Naumann-Godo}, {Nolan}, {Norris}, {Nuss}, {Ohsugi},
  {Okumura}, {Omodei}, {Orlando}, {Ormes}, {Panetta}, {Parent}, {Pelassa},
  {Pepe}, {Pesce-Rollins}, {Piron}, {Porter}, {Rain{\`o}}, {Rando}, {Razzano},
  {Reimer}, {Reimer}, {Reposeur}, {Ripken}, {Ritz}, {Romani}, {Sadrozinski},
  {Sander}, {Saz Parkinson}, {Scargle}, {Sgr{\`o}}, {Siskind}, {Smith},
  {Smith}, {Spandre}, {Spinelli}, {Strickman}, {Strong}, {Suson}, {Takahashi},
  {Takahashi}, {Tanaka}, {Thayer}, {Thayer}, {Thompson}, {Tibaldo}, {Torres},
  {Tosti}, {Tramacere}, {Uchiyama}, {Usher}, {Vandenbroucke}, {Vasileiou},
  {Vilchez}, {Vitale}, {Waite}, {Wang}, {Winer}, {Wood}, {Yang}, {Ylinen}, \&
  {Ziegler}}]{Fermi-LAT:2010fcp}
{Abdo}, A.~A., {Ackermann}, M., {Ajello}, M., {et~al.} 2010{\natexlab{a}},
  \aap, 523, A46

\bibitem[{{Abdo} {et~al.}(2010{\natexlab{b}}){Abdo}, {Ackermann}, {Ajello},
  {Atwood}, {Baldini}, {Ballet}, {Barbiellini}, {Bastieri}, {Baughman},
  {Bechtol}, {Bellazzini}, {Berenji}, {Blandford}, {Bloom}, {Bonamente},
  {Borgland}, {Bregeon}, {Brez}, {Brigida}, {Bruel}, {Burnett}, {Buson},
  {Caliandro}, {Cameron}, {Caraveo}, {Casandjian}, {Cavazzuti}, {Cecchi}, {{\c
  C}elik}, {Chekhtman}, {Cheung}, {Chiang}, {Ciprini}, {Claus}, {Cohen-Tanugi},
  {Colafrancesco}, {Cominsky}, {Conrad}, {Costamante}, {Cutini}, {Davis},
  {Dermer}, {de Angelis}, {de Palma}, {Digel}, {do Couto e Silva}, {Drell},
  {Dubois}, {Dumora}, {Farnier}, {Favuzzi}, {Fegan}, {Finke}, {Focke},
  {Fortin}, {Fukazawa}, {Funk}, {Fusco}, {Gargano}, {Gasparrini}, {Gehrels},
  {Georganopoulos}, {Germani}, {Giebels}, {Giglietto}, {Giordano}, {Giroletti},
  {Glanzman}, {Godfrey}, {Grenier}, {Grove}, {Guillemot}, {Guiriec},
  {Hanabata}, {Harding}, {Hayashida}, {Hays}, {Hughes}, {Jackson},
  {J{\'o}hannesson G.}, {Johnson}, {Johnson}, {Johnson}, {Kamae}, {Katagiri},
  {Kataoka}, {Kawai}, {Kerr}, {Kn{\"o}dlseder}, {Kocian}, {Kuss}, {Lande},
  {Latronico}, {Lemoine-Goumard}, {Longo}, {Loparco}, {Lott}, {Lovellette},
  {Lubrano}, {Madejski}, {Makeev}, {Mazziotta}, {McConville}, {McEnery},
  {Meurer}, {Michelson}, {Mitthumsiri}, {Mizuno}, {Moiseev}, {Monte},
  {Monzani}, {Morselli}, {Moskalenko}, {Murgia}, {Nolan}, {Norris}, {Nuss},
  {Ohsugi}, {Omodei}, {Orlando}, {Ormes}, {Paneque}, {Parent}, {Pelassa},
  {Pepe}, {Pesce-Rollins}, {Piron}, {Porter}, {Rain{\`o}}, {Rando}, {Razzano},
  {Razzaque}, {Reimer}, {Reimer}, {Reposeur}, {Ritz}, {Rochester}, {Rodriguez},
  {Romani}, {Roth}, {Ryde}, {Sadrozinski}, {Sambruna}, {Sanchez}, {Sander},
  {Saz Parkinson}, {Scargle}, {Sgr{\`o}}, {Siskind}, {Smith}, {Smith},
  {Spandre}, {Spinelli}, {Starck}, {Stawarz}, {Strickman}, {Suson}, {Tajima},
  {Takahashi}, {Takahashi}, {Tanaka}, {Thayer}, {Thayer}, {Thompson},
  {Tibaldo}, {Torres}, {Tosti}, {Tramacere}, {Uchiyama}, {Vasileiou},
  {Vilchez}, {Vitale}, {Waite}, {Wallace}, {Wang}, {Winer}, {Wood}, {Ylinen},
  {Ziegler}, {Hardcastle}, {Kazanas}, \& {Fermi LAT
  Collaboration}}]{2010Sci...328..725A}
---. 2010{\natexlab{b}}, Science, 328, 725

\bibitem[{{Abdo} {et~al.}(2010{\natexlab{c}}){Abdo}, {Ackermann}, {Ajello},
  {Allafort}, {Antolini}, {Atwood}, {Axelsson}, {Baldini}, {Ballet},
  {Barbiellini}, \& et~al.}]{1fgl}
---. 2010{\natexlab{c}}, \apjs, 188, 405

\bibitem[{{Abdo} {et~al.}(2010{\natexlab{d}}){Abdo}, {Ackermann}, {Ajello},
  {Allafort}, {Baldini}, {Ballet}, {Barbiellini}, {Baring}, {Bastieri},
  {Baughman}, {Bechtol}, {Bellazzini}, {Berenji}, {Blandford}, {Bloom},
  {Bonamente}, {Borgland}, {Bregeon}, {Brez}, {Brigida}, {Bruel}, {Buehler},
  {Burnett}, {Busetto}, {Caliandro}, {Cameron}, {Caraveo}, {Casandjian},
  {Cecchi}, {{\c C}elik}, {Charles}, {Chaty}, {Chekhtman}, {Cheung}, {Chiang},
  {Cillis}, {Ciprini}, {Claus}, {Cohen-Tanugi}, {Conrad}, {Corbel}, {de Palma},
  {Digel}, {Dormody}, {Silva}, {Drell}, {Dubois}, {Dumora}, {Edmonds},
  {Farnier}, {Favuzzi}, {Fegan}, {Ferrara}, {Focke}, {Fortin}, {Frailis},
  {Fukazawa}, {Funk}, {Fusco}, {Gargano}, {Gasparrini}, {Gehrels}, {Germani},
  {Giavitto}, {Giglietto}, {Giordano}, {Glanzman}, {Godfrey}, {Grenier},
  {Grondin}, {Grove}, {Guillemot}, {Guiriec}, {Hanabata}, {Hays}, {Harding},
  {Hayashida}, {Horan}, {Hughes}, {Jackson}, {Johnson}, {Johnson}, {Johnson},
  {Kamae}, {Katagiri}, {Kataoka}, {Kawai}, {Kerr}, {Kn{\"o}dlseder}, {Kuss},
  {Lande}, {Latronico}, {Lemoine-Goumard}, {Longo}, {Loparco}, {Lott},
  {Lovellette}, {Lubrano}, {Makeev}, {Mazziotta}, {Meurer}, {Michelson},
  {Mitthumsiri}, {Mizuno}, {Monte}, {Monzani}, {Morselli}, {Moskalenko},
  {Murgia}, {Nakamori}, {Nolan}, {Norris}, {Nuss}, {Ohsugi}, {Okumura},
  {Omodei}, {Orlando}, {Ormes}, {Paneque}, {Panetta}, {Pelassa}, {Pepe},
  {Pesce-Rollins}, {Piron}, {Pohl}, {Porter}, {Rain{\`o}}, {Rando}, {Reimer},
  {Reimer}, {Reposeur}, {Ritz}, {Rodriguez}, {Romani}, {Roth}, {Sadrozinski},
  {Sander}, {Saz Parkinson}, {Scargle}, {Sgr{\`o}}, {Siskind}, {Smith},
  {Smith}, {Spinelli}, {Strickman}, {Suson}, {Tajima}, {Takahashi}, {Tanaka},
  {Thayer}, {Thayer}, {Thompson}, {Thorsett}, {Tibaldo}, {Tibolla}, {Torres},
  {Tosti}, {Tramacere}, {Uchiyama}, {Usher}, {Van Etten}, {Vasileiou},
  {Venter}, {Vilchez}, {Vitale}, {Waite}, {Wang}, {Winer}, {Wood}, {Yamazaki},
  {Ylinen}, \& {Ziegler}}]{2010ApJ...710L..92A}
---. 2010{\natexlab{d}}, \apjl, 710, L92

\bibitem[{{Abdo} {et~al.}(2010{\natexlab{e}}){Abdo}, {Ackermann}, {Ajello},
  {Atwood}, {Baldini}, {Ballet}, {Barbiellini}, {Bastieri}, {Baughman},
  {Bechtol}, {Bellazzini}, {Berenji}, {Blandford}, {Bloom}, {Bonamente},
  {Borgland}, {Bregeon}, {Brez}, {Brigida}, {Bruel}, {Burnett}, {Buson},
  {Caliandro}, {Cameron}, {Caraveo}, {Casandjian}, {Cecchi}, {{\c C}elik},
  {Chekhtman}, {Cheung}, {Chiang}, {Ciprini}, {Claus}, {Cohen-Tanugi},
  {Cominsky}, {Conrad}, {Cutini}, {Dermer}, {de Angelis}, {de Palma}, {Digel},
  {Silva}, {Drell}, {Dubois}, {Dumora}, {Farnier}, {Favuzzi}, {Fegan}, {Focke},
  {Fortin}, {Frailis}, {Fukazawa}, {Fusco}, {Gargano}, {Gasparrini}, {Gehrels},
  {Germani}, {Giavitto}, {Giebels}, {Giglietto}, {Giordano}, {Glanzman},
  {Godfrey}, {Gotthelf}, {Grenier}, {Grondin}, {Grove}, {Guillemot}, {Guiriec},
  {Hanabata}, {Harding}, {Hayashida}, {Hays}, {Horan}, {Hughes}, {Jackson},
  {Jean}, {J{\'o}hannesson}, {Johnson}, {Johnson}, {Johnson}, {Johnson},
  {Kamae}, {Katagiri}, {Kataoka}, {Kawai}, {Kerr}, {Kn{\"o}dlseder}, {Kocian},
  {Kuss}, {Lande}, {Latronico}, {Lemoine-Goumard}, {Longo}, {Loparco}, {Lott},
  {Lovellette}, {Lubrano}, {Madejski}, {Makeev}, {Marshall}, {Martin},
  {Mazziotta}, {McConville}, {McEnery}, {Meurer}, {Michelson}, {Mitthumsiri},
  {Mizuno}, {Moiseev}, {Monte}, {Monzani}, {Morselli}, {Moskalenko}, {Murgia},
  {Nolan}, {Norris}, {Nuss}, {Ohsugi}, {Omodei}, {Orlando}, {Ormes}, {Paneque},
  {Parent}, {Pelassa}, {Pepe}, {Pesce-Rollins}, {Piron}, {Porter}, {Rain{\`o}},
  {Rando}, {Razzano}, {Reimer}, {Reimer}, {Reposeur}, {Ritz}, {Rodriguez},
  {Romani}, {Roth}, {Ryde}, {Sadrozinski}, {Sanchez}, {Sander}, {Saz
  Parkinson}, {Scargle}, {Sellerholm}, {Sgr{\`o}}, {Siskind}, {Smith}, {Smith},
  {Spandre}, {Spinelli}, {Starck}, {Strickman}, {Strong}, {Suson}, {Tajima},
  {Takahashi}, {Tanaka}, {Thayer}, {Thayer}, {Thompson}, {Tibaldo}, {Torres},
  {Tosti}, {Tramacere}, {Uchiyama}, {Usher}, {Vasileiou}, {Venter}, {Vilchez},
  {Vitale}, {Waite}, {Wang}, {Weltevrede}, {Winer}, {Wood}, {Ylinen}, \&
  {Ziegler}}]{2010A&A...512A...7A}
---. 2010{\natexlab{e}}, \apj, 512, A7

\bibitem[{{Abergel} {et~al.}(2014){Abergel}, {Ade}, {Aghanim}, {Alves},
  {Aniano}, {Armitage-Caplan}, {Arnaud}, {Ashdown}, {Atrio-Barandela}, \&
  et~al.}]{2014A&A...571A..11P}
{Abergel}, A., {Ade}, P.~A.~R., {Aghanim}, N., {et~al.} 2014, \aap, 571, A11

\bibitem[{{{\'A}brah{\'a}m} {et~al.}(2000){{\'A}brah{\'a}m}, {Bal{\'a}zs}, \&
  {Kun}}]{2000A&A...354..645A}
{{\'A}brah{\'a}m}, P., {Bal{\'a}zs}, L.~G., \& {Kun}, M. 2000, \aap, 354, 645

\bibitem[{Abrahams {et~al.}(2017)Abrahams, Teachey, \&
  Paglione}]{Abrahams:2016chw}
Abrahams, R.~D., Teachey, A., \& Paglione, T. A.~D. 2017, \apj, 834, 91

\bibitem[{{Abramowski} {et~al.}(2010){Abramowski}, {Acero}, {Aharonian},
  {Akhperjanian}, {Anton}, {Barres de Almeida}, {Bazer-Bachi}, {Becherini},
  {Behera}, {Benbow}, {Bernl{\"o}hr}, {Bochow}, {Boisson}, {Bolmont}, {Borrel},
  {Brucker}, {Brun}, {Brun}, {B{\"u}hler}, {Bulik}, {B{\"u}sching},
  {Boutelier}, {Chadwick}, {Charbonnier}, {Chaves}, {Cheesebrough}, {Conrad},
  {Chounet}, {Clapson}, {Coignet}, {Costamante}, {Dalton}, {Daniel}, {Davids},
  {Degrange}, {Deil}, {Dickinson}, {Djannati-Ata{\"i}}, {Domainko},
  {O'C.~Drury}, {Dubois}, {Dubus}, {Dyks}, {Dyrda}, {Egberts}, {Eger},
  {Espigat}, {Fallon}, {Farnier}, {Fegan}, {Feinstein}, {Fernandes}, {Fiasson},
  {F{\"o}rster}, {Fontaine}, {F{\"u}{\ss}ling}, {Gabici}, {Gallant},
  {G{\'e}rard}, {Gerbig}, {Giebels}, {Glicenstein}, {Gl{\"u}ck}, {Goret},
  {G{\"o}ring}, {Hampf}, {Hauser}, {Heinz}, {Heinzelmann}, {Henri}, {Hermann},
  {Hinton}, {Hoffmann}, {Hofmann}, {Hofverberg}, {Holleran}, {Hoppe}, {Horns},
  {Jacholkowska}, {de Jager}, {Jahn}, {Jung}, {Katarzy{\'n}ski}, {Katz},
  {Kaufmann}, {Kerschhaggl}, {Khangulyan}, {Kh{\'e}lifi}, {Keogh}, {Klochkov},
  {Klu{\v z}niak}, {Kneiske}, {Komin}, {Kosack}, {Kossakowski}, {Lamanna},
  {Lenain}, {Lohse}, {Lu}, {Marandon}, {Marcowith}, {Masbou}, {Maurin},
  {McComb}, {Medina}, {M{\'e}hault}, {Moderski}, {Moulin}, {Naumann-Godo}, {de
  Naurois}, {Nedbal}, {Nekrassov}, {Nguyen}, {Nicholas}, {Niemiec}, {Nolan},
  {Ohm}, {Olive}, {de O{\~n}a Wilhelmi}, {Opitz}, {Orford}, {Ostrowski},
  {Panter}, {Paz Arribas}, {Pedaletti}, {Pelletier}, {Petrucci}, {Pita},
  {P{\"u}hlhofer}, {Punch}, {Quirrenbach}, {Raubenheimer}, {Raue}, {Rayner},
  {Reimer}, {Renaud}, {de Los Reyes}, {Rieger}, {Ripken}, {Rob}, {Rosier-Lees},
  {Rowell}, {Rudak}, {Rulten}, {Ruppel}, {Ryde}, {Sahakian}, {Santangelo},
  {Schlickeiser}, {Sch{\"o}ck}, {Sch{\"o}nwald}, {Schwanke}, {Schwarzburg},
  {Schwemmer}, {Shalchi}, {Sushch}, {Sikora}, {Skilton}, {Sol}, {Stawarz},
  {Steenkamp}, {Stegmann}, {Stinzing}, {Szostek}, {Tam}, {Tavernet}, {Terrier},
  {Tibolla}, {Tluczykont}, {Valerius}, {van Eldik}, {Vasileiadis}, {Venter},
  {Venter}, {Vialle}, {Viana}, {Vincent}, {Vivier}, {V{\"o}lk}, {Volpe},
  {Vorobiov}, {Wagner}, {Ward}, {Zdziarski}, {Zech}, \&
  {Zechlin}}]{2010AA...516A..56H}
{Abramowski}, A., {Acero}, F., {Aharonian}, F., {et~al.} 2010, \aap, 516, A56

\bibitem[{{Abramowski} {et~al.}(2012{\natexlab{a}}){Abramowski}, {Acero},
  {Aharonian}, {Akhperjanian}, {Anton}, {Balzer}, {Barnacka}, {Barres de
  Almeida}, {Becherini}, {Becker}, {Behera}, {Bernloehr}, {Birsin}, {Biteau},
  {Bochow}, {Boisson}, {Bolmont}, {Bordas}, {Brucker}, {Brun}, {Brun}, {Bulik},
  {Buesching}, {Carrigan}, {Casanova}, {Cerruti}, {Chadwick}, {Charbonnier},
  {Chaves}, {Cheesebrough}, {Chounet}, {Clapson}, {Coignet}, {Cologna},
  {Conrad}, {Dalton}, {Daniel}, {Davids}, {Degrange}, {Deil}, {Dickinson},
  {Djannati-Ataie}, {Domainko}, {Drury}, {Dubois}, {Dubus}, {Dutson}, {Dyks},
  {Dyrda}, {Egberts}, {Eger}, {Espigat}, {Fallon}, {Farnier}, {Feinstein},
  {Fernandes}, {Fiasson}, {Fontaine}, {Foerster}, {Fuesling}, {Gallant},
  {Gast}, {Gerard}, {Gerbig}, {Giebels}, {Glicenstein}, {Glueck}, {Goret},
  {Goering}, {Haeffner}, {Hague}, {Hampf}, {Hauser}, {Heinz}, {Heinzelmann},
  {Henri}, {Hermann}, {Hinton}, {Hoffmann}, {Hofmann}, {Hofverberg}, {Holler},
  {Horns}, {Jacholkowska}, {de Jager}, {Jahn}, {Jamrozy}, {Jung},
  {Kastendieck}, {Katarzynski}, {Katz}, {Kaufmann}, {Keogh}, {Khangulyan},
  {Khelifi}, {Klochkov}, {Kluzniak}, {Kneiske}, {Komin}, {Kosack},
  {Kossakowski}, {Laffon}, {Lamanna}, {Lennarz}, {Lohse}, {Lopatin}, {Lu},
  {Marandon}, {Marcowith}, {Masbou}, {Maurin}, {Maxted}, {Mayer}, {McComb},
  {Medina}, {Mehault}, {Moderski}, {Moulin}, {Naumann}, {Naumann-Godo}, {de
  Naurois}, {Nedbal}, {Nekrassov}, {Nguyen}, {Nicholas}, {Niemiec}, {Nolan},
  {Ohm}, {de Ona Wilhelmi}, {Opitz}, {Ostrowski}, {Oya}, {Panter}, {Paz
  Arribas}, {Pedaletti}, {Pelletier}, {Petrucci}, {Pita}, {Puehlhofer},
  {Punch}, {Quirrenbach}, {Raue}, {Rayner}, {Reimer}, {Reimer}, {Renaud}, {de
  Los Reyes}, {Rieger}, {Ripken}, {Rob}, {Rosier-Lees}, {Rowell}, {Rudak},
  {Rulten}, {Ruppel}, {Sahakian}, {Sanchez}, {Santangelo}, {Schlickeiser},
  {Schoeck}, {Schulz}, {Schwanke}, {Schwarzburg}, {Schwemmer}, {Sheidaei},
  {Sikora}, {Skilton}, {Sol}, {Spengler}, {Stawarz}, {Steenkamp}, {Stegmann},
  {Stinzing}, {Stycz}, {Sushch}, {Szostek}, {Tavernet}, {Terrier},
  {Tluczykont}, {Valerius}, {van Eldik}, {Vasileiadis}, {Venter}, {Vialle},
  {Viana}, {Vincent}, {Voelk}, {Volpe}, {Vorobiov}, {Vorster}, {Wagner},
  {Ward}, {White}, {Wierzcholska}, {Zacharias}, {Zajczyk}, {Zdziarski}, {Zech},
  {Zechlin}, {Costamante}, {Fegan}, \& {Ajello}}]{2012AA...538A.103H}
---. 2012{\natexlab{a}}, \aap, 538, A103

\bibitem[{{Abramowski} {et~al.}(2012{\natexlab{b}}){Abramowski}, {Acero},
  {Aharonian}, {Akhperjanian}, {Anton}, {Balzer}, {Barnacka}, {Becherini},
  {Becker}, {Bernl{\"o}hr}, {Birsin}, {Biteau}, {Bochow}, {Boisson}, {Bolmont},
  {Bordas}, {Brucker}, {Brun}, {Brun}, {Bulik}, {B{\"u}sching}, {Carrigan},
  {Casanova}, {Cerruti}, {Chadwick}, {Charbonnier}, {Chaves}, {Cheesebrough},
  {Cologna}, {Conrad}, {Dalton}, {Daniel}, {Davids}, {Degrange}, {Deil},
  {Dickinson}, {Djannati-Ata{\"i}}, {Domainko}, {Drury}, {Dubus}, {Dutson},
  {Dyks}, {Dyrda}, {Egberts}, {Eger}, {Espigat}, {Fallon}, {Fegan},
  {Feinstein}, {Fernandes}, {Fiasson}, {Fontaine}, {F{\"o}rster},
  {F{\"u}{\ss}ling}, {Gallant}, {Gast}, {G{\'e}rard}, {Gerbig}, {Giebels},
  {Glicenstein}, {Gl{\"u}ck}, {G{\"o}ring}, {H{\"a}ffner}, {Hague}, {Hahn},
  {Hampf}, {Harris}, {Hauser}, {Heinz}, {Heinzelmann}, {Henri}, {Hermann},
  {Hillert}, {Hinton}, {Hofmann}, {Hofverberg}, {Holler}, {Horns},
  {Jacholkowska}, {de Jager}, {Jahn}, {Jamrozy}, {Jung}, {Kastendieck},
  {Katarzy{\'n}ski}, {Katz}, {Kaufmann}, {Keogh}, {Kh{\'e}lifi}, {Klochkov},
  {Klu{\'z}niak}, {Kneiske}, {Komin}, {Kosack}, {Kossakowski}, {Krayzel},
  {Laffon}, {Lamanna}, {Lenain}, {Lennarz}, {Lohse}, {Lopatin}, {Lu},
  {Marandon}, {Marcowith}, {Masbou}, {Maxted}, {Mayer}, {McComb}, {Medina},
  {M{\'e}hault}, {Moderski}, {Mohamed}, {Moulin}, {Naumann}, {Naumann-Godo},
  {de Naurois}, {Nedbal}, {Nekrassov}, {Nguyen}, {Nicholas}, {Niemiec},
  {Nolan}, {Ohm}, {de O{\~n}a Wilhelmi}, {Opitz}, {Ostrowski}, {Oya}, {Panter},
  {Paz Arribas}, {Pekeur}, {Pelletier}, {Perez}, {Petrucci}, {Peyaud}, {Pita},
  {P{\"u}hlhofer}, {Punch}, {Quirrenbach}, {Raue}, {Rayner}, {Reimer},
  {Reimer}, {Renaud}, {de los Reyes}, {Rieger}, {Ripken}, {Rob}, {Rosier-Lees},
  {Rowell}, {Rudak}, {Rulten}, {Sahakian}, {Sanchez}, {Santangelo},
  {Schlickeiser}, {Schulz}, {Schwanke}, {Schwarzburg}, {Schwemmer}, {Sheidaei},
  {Skilton}, {Sol}, {Spengler}, {Stawarz}, {Steenkamp}, {Stegmann}, {Stinzing},
  {Stycz}, {Sushch}, {Szostek}, {Tavernet}, {Terrier}, {Tluczykont},
  {Valerius}, {van Eldik}, {Vasileiadis}, {Venter}, {Viana}, {Vincent},
  {V{\"o}lk}, {Volpe}, {Vorobiov}, {Vorster}, {Wagner}, {Ward}, {White},
  {Wierzcholska}, {Zacharias}, {Zajczyk}, {Zdziarski}, {Zech}, \&
  {Zechlin}}]{2012AA...542A..94H}
---. 2012{\natexlab{b}}, \aap, 542, A94

\bibitem[{{Abramowski} {et~al.}(2013){Abramowski}, {Acero}, {Aharonian},
  {Akhperjanian}, {Ang{\"u}ner}, {Anton}, {Balenderan}, {Balzer}, {Barnacka},
  \& et~al.}]{2013MNRAS.434.1889H}
---. 2013, \mnras, 434, 1889

\bibitem[{{Abramowski} {et~al.}(2014){Abramowski}, {Aharonian}, {Ait Benkhali},
  {Akhperjanian}, {Ang{\"u}ner}, {Anton}, {Backes}, {Balenderan}, {Balzer}, \&
  et~al.}]{hess2014ph}
{Abramowski}, A., {Aharonian}, F., {Ait Benkhali}, F., {et~al.} 2014, \aap,
  562, A145

\bibitem[{{Abramowski} {et~al.}(2015){Abramowski}, {Aharonian}, {Ait Benkhali},
  {Akhperjanian}, {Ang{\"u}ner}, {Backes}, {Balenderan}, {Balzer}, {Barnacka},
  \& et~al.}]{2015Sci...347..406H}
---. 2015, Science, 347, 406

\bibitem[{{Acciari} {et~al.}(2009){Acciari}, {Aliu}, {Arlen}, {Beilicke},
  {Benbow}, {Bradbury}, {Buckley}, {Bugaev}, {Butt}, {Byrum}, {Celik},
  {Cesarini}, {Ciupik}, {Chow}, {Cogan}, {Colin}, {Cui}, {Daniel}, {Ergin},
  {Falcone}, {Fegan}, {Finley}, {Fortin}, {Fortson}, {Furniss}, {Gillanders},
  {Grube}, {Guenette}, {Gyuk}, {Hanna}, {Hays}, {Holder}, {Horan}, {Hui},
  {Humensky}, {Imran}, {Kaaret}, {Karlsson}, {Kertzman}, {Kieda}, {Kildea},
  {Konopelko}, {Krawczynski}, {Krennrich}, {Lang}, {LeBohec}, {Maier},
  {McCann}, {McCutcheon}, {Moriarty}, {Mukherjee}, {Nagai}, {Niemiec}, {Ong},
  {Pandel}, {Perkins}, {Pohl}, {Quinn}, {Ragan}, {Reyes}, {Reynolds}, {Rose},
  {Schroedter}, {Sembroski}, {Smith}, {Steele}, {Swordy}, {Toner}, {Valcarcel},
  {Vassiliev}, {Wagner}, {Wakely}, {Ward}, {Weekes}, {Weinstein}, {White},
  {Williams}, {Wissel}, {Wood}, \& {Zitzer}}]{2009ApJ...695.1370A}
{Acciari}, V.~A., {Aliu}, E., {Arlen}, T., {et~al.} 2009, \apj, 695, 1370

\bibitem[{{Acciari} {et~al.}(2010{\natexlab{a}}){Acciari}, {Aliu}, {Arlen},
  {Beilicke}, {Benbow}, {Bradbury}, {Buckley}, {Bugaev}, {Butt}, {Byrum},
  {Celik}, {Cesarini}, {Ciupik}, {Chow}, {Cogan}, {Colin}, {Cui}, {Daniel},
  {Ergin}, {Falcone}, {Fegan}, {Finley}, {Fortin}, {Fortson}, {Furniss},
  {Gillanders}, {Grube}, {Guenette}, {Gyuk}, {Hanna}, {Hays}, {Holder},
  {Horan}, {Hui}, {Humensky}, {Imran}, {Kaaret}, {Karlsson}, {Kertzman},
  {Kieda}, {Kildea}, {Konopelko}, {Krawczynski}, {Krennrich}, {Lang},
  {LeBohec}, {Maier}, {McCann}, {McCutcheon}, {Moriarty}, {Mukherjee}, {Nagai},
  {Niemiec}, {Ong}, {Pandel}, {Perkins}, {Pohl}, {Quinn}, {Ragan}, {Reyes},
  {Reynolds}, {Rose}, {Schroedter}, {Sembroski}, {Smith}, {Steele}, {Swordy},
  {Toner}, {Valcarcel}, {Vassiliev}, {Wagner}, {Wakely}, {Ward}, {Weekes},
  {Weinstein}, {White}, {Williams}, {Wissel}, {Wood}, \&
  {Zitzer}}]{2010ApJ...709L.163A}
---. 2010{\natexlab{a}}, \apj, 709, L163

\bibitem[{{Acciari} {et~al.}(2010{\natexlab{b}}){Acciari}, {Aliu}, {Arlen},
  {Aune}, {Bautista}, {Beilicke}, {Benbow}, {B{\"o}ttcher}, {Boltuch},
  {Bradbury}, \& et~al.}]{2010ApJ...715L..49A}
---. 2010{\natexlab{b}}, \apj, 715, L49

\bibitem[{{Acero} {et~al.}(2011){Acero}, {Aharonian}, {Akhperjanian}, {Anton},
  {Barres de Almeida}, {Bazer-Bachi}, {Becherini}, {Behera}, {Bernl{\"o}hr},
  {Bochow}, {Boisson}, {Bolmont}, {Borrel}, {Brucker}, {Brun}, {Brun},
  {B{\"u}hler}, {Bulik}, {B{\"u}sching}, {Boutelier}, {Chadwick},
  {Charbonnier}, {Chaves}, {Cheesebrough}, {Chounet}, {Clapson}, {Coignet},
  {Dalton}, {Daniel}, {Davids}, {Degrange}, {Deil}, {Dickinson},
  {Djannati-Ata{\"i}}, {Domainko}, {Drury}, {Dubois}, {Dubus}, {Dyks}, {Dyrda},
  {Egberts}, {Emmanoulopoulos}, {Espigat}, {Farnier}, {Feinstein}, {Fiasson},
  {F{\"o}rster}, {Fontaine}, {F{\"u}{\ss}ling}, {Gabici}, {Gallant},
  {G{\'e}rard}, {Gerbig}, {Giebels}, {Glicenstein}, {Gl{\"u}ck}, {Goret},
  {G{\"o}ring}, {Hauser}, {Hauser}, {Heinz}, {Heinzelmann}, {Henri}, {Hermann},
  {Hinton}, {Hoffmann}, {Hofmann}, {Holleran}, {Hoppe}, {Horns},
  {Jacholkowska}, {de Jager}, {Jahn}, {Jung}, {Katarzy{\'n}ski}, {Katz},
  {Kaufmann}, {Kerschhaggl}, {Khangulyan}, {Kh{\'e}lifi}, {Keogh}, {Klochkov},
  {Klu{\'z}niak}, {Kneiske}, {Komin}, {Kosack}, {Kossakowski}, {Lamanna},
  {Lenain}, {Lohse}, {Marandon}, {Martineau-Huynh}, {Marcowith}, {Masbou},
  {Maurin}, {McComb}, {Medina}, {M{\'e}hault}, {Moderski}, {Moulin},
  {Naumann-Godo}, {de Naurois}, {Nedbal}, {Nekrassov}, {Nicholas}, {Niemiec},
  {Nolan}, {Ohm}, {Olive}, {de O{\~n}a Wilhelmi}, {Orford}, {Ostrowski},
  {Panter}, {Paz Arribas}, {Pedaletti}, {Pelletier}, {Petrucci}, {Pita},
  {P{\"u}hlhofer}, {Punch}, {Quirrenbach}, {Raubenheimer}, {Raue}, {Rayner},
  {Reimer}, {Renaud}, {Rieger}, {Ripken}, {Rob}, {Rosier-Lees}, {Rowell},
  {Rudak}, {Rulten}, {Ruppel}, {Sahakian}, {Santangelo}, {Schlickeiser},
  {Sch{\"o}ck}, {Schwanke}, {Schwarzburg}, {Schwemmer}, {Shalchi}, {Sikora},
  {Skilton}, {Sol}, {Stawarz}, {Steenkamp}, {Stegmann}, {Stinzing}, {Superina},
  {Szostek}, {Tam}, {Tavernet}, {Terrier}, {Tibolla}, {Tluczykont}, {van
  Eldik}, {Vasileiadis}, {Venter}, {Venter}, {Vialle}, {Vincent}, {Vivier},
  {V{\"o}lk}, {Volpe}, {Wagner}, {Ward}, {Zdziarski}, {Zech}, \&
  {H.E.S.S.~Collaboration}}]{2011A&A...525A..45H}
{Acero}, F., {Aharonian}, F., {Akhperjanian}, A.~G., {et~al.} 2011, \aap, 525,
  A45

\bibitem[{{Acero} {et~al.}(2013){Acero}, {Ackermann}, {Ajello}, {Allafort},
  {Baldini}, {Ballet}, {Barbiellini}, {Bastieri}, {Bechtol}, {Bellazzini},
  {Blandford}, {Bloom}, {Bonamente}, {Bottacini}, {Brandt}, {Bregeon},
  {Brigida}, {Bruel}, {Buehler}, {Buson}, {Caliandro}, {Cameron}, {Caraveo},
  {Cecchi}, {Charles}, {Chaves}, {Chekhtman}, {Chiang}, {Chiaro}, {Ciprini},
  {Claus}, {Cohen-Tanugi}, {Conrad}, {Cutini}, {Dalton}, {D'Ammando}, {de
  Palma}, {Dermer}, {Di Venere}, {Silva}, {Drell}, {Drlica-Wagner}, {Falletti},
  {Favuzzi}, {Fegan}, {Ferrara}, {Focke}, {Franckowiak}, {Fukazawa}, {Funk},
  {Fusco}, {Gargano}, {Gasparrini}, {Giglietto}, {Giordano}, {Giroletti},
  {Glanzman}, {Godfrey}, {Gr{\'e}goire}, {Grenier}, {Grondin}, {Grove},
  {Guiriec}, {Hadasch}, {Hanabata}, {Harding}, {Hayashida}, {Hayashi}, {Hays},
  {Hewitt}, {Hill}, {Horan}, {Hou}, {Hughes}, {Inoue}, {Jackson}, {Jogler},
  {J{\'o}hannesson}, {Johnson}, {Kamae}, {Kawano}, {Kerr}, {Kn{\"o}dlseder},
  {Kuss}, {Lande}, {Larsson}, {Latronico}, {Lemoine-Goumard}, {Longo},
  {Loparco}, {Lovellette}, {Lubrano}, {Marelli}, {Massaro}, {Mayer},
  {Mazziotta}, {McEnery}, {Mehault}, {Michelson}, {Mitthumsiri}, {Mizuno},
  {Monte}, {Monzani}, {Morselli}, {Moskalenko}, {Murgia}, {Nakamori}, {Nemmen},
  {Nuss}, {Ohsugi}, {Okumura}, {Orienti}, {Orlando}, {Ormes}, {Paneque},
  {Panetta}, {Perkins}, {Pesce-Rollins}, {Piron}, {Pivato}, {Porter},
  {Rain{\`o}}, {Rando}, {Razzano}, {Reimer}, {Reimer}, {Reposeur}, {Ritz},
  {Roth}, {Rousseau}, {Saz Parkinson}, {Schulz}, {Sgr{\`o}}, {Siskind},
  {Smith}, {Spandre}, {Spinelli}, {Suson}, {Takahashi}, {Takeuchi}, {Thayer},
  {Thayer}, {Thompson}, {Tibaldo}, {Tibolla}, {Tinivella}, {Torres}, {Tosti},
  {Troja}, {Uchiyama}, {Vandenbroucke}, {Vasileiou}, {Vianello}, {Vitale},
  {Werner}, {Winer}, {Wood}, \& {Yang}}]{2013ApJ...773...77A}
{Acero}, F., {Ackermann}, M., {Ajello}, M., {et~al.} 2013, \apj, 773, 77

\bibitem[{{Acero} {et~al.}(2015){Acero}, {Ackermann}, {Ajello}, {Albert},
  {Atwood}, {Axelsson}, {Baldini}, {Ballet}, {Barbiellini}, {Bastieri},
  {Belfiore}, {Bellazzini}, {Bissaldi}, {Blandford}, {Bloom}, {Bogart},
  {Bonino}, {Bottacini}, {Bregeon}, {Britto}, {Bruel}, {Buehler}, {Burnett},
  {Buson}, {Caliandro}, {Cameron}, {Caputo}, {Caragiulo}, {Caraveo},
  {Casandjian}, {Cavazzuti}, {Charles}, {Chaves}, {Chekhtman}, {Cheung},
  {Chiang}, {Chiaro}, {Ciprini}, {Claus}, {Cohen-Tanugi}, {Cominsky}, {Conrad},
  {Cutini}, {D'Ammando}, {de Angelis}, {DeKlotz}, {de Palma}, {Desiante},
  {Digel}, {Di Venere}, {Drell}, {Dubois}, {Dumora}, {Favuzzi}, {Fegan},
  {Ferrara}, {Finke}, {Franckowiak}, {Fukazawa}, {Funk}, {Fusco}, {Gargano},
  {Gasparrini}, {Giebels}, {Giglietto}, {Giommi}, {Giordano}, {Giroletti},
  {Glanzman}, {Godfrey}, {Grenier}, {Grondin}, {Grove}, {Guillemot}, {Guiriec},
  {Hadasch}, {Harding}, {Hays}, {Hewitt}, {Hill}, {Horan}, {Iafrate}, {Jogler},
  {J{\'o}hannesson}, {Johnson}, {Johnson}, {Johnson}, {Johnson}, {Kamae},
  {Kataoka}, {Katsuta}, {Kuss}, {La Mura}, {Landriu}, {Larsson}, {Latronico},
  {Lemoine-Goumard}, {Li}, {Li}, {Longo}, {Loparco}, {Lott}, {Lovellette},
  {Lubrano}, {Madejski}, {Massaro}, {Mayer}, {Mazziotta}, {McEnery},
  {Michelson}, {Mirabal}, {Mizuno}, {Moiseev}, {Mongelli}, {Monzani},
  {Morselli}, {Moskalenko}, {Murgia}, {Nuss}, {Ohno}, {Ohsugi}, {Omodei},
  {Orienti}, {Orlando}, {Ormes}, {Paneque}, {Panetta}, {Perkins},
  {Pesce-Rollins}, {Piron}, {Pivato}, {Porter}, {Racusin}, {Rando}, {Razzano},
  {Razzaque}, {Reimer}, {Reimer}, {Reposeur}, {Rochester}, {Romani},
  {Salvetti}, {S{\'a}nchez-Conde}, {Saz Parkinson}, {Schulz}, {Siskind},
  {Smith}, {Spada}, {Spandre}, {Spinelli}, {Stephens}, {Strong}, {Suson},
  {Takahashi}, {Takahashi}, {Tanaka}, {Thayer}, {Thayer}, {Thompson},
  {Tibaldo}, {Tibolla}, {Torres}, {Torresi}, {Tosti}, {Troja}, {Van Klaveren},
  {Vianello}, {Winer}, {Wood}, {Wood}, {Zimmer}, \& {Fermi-LAT
  Collaboration}}]{3fgl}
---. 2015, \apjs, 218, 23

\bibitem[{{Acero} {et~al.}(2016{\natexlab{a}}){Acero}, {Ackermann}, {Ajello},
  {Albert}, {Baldini}, {Ballet}, {Barbiellini}, {Bastieri}, {Bellazzini},
  {Bissaldi}, {Bloom}, {Bonino}, {Bottacini}, {Brandt}, {Bregeon}, {Bruel},
  {Buehler}, {Buson}, {Caliandro}, {Cameron}, {Caragiulo}, {Caraveo},
  {Casandjian}, {Cavazzuti}, {Cecchi}, {Charles}, {Chekhtman}, {Chiang},
  {Chiaro}, {Ciprini}, {Claus}, {Cohen-Tanugi}, {Conrad}, {Cuoco}, {Cutini},
  {D'Ammando}, {de Angelis}, {de Palma}, {Desiante}, {Digel}, {Di Venere},
  {Drell}, {Favuzzi}, {Fegan}, {Ferrara}, {Focke}, {Franckowiak}, {Funk},
  {Fusco}, {Gargano}, {Gasparrini}, {Giglietto}, {Giordano}, {Giroletti},
  {Glanzman}, {Godfrey}, {Grenier}, {Guiriec}, {Hadasch}, {Harding}, {Hayashi},
  {Hays}, {Hewitt}, {Hill}, {Horan}, {Hou}, {Jogler}, {J{\'o}hannesson},
  {Kamae}, {Kuss}, {Landriu}, {Larsson}, {Latronico}, {Li}, {Li}, {Longo},
  {Loparco}, {Lovellette}, {Lubrano}, {Maldera}, {Malyshev}, {Manfreda},
  {Martin}, {Mayer}, {Mazziotta}, {McEnery}, {Michelson}, {Mirabal}, {Mizuno},
  {Monzani}, {Morselli}, {Nuss}, {Ohsugi}, {Omodei}, {Orienti}, {Orlando},
  {Ormes}, {Paneque}, {Pesce-Rollins}, {Piron}, {Pivato}, {Rain{\`o}}, {Rando},
  {Razzano}, {Razzaque}, {Reimer}, {Reimer}, {Remy}, {Renault},
  {S{\'a}nchez-Conde}, {Schaal}, {Schulz}, {Sgr{\`o}}, {Siskind}, {Spada},
  {Spandre}, {Spinelli}, {Strong}, {Suson}, {Tajima}, {Takahashi}, {Thayer},
  {Thompson}, {Tibaldo}, {Tinivella}, {Torres}, {Tosti}, {Troja}, {Vianello},
  {Werner}, {Wood}, {Wood}, {Zaharijas}, \& {Zimmer}}]{2016ApJS..223...26A}
---. 2016{\natexlab{a}}, \apjs, 223, 26

\bibitem[{{Acero} {et~al.}(2016{\natexlab{b}}){Acero}, {Ackermann}, {Ajello},
  {Baldini}, {Ballet}, {Barbiellini}, {Bastieri}, {Bellazzini}, {Bissaldi},
  {Blandford}, {Bloom}, {Bonino}, {Bottacini}, {Brandt}, {Bregeon}, {Bruel},
  {Buehler}, {Buson}, {Caliandro}, {Cameron}, {Caputo}, {Caragiulo}, {Caraveo},
  {Casandjian}, {Cavazzuti}, {Cecchi}, {Chekhtman}, {Chiang}, {Chiaro},
  {Ciprini}, {Claus}, {Cohen}, {Cohen-Tanugi}, {Cominsky}, {Condon}, {Conrad},
  {Cutini}, {D'Ammando}, {de Angelis}, {de Palma}, {Desiante}, {Digel}, {Di
  Venere}, {Drell}, {Drlica-Wagner}, {Favuzzi}, {Ferrara}, {Franckowiak},
  {Fukazawa}, {Funk}, {Fusco}, {Gargano}, {Gasparrini}, {Giglietto}, {Giommi},
  {Giordano}, {Giroletti}, {Glanzman}, {Godfrey}, {Gomez-Vargas}, {Grenier},
  {Grondin}, {Guillemot}, {Guiriec}, {Gustafsson}, {Hadasch}, {Harding},
  {Hayashida}, {Hays}, {Hewitt}, {Hill}, {Horan}, {Hou}, {Iafrate}, {Jogler},
  {J{\'o}hannesson}, {Johnson}, {Kamae}, {Katagiri}, {Kataoka}, {Katsuta},
  {Kerr}, {Kn{\"o}dlseder}, {Kocevski}, {Kuss}, {Laffon}, {Lande}, {Larsson},
  {Latronico}, {Lemoine-Goumard}, {Li}, {Li}, {Longo}, {Loparco}, {Lovellette},
  {Lubrano}, {Magill}, {Maldera}, {Marelli}, {Mayer}, {Mazziotta}, {Michelson},
  {Mitthumsiri}, {Mizuno}, {Moiseev}, {Monzani}, {Moretti}, {Morselli},
  {Moskalenko}, {Murgia}, {Nemmen}, {Nuss}, {Ohsugi}, {Omodei}, {Orienti},
  {Orlando}, {Ormes}, {Paneque}, {Perkins}, {Pesce-Rollins}, {Petrosian},
  {Piron}, {Pivato}, {Porter}, {Rain{\`o}}, {Rando}, {Razzano}, {Razzaque},
  {Reimer}, {Reimer}, {Renaud}, {Reposeur}, {Rousseau}, {Saz Parkinson},
  {Schmid}, {Schulz}, {Sgr{\`o}}, {Siskind}, {Spada}, {Spandre}, {Spinelli},
  {Strong}, {Suson}, {Tajima}, {Takahashi}, {Tanaka}, {Thayer}, {Thompson},
  {Tibaldo}, {Tibolla}, {Torres}, {Tosti}, {Troja}, {Uchiyama}, {Vianello},
  {Wells}, {Wood}, {Wood}, {Yassine}, {den Hartog}, \&
  {Zimmer}}]{Acero:2015prw}
---. 2016{\natexlab{b}}, \apjs, 224, 8

\bibitem[{{Ackermann} {et~al.}(2011){Ackermann}, {Ajello}, {Allafort},
  {Baldini}, {Ballet}, {Barbiellini}, {Bastieri}, {Belfiore}, {Bellazzini},
  {Berenji}, {Blandford}, {Bloom}, {Bonamente}, {Borgland}, {Bottacini},
  {Brigida}, {Bruel}, {Buehler}, {Buson}, {Caliandro}, {Cameron}, {Caraveo},
  {Casandjian}, {Cecchi}, {Chekhtman}, {Cheung}, {Chiang}, {Ciprini}, {Claus},
  {Cohen-Tanugi}, {de Angelis}, {de Palma}, {Dermer}, {do Couto e Silva},
  {Drell}, {Dumora}, {Favuzzi}, {Fegan}, {Focke}, {Fortin}, {Fukazawa},
  {Fusco}, {Gargano}, {Germani}, {Giglietto}, {Giordano}, {Giroletti},
  {Glanzman}, {Godfrey}, {Grenier}, {Guillemot}, {Guiriec}, {Hadasch},
  {Hanabata}, {Harding}, {Hayashida}, {Hayashi}, {Hays}, {J{\'o}hannesson},
  {Johnson}, {Kamae}, {Katagiri}, {Kataoka}, {Kerr}, {Kn{\"o}dlseder}, {Kuss},
  {Lande}, {Latronico}, {Lee}, {Longo}, {Loparco}, {Lott}, {Lovellette},
  {Lubrano}, {Martin}, {Mazziotta}, {McEnery}, {Mehault}, {Michelson},
  {Mitthumsiri}, {Mizuno}, {Monte}, {Monzani}, {Morselli}, {Moskalenko},
  {Murgia}, {Naumann-Godo}, {Nolan}, {Norris}, {Nuss}, {Ohsugi}, {Okumura},
  {Orlando}, {Ormes}, {Ozaki}, {Paneque}, {Parent}, {Pesce-Rollins},
  {Pierbattista}, {Piron}, {Pohl}, {Prokhorov}, {Rain{\`o}}, {Rando},
  {Razzano}, {Reposeur}, {Ritz}, {Parkinson}, {Sgr{\`o}}, {Siskind}, {Smith},
  {Spinelli}, {Strong}, {Takahashi}, {Tanaka}, {Thayer}, {Thayer}, {Thompson},
  {Tibaldo}, {Torres}, {Tosti}, {Tramacere}, {Troja}, {Uchiyama},
  {Vandenbroucke}, {Vasileiou}, {Vianello}, {Vitale}, {Waite}, {Wang}, {Winer},
  {Wood}, {Yang}, {Zimmer}, \& {Bontemps}}]{2011Sci...334.1103A}
{Ackermann}, M., {Ajello}, M., {Allafort}, A., {et~al.} 2011, Science, 334,
  1103

\bibitem[{{Ackermann} {et~al.}(2012){Ackermann}, {Albert}, {Baldini}, {Ballet},
  {Barbiellini}, {Bastieri}, {Bechtol}, {Bellazzini}, {Blandford}, {Bloom},
  {Bonamente}, {Borgland}, {Bottacini}, {Brandt}, {Bregeon}, {Brigida},
  {Bruel}, {Buehler}, {Burnett}, {Caliandro}, {Cameron}, {Caraveo},
  {Casandjian}, {Cecchi}, {Charles}, {Chiang}, {Ciprini}, {Claus},
  {Cohen-Tanugi}, {Conrad}, {Cutini}, {de Palma}, {Dermer}, {Digel}, {Silva},
  {Drell}, {Drlica-Wagner}, {Essig}, {Falletti}, {Favuzzi}, {Fegan}, {Focke},
  {Fukazawa}, {Funk}, {Fusco}, {Gargano}, {Germani}, {Giglietto}, {Giordano},
  {Giroletti}, {Glanzman}, {Godfrey}, {Grenier}, {Guiriec}, {Gustafsson},
  {Hadasch}, {Hayashida}, {Hou}, {Hughes}, {Johnson}, {Johnson}, {Kamae},
  {Katagiri}, {Kataoka}, {Kn{\"o}dlseder}, {Kuss}, {Lande}, {Latronico}, {Lee},
  {Lionetto}, {Llena Garde}, {Longo}, {Loparco}, {Lovellette}, {Lubrano},
  {Mazziotta}, {McEnery}, {Michelson}, {Mitthumsiri}, {Mizuno}, {Moiseev},
  {Monte}, {Monzani}, {Morselli}, {Moskalenko}, {Murgia}, {Naumann-Godo},
  {Norris}, {Nuss}, {Ohsugi}, {Okumura}, {Orlando}, {Ormes}, {Ozaki},
  {Paneque}, {Pelassa}, {Pierbattista}, {Piron}, {Pivato}, {Porter},
  {Rain{\`o}}, {Rando}, {Razzano}, {Reimer}, {Reimer}, {Ritz}, {Sadrozinski},
  {Sehgal}, {Sgr{\`o}}, {Siskind}, {Spinelli}, {Strigari}, {Suson}, {Tajima},
  {Takahashi}, {Tanaka}, {Thayer}, {Thayer}, {Tibaldo}, {Tinivella}, {Torres},
  {Troja}, {Uchiyama}, {Usher}, {Vandenbroucke}, {Vasileiou}, {Vianello},
  {Vitale}, {Waite}, {Wang}, {Winer}, {Wood}, {Yang}, {Zalewski}, \&
  {Zimmer}}]{Ackermann:2012nb}
{Ackermann}, M., {Albert}, A., {Baldini}, L., {et~al.} 2012, \apj, 747, 121

\bibitem[{{Ackermann} {et~al.}(2013){Ackermann}, {Ajello}, {Allafort}, {Asano},
  {Atwood}, {Baldini}, {Ballet}, {Barbiellini}, {Bastieri}, {Bechtol},
  {Bellazzini}, {Bloom}, {Bonamente}, {Borgland}, {Bottacini}, {Brandt},
  {Bregeon}, {Brigida}, {Bruel}, {Buehler}, {Burnett}, {Busetto}, {Buson},
  {Caliandro}, {Cameron}, {Caraveo}, {Casandjian}, {Cecchi}, {Charles},
  {Chaty}, {Chekhtman}, {Cheung}, {Chiang}, {Cillis}, {Ciprini}, {Claus},
  {Cohen-Tanugi}, {Colafrancesco}, {Conrad}, {Cutini}, {D'Ammando}, {de Palma},
  {Dermer}, {Silva}, {Drell}, {Drlica-Wagner}, {Dubois}, {Favuzzi}, {Fegan},
  {Ferrara}, {Focke}, {Fortin}, {Fukazawa}, {Funk}, {Fusco}, {Gargano},
  {Gasparrini}, {Gehrels}, {Germani}, {Giglietto}, {Giordano}, {Giroletti},
  {Glanzman}, {Godfrey}, {Grandi}, {Grenier}, {Grove}, {Guiriec}, {Hadasch},
  {Hayashida}, {Hays}, {Horan}, {Hou}, {Hughes}, {Jackson}, {Jogler},
  {J{\'o}hannesson}, {Johnson}, {Johnson}, {Kamae}, {Kataoka}, {Kerr},
  {Kn{\"o}dlseder}, {Kuss}, {Lande}, {Larsson}, {Latronico}, {Lavalley}, {Lee},
  {Longo}, {Loparco}, {Lott}, {Lovellette}, {Lubrano}, {Mazziotta},
  {McConville}, {McEnery}, {Mehault}, {Michelson}, {Mignani}, {Mitthumsiri},
  {Mizuno}, {Moiseev}, {Monte}, {Monzani}, {Morselli}, {Moskalenko}, {Murgia},
  {Naumann-Godo}, {Nemmen}, {Nishino}, {Norris}, {Nuss}, {Ohsugi}, {Omodei},
  {Orienti}, {Orlando}, {Ormes}, {Paneque}, {Panetta}, {Pelassa}, {Perkins},
  {Pesce-Rollins}, {Pierbattista}, {Piron}, {Pivato}, {Poon}, {Porter},
  {Rain{\`o}}, {Rando}, {Razzano}, {Razzaque}, {Reimer}, {Reimer}, {Reyes},
  {Ritz}, {Rochester}, {Romoli}, {Roth}, {Sanchez}, {Saz Parkinson}, {Scargle},
  {Sgr{\`o}}, {Siskind}, {Snyder}, {Spandre}, {Spinelli}, {Stephens}, {Suson},
  {Tajima}, {Takahashi}, {Tanaka}, {Thayer}, {Thayer}, {Thompson}, {Tibaldo},
  {Tibolla}, {Tinivella}, {Tosti}, {Troja}, {Usher}, {Vandenbroucke},
  {Vasileiou}, {Vianello}, {Vitale}, {von Kienlin}, {Waite}, {Wallace},
  {Weltevrede}, {Winer}, {Wood}, {Wood}, {Yang}, \&
  {Zimmer}}]{Ackermann:2013yma}
{Ackermann}, M., {Ajello}, M., {Allafort}, A., {et~al.} 2013, \apj, 765, 54

\bibitem[{{Ackermann} {et~al.}(2015){Ackermann}, {Ajello}, {Atwood}, {Baldini},
  {Ballet}, {Barbiellini}, {Bastieri}, {Becerra Gonzalez}, {Bellazzini},
  {Bissaldi}, {Blandford}, {Bloom}, {Bonino}, {Bottacini}, {Brandt}, {Bregeon},
  {Britto}, {Bruel}, {Buehler}, {Buson}, {Caliandro}, {Cameron}, {Caragiulo},
  {Caraveo}, {Carpenter}, {Casandjian}, {Cavazzuti}, {Cecchi}, {Charles},
  {Chekhtman}, {Cheung}, {Chiang}, {Chiaro}, {Ciprini}, {Claus},
  {Cohen-Tanugi}, {Cominsky}, {Conrad}, {Cutini}, {D'Abrusco}, {D'Ammando}, {de
  Angelis}, {Desiante}, {Digel}, {Di Venere}, {Drell}, {Favuzzi}, {Fegan},
  {Ferrara}, {Finke}, {Focke}, {Franckowiak}, {Fuhrmann}, {Fukazawa},
  {Furniss}, {Fusco}, {Gargano}, {Gasparrini}, {Giglietto}, {Giommi},
  {Giordano}, {Giroletti}, {Glanzman}, {Godfrey}, {Grenier}, {Grove},
  {Guiriec}, {Hewitt}, {Hill}, {Horan}, {Itoh}, {J{\'o}hannesson}, {Johnson},
  {Johnson}, {Kataoka}, {Kawano}, {Krauss}, {Kuss}, {La Mura}, {Larsson},
  {Latronico}, {Leto}, {Li}, {Li}, {Longo}, {Loparco}, {Lott}, {Lovellette},
  {Lubrano}, {Madejski}, {Mayer}, {Mazziotta}, {McEnery}, {Michelson},
  {Mizuno}, {Moiseev}, {Monzani}, {Morselli}, {Moskalenko}, {Murgia}, {Nuss},
  {Ohno}, {Ohsugi}, {Ojha}, {Omodei}, {Orienti}, {Orlando}, {Paggi}, {Paneque},
  {Perkins}, {Pesce-Rollins}, {Piron}, {Pivato}, {Porter}, {Rain{\`o}},
  {Rando}, {Razzano}, {Razzaque}, {Reimer}, {Reimer}, {Romani}, {Salvetti},
  {Schaal}, {Schinzel}, {Schulz}, {Sgr{\`o}}, {Siskind}, {Sokolovsky}, {Spada},
  {Spandre}, {Spinelli}, {Stawarz}, {Suson}, {Takahashi}, {Takahashi},
  {Tanaka}, {Thayer}, {Thayer}, {Tibaldo}, {Torres}, {Torresi}, {Tosti},
  {Troja}, {Uchiyama}, {Vianello}, {Winer}, {Wood}, \&
  {Zimmer}}]{2015ApJ...810...14A}
{Ackermann}, M., {Ajello}, M., {Atwood}, W.~B., {et~al.} 2015, \apj, 810, 14

\bibitem[{{Ackermann} {et~al.}(2016{\natexlab{a}}){Ackermann}, {Ajello},
  {Atwood}, {Baldini}, {Ballet}, {Barbiellini}, {Bastieri}, {Becerra Gonzalez},
  {Bellazzini}, {Bissaldi}, {Blandford}, {Bloom}, {Bonino}, {Bottacini},
  {Brandt}, {Bregeon}, {Bruel}, {Buehler}, {Buson}, {Caliandro}, {Cameron},
  {Caputo}, {Caragiulo}, {Caraveo}, {Cavazzuti}, {Cecchi}, {Charles},
  {Chekhtman}, {Cheung}, {Chiang}, {Chiaro}, {Ciprini}, {Cohen},
  {Cohen-Tanugi}, {Cominsky}, {Conrad}, {Cuoco}, {Cutini}, {D'Ammando}, {de
  Angelis}, {de Palma}, {Desiante}, {Di Mauro}, {Di Venere},
  {Dom{\'{\i}}nguez}, {Drell}, {Favuzzi}, {Fegan}, {Ferrara}, {Focke},
  {Fortin}, {Franckowiak}, {Fukazawa}, {Funk}, {Furniss}, {Fusco}, {Gargano},
  {Gasparrini}, {Giglietto}, {Giommi}, {Giordano}, {Giroletti}, {Glanzman},
  {Godfrey}, {Grenier}, {Grondin}, {Guillemot}, {Guiriec}, {Harding}, {Hays},
  {Hewitt}, {Hill}, {Horan}, {Iafrate}, {Hartmann}, {Jogler},
  {J{\'o}hannesson}, {Johnson}, {Kamae}, {Kataoka}, {Kn{\"o}dlseder}, {Kuss},
  {La Mura}, {Larsson}, {Latronico}, {Lemoine-Goumard}, {Li}, {Li}, {Longo},
  {Loparco}, {Lott}, {Lovellette}, {Lubrano}, {Madejski}, {Maldera},
  {Manfreda}, {Mayer}, {Mazziotta}, {Michelson}, {Mirabal}, {Mitthumsiri},
  {Mizuno}, {Moiseev}, {Monzani}, {Morselli}, {Moskalenko}, {Murgia}, {Nuss},
  {Ohsugi}, {Omodei}, {Orienti}, {Orlando}, {Ormes}, {Paneque}, {Perkins},
  {Pesce-Rollins}, {Petrosian}, {Piron}, {Pivato}, {Porter}, {Rain{\`o}},
  {Rando}, {Razzano}, {Razzaque}, {Reimer}, {Reimer}, {Reposeur}, {Romani},
  {S{\'a}nchez-Conde}, {Saz Parkinson}, {Schmid}, {Schulz}, {Sgr{\`o}},
  {Siskind}, {Spada}, {Spandre}, {Spinelli}, {Suson}, {Tajima}, {Takahashi},
  {Takahashi}, {Takahashi}, {Thayer}, {Thompson}, {Tibaldo}, {Torres}, {Tosti},
  {Troja}, {Vianello}, {Wood}, {Wood}, {Yassine}, {Zaharijas}, \&
  {Zimmer}}]{Ackermann:2015uya}
---. 2016{\natexlab{a}}, \apjs, 222, 5

\bibitem[{{Ackermann} {et~al.}(2016{\natexlab{b}}){Ackermann}, {Albert},
  {Atwood}, {Baldini}, {Ballet}, {Barbiellini}, {Bastieri}, {Bellazzini},
  {Bissaldi}, {Bloom}, {Bonino}, {Brandt}, {Bregeon}, {Bruel}, {Buehler},
  {Caliandro}, {Cameron}, {Caragiulo}, {Caraveo}, {Cavazzuti}, {Cecchi},
  {Charles}, {Chekhtman}, {Chiang}, {Chiaro}, {Ciprini}, {Cohen-Tanugi},
  {Cutini}, {D'Ammando}, {de Angelis}, {de Palma}, {Desiante}, {Digel},
  {Drell}, {Favuzzi}, {Ferrara}, {Focke}, {Franckowiak}, {Fusco}, {Gargano},
  {Gasparrini}, {Giglietto}, {Giordano}, {Godfrey}, {Grenier}, {Grondin},
  {Guillemot}, {Guiriec}, {Harding}, {Hill}, {Horan}, {J{\'o}hannesson},
  {Kn{\"o}dlseder}, {Kuss}, {Larsson}, {Latronico}, {Li}, {Li}, {Longo},
  {Loparco}, {Lubrano}, {Maldera}, {Martin}, {Mayer}, {Mazziotta}, {Michelson},
  {Mizuno}, {Monzani}, {Morselli}, {Murgia}, {Nuss}, {Ohsugi}, {Orienti},
  {Orlando}, {Ormes}, {Paneque}, {Pesce-Rollins}, {Piron}, {Pivato}, {Porter},
  {Rain{\`o}}, {Rando}, {Razzano}, {Reimer}, {Reimer}, {Romani},
  {S{\'a}nchez-Conde}, {Schulz}, {Sgr{\`o}}, {Siskind}, {Smith}, {Spada},
  {Spandre}, {Spinelli}, {Suson}, {Takahashi}, {Thayer}, {Tibaldo}, {Torres},
  {Tosti}, {Troja}, {Vianello}, {Wood}, \& {Zimmer}}]{TheFermi-LAT:2015lxa}
{Ackermann}, M., {Albert}, A., {Atwood}, W.~B., {et~al.} 2016{\natexlab{b}},
  \aap, 586, A71

\bibitem[{{Ackermann} {et~al.}(2016{\natexlab{c}}){Ackermann}, {Ajello},
  {Baldini}, {Ballet}, {Barbiellini}, {Bastieri}, {Bellazzini}, {Bissaldi},
  {Blandford}, {Bloom}, {Bonino}, {Brandt}, {Bregeon}, {Bruel}, {Buehler},
  {Buson}, {Caliandro}, {Cameron}, {Caragiulo}, {Caraveo}, {Cavazzuti},
  {Cecchi}, {Charles}, {Chekhtman}, {Cheung}, {Chiaro}, {Ciprini}, {Cohen},
  {Cohen-Tanugi}, {Costanza}, {Cutini}, {D'Ammando}, {Davis}, {de Angelis}, {de
  Palma}, {Desiante}, {Digel}, {Di Lalla}, {Di Mauro}, {Di Venere}, {Favuzzi},
  {Fegan}, {Ferrara}, {Focke}, {Fukazawa}, {Funk}, {Fusco}, {Gargano},
  {Gasparrini}, {Georganopoulos}, {Giglietto}, {Giordano}, {Giroletti},
  {Godfrey}, {Green}, {Grenier}, {Guiriec}, {Hays}, {Hewitt}, {Hill}, {Jogler},
  {J{\'o}hannesson}, {Kensei}, {Kuss}, {Larsson}, {Latronico}, {Li}, {Li},
  {Longo}, {Loparco}, {Lubrano}, {Magill}, {Maldera}, {Manfreda}, {Mayer},
  {Mazziotta}, {McConville}, {McEnery}, {Michelson}, {Mitthumsiri}, {Mizuno},
  {Monzani}, {Morselli}, {Moskalenko}, {Murgia}, {Negro}, {Nuss}, {Ohno},
  {Ohsugi}, {Orienti}, {Orlando}, {Ormes}, {Paneque}, {Perkins},
  {Pesce-Rollins}, {Piron}, {Pivato}, {Porter}, {Rain{\`o}}, {Rando},
  {Razzano}, {Reimer}, {Reimer}, {Schmid}, {Sgr{\`o}}, {Simone}, {Siskind},
  {Spada}, {Spandre}, {Spinelli}, {Stawarz}, {Takahashi}, {Thayer}, {Thompson},
  {Torres}, {Tosti}, {Troja}, {Vianello}, {Wood}, {Wood}, {Zimmer}, \& {Fermi
  LAT Collaboration}}]{Ackermann:2016arn}
{Ackermann}, M., {Ajello}, M., {Baldini}, L., {et~al.} 2016{\natexlab{c}},
  \apj, 826, 1

\bibitem[{{Ackermann} {et~al.}(2017{\natexlab{a}}){Ackermann}, {Ajello},
  {Albert}, {Baldini}, {Ballet}, {Barbiellini}, {Bastieri}, {Bellazzini},
  {Bissaldi}, {Bloom}, {Bonino}, {Bottacini}, {Brandt}, {Bregeon}, {Bruel},
  {Buehler}, {Cameron}, {Caputo}, {Caragiulo}, {Caraveo}, {Cavazzuti},
  {Cecchi}, {Charles}, {Chekhtman}, {Chiaro}, {Ciprini}, {Costanza}, {Cutini},
  {D'Ammando}, {de Palma}, {Desiante}, {Digel}, {Di Lalla}, {Di Mauro}, {Di
  Venere}, {Favuzzi}, {Funk}, {Fusco}, {Gargano}, {Giglietto}, {Giordano},
  {Giroletti}, {Glanzman}, {Green}, {Grenier}, {Guillemot}, {Guiriec},
  {Hayashi}, {Hou}, {J{\'o}hannesson}, {Kamae}, {Kn{\"o}dlseder}, {Kong},
  {Kuss}, {La Mura}, {Larsson}, {Latronico}, {Li}, {Longo}, {Loparco},
  {Lubrano}, {Maldera}, {Malyshev}, {Manfreda}, {Martin}, {Mazziotta},
  {Michelson}, {Mirabal}, {Mitthumsiri}, {Mizuno}, {Monzani}, {Morselli},
  {Moskalenko}, {Negro}, {Nuss}, {Ohsugi}, {Omodei}, {Orlando}, {Ormes},
  {Paneque}, {Persic}, {Pesce-Rollins}, {Piron}, {Porter}, {Principe},
  {Rain{\`o}}, {Rando}, {Razzano}, {Reimer}, {S{\'a}nchez-Conde}, {Sgr{\`o}},
  {Simone}, {Siskind}, {Spada}, {Spandre}, {Spinelli}, {Tanaka}, {Tibaldo},
  {Torres}, {Troja}, {Uchiyama}, {Wang}, {Wood}, {Wood}, {Zaharijas}, \&
  {Zhou}}]{Ackermann:2017nya}
{Ackermann}, M., {Ajello}, M., {Albert}, A., {et~al.} 2017{\natexlab{a}}, \apj,
  836, 208

\bibitem[{{Ackermann} {et~al.}(2017{\natexlab{b}}){Ackermann}, {Ajello},
  {Baldini}, {Ballet}, {Barbiellini}, {Bastieri}, {Bellazzini}, {Bissaldi},
  {Bloom}, {Bonino}, {Bottacini}, {Brandt}, {Bregeon}, {Bruel}, {Buehler},
  {Cameron}, {Caragiulo}, {Caraveo}, {Castro}, {Cavazzuti}, {Cecchi},
  {Charles}, {Chekhtman}, {Cheung}, {Chiaro}, {Ciprini}, {Cohen}, {Costantin},
  {Costanza}, {Cutini}, {D'Ammando}, {de Palma}, {Desiante}, {Digel}, {Di
  Lalla}, {Di Mauro}, {Di Venere}, {Favuzzi}, {Fegan}, {Ferrara},
  {Franckowiak}, {Fukazawa}, {Funk}, {Fusco}, {Gargano}, {Gasparrini},
  {Giglietto}, {Giordano}, {Giroletti}, {Green}, {Grenier}, {Grondin},
  {Guillemot}, {Guiriec}, {Harding}, {Hays}, {Hewitt}, {Horan}, {Hou},
  {J{\'o}hannesson}, {Kamae}, {Kuss}, {La Mura}, {Larsson}, {Lemoine-Goumard},
  {Li}, {Longo}, {Loparco}, {Lubrano}, {Magill}, {Maldera}, {Malyshev},
  {Manfreda}, {Mazziotta}, {Michelson}, {Mitthumsiri}, {Mizuno}, {Monzani},
  {Morselli}, {Moskalenko}, {Negro}, {Nuss}, {Ohsugi}, {Omodei}, {Orienti},
  {Orlando}, {Ormes}, {Paliya}, {Paneque}, {Perkins}, {Persic},
  {Pesce-Rollins}, {Petrosian}, {Piron}, {Porter}, {Principe}, {Raino},
  {Rando}, {Razzano}, {Razzaque}, {Reimer}, {Reimer}, {Reposeur}, {Sgro},
  {Simone}, {Siskind}, {Spada}, {Spandre}, {Spinelli}, {Suson}, {Tak},
  {Thayer}, {Thompson}, {Torres}, {Tosti}, {Troja}, {Vianello}, {Wood}, \&
  {Wood}}]{2017arXiv170200476T}
{Ackermann}, M., {Ajello}, M., {Baldini}, L., {et~al.} 2017{\natexlab{b}},
  \apj, 843, 139

\bibitem[{{Ackermann} {et~al.}(2017{\natexlab{c}}){Ackermann}, {Ajello},
  {Albert}, {Atwood}, {Baldini}, {Ballet}, {Barbiellini}, {Bastieri},
  {Bellazzini}, {Bissaldi}, {Blandford}, {Bloom}, {Bonino}, {Bottacini},
  {Brandt}, {Bregeon}, {Bruel}, {Buehler}, {Burnett}, {Cameron}, {Caputo},
  {Caragiulo}, {Caraveo}, {Cavazzuti}, {Cecchi}, {Charles}, {Chekhtman},
  {Chiang}, {Chiappo}, {Chiaro}, {Ciprini}, {Conrad}, {Costanza}, {Cuoco},
  {Cutini}, {D'Ammando}, {de Palma}, {Desiante}, {Digel}, {Di Lalla}, {Di
  Mauro}, {Di Venere}, {Drell}, {Favuzzi}, {Fegan}, {Ferrara}, {Focke},
  {Franckowiak}, {Fukazawa}, {Funk}, {Fusco}, {Gargano}, {Gasparrini},
  {Giglietto}, {Giordano}, {Giroletti}, {Glanzman}, {Gomez-Vargas}, {Green},
  {Grenier}, {Grove}, {Guillemot}, {Guiriec}, {Gustafsson}, {Harding}, {Hays},
  {Hewitt}, {Horan}, {Jogler}, {Johnson}, {Kamae}, {Kocevski}, {Kuss}, {La
  Mura}, {Larsson}, {Latronico}, {Li}, {Longo}, {Loparco}, {Lovellette},
  {Lubrano}, {Magill}, {Maldera}, {Malyshev}, {Manfreda}, {Martin},
  {Mazziotta}, {Michelson}, {Mirabal}, {Mitthumsiri}, {Mizuno}, {Moiseev},
  {Monzani}, {Morselli}, {Negro}, {Nuss}, {Ohsugi}, {Orienti}, {Orlando},
  {Ormes}, {Paneque}, {Perkins}, {Persic}, {Pesce-Rollins}, {Piron},
  {Principe}, {Rain{\`o}}, {Rando}, {Razzano}, {Razzaque}, {Reimer}, {Reimer},
  {S{\'a}nchez-Conde}, {Sgr{\`o}}, {Simone}, {Siskind}, {Spada}, {Spandre},
  {Spinelli}, {Suson}, {Tajima}, {Tanaka}, {Thayer}, {Tibaldo}, {Torres},
  {Troja}, {Uchiyama}, {Vianello}, {Wood}, {Wood}, {Zaharijas}, {Zimmer}, \&
  {Fermi LAT Collaboration}}]{TheFermi-LAT:2017vmf}
{Ackermann}, M., {Ajello}, M., {Albert}, A., {et~al.} 2017{\natexlab{c}}, \apj,
  840, 43

\bibitem[{{Aharonian} {et~al.}(2003){Aharonian}, {Akhperjanian}, {Beilicke},
  {Bernl{\"o}hr}, {B{\"o}rst}, {Bojahr}, {Bolz}, {Coarasa}, {Contreras},
  {Cortina}, {Costamante}, {Denninghoff}, {Fonseca}, {Girma}, {G{\"o}tting},
  {Heinzelmann}, {Hermann}, {Heusler}, {Hofmann}, {Horns}, {Jung}, {Kankanyan},
  {Kestel}, {Kohnle}, {Konopelko}, {Kornmeyer}, {Kranich}, {Lampeitl}, {Lopez},
  {Lorenz}, {Lucarelli}, {Mang}, {Mazine}, {Meyer}, {Mirzoyan}, {Moralejo},
  {Ona-Wilhelmi}, {Panter}, {Plyasheshnikov}, {Prahl}, {P{\"u}hlhofer}, {de los
  Reyes}, {Rhode}, {Ripken}, {Rowell}, {Sahakian}, {Samorski}, {Schilling},
  {Siems}, {Sobzynska}, {Stamm}, {Tluczykont}, {Vitale}, {V{\"o}lk}, {Wiedner},
  \& {Wittek}}]{2003A&A...403..523A}
{Aharonian}, F., {Akhperjanian}, A., {Beilicke}, M., {et~al.} 2003, \aap, 403,
  523

\bibitem[{{Aharonian} {et~al.}(2006){Aharonian}, {Akhperjanian}, {Bazer-Bachi},
  {Beilicke}, {Benbow}, {Berge}, {Bernl{\"o}hr}, {Boisson}, {Bolz}, {Borrel},
  {Braun}, {Breitling}, {Brown}, {Chadwick}, {Chounet}, {Cornils},
  {Costamante}, {Degrange}, {Dickinson}, {Djannati-Ata{\"i}}, {O'C.~Drury},
  {Dubus}, {Emmanoulopoulos}, {Espigat}, {Feinstein}, {Fontaine}, {Fuchs},
  {Funk}, {Gallant}, {Giebels}, {Glicenstein}, {Goret}, {Hadjichristidis},
  {Hauser}, {Hauser}, {Heinzelmann}, {Henri}, {Hermann}, {Hinton}, {Hofmann},
  {Holleran}, {Horns}, {Jacholkowska}, {de Jager}, {Kh{\'e}lifi}, {Klages},
  {Komin}, {Konopelko}, {Latham}, {Le Gallou}, {Lemi{\`e}re},
  {Lemoine-Goumard}, {Lohse}, {Martin}, {Martineau-Huynh}, {Marcowith},
  {Masterson}, {McComb}, {de Naurois}, {Nedbal}, {Nolan}, {Noutsos}, {Orford},
  {Osborne}, {Ouchrif}, {Panter}, {Pelletier}, {Pita}, {P{\"u}hlhofer},
  {Punch}, {Raubenheimer}, {Raue}, {Rayner}, {Reimer}, {Reimer}, {Ripken},
  {Rob}, {Rolland}, {Rowell}, {Sahakian}, {Saug{\'e}}, {Schlenker},
  {Schlickeiser}, {Schuster}, {Schwanke}, {Siewert}, {Sol}, {Spangler},
  {Steenkamp}, {Stegmann}, {Superina}, {Tavernet}, {Terrier}, {Th{\'e}oret},
  {Tluczykont}, {van Eldik}, {Vasileiadis}, {Venter}, {Vincent}, {V{\"o}lk}, \&
  {Wagner}}]{2006A&A...449..223A}
{Aharonian}, F., {Akhperjanian}, A.~G., {Bazer-Bachi}, A.~R., {et~al.} 2006,
  \aap, 449, 223

\bibitem[{{Aharonian} {et~al.}(2007{\natexlab{a}}){Aharonian}, {Akhperjanian},
  {Bazer-Bachi}, {Beilicke}, {Benbow}, {Berge}, {Bernl{\"o}hr}, {Boisson},
  {Bolz}, {Borrel}, {Braun}, {Brion}, {Brown}, {B{\"u}hler}, {B{\"u}sching},
  {Boutelier}, {Carrigan}, {Chadwick}, {Chounet}, {Coignet}, {Cornils},
  {Costamante}, {Degrange}, {Dickinson}, {Djannati-Ata{\"i}}, {O'C.~Drury},
  {Dubus}, {Egberts}, {Emmanoulopoulos}, {Espigat}, {Farnier}, {Feinstein},
  {Ferrero}, {Fiasson}, {Fontaine}, {Funk}, {Funk}, {F{\"u}{\ss}ling},
  {Gallant}, {Giebels}, {Glicenstein}, {Gl{\"u}ck}, {Goret}, {Hadjichristidis},
  {Hauser}, {Hauser}, {Heinzelmann}, {Henri}, {Hermann}, {Hinton}, {Hoffmann},
  {Hofmann}, {Holleran}, {Hoppe}, {Horns}, {Jacholkowska}, {de Jager},
  {Kendziorra}, {Kerschhaggl}, {Kh{\'e}lifi}, {Komin}, {Kosack}, {Lamanna},
  {Latham}, {Le Gallou}, {Lemi{\`e}re}, {Lemoine-Goumard}, {Lohse}, {Martin},
  {Martineau-Huynh}, {Marcowith}, {Masterson}, {Maurin}, {McComb}, {Moulin},
  {de Naurois}, {Nedbal}, {Nolan}, {Noutsos}, {Olive}, {Orford}, {Osborne},
  {Panter}, {Pelletier}, {Petrucci}, {Pita}, {P{\"u}hlhofer}, {Punch},
  {Ranchon}, {Raubenheimer}, {Raue}, {Rayner}, {Ripken}, {Rob}, {Rolland},
  {Rosier-Lees}, {Rowell}, {Sahakian}, {Santangelo}, {Saug{\'e}}, {Schlenker},
  {Schlickeiser}, {Schr{\"o}der}, {Schwanke}, {Schwarzburg}, {Schwemmer},
  {Shalchi}, {Sol}, {Spangler}, {Spanier}, {Steenkamp}, {Stegmann}, {Superina},
  {Tam}, {Tavernet}, {Terrier}, {Tluczykont}, {van Eldik}, {Vasileiadis},
  {Venter}, {Vialle}, {Vincent}, {V{\"o}lk}, {Wagner}, \&
  {Ward}}]{2007AA...470..475A}
---. 2007{\natexlab{a}}, \aap, 470, 475

\bibitem[{{Aharonian} {et~al.}(2007{\natexlab{b}}){Aharonian}, {Akhperjanian},
  {Barres de Almeida}, {Bazer-Bachi}, {Behera}, {Beilicke}, {Benbow},
  {Bernl{\"o}hr}, {Boisson}, {Bolz}, {Borrel}, {Braun}, {Brion}, {Brown},
  {B{\"u}hler}, {Bulik}, {B{\"u}sching}, {Boutelier}, {Carrigan}, {Chadwick},
  {Chounet}, {Clapson}, {Coignet}, {Cornils}, {Costamante}, {Dalton},
  {Degrange}, {Dickinson}, {Djannati-Ata{\"i}}, {Domainko}, {O'C.~Drury},
  {Dubois}, {Dubus}, {Dyks}, {Egberts}, {Emmanoulopoulos}, {Espigat},
  {Farnier}, {Feinstein}, {Fiasson}, {F{\"o}rster}, {Fontaine}, {Funk},
  {F{\"u}{\ss}ling}, {Gallant}, {Giebels}, {Glicenstein}, {Gl{\"u}ck}, {Goret},
  {Hadjichristidis}, {Hauser}, {Hauser}, {Heinzelmann}, {Henri}, {Hermann},
  {Hinton}, {Hoffmann}, {Hofmann}, {Holleran}, {Hoppe}, {Horns},
  {Jacholkowska}, {de Jager}, {Jung}, {Katarzy{\'n}ski}, {Kendziorra},
  {Kerschhaggl}, {Kh{\'e}lifi}, {Keogh}, {Komin}, {Kosack}, {Lamanna},
  {Latham}, {Lemi{\`e}re}, {Lemoine-Goumard}, {Lenain}, {Lohse}, {Martin},
  {Martineau-Huynh}, {Marcowith}, {Masterson}, {Maurin}, {Maurin}, {McComb},
  {Moderski}, {Moulin}, {de Naurois}, {Nedbal}, {Nolan}, {Ohm}, {Olive}, {de
  O{\~n}a Wilhelmi}, {Orford}, {Osborne}, {Ostrowski}, {Panter}, {Pedaletti},
  {Pelletier}, {Petrucci}, {Pita}, {P{\"u}hlhofer}, {Punch}, {Ranchon},
  {Raubenheimer}, {Raue}, {Rayner}, {Renaud}, {Ripken}, {Rob}, {Rolland},
  {Rosier-Lees}, {Rowell}, {Rudak}, {Ruppel}, {Sahakian}, {Santangelo},
  {Schlickeiser}, {Sch{\"o}ck}, {Schr{\"o}der}, {Schwanke}, {Schwarzburg},
  {Schwemmer}, {Shalchi}, {Sol}, {Spangler}, {Stawarz}, {Steenkamp},
  {Stegmann}, {Superina}, {Tam}, {Tavernet}, {Terrier}, {van Eldik},
  {Vasileiadis}, {Venter}, {Vialle}, {Vincent}, {Vivier}, {V{\"o}lk}, {Volpe},
  {Wagner}, {Ward}, {Zdziarski}, \& {Zech}}]{2007AA...473L..25A}
{Aharonian}, F., {Akhperjanian}, A.~G., {Barres de Almeida}, U., {et~al.}
  2007{\natexlab{b}}, \aap, 473, L25

\bibitem[{{Aharonian} {et~al.}(2007{\natexlab{c}}){Aharonian}, {Akhperjanian},
  {Barres de Almeida}, {Bazer-Bachi}, {Behera}, {Beilicke}, {Benbow},
  {Bernl{\"o}hr}, {Boisson}, {Bolz}, {Borrel}, {Braun}, {Brion}, {Brown},
  {B{\"u}hler}, {Bulik}, {B{\"u}sching}, {Boutelier}, {Carrigan}, {Chadwick},
  {Chounet}, {Clapson}, {Coignet}, {Cornils}, {Costamante}, {Dalton},
  {Degrange}, {Dickinson}, {Djannati-Ata{\"i}}, {Domainko}, {O'C.~Drury},
  {Dubois}, {Dubus}, {Dyks}, {Egberts}, {Emmanoulopoulos}, {Espigat},
  {Farnier}, {Feinstein}, {Fiasson}, {F{\"o}rster}, {Fontaine}, {Funk},
  {F{\"u}{\ss}ling}, {Gallant}, {Giebels}, {Glicenstein}, {Gl{\"u}ck}, {Goret},
  {Hadjichristidis}, {Hauser}, {Hauser}, {Heinzelmann}, {Henri}, {Hermann},
  {Hinton}, {Hoffmann}, {Hofmann}, {Holleran}, {Hoppe}, {Horns},
  {Jacholkowska}, {de Jager}, {Jung}, {Katarzy{\'n}ski}, {Kendziorra},
  {Kerschhaggl}, {Kh{\'e}lifi}, {Keogh}, {Komin}, {Kosack}, {Lamanna},
  {Latham}, {Lemi{\`e}re}, {Lemoine-Goumard}, {Lenain}, {Lohse}, {Martin},
  {Martineau-Huynh}, {Marcowith}, {Masterson}, {Maurin}, {Maurin}, {McComb},
  {Moderski}, {Moulin}, {de Naurois}, {Nedbal}, {Nolan}, {Ohm}, {Olive}, {de
  O{\~n}a Wilhelmi}, {Orford}, {Osborne}, {Ostrowski}, {Panter}, {Pedaletti},
  {Pelletier}, {Petrucci}, {Pita}, {P{\"u}hlhofer}, {Punch}, {Ranchon},
  {Raubenheimer}, {Raue}, {Rayner}, {Renaud}, {Ripken}, {Rob}, {Rolland},
  {Rosier-Lees}, {Rowell}, {Rudak}, {Ruppel}, {Sahakian}, {Santangelo},
  {Schlickeiser}, {Sch{\"o}ck}, {Schr{\"o}der}, {Schwanke}, {Schwarzburg},
  {Schwemmer}, {Shalchi}, {Sol}, {Spangler}, {Stawarz}, {Steenkamp},
  {Stegmann}, {Superina}, {Tam}, {Tavernet}, {Terrier}, {van Eldik},
  {Vasileiadis}, {Venter}, {Vialle}, {Vincent}, {Vivier}, {V{\"o}lk}, {Volpe},
  {Wagner}, {Ward}, {Zdziarski}, \& {Zech}}]{2007AA...475L...9A}
---. 2007{\natexlab{c}}, \aap, 475, L9

\bibitem[{{Aharonian} {et~al.}(2008){Aharonian}, {Akhperjanian}, {Bazer-Bachi},
  {Behera}, {Beilicke}, {Benbow}, {Berge}, {Bernl{\"o}hr}, {Boisson}, {Bolz},
  {Borrel}, {Braun}, {Brion}, {Brown}, {B{\"u}hler}, {Bulik}, {B{\"u}sching},
  {Boutelier}, {Carrigan}, {Chadwick}, {Chounet}, {Clapson}, {Coignet},
  {Cornils}, {Costamante}, {Degrange}, {Dickinson}, {Djannati-Ata{\"i}},
  {Domainko}, {O'C.~Drury}, {Dubus}, {Dyks}, {Egberts}, {Emmanoulopoulos},
  {Espigat}, {Farnier}, {Feinstein}, {Fiasson}, {F{\"o}rster}, {Fontaine},
  {Fukui}, {Funk}, {Funk}, {F{\"u}{\ss}ling}, {Gallant}, {Giebels},
  {Glicenstein}, {Gl{\"u}ck}, {Goret}, {Hadjichristidis}, {Hauser}, {Hauser},
  {Heinzelmann}, {Henri}, {Hermann}, {Hinton}, {Hoffmann}, {Hofmann},
  {Holleran}, {Hoppe}, {Horns}, {Jacholkowska}, {de Jager}, {Kendziorra},
  {Kerschhaggl}, {Kh{\'e}lifi}, {Komin}, {Kosack}, {Lamanna}, {Latham}, {Le
  Gallou}, {Lemi{\`e}re}, {Lemoine-Goumard}, {Lenain}, {Lohse}, {Martin},
  {Martineau-Huynh}, {Marcowith}, {Masterson}, {Maurin}, {McComb}, {Moderski},
  {Moriguchi}, {Moulin}, {de Naurois}, {Nedbal}, {Nolan}, {Olive}, {Orford},
  {Osborne}, {Ostrowski}, {Panter}, {Pedaletti}, {Pelletier}, {Petrucci},
  {Pita}, {P{\"u}hlhofer}, {Punch}, {Ranchon}, {Raubenheimer}, {Raue},
  {Rayner}, {Reimer}, {Renaud}, {Ripken}, {Rob}, {Rolland}, {Rosier-Lees},
  {Rowell}, {Rudak}, {Ruppel}, {Sahakian}, {Santangelo}, {Saug{\'e}},
  {Schlenker}, {Schlickeiser}, {Schr{\"o}der}, {Schwanke}, {Schwarzburg},
  {Schwemmer}, {Shalchi}, {Sol}, {Spangler}, {Stawarz}, {Steenkamp},
  {Stegmann}, {Superina}, {Takeuchi}, {Tam}, {Tavernet}, {Terrier}, {van
  Eldik}, {Vasileiadis}, {Venter}, {Vialle}, {Vincent}, {Vivier}, {V{\"o}lk},
  {Volpe}, {Wagner}, \& {Ward}}]{2008A&A...481..401A}
{Aharonian}, F., {Akhperjanian}, A.~G., {Bazer-Bachi}, A.~R., {et~al.} 2008,
  \aap, 481, 401

\bibitem[{{Aharonian} {et~al.}(2009){Aharonian}, {Akhperjanian}, {Anton}, {de
  Almeida}, {Bazer-Bachi}, {Becherini}, {Behera}, {Benbow}, {Bernl{\"o}hr},
  {Boisson}, {Bochow}, {Borrel}, {Brion}, {Brucker}, {Brun}, {B{\"u}hler},
  {Bulik}, {B{\"u}sching}, {Boutelier}, {Chadwick}, {Charbonnier}, {Chaves},
  {Cheesebrough}, {Chounet}, {Clapson}, {Coignet}, {Dalton}, {Daniel},
  {Davids}, {Degrange}, {Deil}, {Dickinson}, {Djannati-Ata{\"i}}, {Domainko},
  {Drury}, {Dubois}, {Dubus}, {Dyks}, {Dyrda}, {Egberts}, {Emmanoulopoulos},
  {Espigat}, {Farnier}, {Feinstein}, {Fiasson}, {F{\"o}rster}, {Fontaine},
  {F{\"u}{\ss}ling}, {Gabici}, {Gallant}, {G{\'e}rard}, {Giebels},
  {Glicenstein}, {Gl{\"u}ck}, {Goret}, {G{\"o}hring}, {Hauser}, {Hauser},
  {Heinz}, {Heinzelmann}, {Henri}, {Hermann}, {Hinton}, {Hoffmann}, {Hofmann},
  {Holleran}, {Hoppe}, {Horns}, {Jacholkowska}, {de Jager}, {Jahn}, {Jung},
  {Katarzy{\'n}ski}, {Katz}, {Kaufmann}, {Kendziorra}, {Kerschhaggl},
  {Khangulyan}, {Kh{\'e}lifi}, {Keogh}, {Klu{\'z}niak}, {Kneiske}, {Komin},
  {Kosack}, {Lamanna}, {Latham}, {Lenain}, {Lohse}, {Marandon}, {Martin},
  {Martineau-Huynh}, {Marcowith}, {Maurin}, {McComb}, {Medina}, {Moderski},
  {Moulin}, {Naumann-Godo}, {de Naurois}, {Nedbal}, {Nekrassov}, {Niemiec},
  {Nolan}, {Ohm}, {Olive}, {de O{\~n}a Wilhelmi}, {Orford}, {Ostrowski},
  {Panter}, {Arribas}, {Pedaletti}, {Pelletier}, {Petrucci}, {Pita},
  {P{\"u}hlhofer}, {Punch}, {Quirrenbach}, {Raubenheimer}, {Raue}, {Rayner},
  {Renaud}, {Rieger}, {Ripken}, {Rob}, {Rosier-Lees}, {Rowell}, {Rudak},
  {Rulten}, {Ruppel}, {Sahakian}, {Santangelo}, {Schlickeiser}, {Sch{\"o}ck},
  {Schr{\"o}der}, {Schwanke}, {Schwarzburg}, {Schwemmer}, {Shalchi}, {Sikora},
  {Skilton}, {Sol}, {Spangler}, {Stawarz}, {Steenkamp}, {Stegmann}, {Superina},
  {Szostek}, {Tam}, {Tavernet}, {Terrier}, {Tibolla}, {Tluczykont}, {van
  Eldik}, {Vasileiadis}, {Venter}, {Venter}, {Vialle}, {Vincent}, {Vink},
  {Vivier}, {V{\"o}lk}, {Volpe}, {Wagner}, {Ward}, {Zdziarski}, \&
  {Zech}}]{2009ApJ...695L..40A}
{Aharonian}, F., {Akhperjanian}, A.~G., {Anton}, G., {et~al.} 2009, \apjl, 695,
  L40

\bibitem[{{Aharonian} {et~al.}(1994){Aharonian}, {Coppi}, \&
  {Voelk}}]{aharonian1994}
{Aharonian}, F.~A., {Coppi}, P.~S., \& {Voelk}, H.~J. 1994, \apjl, 423, L5

\bibitem[{{Ajello} {et~al.}(2017){Ajello}, {Atwood}, {Baldini}, {Ballet},
  {Barbiellini}, {Bastieri}, {Bellazzini}, {Bissaldi}, {Blandford}, {Bloom},
  {Bonino}, {Bregeon}, {Britto}, {Bruel}, {Buehler}, {Buson}, {Cameron},
  {Caputo}, {Caragiulo}, {Caraveo}, {Cavazzuti}, {Cecchi}, {Charles},
  {Chekhtman}, {Cheung}, {Chiaro}, {Ciprini}, {Cohen}, {Costantin}, {Costanza},
  {Cuoco}, {Cutini}, {D'Ammando}, {de Palma}, {Desiante}, {Digel}, {Di Lalla},
  {Di Mauro}, {Di Venere}, {Dom{\'{\i}}nguez}, {Drell}, {Dumora}, {Favuzzi},
  {Fegan}, {Ferrara}, {Fortin}, {Franckowiak}, {Fukazawa}, {Funk}, {Fusco},
  {Gargano}, {Gasparrini}, {Giglietto}, {Giommi}, {Giordano}, {Giroletti},
  {Glanzman}, {Green}, {Grenier}, {Grondin}, {Grove}, {Guillemot}, {Guiriec},
  {Harding}, {Hays}, {Hewitt}, {Horan}, {J{\'o}hannesson}, {Kensei}, {Kuss},
  {La Mura}, {Larsson}, {Latronico}, {Lemoine-Goumard}, {Li}, {Longo},
  {Loparco}, {Lott}, {Lubrano}, {Magill}, {Maldera}, {Manfreda}, {Mazziotta},
  {McEnery}, {Meyer}, {Michelson}, {Mirabal}, {Mitthumsiri}, {Mizuno},
  {Moiseev}, {Monzani}, {Morselli}, {Moskalenko}, {Negro}, {Nuss}, {Ohsugi},
  {Omodei}, {Orienti}, {Orlando}, {Palatiello}, {Paliya}, {Paneque}, {Perkins},
  {Persic}, {Pesce-Rollins}, {Piron}, {Porter}, {Principe}, {Rain{\`o}},
  {Rando}, {Razzano}, {Razzaque}, {Reimer}, {Reimer}, {Reposeur}, {Saz
  Parkinson}, {Sgr{\`o}}, {Simone}, {Siskind}, {Spada}, {Spandre}, {Spinelli},
  {Stawarz}, {Suson}, {Takahashi}, {Tak}, {Thayer}, {Thayer}, {Thompson},
  {Torres}, {Torresi}, {Troja}, {Vianello}, {Wood}, \&
  {Wood}}]{2017arXiv170200664T}
{Ajello}, M., {Atwood}, W.~B., {Baldini}, L., {et~al.} 2017, \apjs, 232, 18

\bibitem[{Akaike(1974)}]{Akaike}
Akaike, H. 1974, IEEE Transactions on Automatic Control, 19, 716

\bibitem[{{Aliu} {et~al.}(2012){Aliu}, {Archambault}, {Arlen}, {Aune},
  {Beilicke}, {Benbow}, {B{\"o}ttcher}, {Bouvier}, {Bugaev}, {Cannon},
  {Cesarini}, {Ciupik}, {Collins-Hughes}, {Connolly}, {Cui}, {Dickherber},
  {Dumm}, {Errando}, {Falcone}, {Federici}, {Feng}, {Finley}, {Finnegan},
  {Fortson}, {Furniss}, {Galante}, {Gall}, {Godambe}, {Griffin}, {Grube},
  {Gyuk}, {Hanna}, {Holder}, {Huan}, {Hughes}, {Hui}, {Imran}, {Jameil},
  {Kaaret}, {Karlsson}, {Kertzman}, {Kerr}, {Khassen}, {Kieda}, {Krawczynski},
  {Krennrich}, {Lang}, {Lee}, {Madhavan}, {Majumdar}, {McArthur}, {McCann},
  {Moriarty}, {Mukherjee}, {Nelson}, {O'Faol{\'a}in de Bhr{\'o}ithe}, {Ong},
  {Orr}, {Otte}, {Park}, {Perkins}, {Pichel}, {Pohl}, {Quinn}, {Ragan},
  {Reynolds}, {Roache}, {Ruppel}, {Saxon}, {Schroedter}, {Sembroski}, {{\c
  S}ent{\"u}rk}, {Smith}, {Staszak}, {Stroh}, {Telezhinsky}, {Te{\v s}i{\'c}},
  {Theiling}, {Thibadeau}, {Tsurusaki}, {Varlotta}, {Vassiliev}, {Vivier},
  {Wakely}, {Ward}, {Weinstein}, {Welsing}, {Williams}, \&
  {Zitzer}}]{2012ApJ...755..118A}
{Aliu}, E., {Archambault}, S., {Arlen}, T., {et~al.} 2012, \apj, 755, 118

\bibitem[{{Aliu} {et~al.}(2013){Aliu}, {Archambault}, {Arlen}, {Aune},
  {Beilicke}, {Benbow}, {Bouvier}, {Buckley}, {Bugaev}, {Cesarini}, {Ciupik},
  {Collins-Hughes}, {Connolly}, {Cui}, {Dickherber}, {Duke}, {Dumm},
  {Dwarkadas}, {Errando}, {Falcone}, {Federici}, {Feng}, {Finley}, {Finnegan},
  {Fortson}, {Furniss}, {Galante}, {Gall}, {Gillanders}, {Godambe}, {Gotthelf},
  {Griffin}, {Grube}, {Gyuk}, {Hanna}, {Holder}, {Hughes}, {Humensky},
  {Kaaret}, {Kargaltsev}, {Karlsson}, {Khassen}, {Kieda}, {Krawczynski},
  {Krennrich}, {Lang}, {Lee}, {Madhavan}, {Maier}, {Majumdar}, {McArthur},
  {McCann}, {Moriarty}, {Mukherjee}, {Nelson}, {O'Faol{\'a}in de Bhr{\'o}ithe},
  {Ong}, {Orr}, {Otte}, {Park}, {Perkins}, {Pohl}, {Prokoph}, {Quinn}, {Ragan},
  {Reyes}, {Reynolds}, {Roache}, {Roberts}, {Saxon}, {Schroedter}, {Sembroski},
  {Slane}, {Smith}, {Staszak}, {Telezhinsky}, {Te{\v s}i{\'c}}, {Theiling},
  {Thibadeau}, {Tsurusaki}, {Tyler}, {Varlotta}, {Vassiliev}, {Vincent},
  {Vivier}, {Wakely}, {Weekes}, {Weinstein}, {Welsing}, {Williams}, \&
  {Zitzer}}]{2013ApJ...764...38A}
---. 2013, \apj, 764, 38

\bibitem[{{Aliu} {et~al.}(2014){Aliu}, {Archambault}, {Arlen}, {Aune},
  {Behera}, {Beilicke}, {Benbow}, {Berger}, {Bird}, {Bouvier}, {Buckley},
  {Bugaev}, {Byrum}, {Cerruti}, {Chen}, {Ciupik}, {Connolly}, {Cui}, {Duke},
  {Dumm}, {Errando}, {Falcone}, {Federici}, {Feng}, {Finley}, {Fleischhack},
  {Fortin}, {Fortson}, {Furniss}, {Galante}, {Gillanders}, {Griffin},
  {Griffiths}, {Grube}, {Gyuk}, {Hanna}, {Holder}, {Hughes}, {Humensky},
  {Johnson}, {Kaaret}, {Kertzman}, {Khassen}, {Kieda}, {Krawczynski},
  {Krennrich}, {Lang}, {Madhavan}, {Maier}, {Majumdar}, {McArthur}, {McCann},
  {Meagher}, {Millis}, {Moriarty}, {Mukherjee}, {Nieto}, {O'Faol{\'a}in de
  Bhr{\'o}ithe}, {Ong}, {Otte}, {Park}, {Perkins}, {Pohl}, {Popkow}, {Prokoph},
  {Quinn}, {Ragan}, {Reyes}, {Reynolds}, {Richards}, {Roache}, {Sembroski},
  {Smith}, {Staszak}, {Telezhinsky}, {Theiling}, {Varlotta}, {Vassiliev},
  {Vincent}, {Wakely}, {Weekes}, {Weinstein}, {Welsing}, {Williams}, {Zajczyk},
  \& {Zitzer}}]{2014ApJ...782...13A}
---. 2014, \apj, 782, 13

\bibitem[{{Alves Batista} {et~al.}(2016){Alves Batista}, {Saveliev}, {Sigl}, \&
  {Vachaspati}}]{2016PhRvD..94h3005A}
{Alves Batista}, R., {Saveliev}, A., {Sigl}, G., \& {Vachaspati}, T. 2016,
  \prd, 94, 083005

\bibitem[{{Ando} \& {Kusenko}(2010)}]{2010ApJ...722L..39A}
{Ando}, S., \& {Kusenko}, A. 2010, \apjl, 722, L39

\bibitem[{Araya(2018)}]{Araya:2016cbn}
Araya, M. 2018, \mnras, 474, 102

\bibitem[{{Archambault} {et~al.}(2017{\natexlab{a}}){Archambault}, {Archer},
  {Benbow}, {Bird}, {Bourbeau}, {Buchovecky}, {Buckley}, {Bugaev}, {Cerruti},
  {Connolly}, {Cui}, {Dwarkadas}, {Errando}, {Falcone}, {Feng}, {Finley},
  {Fleischhack}, {Fortson}, {Furniss}, {Griffin}, {H{\"u}tten}, {Hanna},
  {Holder}, {Johnson}, {Kaaret}, {Kar}, {Kelley-Hoskins}, {Kertzman}, {Kieda},
  {Krause}, {Kumar}, {Lang}, {Maier}, {McArthur}, {McCann}, {Moriarty},
  {Mukherjee}, {Nieto}, {O'Brien}, {Ong}, {Otte}, {Park}, {Pohl}, {Popkow},
  {Pueschel}, {Quinn}, {Ragan}, {Reynolds}, {Richards}, {Roache}, {Sadeh},
  {Santander}, {Sembroski}, {Shahinyan}, {Slane}, {Staszak}, {Telezhinsky},
  {Trepanier}, {Tyler}, {Wakely}, {Weinstein}, {Weisgarber}, {Wilcox},
  {Wilhelm}, {Williams}, \& {Zitzer}}]{2017ApJ...836...23A}
{Archambault}, S., {Archer}, A., {Benbow}, W., {et~al.} 2017{\natexlab{a}},
  \apj, 836, 23

\bibitem[{{Archambault} {et~al.}(2017{\natexlab{b}}){Archambault}, {Archer},
  {Benbow}, {Buchovecky}, {Bugaev}, {Cerruti}, {Connolly}, {Cui}, {Falcone},
  {Fern{\'a}ndez Alonso}, {Finley}, {Fleischhack}, {Fortson}, {Furniss},
  {Griffin}, {H{\"u}tten}, {Hervet}, {Holder}, {Humensky}, {Johnson}, {Kaaret},
  {Kar}, {Kieda}, {Krause}, {Krennrich}, {Lang}, {Lin}, {Maier}, {McArthur},
  {Moriarty}, {Nieto}, {O'Brien}, {Ong}, {Otte}, {Pohl}, {Popkow}, {Pueschel},
  {Quinn}, {Ragan}, {Reynolds}, {Richards}, {Roache}, {Rovero}, {Sadeh},
  {Shahinyan}, {Staszak}, {Telezhinsky}, {Tyler}, {Wakely}, {Weinstein},
  {Weisgarber}, {Wilcox}, {Wilhelm}, {Williams}, \&
  {Zitzer}}]{2017ApJ...835..288A}
---. 2017{\natexlab{b}}, \apj, 835, 288

\bibitem[{{Arlen} {et~al.}(2014){Arlen}, {Vassilev}, {Weisgarber}, {Wakely}, \&
  {Yusef Shafi}}]{arlen2014}
{Arlen}, T.~C., {Vassilev}, V.~V., {Weisgarber}, T., {Wakely}, S.~P., \& {Yusef
  Shafi}, S. 2014, \apj, 796, 18

\bibitem[{{Atwood} {et~al.}(2012){Atwood}, {Albert}, {Baldini}, {Tinivella},
  {Bregeon}, {Pesce-Rollins}, {Sgr{\`o}}, {Bruel}, {Charles}, {Drlica-Wagner},
  {Franckowiak}, {Jogler}, {Rochester}, {Usher}, {Wood}, {Cohen-Tanugi}, \&
  {S.~Zimmer for the Fermi-LAT Collaboration}}]{2013arXiv1303.3514A}
{Atwood}, W., {Albert}, A., {Baldini}, L., {et~al.} 2012, in Fourth
  International Fermi Symposium, Monterrey, California, USA

\bibitem[{{Atwood} {et~al.}(2009){Atwood}, {Abdo}, {Ackermann}, {Althouse},
  {Anderson}, {Axelsson}, {Baldini}, {Ballet}, {Band}, {Barbiellini}, \&
  et~al.}]{Atwood:2009ez}
{Atwood}, W.~B., {Abdo}, A.~A., {Ackermann}, M., {et~al.} 2009, \apj., 697,
  1071

\bibitem[{{Bartoli} {et~al.}(2014){Bartoli}, {Bernardini}, {Bi},
  {et~al.}}]{2014ApJ...790..152B}
{Bartoli}, B., {Bernardini}, P., {Bi}, X.~J., {et~al.} 2014, \apj, 790, 152

\bibitem[{{Bertoni} {et~al.}(2016){Bertoni}, {Hooper}, \&
  {Linden}}]{2016JCAP...05..049B}
{Bertoni}, B., {Hooper}, D., \& {Linden}, T. 2016, \jcap, 5, 049

\bibitem[{{Biteau} \& {Williams}(2015)}]{2015ApJ...812...60B}
{Biteau}, J., \& {Williams}, D.~A. 2015, \apj, 812, 60

\bibitem[{Bonnoli {et~al.}(2015)Bonnoli, Tavecchio, Ghisellini, \&
  Sbarrato}]{Bonnoli:2015yia}
Bonnoli, G., Tavecchio, F., Ghisellini, G., \& Sbarrato, T. 2015, \mnras, 451,
  611

\bibitem[{{Broderick} {et~al.}(2012){Broderick}, {Chang}, \&
  {Pfrommer}}]{broderick2012}
{Broderick}, A.~E., {Chang}, P., \& {Pfrommer}, C. 2012, \apj, 752, 22

\bibitem[{{Buckley} {et~al.}(2015){Buckley}, {Charles}, {Gaskins}, {Brooks},
  {Drlica-Wagner}, {Martin}, \& {Zhao}}]{2015PhRvD..91j2001B}
{Buckley}, M.~R., {Charles}, E., {Gaskins}, J.~M., {et~al.} 2015, \prd, 91,
  102001

\bibitem[{Buckley \& Hooper(2010)}]{Buckley:2010vg}
Buckley, M.~R., \& Hooper, D. 2010, Phys. Rev., D82, 063501

\bibitem[{{Bykov}(2014)}]{2014A&ARv..22...77B}
{Bykov}, A.~M. 2014, \aapr, 22, 77

\bibitem[{{Bykov} \& {Toptygin}(2001)}]{2001AstL...27..625B}
{Bykov}, A.~M., \& {Toptygin}, I.~N. 2001, Astronomy Letters, 27, 625

\bibitem[{{Caprini} \& {Gabici}(2015)}]{caprini2015}
{Caprini}, C., \& {Gabici}, S. 2015, \prd, 91, 123514

\bibitem[{{Caprioli}(2015)}]{2015ICRC...34....8C}
{Caprioli}, D. 2015, in International Cosmic Ray Conference, Vol.~34, 34th
  International Cosmic Ray Conference (ICRC2015), 8

\bibitem[{{Caputo} {et~al.}(2016){Caputo}, {Buckley}, {Martin}, {Charles},
  {Brooks}, {Drlica-Wagner}, {Gaskins}, \& {Wood}}]{2016PhRvD..93f2004C}
{Caputo}, R., {Buckley}, M.~R., {Martin}, P., {et~al.} 2016, \prd, 93, 062004

\bibitem[{Chang {et~al.}(2016)Chang, Broderick, Pfrommer, Puchwein, Lamberts,
  Shalaby, \& Vasil}]{chang2016}
Chang, P., Broderick, A.~E., Pfrommer, C., {et~al.} 2016, \apj, 833, 118

\bibitem[{{Chaves} {et~al.}(2008){Chaves}, {Renaud}, {Lemoine-Goumard}, \&
  {Goret}}]{2008AIPC.1085..372C}
{Chaves}, R.~C.~G., {Renaud}, M., {Lemoine-Goumard}, M., \& {Goret}, P. 2008,
  in American Institute of Physics Conference Series, Vol. 1085, 372--375

\bibitem[{Chen {et~al.}(2015{\natexlab{a}})Chen, Buckley, \&
  Ferrer}]{Chen:2014rsa}
Chen, W., Buckley, J.~H., \& Ferrer, F. 2015{\natexlab{a}}, Phys. Rev. Lett.,
  115, 211103

\bibitem[{Chen {et~al.}(2015{\natexlab{b}})Chen, Chowdhury, Ferrer, Tashiro, \&
  Vachaspati}]{Chen:2014qva}
Chen, W., Chowdhury, B.~D., Ferrer, F., Tashiro, H., \& Vachaspati, T.
  2015{\natexlab{b}}, \mnras, 450, 3371

\bibitem[{{Chen} {et~al.}(2015){Chen}, {Chowdhury}, {Ferrer}, {Tashiro}, \&
  {Vachaspati}}]{chen2015helical}
{Chen}, W., {Chowdhury}, B.~D., {Ferrer}, F., {Tashiro}, H., \& {Vachaspati},
  T. 2015, \mnras, 450, 3371

\bibitem[{Chernoff(1954)}]{Chernoff}
Chernoff, H. 1954, Ann. Math. Statist., 25, 573

\bibitem[{{Condon} {et~al.}(2017){Condon}, {Lemoine-Goumard}, {Acero}, \&
  {Katagiri}}]{2017ApJ...851..100C}
{Condon}, B., {Lemoine-Goumard}, M., {Acero}, F., \& {Katagiri}, H. 2017, \apj,
  851, 100

\bibitem[{{Condon} {et~al.}(1998){Condon}, {Cotton}, {Greisen}, {Yin},
  {Perley}, {Taylor}, \& {Broderick}}]{1998AJ....115.1693C}
{Condon}, J.~J., {Cotton}, W.~D., {Greisen}, E.~W., {et~al.} 1998, \aj, 115,
  1693

\bibitem[{{de Gasperin} {et~al.}(2014){de Gasperin}, {Evoli}, {Br{\"u}ggen},
  {Hektor}, {Cardillo}, {Thorman}, {Dawson}, \&
  {Morrison}}]{deGasperin:2014caa}
{de Gasperin}, F., {Evoli}, C., {Br{\"u}ggen}, M., {et~al.} 2014, \aap, 568,
  A107

\bibitem[{{de Naurois} \& {H.E.S.S.~Collaboration}(2013)}]{2013AdSpR..51..258D}
{de Naurois}, M., \& {H.E.S.S.~Collaboration}. 2013, Advances in Space
  Research, 51, 258

\bibitem[{{Dermer} {et~al.}(2011){Dermer}, {Cavadini}, {Razzaque},
  {et~al.}}]{dermer2011}
{Dermer}, C.~D., {Cavadini}, M., {Razzaque}, S., {et~al.} 2011, \apjl, 733,
  L21+

\bibitem[{{Dolag} {et~al.}(2009){Dolag}, {Kachelrie{\ss}}, {Ostapchenko}, \&
  {Tom{\`a}s}}]{2009ApJ...703.1078D}
{Dolag}, K., {Kachelrie{\ss}}, M., {Ostapchenko}, S., \& {Tom{\`a}s}, R. 2009,
  \apj, 703, 1078

\bibitem[{{Dolag} {et~al.}(2011){Dolag}, {Kachelriess}, {Ostapchenko}, \&
  {Tom{\`a}s}}]{dolag2011}
{Dolag}, K., {Kachelriess}, M., {Ostapchenko}, S., \& {Tom{\`a}s}, R. 2011,
  \apjl, 727, L4

\bibitem[{{Dom{\'{\i}}nguez} {et~al.}(2011){Dom{\'{\i}}nguez}, {Primack},
  {Rosario}, {Prada}, {Gilmore}, {Faber}, {Koo}, {Somerville},
  {P{\'e}rez-Torres}, {P{\'e}rez-Gonz{\'a}lez}, {Huang}, {Davis},
  {Guhathakurta}, {Barmby}, {Conselice}, {Lozano}, {Newman}, \&
  {Cooper}}]{Dominguez:2010bv}
{Dom{\'{\i}}nguez}, A., {Primack}, J.~R., {Rosario}, D.~J., {et~al.} 2011,
  \mnras, 410, 2556

\bibitem[{Dubner {et~al.}(2017)Dubner, Castelletti, Kargaltsev, Pavlov,
  Bietenholz, \& Talavera}]{2017arXiv170402968D}
Dubner, G., Castelletti, G., Kargaltsev, O., {et~al.} 2017, \apj, 840, 82

\bibitem[{Durrer \& Neronov(2013)}]{Durrer:2013pga}
Durrer, R., \& Neronov, A. 2013, \aapr, 21, 62

\bibitem[{{Dwek} \& {Krennrich}(2013)}]{dwek2013}
{Dwek}, E., \& {Krennrich}, F. 2013, Astroparticle Physics, 43, 112

\bibitem[{Finke {et~al.}(2010)Finke, Razzaque, \& Dermer}]{Finke:2009xi}
Finke, J.~D., Razzaque, S., \& Dermer, C.~D. 2010, \apj, 712, 238

\bibitem[{{Finke} {et~al.}(2015){Finke}, {Reyes}, {Georganopoulos}, {Reynolds},
  {Ajello}, {Fegan}, \& {McCann}}]{finke2015}
{Finke}, J.~D., {Reyes}, L.~C., {Georganopoulos}, M., {et~al.} 2015, \apj, 814,
  20

\bibitem[{{Fitoussi} {et~al.}(2017){Fitoussi}, {Belmont}, {Malzac},
  {Marcowith}, {Cohen-Tanugi}, \& {Jean}}]{2017MNRAS.466.3472F}
{Fitoussi}, T., {Belmont}, R., {Malzac}, J., {et~al.} 2017, \mnras, 466, 3472

\bibitem[{Ghisellini {et~al.}(2017)Ghisellini, Righi, Costamante, \&
  Tavecchio}]{Ghisellini:2017ico}
Ghisellini, G., Righi, C., Costamante, L., \& Tavecchio, F. 2017, \mnras, 469,
  255

\bibitem[{{Gould} \& {Schr{\'e}der}(1967{\natexlab{a}})}]{gould1967a}
{Gould}, R.~J., \& {Schr{\'e}der}, G.~P. 1967{\natexlab{a}}, Physical Review,
  155, 1408

\bibitem[{{Gould} \& {Schr{\'e}der}(1967{\natexlab{b}})}]{gould1967}
---. 1967{\natexlab{b}}, Physical Review, 155, 1404

\bibitem[{{Grondin} {et~al.}(2013){Grondin}, {Romani}, {Lemoine-Goumard},
  {Guillemot}, {Harding}, \& {Reposeur}}]{2013ApJ...774..110G}
{Grondin}, M.-H., {Romani}, R.~W., {Lemoine-Goumard}, M., {et~al.} 2013, \apj,
  774, 110

\bibitem[{Haffner {et~al.}(2003)Haffner, Reynolds, Tufte,
  {et~al.}}]{0067-0049-149-2-405}
Haffner, L.~M., Reynolds, R.~J., Tufte, S.~L., {et~al.} 2003, \apjs, 149, 405

\bibitem[{Haffner {et~al.}(\noop{3001} 2018 in prep.)}]{whaminprep}
Haffner, L.~M., {et~al.} \noop{3001} 2018 in prep.

\bibitem[{{Hauser} \& {Dwek}(2001)}]{hauser2001}
{Hauser}, M.~G., \& {Dwek}, E. 2001, \araa, 39, 249

\bibitem[{{Hester}(2008)}]{2008ARA&A..46..127H}
{Hester}, J.~J. 2008, \araa, 46, 127

\bibitem[{Hillas {et~al.}(1998)Hillas, Akerlof, Biller, Buckley, Carter-Lewis,
  Catanese, Cawley, Fegan, Finley, Gaidos, Krennrich, Lamb, Lang, Mohanty,
  Punch, Reynolds, Rodgers, Rose, Rovero, Schubnell, Sembroski, Vacanti,
  Weekes, West, \& Zweerink}]{1998ApJ...503..744H}
Hillas, A.~M., Akerlof, C.~W., Biller, S.~D., {et~al.} 1998, \apj, 503, 744

\bibitem[{Holler {et~al.}(2017)Holler, Berge, Hahn, Khangulyan, \&
  Parsons}]{Holler:2017dtw}
Holler, M., Berge, D., Hahn, J., Khangulyan, D., \& Parsons, R.~D. 2017, in
  {35th International Cosmic Ray Conference (ICRC 2017) Bexco, Busan, Korea}

\bibitem[{Israel(1998)}]{Israel1998}
Israel, F. 1998, \aapr, 8, 237

\bibitem[{{Kachelrie{\ss}} {et~al.}(2012){Kachelrie{\ss}}, {Ostapchenko}, \&
  {Tom{\`a}s}}]{2012CoPhC.183.1036K}
{Kachelrie{\ss}}, M., {Ostapchenko}, S., \& {Tom{\`a}s}, R. 2012, Computer
  Physics Communications, 183, 1036

\bibitem[{{Kashlinsky}(2005)}]{kashlinsky2005}
{Kashlinsky}, A. 2005, \physrep, 409, 361

\bibitem[{Katsuta {et~al.}(2017)Katsuta, Uchiyama, \& Funk}]{Katsuta:2017ctk}
Katsuta, J., Uchiyama, Y., \& Funk, S. 2017, \apj, 839, 129

\bibitem[{Kerr {et~al.}(2015)Kerr, Ray, Johnston, Shannon, \&
  Camilo}]{Kerr:2015tva}
Kerr, M., Ray, P., Johnston, S., Shannon, R., \& Camilo, F. 2015, \apj, 814,
  128

\bibitem[{{Kharchenko} {et~al.}(2013){Kharchenko}, {Piskunov}, {Schilbach},
  {R{\"o}ser}, \& {Scholz}}]{2013A&A...558A..53K}
{Kharchenko}, N.~V., {Piskunov}, A.~E., {Schilbach}, E., {R{\"o}ser}, S., \&
  {Scholz}, R.-D. 2013, \aap, 558, A53

\bibitem[{{Kinzer} {et~al.}(1995){Kinzer}, {Johnson}, {Dermer}, {Kurfess},
  {Strickman}, {Grove}, {Kroeger}, {Grabelsky}, {Purcell}, {Ulmer}, {Jung}, \&
  {McNaron-Brown}}]{1995ApJ...449..105K}
{Kinzer}, R.~L., {Johnson}, W.~N., {Dermer}, C.~D., {et~al.} 1995, \apj, 449,
  105

\bibitem[{Kotelnikov {et~al.}(2015)Kotelnikov, Rubtsov, \&
  Troitsky}]{Kotelnikov:2014vda}
Kotelnikov, E., Rubtsov, G., \& Troitsky, S. 2015, \mnras, 450, L44

\bibitem[{Kuhlen {et~al.}(2008)Kuhlen, Diemand, \& Madau}]{Kuhlen:2008aw}
Kuhlen, M., Diemand, J., \& Madau, P. 2008, \apj, 686, 262

\bibitem[{{Kun} {et~al.}(2008){Kun}, {Kiss}, \& {Balog}}]{2008hsf1.book..136K}
{Kun}, M., {Kiss}, Z.~T., \& {Balog}, Z. 2008, {Star Forming Regions in
  Cepheus}, ed. B.~{Reipurth}, 136

\bibitem[{{Lande} {et~al.}(2012){Lande}, {Ackermann}, {Allafort}, {Ballet},
  {Bechtol}, {Burnett}, {Cohen-Tanugi}, {Drlica-Wagner}, {Funk}, {Giordano},
  {Grondin}, {Kerr}, \& {Lemoine-Goumard}}]{2012ApJ...756....5L}
{Lande}, J., {Ackermann}, M., {Allafort}, A., {et~al.} 2012, \apj, 756, 5

\bibitem[{{Lang} {et~al.}(2005){Lang}, {Johnson}, {Goss}, \&
  {Rodr{\'{\i}}guez}}]{2005AJ....130.2185L}
{Lang}, C.~C., {Johnson}, K.~E., {Goss}, W.~M., \& {Rodr{\'{\i}}guez}, L.~F.
  2005, \aj, 130, 2185

\bibitem[{{Li} {et~al.}(2016){Li}, {Torres}, {de O{\~n}a Wilhelmi}, {Rea}, \&
  {Martin}}]{2016ApJ...831...19L}
{Li}, J., {Torres}, D.~F., {de O{\~n}a Wilhelmi}, E., {Rea}, N., \& {Martin},
  J. 2016, \apj, 831, 19

\bibitem[{{Lyne} {et~al.}(1993){Lyne}, {Pritchard}, \&
  {Graham-Smith}}]{1993MNRAS.265.1003L}
{Lyne}, A.~G., {Pritchard}, R.~S., \& {Graham-Smith}, F. 1993, \mnras, 265,
  1003

\bibitem[{{Maurin} {et~al.}(2016){Maurin}, {Marcowith}, {Komin}, {Krayzel}, \&
  {Lamanna}}]{2016A&A...591A..71M}
{Maurin}, G., {Marcowith}, A., {Komin}, N., {Krayzel}, F., \& {Lamanna}, G.
  2016, \aap, 591, A71

\bibitem[{{Mayer-Hasselwander} {et~al.}(1980){Mayer-Hasselwander}, {Buccheri},
  {Kanbach}, {Bennett}, {Lichti}, {Bignami}, {Damico}, {Hermsen}, {Lebrun}, \&
  {Masnou}}]{1980NYASA.336..211M}
{Mayer-Hasselwander}, H.~A., {Buccheri}, R., {Kanbach}, G., {et~al.} 1980, in
  Annals of the New York Academy of Sciences, Vol. 336, Ninth Texas Symposium
  on Relativistic Astrophysics, ed. J.~{Ehlers}, J.~J. {Perry}, \& M.~{Walker},
  211--222

\bibitem[{{Menzler} \& {Schlickeiser}(2015)}]{menzler2015}
{Menzler}, U., \& {Schlickeiser}, R. 2015, \mnras, 448, 3405

\bibitem[{Meyer {et~al.}(2016)Meyer, Conrad, \& Dickinson}]{Meyer:2016xvq}
Meyer, M., Conrad, J., \& Dickinson, H. 2016, \apj, 827, 147

\bibitem[{{Meyer} {et~al.}(2010){Meyer}, {Horns}, \&
  {Zechlin}}]{2010A&A...523A...2M}
{Meyer}, M., {Horns}, D., \& {Zechlin}, H.-S. 2010, \aap, 523, A2

\bibitem[{{Meyer} {et~al.}(2012){Meyer}, {Raue}, {Mazin}, \&
  {Horns}}]{meyer2012}
{Meyer}, M., {Raue}, M., {Mazin}, D., \& {Horns}, D. 2012, \aap, 542, A59

\bibitem[{Mirabal {et~al.}(2012)Mirabal, Frias-Martinez, Hassan, \&
  Frias-Martinez}]{Mirabal:2012em}
Mirabal, N., Frias-Martinez, V., Hassan, T., \& Frias-Martinez, E. 2012,
  \mnras, 424, L64

\bibitem[{{Neronov} {et~al.}(2010){Neronov}, {Semikoz}, {Kachelriess},
  {Ostapchenko}, \& {Elyiv}}]{2010ApJ...719L.130N}
{Neronov}, A., {Semikoz}, D., {Kachelriess}, M., {Ostapchenko}, S., \& {Elyiv},
  A. 2010, \apjl, 719, L130

\bibitem[{{Neronov} \& {Semikoz}(2009)}]{2009PhRvD..80l3012N}
{Neronov}, A., \& {Semikoz}, D.~V. 2009, \prd, 80, 123012

\bibitem[{{Neronov} {et~al.}(2011){Neronov}, {Semikoz}, {Tinyakov}, \&
  {Tkachev}}]{2011A&A...526A..90N}
{Neronov}, A., {Semikoz}, D.~V., {Tinyakov}, P.~G., \& {Tkachev}, I.~I. 2011,
  \aap, 526, A90

\bibitem[{{Neronov} \& {Vovk}(2010)}]{neronov2010}
{Neronov}, A., \& {Vovk}, I. 2010, Science, 328, 73

\bibitem[{{Nikishov}(1962)}]{nikishov1962}
{Nikishov}, A.~I. 1962, Sov. Phys. JETP, 14, 393

\bibitem[{{Ohm} {et~al.}(2013){Ohm}, {Hinton}, \&
  {White}}]{2013MNRAS.434.2289O}
{Ohm}, S., {Hinton}, J.~A., \& {White}, R. 2013, \mnras, 434, 2289

\bibitem[{Olive(2003)}]{Olive:2003iq}
Olive, K.~A. 2003, ArXiv e-prints, arXiv:astro-ph/0301505

\bibitem[{{Parma} {et~al.}(2002){Parma}, {Murgia}, {de Ruiter}, \&
  {Fanti}}]{parma2002}
{Parma}, P., {Murgia}, M., {de Ruiter}, H.~R., \& {Fanti}, R. 2002, \nar, 46,
  313

\bibitem[{Pieri {et~al.}(2008)Pieri, Bertone, \& Branchini}]{Pieri:2007ir}
Pieri, L., Bertone, G., \& Branchini, E. 2008, \mnras, 384, 1627

\bibitem[{{Pineault} {et~al.}(1997){Pineault}, {Landecker}, {Swerdlyk}, \&
  {Reich}}]{1997A&A...324.1152P}
{Pineault}, S., {Landecker}, T.~L., {Swerdlyk}, C.~M., \& {Reich}, W. 1997,
  \aap, 324, 1152

\bibitem[{Plaga(1995)}]{plaga1995}
Plaga, R. 1995, Nature, 374, 430

\bibitem[{{Protheroe} \& {Stanev}(1993)}]{protheroe1993}
{Protheroe}, R.~J., \& {Stanev}, T. 1993, \mnras, 264, 191

\bibitem[{Quireza {et~al.}(2006)Quireza, Rood, Bania, Balser, \&
  Maciel}]{Quireza:2006sn}
Quireza, C., Rood, R.~T., Bania, T.~M., Balser, D.~S., \& Maciel, W.~J. 2006,
  \apj, 653, 1226

\bibitem[{{Remy} {et~al.}(2017){Remy}, {Grenier}, {Marshall}, \&
  {Casandjian}}]{remy2017}
{Remy}, Q., {Grenier}, I.~A., {Marshall}, D.~J., \& {Casandjian}, J.~M. 2017,
  \aap, 601, A78

\bibitem[{Reynoso \& Green(2007)}]{Reynoso:2006zn}
Reynoso, E.~M., \& Green, A.~J. 2007, \mnras, 375, 92

\bibitem[{Rubin {et~al.}(1980)Rubin, Thonnard, \& Ford}]{Rubin:1980zd}
Rubin, V., Thonnard, N., \& Ford, W.K., J. 1980, \apj, 238, 471

\bibitem[{{Schlegel} {et~al.}(1998){Schlegel}, {Finkbeiner}, \&
  {Davis}}]{1998ApJ...500..525S}
{Schlegel}, D.~J., {Finkbeiner}, D.~P., \& {Davis}, M. 1998, \apj, 500, 525

\bibitem[{{Seward} {et~al.}(1995){Seward}, {Schmidt}, \&
  {Slane}}]{1995ApJ...453..284S}
{Seward}, F.~D., {Schmidt}, B., \& {Slane}, P. 1995, \apj, 453, 284

\bibitem[{{Sironi} \& {Giannios}(2014)}]{sironi2014}
{Sironi}, L., \& {Giannios}, D. 2014, \apj, 787, 49

\bibitem[{{Slane} {et~al.}(1997){Slane}, {Seward}, {Bandiera}, {Torii}, \&
  {Tsunemi}}]{1997ApJ...485..221S}
{Slane}, P., {Seward}, F.~D., {Bandiera}, R., {Torii}, K., \& {Tsunemi}, H.
  1997, \apj, 485, 221

\bibitem[{Slane {et~al.}(2004)Slane, Zimmerman, Hughes, Seward, Gaensler, \&
  Clarke}]{Slane:2003du}
Slane, P., Zimmerman, E.~R., Hughes, J.~P., {et~al.} 2004, \apj, 601, 1045

\bibitem[{{Strong} {et~al.}(1982){Strong}, {Bignami}, {Bloemen}, {Buccheri},
  {Caraveo}, {Hermsen}, {Kanbach}, {Lebrun}, {Mayer-Hasselwander}, {Paul}, \&
  {Wills}}]{strong1982}
{Strong}, A.~W., {Bignami}, G.~F., {Bloemen}, J.~B.~G.~M., {et~al.} 1982, \aap,
  115, 404

\bibitem[{Su\'{a}rez {et~al.}(2015)Su\'{a}rez, Combi, Albacete-Colombo, Paron,
  García, \& Miceli}]{Suarez:2015hga}
Su\'{a}rez, A.~E., Combi, J.~A., Albacete-Colombo, J.~F., {et~al.} 2015, \aap,
  583, A84

\bibitem[{Sun {et~al.}(2016)Sun, Yang, Mckinley, \& Aharonian}]{Sun:2016ibh}
Sun, X.-n., Yang, R.-z., Mckinley, B., \& Aharonian, F. 2016, \aap, 595, A29

\bibitem[{{Swanenburg} {et~al.}(1981){Swanenburg}, {Bennett}, {Bignami},
  {Buccheri}, {Caraveo}, {Hermsen}, {Kanbach}, {Lichti}, {Masnou},
  {Mayer-Hasselwander}, {Paul}, {Sacco}, {Scarsi}, \&
  {Wills}}]{1981ApJ...243L..69S}
{Swanenburg}, B.~N., {Bennett}, K., {Bignami}, G.~F., {et~al.} 1981, \apjl,
  243, L69

\bibitem[{{Tavecchio} {et~al.}(2011){Tavecchio}, {Ghisellini}, {Bonnoli}, \&
  {Foschini}}]{tavecchio2011}
{Tavecchio}, F., {Ghisellini}, G., {Bonnoli}, G., \& {Foschini}, L. 2011,
  \mnras, 414, 3566

\bibitem[{{Taylor} {et~al.}(2011{\natexlab{a}}){Taylor}, {Vovk}, \&
  {Neronov}}]{taylor2011}
{Taylor}, A.~M., {Vovk}, I., \& {Neronov}, A. 2011{\natexlab{a}}, \aap, 529,
  A144

\bibitem[{{Taylor} {et~al.}(2011{\natexlab{b}}){Taylor}, {Vovk}, \&
  {Neronov}}]{2011A&A...529A.144T}
---. 2011{\natexlab{b}}, \aap, 529, A144

\bibitem[{{Thompson} {et~al.}(1995){Thompson}, {Bertsch}, {Dingus}, {Esposito},
  {Etienne}, {Fichtel}, {Friedlander}, {Hartman}, {Hunter}, {Kendig}, {Mattox},
  {McDonald}, {von Montigny}, {Mukherjee}, {Ramanamurthy}, {Sreekumar},
  {Fierro}, {Lin}, {Michelson}, {Nolan}, {Shriver}, {Willis}, {Kanbach},
  {Mayer-Hasselwander}, {Merck}, {Radecke}, {Kniffen}, \&
  {Schneid}}]{1995ApJS..101..259T}
{Thompson}, D.~J., {Bertsch}, D.~L., {Dingus}, B.~L., {et~al.} 1995, \apjs,
  101, 259

\bibitem[{{Vovk} {et~al.}(2012){Vovk}, {Taylor}, {Semikoz}, \&
  {Neronov}}]{2012ApJ...747L..14V}
{Vovk}, I., {Taylor}, A.~M., {Semikoz}, D., \& {Neronov}, A. 2012, \apjl, 747,
  L14

\bibitem[{{Wood} {et~al.}(2017){Wood}, {Caputo}, {Charles}, {Di Mauro},
  {Magill}, \& {Jeremy Perkins for the Fermi-LAT
  Collaboration}}]{2017arXiv170709551W}
{Wood}, M., {Caputo}, R., {Charles}, E., {et~al.} 2017, ArXiv e-prints,
  arXiv:1707.09551

\bibitem[{{Xia} {et~al.}(2017){Xia}, {Duan}, {Li}, {Liang}, {Shen}, {Yue},
  {Wang}, {Yuan}, {Fan}, {Wu}, \& {Chang}}]{2017PhRvD..95j2001X}
{Xia}, Z.-Q., {Duan}, K.-K., {Li}, S., {et~al.} 2017, \prd, 95, 102001

\bibitem[{{Yang} \& {Aharonian}(2017)}]{2017A&A...600A.107Y}
{Yang}, R.-Z., \& {Aharonian}, F. 2017, \aap, 600, A107

\bibitem[{{Yang} {et~al.}(2017){Yang}, {de O{\~n}a Wilhelmi}, \&
  {Aharonian}}]{2017arXiv171002803Y}
{Yang}, R.-Z., {de O{\~n}a Wilhelmi}, E., \& {Aharonian}, F. 2017, ArXiv
  e-prints, arXiv:1710.02803

\bibitem[{Zechlin \& Horns(2012)}]{Zechlin:2012by}
Zechlin, H.-S., \& Horns, D. 2012, JCAP, 1211, 050, [Erratum:
  JCAP1502,no.02,E01(2015)]

\bibitem[{Zhu {et~al.}(2015)Zhu, Tian, \& Wu}]{Zhu:2015fwa}
Zhu, H., Tian, W.~W., \& Wu, D. 2015, \mnras, 452, 3470

\bibitem[{Zwicky(1933)}]{Zwicky:1933gu}
Zwicky, F. 1933, Helv. Phys. Acta, 6, 110

\end{thebibliography}

\end{document}